\documentclass[twocolumn]{aastex63}

\graphicspath{{./}{figures/}}

\shorttitle{$\alpha_\mathrm{CO}$ in Barred Centers}
\shortauthors{Teng et al.}

\usepackage{amsmath}
\usepackage{bm}
\usepackage{lineno}
\usepackage{xparse}
\newcommand*\species[1]{%
    \ensuremath{\IfBold{\bm{\mathrm{#1}}}{\mathrm{#1}}}%
}

\newcommand*{\trans}[1]{%
    \ensuremath{\IfBold{\bm{\mytrans{#1}}}{\mytrans{#1}}}%
}

\newcommand*{\delim}{\text{\scalebox{0.7}[1.0]{\ensuremath{-}}}}
\newcommand*\mytrans[1]{%
    \ifnum#1=10 (1\delim0)\else%
    \ifnum#1=21 (2\delim1)\else%
    \ifnum#1=32 (3\delim2)\else%
    \ifnum#1=43 (4\delim3)\else%
    \ifnum#1=54 (5\delim4)\else%
    \ifnum#1=65 (6\delim5)\else%
    \ifnum#1=76 (7\delim6)\else%
    \ifnum#1=87 (8\delim7)\else%
    \ifnum#1=98 (9\delim8)\else%
    \ifnum#1=109 (10\delim9)\else#1%
    \fi\fi\fi\fi\fi\fi\fi\fi\fi\fi%
}

\makeatletter
\newcommand*{\IfBold}{%
  \ifx\f@series\my@test@bx
    \expandafter\@firstoftwo
  \else
    \expandafter\@secondoftwo
  \fi
}
\newcommand*{\my@test@bx}{bx}
\makeatother

\DeclareDocumentCommand\chem{ m g }{%
    {\species{#1}%
        \IfNoValueF {#2} {\,\trans{#2}}%
    }%
}

\begin{document}

\title{The Physical Drivers and Observational Tracers of CO-to-$\bm{\mathrm{H_2}}$ Conversion Factor Variations in Nearby Barred Galaxy Centers}

\newcommand{\UCSD}{\affiliation{Center for Astrophysics and Space Sciences, Department of Physics, University of California San Diego, \\ 9500 Gilman Drive, La Jolla, CA 92093, USA}}

\newcommand{\McMaster}{\affiliation{Department of Physics and Astronomy, McMaster University, 1280 Main Street West, Hamilton, ON L8S 4M1, Canada}}

\newcommand{\CITA}{\affiliation{Canadian Institute for Theoretical Astrophysics (CITA), University of Toronto, 60 St George Street, Toronto, ON M5S 3H8, Canada}}

\newcommand{\OSU}{\affiliation{Department of Astronomy, The Ohio State University, 140 West 18th Avenue, Columbus, OH 43210, USA}}

\newcommand{\ANU}{\affiliation{Research School of Astronomy and Astrophysics, Australian National University, Canberra, ACT 2611, Australia}}

\newcommand{\ASIAA}{\affiliation{Institute of Astronomy and Astrophysics, Academia Sinica, No. 1, Sec. 4, Roosevelt Road, Taipei 10617, Taiwan}}

\newcommand{\Bonn}{\affiliation{Argelander-Institut f\"ur Astronomie, Universit\"at Bonn, Auf dem H\"ugel 71, 53121 Bonn, Germany}}

\newcommand{\Heidelberg}{\affiliation{Astronomisches Rechen-Institut, Zentrum f\"{u}r Astronomie der Universit\"{a}t Heidelberg, M\"{o}nchhofstra\ss e 12-14, D-69120 Heidelberg, Germany}}

\newcommand{\ITA}{\affiliation{Universit\"{a}t Heidelberg, Zentrum f\"{u}r Astronomie, Institut f\"{u}r Theoretische Astrophysik, \\ Albert-Ueberle-Str 2, D-69120 Heidelberg, Germany}}

\newcommand{\Leiden}{\affiliation{Leiden Observatory, Leiden University, P.O. Box 9513, 2300 RA Leiden, The Netherlands}}

\newcommand{\Maryland}{\affiliation{Department of Astronomy, University of Maryland, College Park, MD 20742, USA}}

\newcommand{\MPE}{\affiliation{Max-Planck-Institut f\"{u}r extraterrestrische Physik, Giessenbachstra{\ss}e 1, D-85748 Garching, Germany}}

\newcommand{\MPIA}{\affiliation{Max-Planck-Institut f\"{u}r Astronomie, K\"{o}nigstuhl 17, D-69117, Heidelberg, Germany}}

\newcommand{\NRAO}{\affiliation{National Radio Astronomy Observatory, 520 Edgemont Road, Charlottesville, VA 22903-2475, USA}}

\newcommand{\OAN}{\affiliation{Observatorio Astron\'{o}mico Nacional (IGN), C/Alfonso XII, 3, E-28014 Madrid, Spain}}

\newcommand{\Tamkang}{\affiliation{Department of Physics, Tamkang University, No.151, Yingzhuan Rd., Tamsui Dist., New Taipei City 251301, Taiwan}}

\newcommand{\Toledo}{\affiliation{Department of Physics and Astronomy, University of Toledo, Ritter Obs., MS \#113, Toledo, OH 43606, USA}}

\correspondingauthor{Yu-Hsuan Teng}
\email{yuteng@ucsd.edu}

\author[0000-0003-4209-1599]{Yu-Hsuan Teng}
\UCSD

\author[0000-0002-4378-8534]{Karin M. Sandstrom}
\UCSD

\author[0000-0003-0378-4667]{Jiayi Sun}
\McMaster
\CITA

\author[0000-0003-1613-6263]{Munan Gong}
\MPE

\author[0000-0002-5480-5686]{Alberto D. Bolatto}
\Maryland

\author[0000-0003-2551-7148]{I-Da Chiang}
\ASIAA

\author[0000-0002-2545-1700]{Adam K. Leroy}
\OSU

\author[0000-0003-1242-505X]{Antonio Usero}
\OAN

\author[0000-0001-6708-1317]{Simon C.~O. Glover}
\ITA

\author[0000-0002-0560-3172]{Ralf S.\ Klessen}
\ITA
\affiliation{Universit\"{a}t Heidelberg, Interdisziplin\"{a}res Zentrum f\"{u}r Wissenschaftliches Rechnen, \\ Im Neuenheimer Feld 205, D-69120 Heidelberg, Germany}

\author[0000-0001-9773-7479]{Daizhong Liu}
\MPE

\author[0000-0002-0472-1011]{Miguel Querejeta}
\OAN

\author[0000-0002-3933-7677]{Eva Schinnerer}
\MPIA

\author[0000-0003-0166-9745]{Frank Bigiel}
\Bonn

\author[0000-0001-5301-1326]{Yixian Cao}
\MPE

\author[0000-0002-5635-5180]{M\'elanie Chevance}
\ITA
\affiliation{Cosmic Origins Of Life (COOL) Research DAO, coolresearch.io}

\author[0000-0002-1185-2810]{Cosima Eibensteiner}
\Bonn

\author[0000-0002-3247-5321]{Kathryn Grasha}
\ANU

\author[0000-0002-6760-9449]{Frank P. Israel}
\Leiden

\author[0000-0001-7089-7325]{Eric J. Murphy}
\NRAO

\author[0000-0001-9793-6400]{Lukas Neumann}
\Bonn

\author[0000-0002-1370-6964]{Hsi-An Pan}
\Tamkang

\author[0000-0001-5965-3530]{Francesca~Pinna}
\MPIA

\author[0000-0001-6113-6241]{Mattia C.\ Sormani}
\ITA
\affiliation{Department of Physics, University of Surrey, Guildford GU2 7XH, UK}

\author[0000-0003-1545-5078]{J.~D. Smith}
\Toledo

\author[0000-0003-4793-7880]{Fabian Walter}
\MPIA

\author[0000-0002-0012-2142]{Thomas G. Williams}
\affiliation{Sub-department of Astrophysics, Department of Physics, University of Oxford, Keble Road, Oxford OX1 3RH, UK}

\begin{abstract}
The CO-to-H$_2$ conversion factor ($\alpha_\mathrm{CO}$) is central to measuring the amount and properties of molecular gas. It is known to vary with environmental conditions, and previous studies have revealed lower $\alpha_\mathrm{CO}$ in the centers of some barred galaxies on kpc scales. To unveil the physical drivers of such variations, we obtained ALMA Band~3, 6, and~7 observations toward the inner $\sim$2~kpc of NGC~3627 and NGC~4321 tracing $^{12}$CO, $^{13}$CO, and C$^{18}$O lines on $\sim$100~pc scales. Our multi-line modeling and Bayesian likelihood analysis of these datasets reveal variations of molecular gas density, temperature, optical depth, and velocity dispersion, which are among the key drivers of $\alpha_\mathrm{CO}$. The central 300~pc nuclei in both galaxies show strong enhancement of temperature $T_\mathrm{k}\gtrsim100$~K and density $n_\mathrm{H_2}>10^3$~$\mathrm{cm^{-3}}$.
Assuming a CO-to-H$_2$ abundance of $3\times10^{-4}$, we derive 4--15 times lower $\alpha_\mathrm{CO}$ than the Galactic value across our maps, which agrees well with previous kpc-scale measurements. 
Combining the results with our previous work on NGC~3351, we find a strong correlation of $\alpha_\mathrm{CO}$ with low-$J$ $^{12}$CO optical depths ($\tau_\mathrm{CO}$), as well as an anti-correlation with $T_\mathrm{k}$. The $\tau_\mathrm{CO}$ correlation explains most of the $\alpha_\mathrm{CO}$ variation in the three galaxy centers, whereas changes in $T_\mathrm{k}$ influence $\alpha_\mathrm{CO}$ to second order. Overall, the observed line width and $^{12}$CO/$^{13}$CO~2--1 line ratio correlate with $\tau_\mathrm{CO}$ variation in these centers, and thus they are useful observational indicators for $\alpha_\mathrm{CO}$ variation. We also test current simulation-based $\alpha_\mathrm{CO}$ prescriptions and find a systematic overprediction, which likely originates from the mismatch of gas conditions between our data and the simulations.   

\end{abstract}

\keywords{Barred spiral galaxies (136); CO line emission (262); Galaxy nuclei (609); Molecular gas (1073); Star forming regions (1565)}

\section{Introduction} \label{sec:intro}

The cold and dense molecular gas in the interstellar medium (ISM) is the direct fuel for current and future star formation. Measuring the amount and properties of molecular gas is crucial for understanding star formation, the ISM, and their relations with galaxy evolution. While molecular hydrogen (H$_2$) is the primary constituent of molecular gas, it is difficult to be directly observed in the cold ($T \lesssim 100$~K) phase where stars are formed \citep{2010pcim.book.....T,2011piim.book.....D}. Instead, molecular gas mass is often measured with the low-$J$ rotational lines of carbon monoxide ($^{12}$C$^{16}$O, hereafter CO) by applying a CO-to-H$_2$ conversion factor \citep{1987ApJ...319..730S,co-to-h2}. This conversion factor ($\alpha_\mathrm{CO}$) is often defined for the $J$=1--0 line as the ratio of total molecular gas mass ($M_\mathrm{mol}$ in $\mathrm{M_\odot}$) to the CO $J$=1--0 luminosity ($L_\mathrm{CO(1-0)}$ in K~km~s$^{-1}$~pc$^2$), or equivalently, the ratio of molecular gas surface density ($\Sigma_\mathrm{mol}$ in $\mathrm{M_\odot}$~pc$^{-2}$) to the CO 1--0 intensity ($I_\mathrm{CO(1-0)}$ in K~km~s$^{-1}$):
\begin{equation}
\alpha_\mathrm{CO} = \frac{M_\mathrm{mol}}{L_\mathrm{CO(1-0)}} = \frac{\Sigma_\mathrm{mol}}{I_\mathrm{CO(1-0)}}\ \rm \left[\frac{M_\odot}{K\ km~s^{-1}\ pc^2} \right]~.
\label{def_alpha_intro}
\end{equation}
Another common way to express the conversion factor is to quote the ratio between H$_2$ column density and CO intensity, $X_\mathrm{CO} \equiv  N_\mathrm{H_2} / I_\mathrm{CO(1-0)}$, which is related to $\alpha_\mathrm{CO}$ via $X_\mathrm{CO} \left[ \frac{\mathrm{cm^{-2}}}{\mathrm{K\ km~s^{-1}}} \right] = 4.5\times10^{19}\ \alpha_\mathrm{CO} \left[ \rm \frac{M_\odot}{K\ km~s^{-1}\ pc^2} \right]$, where the $4.5\times10^{19}$ factor includes the mass contribution from Helium to $M_\mathrm{mol}$. 

$\alpha_\mathrm{CO}$ can be measured by estimating $M_\mathrm{mol}$ using virial methods, $\gamma$-ray emission, or optically-thin tracers like dust or CO isotopologues \citep[e.g.,][]{2008ApJ...686..948B,2011ApJ...737...12L,2012A&A...538A..71A,2012ApJ...750....3A,2013ApJ...777....5S,2017A&A...601A..78R,2020A&A...635A.131I,2022ApJ...925...72T}. Previous $\alpha_\mathrm{CO}$ measurements toward molecular clouds in the disks of the Milky Way or other nearby spiral galaxies have reported relatively consistent $\alpha_\mathrm{CO}$ values around 4.4~$\mathrm{M_\odot\ (K~km~s^{-1}~pc^2)^{-1}}$ (or $2\times10^{20}$ $\mathrm{cm^{-2}\ (K~km~s^{-1})^{-1}}$ in $X_\mathrm{CO}$) within a factor of $\sim$2 \citep[see the review by][and references therein]{co-to-h2}. This also includes studies across various Galactic disk GMCs using CO and $^{13}$CO observations together with radiative transfer modeling \citep{2008ApJ...680..428G,2013ApJ...775L...2L,2015ApJS..216...18N}, which is similar to the methodology we use in this paper.  
Therefore, many studies assume a constant, Galactic-like $\alpha_\mathrm{CO}$ value when inferring molecular gas mass from CO observations. However, recent theoretical studies have shown that $\alpha_\mathrm{CO}$ can vary by up to one or two orders of magnitude in different environments, and it is known to depend on gas properties including metallicity, temperature, column and volume densities, velocity dispersion, as well as the nature of excitation \citep[e.g.,][]{2010ApJ...716.1191W,2012ApJ...747..124F,2012MNRAS.426..377G,2012MNRAS.421.3127N,2012A&A...542A..65K,2015A&A...574A.127K,co-to-h2,2019A&A...621A.104R,2020ApJ...903..142G}. Such environmental dependence can explain why $\alpha_\mathrm{CO}$ has been found in observations to deviate from the Galactic disk value in various galaxy centers \citep{2009A&A...493..525I,2009A&A...506..689I,2020A&A...635A.131I,2013ApJ...777....5S,2022ApJ...925...72T}, (ultra-)luminous infrared galaxies \citep[U/LIRGs;][]{1998ApJ...507..615D,2014ApJ...795..174K,2017MNRAS.471.2917K,2014ApJ...796L..15S,2017ApJ...840....8S,2019A&A...628A..71H}, or low-metallicity galaxies \citep{1997A&A...328..471I,2000mhs..conf..293I,2018MNRAS.478.1716P,2020A&A...643A.141M}.

The variation of $\alpha_\mathrm{CO}$ within and among galaxies has a direct impact on many important quantities and relations that are widely used in current studies, because of their dependence on molecular gas mass estimation. This includes the molecular gas depletion time (which depends on $M_\mathrm{mol}$ and star formation rate), the cloud free-fall time (which depends on $M_\mathrm{mol}$ and cloud size), the virial parameter and turbulent pressure (both of which depend on $M_\mathrm{mol}$, cloud size, and velocity dispersion), and the gas inflow rates in barred galaxy centers, to name only a few. For instance, \citet{2013AJ....146...19L} and \citet{2023arXiv230203044D} showed that the molecular cloud depletion time in galaxy centers will become significantly shorter if $\alpha_\mathrm{CO}$ depression is considered. 
\citet{2020ApJ...892..148S,2022AJ....164...43S} demonstrated how cloud virial parameter, turbulent pressure, and ISM dynamical equilibrium pressure would vary with different choices of $\alpha_\mathrm{CO}$. 
$\alpha_\mathrm{CO}$ is also the dominant source of uncertainty in estimating the bar-driven mass inflow rates in the Central Molecular Zone \citep{2019MNRAS.484.1213S}.    
Furthermore, $\alpha_\mathrm{CO}$ variation can change the slopes of star formation scaling relations \citep[e.g.,][]{2012ApJ...758..127F,2012MNRAS.421.3127N,2021A&A...650A.134P,2023arXiv230203044D,2023ApJ...945L..19S}, such as the Kennicutt--Schmidt \citep{1998ApJ...498..541K,stacking} and molecular gas main sequence relations \citep{2019ApJ...884L..33L}. 
Therefore, it is critical to understand the physical drivers of $\alpha_\mathrm{CO}$ and establish how $\alpha_\mathrm{CO}$ behaves in different environmental regimes.

Recent years have seen progress on the development of a metallicity-dependent $\alpha_\mathrm{CO}$ prescription \citep{2012AJ....143..138S,2016A&A...588A..23A,2017MNRAS.470.4750A}, which has been applied in several recent works \citep[e.g.,][]{2020ApJ...892..148S,2020ApJ...901L...8S,2021A&A...650A.134P}. In terms of the emissivity dependence, many studies adopt a bimodal $\alpha_\mathrm{CO}$ with $\sim$0.8~$\mathrm{M_\odot\ (K~km~s^{-1}~pc^2)^{-1}}$ in (U)LIRGs or starburst regions \citep{1998ApJ...507..615D} and the Galactic-like 4.4~$\mathrm{M_\odot\ (K~km~s^{-1}~pc^2)^{-1}}$ elsewhere.
However, recent theoretical studies and simulations suggest that $\alpha_\mathrm{CO}$ is not simply bimodal or metallicity dependent. Instead, it is likely to vary continuously with local environmental conditions in addition to metallicity \citep{2012MNRAS.421.3127N,co-to-h2}. Theoretical and observational works have also shown that emissivity-related terms such as temperature, density, and opacity are important drivers of $\alpha_\mathrm{CO}$ variation, especially in actively star-forming galaxies including mergers and galaxy centers \citep{2011MNRAS.418..664N,2012MNRAS.421.3127N,2012ApJ...751...10P,2018ApJ...863..143C,2020ApJ...903..142G,2022ApJ...925...72T}. Therefore, a crucial next step would be to identify observational tracers and establish a robust prescription that can predict the effects of emissivity-related terms on $\alpha_\mathrm{CO}$.

Compared to observational studies, simulations can give direct $\alpha_\mathrm{CO}$ predictions from sophisticated modeling of gas dynamics, chemistry, and radiative transfer, allowing the development of prescriptions useful for observations. Thus, significant efforts have been made to investigate $\alpha_\mathrm{CO}$ variations using numerical simulations \citep{2011MNRAS.412.1686S,2011MNRAS.415.3253S,2011MNRAS.418..664N,2012MNRAS.421.3127N,2012ApJ...747..124F,2015A&A...575A..56B,2015MNRAS.447.2144D,2018MNRAS.475.1508P,2018ApJ...858...16G,2020ApJ...903..142G,2019A&A...621A.104R,2020MNRAS.492.1465S,2021MNRAS.502.2701B,2022ApJ...931...28H}. In particular, \citet{2012MNRAS.421.3127N} proposed a functional prediction of $\alpha_\mathrm{CO}$ from metallicity and $I_\mathrm{CO(1-0)}$ based on low-redshift mergers and high-redshift disks in their simulation. Some studies focusing on starburst mergers also found correlations between $\alpha_\mathrm{CO}$ and star formation rate or molecular gas depletion time \citep{2015A&A...575A..56B,2019A&A...621A.104R}. More recently, (magneto-)hydrodynamical simulations resolving down to pc scales further explored how $\alpha_\mathrm{CO}$ may vary with observational beam size \citep{2020ApJ...903..142G,2022ApJ...931...28H}. 
Both studies have suggested $\alpha_\mathrm{CO}$ dependence on $I_\mathrm{CO(1-0)}$, and \citet{2020ApJ...903..142G} also found $\alpha_\mathrm{CO}$ correlations with the CO 2--1/1--0 line ratio ($R_\mathrm{21}$) and CO line peak temperature. 
While these simulations are limited to Galactic disk-like environments with much lower CO intensity ($< 200$~K km s$^{-1}$) and surface density ($< 100$~$\mathrm{M_\odot}$~pc$^{-2}$) than in galaxy centers, it is important to test the simulation-based predictions and understand if and where they can accurately predict $\alpha_\mathrm{CO}$.   

In this work, we study the spatial variations of molecular gas properties and $\alpha_\mathrm{CO}$ in nearby galaxy centers at 100~pc scales, using observations of multiple CO, $^{13}$CO, and C$^{18}$O rotational transitions with the Atacama Large Millimeter/submillimeter Array (ALMA). We target nearby barred galaxies that were found by previous kpc-scale observations to have $\alpha_\mathrm{CO}$ depression in their central few kpc, including NGC~3351, NGC~3627, and NGC~4321 \citep{2013ApJ...777....5S,2015PASJ...67....2M,2020A&A...635A.131I,2021MNRAS.504.2360J}. These galaxies were also found to have a near-solar gas-phase metallicity \citep{2019ApJ...887...80K,2020MNRAS.499..193K,2022A&A...658A.188S,2022MNRAS.509.1303W}. Following our previous work on the central kpc of NGC~3351 \citep{2022ApJ...925...72T}, here we present an extension towards the centers of NGC~3627 and NGC~4321. In this paper, we discuss the implications of the combined results for all three galaxy centers. The basic information for these galaxies is provided in Table~\ref{tab:source}. 

This paper is structured as follows. Section~\ref{sec:obs} describes the observations and data reduction. Section~\ref{sec:result} presents the results of integrated intensity, line ratios, and the regional statistics. Our multi-line modeling setup and results are presented in Section~\ref{sec:model}. In Section~\ref{sec:discussion}, we discuss implications from our modeling and $\alpha_\mathrm{CO}$ solutions and compare with results from the literature. The conclusions are summarized in Section~\ref{sec:conclusion}.

\begin{table}
\footnotesize
\begin{center}
\caption{Source Information \label{tab:source}}
\begin{tabular}{lcccc}
  \hline\hline
  Property & NGC~3627 & NGC~4321 & NGC~3351 \\
  \hline
  R.A. (J2000) &$11^\mathrm{h}20^\mathrm{m}15\fs0$ &$12^\mathrm{h}22^\mathrm{m}54\fs9$ &$10^\mathrm{h}43^\mathrm{m}57\fs8$\\
  Decl. (J2000) &$+12^\circ59\arcmin29\arcsec$ &$+15^\circ49\arcmin20\arcsec$ &$+11^\circ42\arcmin13\arcsec$\\
  Hubble Type &SABb &SABbc & SBb\\
  Nuclear Type &LINER/AGN &H\,{\scriptsize II}/LINER &H\,{\scriptsize II}\\
  Distance (Mpc) &$11.32\pm0.48$ &$15.21\pm0.49$ &$9.96\pm0.33$\\
  Linear Scale (pc/$\arcsec$) &54.9 &73.7 &48.3 \\
  Matched Beam ($\arcsec$) &2.0 &1.7 &2.1 \\
  Inclination ($^\circ$) &$57.3\pm1.0$ &$38.5\pm2.4$ &$45.1\pm6.0$\\
  Position Angle ($^\circ$) &$173.1\pm3.6$ &$156.2\pm1.7$ &$192.7\pm0.4$\\
  $\log_{10} M_*$ (M$_\odot$) &10.84 &10.75 &10.37 \\
  SFR (M$_\odot$/yr) &3.89 &3.55 &1.32 \\
  \hline
\end{tabular} 
\end{center}
\tablecomments{Positions, stellar masses, and star formation rates from \citet{2021ApJS..257...43L}. Nuclear types suggested by \citet{1997ApJS..112..315H,2000ApJS..129...93F,2010ApJS..190..233M,2022A&A...659A..26B}. Distances from \citet{2021MNRAS.501.3621A}. Inclinations and position angles from \citet{2020ApJ...897..122L}.}
\end{table}

\section{Observations and Data} \label{sec:obs}

We obtained ALMA observations of six low-$J$ CO, $^{13}$CO, and C$^{18}$O lines in Band 3, 6, and 7, covering at least the central $35\arcsec \times 35\arcsec$ (1.5--2~kpc) area in NGC~3627 and NGC~4321. The achieved angular resolutions of 1--2$\arcsec$ (or ${\lesssim}100$~pc in physical scale) allow us to probe molecular gas conditions approaching typical giant molecular cloud (GMC) scales of a few tens of pc \citep[e.g.,][]{1987ApJS...63..821S}. These observations were planned together with and set up similarly to those described in \citet[which cover the central ${\sim}30\arcsec$ of NGC~3351]{2022ApJ...925...72T}. We briefly summarize the data characteristics below, and refer interested readers to \citet{2022ApJ...925...72T} for more details.

Our Band~3 observations (projects 2015.1.00978.S and 2016.1.00972.S) captured the $J$=1--0 line of CO with the 12-m array in the C36-2/3 and C40-4 configurations for NGC~3627 and NGC~4321, respectively. The native beam sizes are accordingly $1.8\arcsec \times 1.7\arcsec$ and $1.3\arcsec \times 1.0\arcsec$. We use a three-pointing mosaic to cover the central $60\arcsec \times 60\arcsec$ area in each galaxy. The rms noise level is 0.16~K (for NGC~3627) and 0.09~K (for NGC~4321) per 2.5~km~s$^{-1}$ velocity channel. 

The Band~6 observations come from two separate projects and cover the $J$=2--1 transitions of CO, $^{13}$CO, and C$^{18}$O in two distinct spectral tunings. Observations of the $^{13}$CO and C$^{18}$O 2--1 (project 2015.1.00978.S) were carried out in the C36-1 and C36-2/3 configurations for NGC~3627 and NGC~4321, respectively. We use a seven-pointing mosaic to cover the central $40\arcsec \times 40\arcsec$ area for each target. The native beam size is $1.5\arcsec \times 1.2\arcsec$ ($1.1\arcsec \times 0.9 \arcsec$) and the rms noise level is 15~mK (9~mK) per 2.5~km~s$^{-1}$ velocity channel for NGC~3627 (NGC~4321). The CO 2--1 data were instead obtained from the PHANGS--ALMA survey (project 2015.1.00956.S) and reach an angular resolution of ${\sim}1.6\arcsec$ and an rms level of ${\sim}0.08$~K (for more details, see \citealt{2021ApJS..257...43L,2021ApJS..255...19L}).

The Band~7 observations cover the $J$=3--2 lines of $^{13}$CO and C$^{18}$O (project 2016.1.00972.S) with a mixture of C40-1, C43-1, and C43-2 configurations for either target. The central $35\arcsec \times 35\arcsec$ area in each galaxy is covered by a 14-pointing mosaic, and the achieved native beam sizes are $1.2\arcsec \times 1.0\arcsec$ for NGC~3627 and $1.1\arcsec \times 0.9\arcsec$ for NGC~4321. The rms noise level is 10~mK (7~mK) per 2.5~km~s$^{-1}$ velocity channel for NGC~3627 (NGC~4321).   

We follow the same calibration and imaging process as described in full detail in \citet{2022ApJ...925...72T}. In short, we calibrated the raw data with scripts provided by the observatory and imaged all the lines in a similar way adapting the PHANGS--ALMA pipeline \citep{2021ApJS..255...19L}. We then convolved all the data cubes to a matched round beam of 2.0$\arcsec$ (110~pc) for NGC~3627 and 1.7$\arcsec$ (125~pc) for NGC~4321, and produced a set of moment maps and effective line width ($\Delta v$) maps\footnote{The effective line width is defined as $I_\mathrm{CO}/(\sqrt{2\pi}\,T_\mathrm{peak})$, which is identical to moment 2 for a Gaussian line profile. See \citet{2001ApJ...551..852H} and \citet{2018ApJ...860..172S,2020ApJ...892..148S} for more details.} for all six lines at the common resolution. The map creation scheme is also similar to that implemented in the PHANGS--ALMA pipeline, except that we start from a high confidence mask including at least two consecutive channels above 5$\sigma$, and then expand into a more inclusive mask with at least two consecutive channels above 2$\sigma$.
Finally, we regridded all data products such that the pixel scale matches 1/2 the beam size (i.e., Nyquist sampling). These beam-matched, Nyquist-sampled data products include a set of moment maps and uncertainty maps for all the lines, where the uncertainty maps were derived from the noise measured in the data cubes propagated through the steps of creating the moment maps. The data products are used throughout this work, and many of the maps are presented in Section~\ref{sec:result} and Appendix~\ref{sec:vel_maps}.   

We note that most of the observations presented here only used the 12-m array, except for the CO 2--1 observations from PHANGS--ALMA \citep[which combines ALMA 12-m, 7-m, and total-power observations to ensure flux recovery on all scales; see][]{2021ApJS..257...43L}. To minimize impacts from the lack of short-spacing data, we adopt the same method introduced in \citet{2022ApJ...925...72T}.
Namely, we estimate the flux recovery ratio by creating a new CO 2--1 image from only the PHANGS 12-m observations and measuring the (pixel-by-pixel) ratio of the moment~0 maps made from the 12-m only image and the combined 12-m+7-m+TP image. Then, we mask out all pixels with that ratio lower than $70\%$ throughout our analysis. For NGC~3627 and NGC~4321, this results in ${\sim}1\%$ and $12\%$ of the number of pixels being masked, respectively, after applying the signal-to-noise (S/N) cuts described in Section~\ref{sec:result}. Thus, we expect individual line intensity errors due to incomplete $u$--$v$ sampling to be less than 30\%, assuming that CO 1--0 and 2--1 emission shows the same distribution. With the above procedure, we make sure that our analysis avoids regions where there can be significant missing flux due to the lack of short-spacing information.

\section{Results} \label{sec:result}

\begin{figure*} \centering
\includegraphics[width=\linewidth]{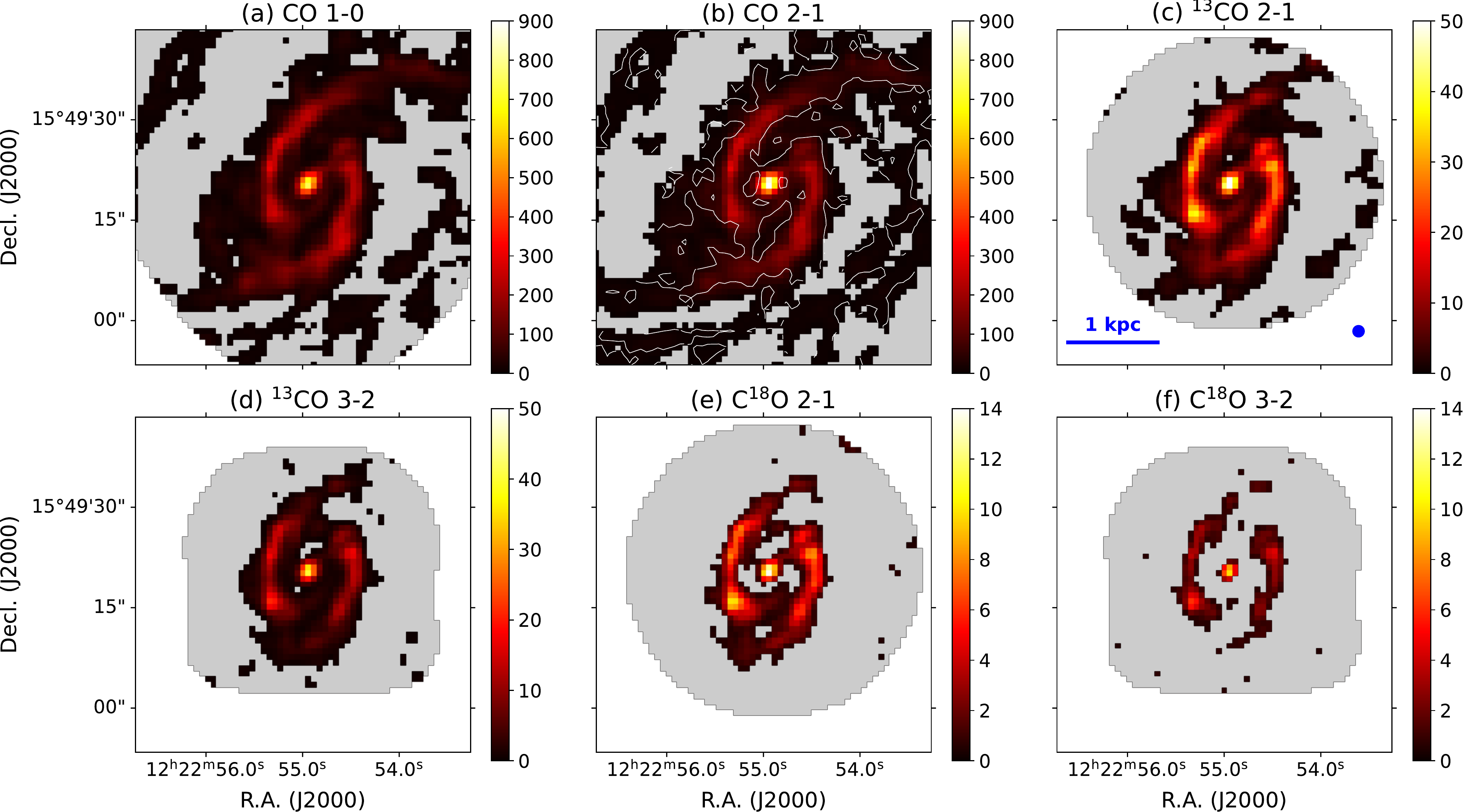} 
\caption{Integrated intensity maps of NGC~4321 (in units of $\mathrm{K~km~s^{-1}}$). The white areas lie outside the field of view of ALMA observations, while the gray areas show the pixels with $< 3\sigma$ detection. The overlaid contour in panel (b) represents a 70\% flux recovery rate (12-m/combined). The matched beam size of 1.7$\arcsec$ and a scale bar of 1~kpc are shown in panel (c). A bright nucleus and the inner spiral arms are securely detected in all six lines. The pixels in the gap between the nucleus and arms generally have low flux recovery rate with the 12-m array alone, and thus most of that region will be excluded from our analysis.}
\label{fig:mom0_4321}
\end{figure*}

\begin{figure*} \centering
\includegraphics[width=\linewidth]{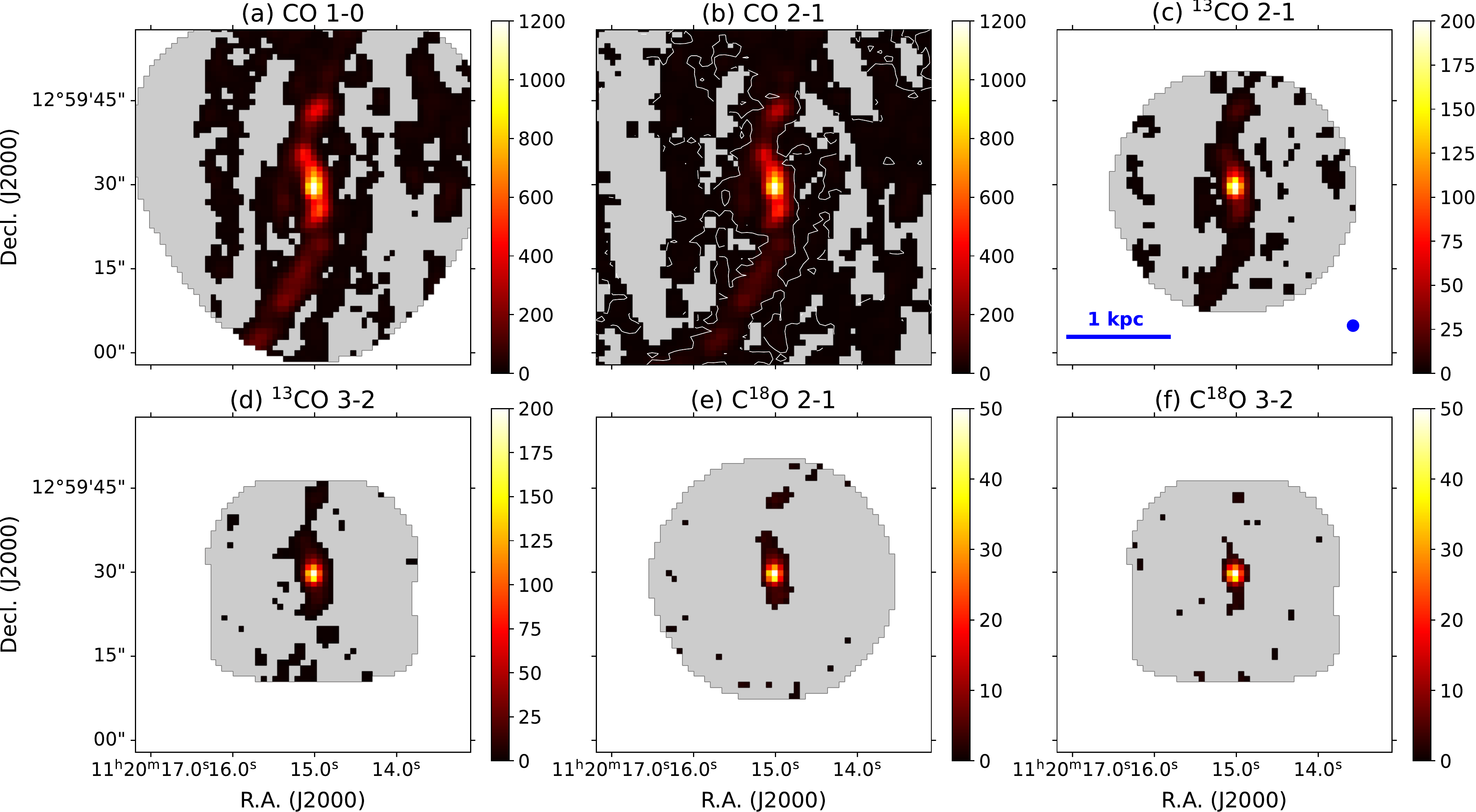} 
\caption{Same as Figure~\ref{fig:mom0_4321}, but for NGC~3627. The matched beam size of 2.0$\arcsec$ and a scale bar of 1~kpc are shown in panel (c). The central nucleus with a size of $\sim$300~pc is securely detected in all six lines, while the inner spiral arms are not bright enough to be detected in C$^{18}$O.}
\label{fig:mom0_3627}
\end{figure*}

Figures~\ref{fig:mom0_4321} and~\ref{fig:mom0_3627} show the integrated intensity (moment~0) maps of the six observed lines for NGC~4321 and NGC~3627, respectively. In these figures, we mask out regions with S/N $< 3$ for the line integrated intensity based on the ratio of moment~0 map and its associated uncertainty map (Section~\ref{sec:obs}). This masking is applied to the 2D moment maps and is distinct from the masking done on the 3D data cubes when creating the moment maps as described in Section~\ref{sec:obs}. 

In NGC~4321, the observations of all six lines capture a bright and compact (${\sim}300$~pc) nucleus surrounded by two inner spiral arms or bar lanes at ${\sim}1$ kpc galactocentric diameter. We note that the regions between the nucleus and the arms generally have $< 70\%$ flux recovery rate (based on CO 2--1), and thus they will be excluded from our analysis. The moment 0 images of NGC~3627 also reveal a ${\sim}300$~pc nucleus as well as bar lanes connected to the center. The nucleus of NGC~3627 is over two times brighter than that of NGC~4321 and is detected in all six lines, while the outer lanes in NGC~3627 are not bright enough to be detected in C$^{18}$O. 
We note that the bar-ends of NGC~3627, which are known to have high star formation rate likely due to interactions between the spiral arms and the bar \citep{2011MNRAS.411.1409W,2015ApJ...813..118M,2017A&A...597A..85B,2020MNRAS.493.2872C,2021MNRAS.506..963B}, are just outside our common field of view but can be slightly seen near the south-east edge of Figure~\ref{fig:mom0_3627}(b).

\begin{figure*}
\hspace{2em}
\begin{minipage}{.4\linewidth}
\centering
\includegraphics[width=\linewidth]{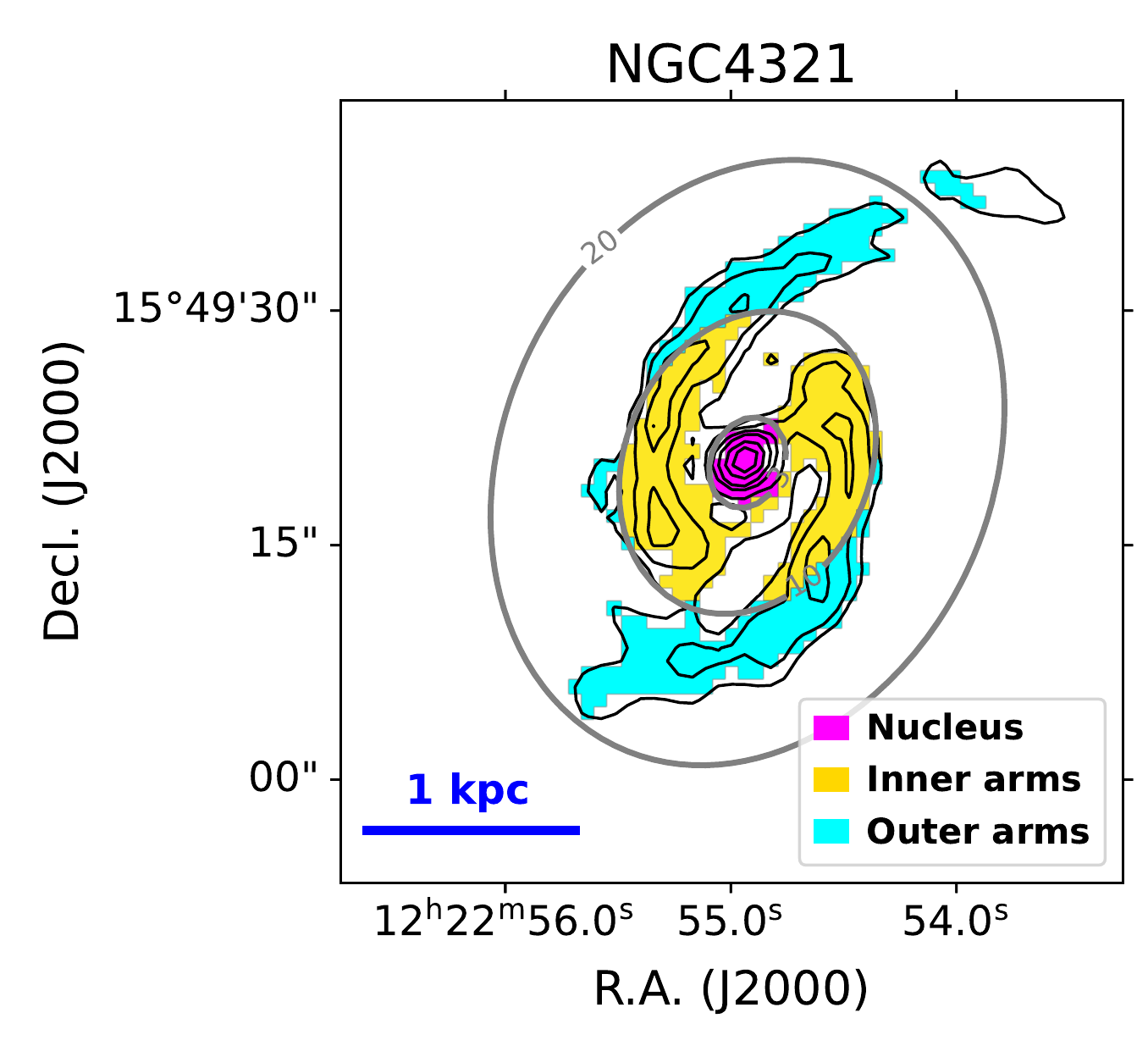}
\end{minipage}
\hspace{2em}
\begin{minipage}{.4\linewidth}
\centering
\includegraphics[width=\linewidth]{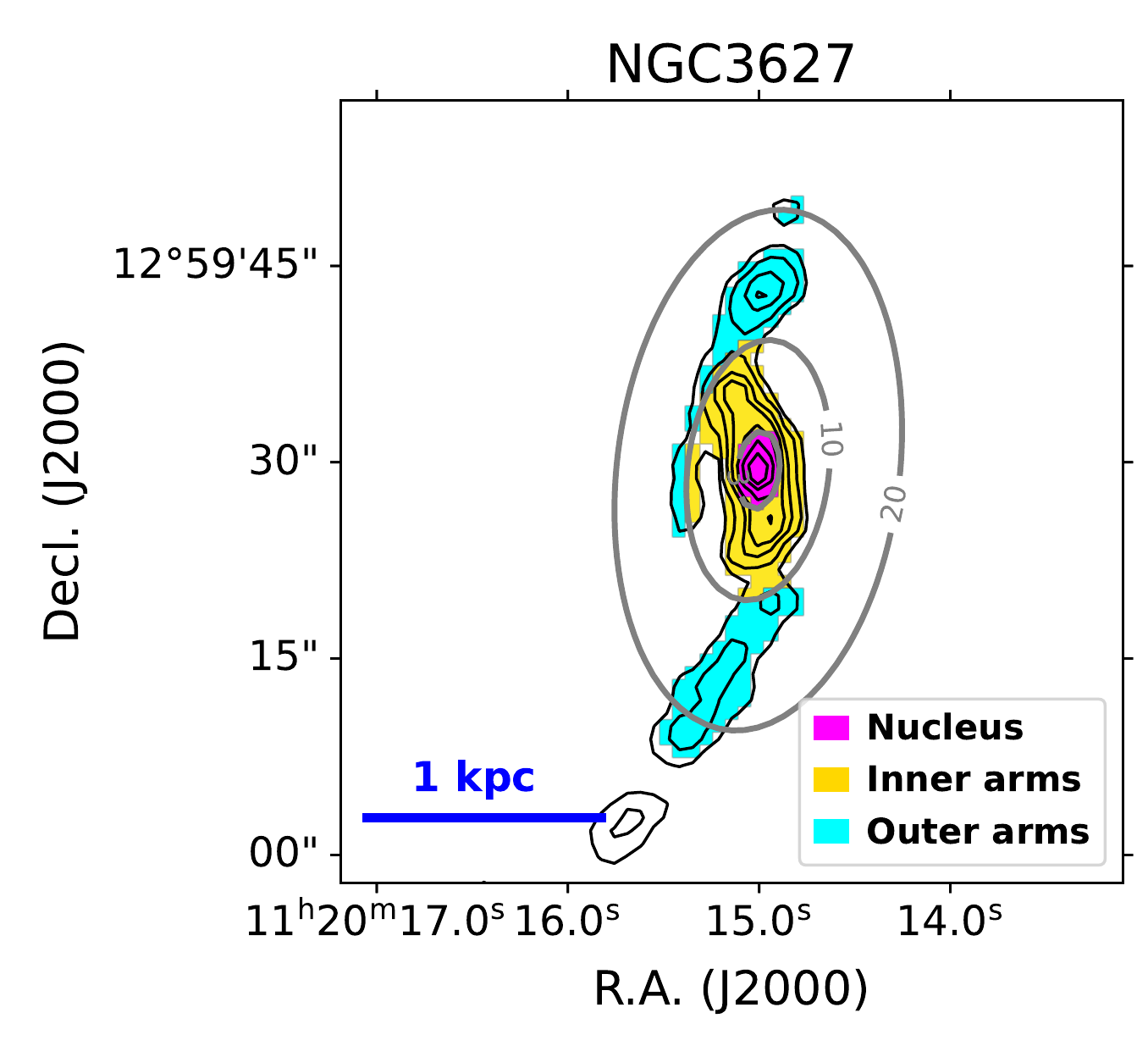}
\end{minipage}
\caption{Definition of the nucleus, inner arms, and outer arms regions based on the galactocentric radius, which will be used for regional statistics and analysis. The black contours represent the CO 2--1 integrated intensity at 50, 100, 200, 300, 500, 700 (and 900 for NGC~3627) K~km~$\mathrm{s^{-1}}$. The gray contours show the projected galactocentric radius of 3, 10, and 20$''$, respectively.}
\label{fig:def_regions}
\end{figure*}

As shown in Figure~\ref{fig:def_regions}, we define three different regions in both galaxies for further analysis. The ``nucleus'' is defined as the central $6''$ (300--450~pc in diameter) region. The ``inner arms'' cover the inner $20''$ region in galactocentric diameter but excludes the nucleus region, and the ``outer arms'' refer to pixels outside a diameter of $20''$ that are connected to the inner arms\footnote{The nomenclature of ``arms'' in this paper is simply based on gas morphology and has no implications on the dynamical driver of this feature. These regions are bar lanes or inner spiral arms within the main galactic bar, which are different from spiral arms seen in the outer disks.}. All the pixels included in our analysis have S/N $>3$ in $^{13}$CO and a flux recovery rate of $> 70\%$. To further ensure a reasonable uncertainty range for the parameters estimated from our modeling (Section~\ref{sec:model}), we only consider pixels with S/N $\gtrsim$5 and 50 in $^{13}$CO and CO lines, respectively, which corresponds to $I_\mathrm{CO(2-1)} > 50$~K~km~$\mathrm{s^{-1}}$. This intensity cutoff ensures a $< 0.5$~dex uncertainty in our $\alpha_\mathrm{CO}$ estimates for every included pixel, and it is applied in addition to the S/N $> 3$ criterion for $^{13}$CO.

\begin{figure*} 
\centering
\includegraphics[width=0.32\linewidth]{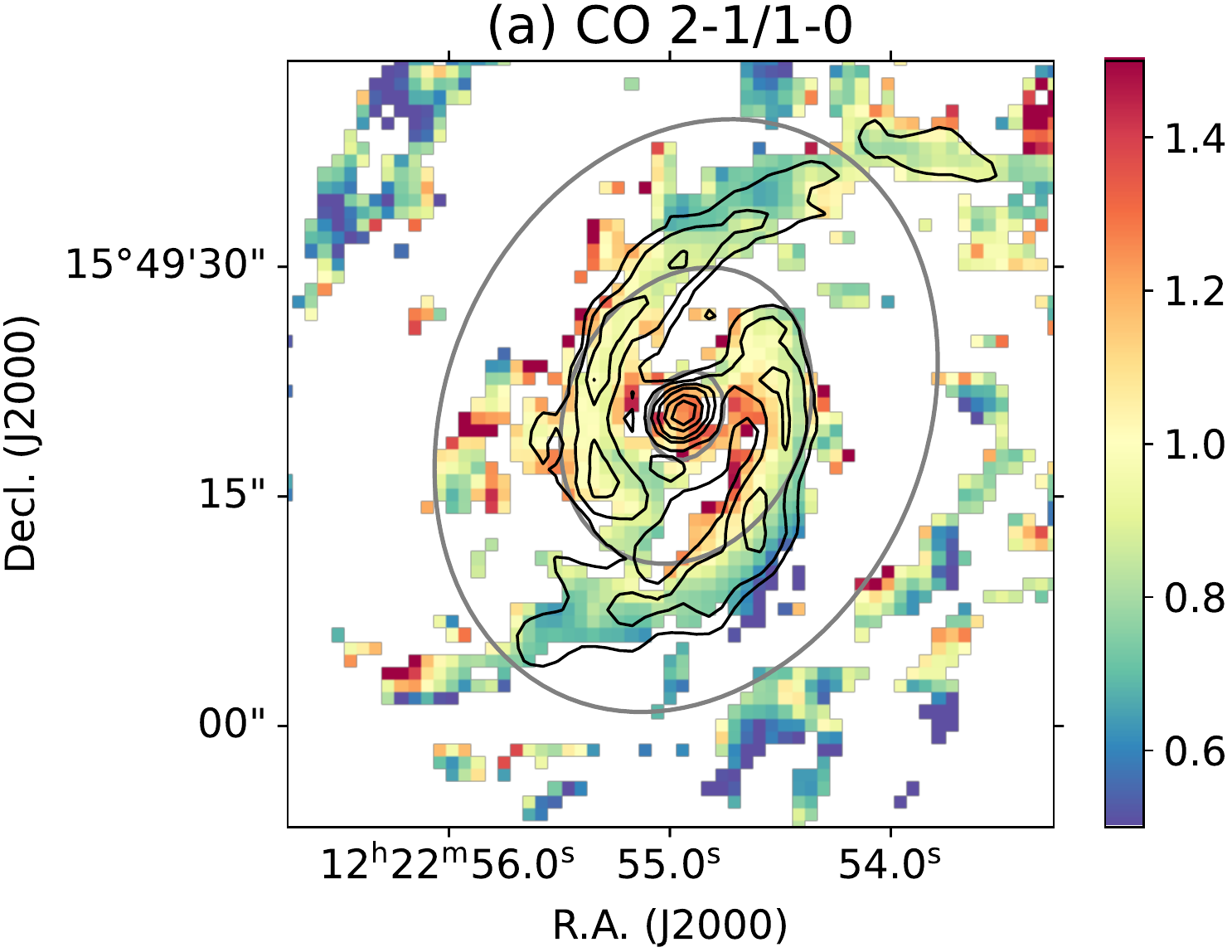} \includegraphics[width=0.32\linewidth]{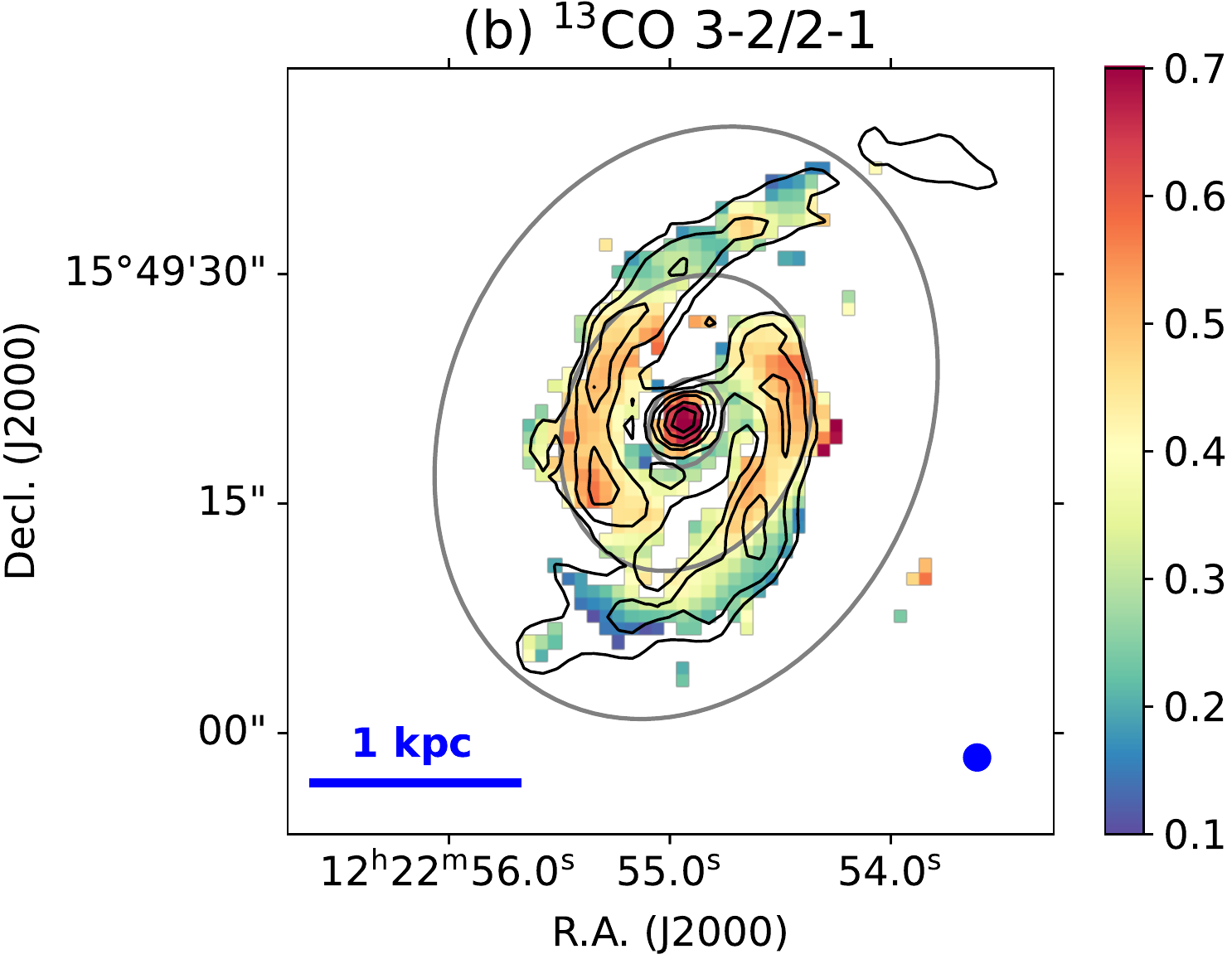} \includegraphics[width=0.32\linewidth]{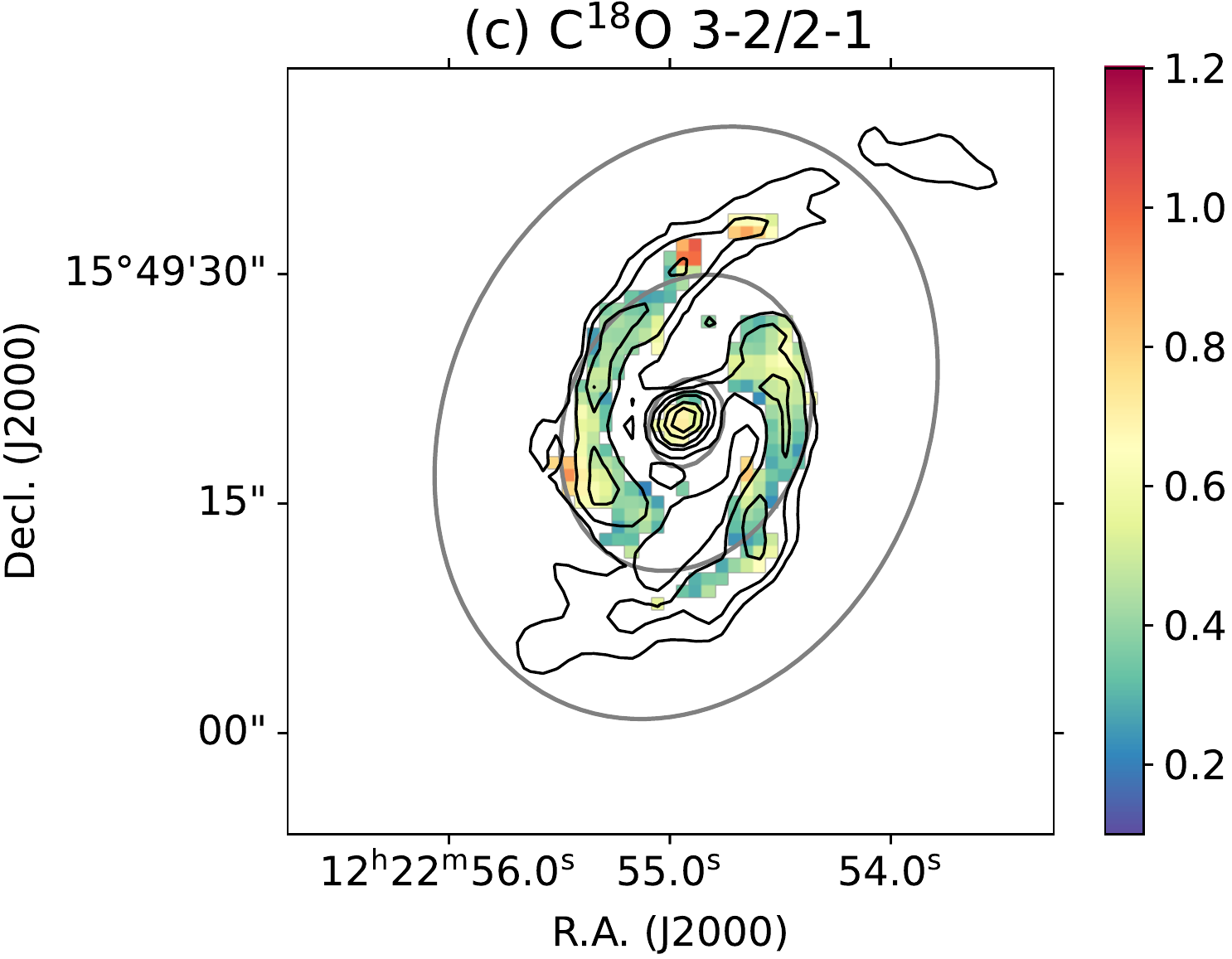} \\
\includegraphics[width=0.33\linewidth]{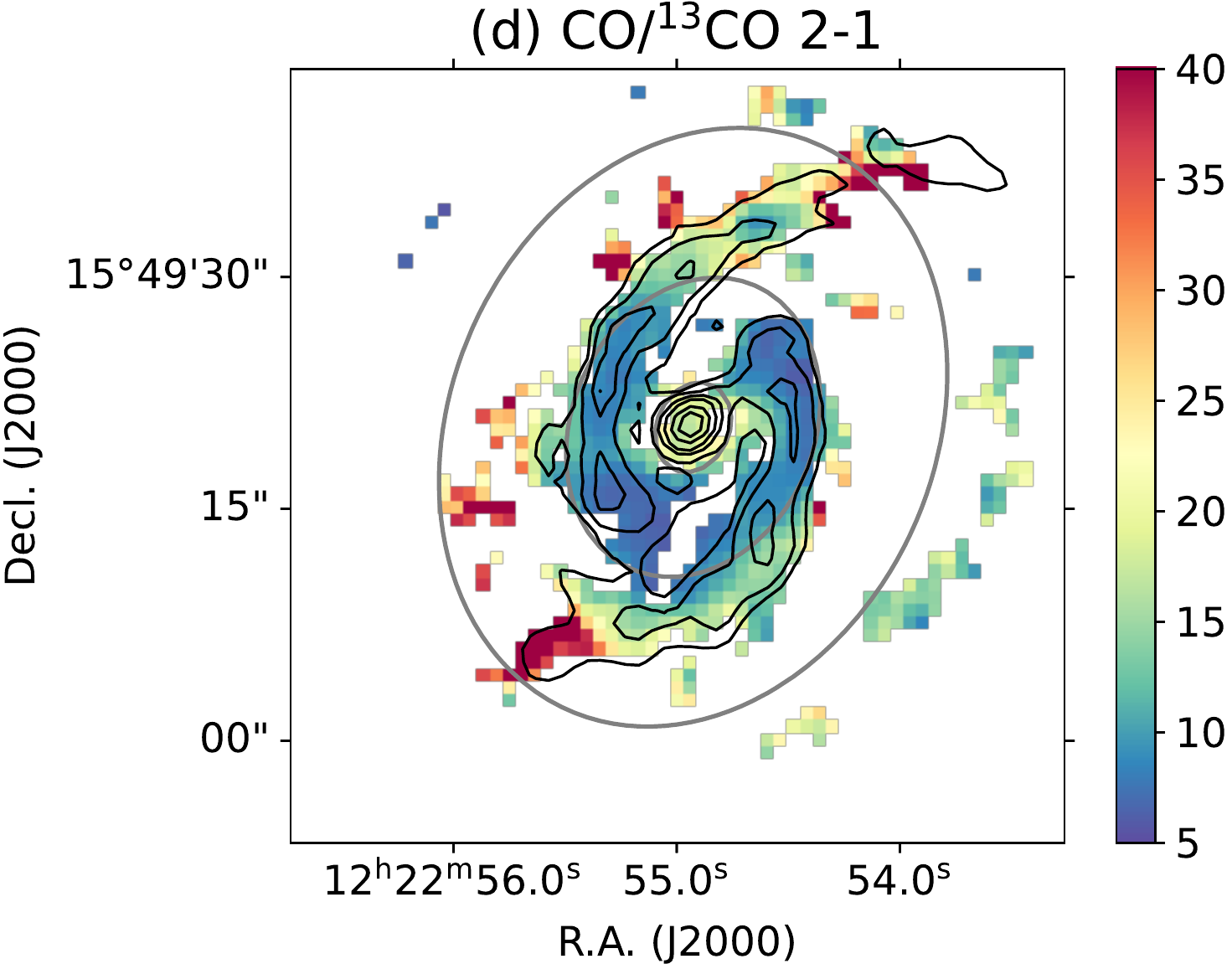} 
\includegraphics[width=0.33\linewidth]{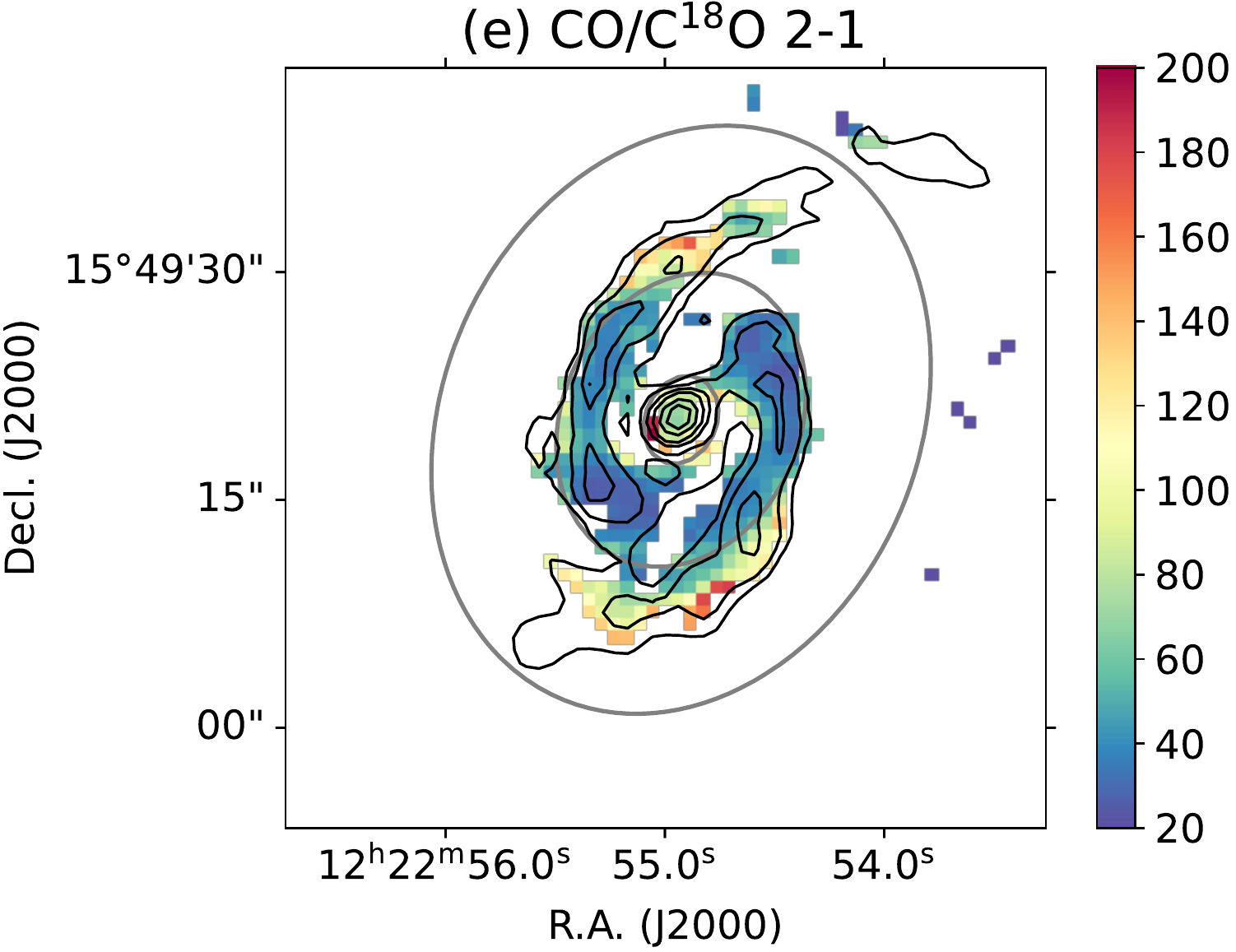} \\ \includegraphics[width=0.33\linewidth]{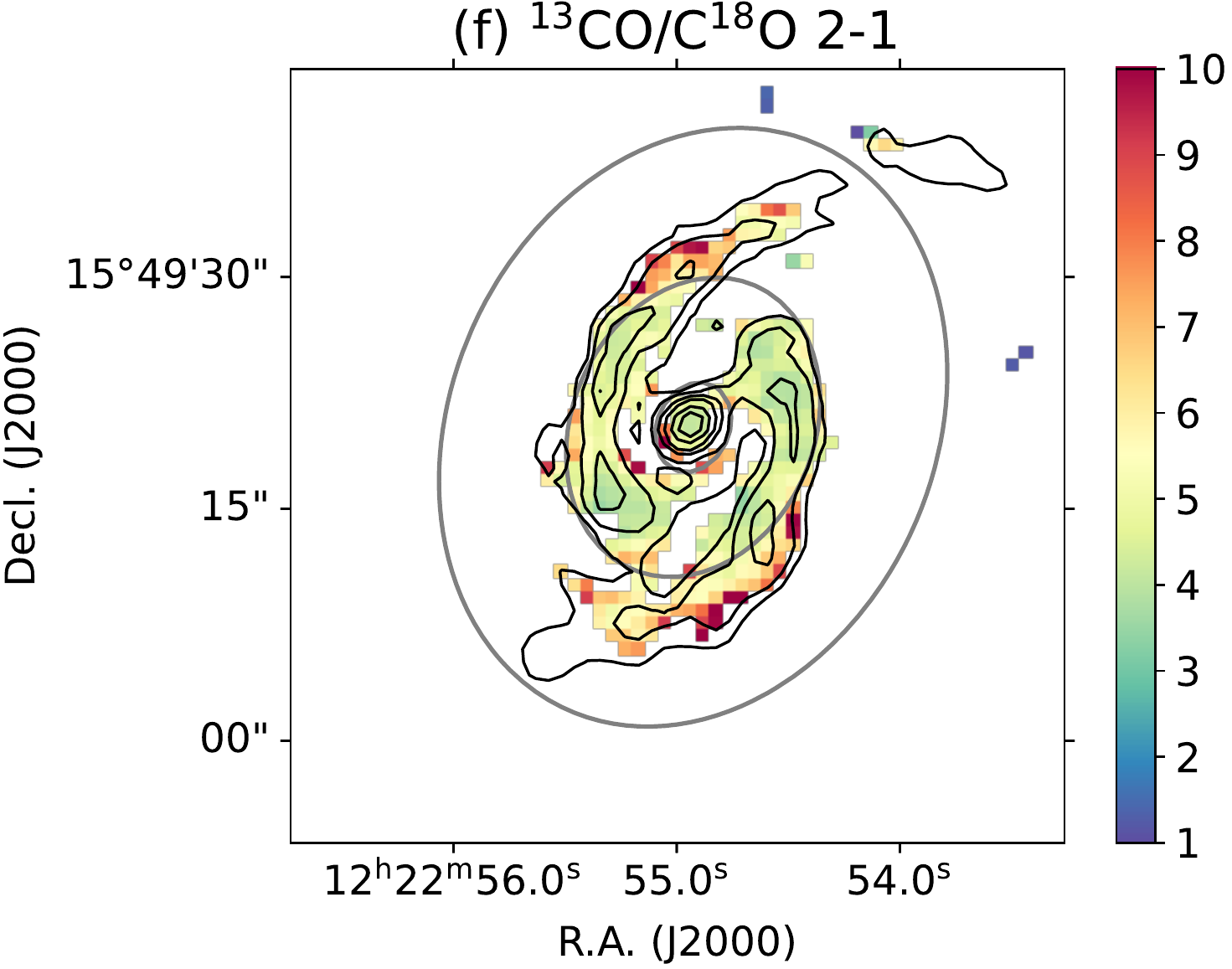} \includegraphics[width=0.33\linewidth]{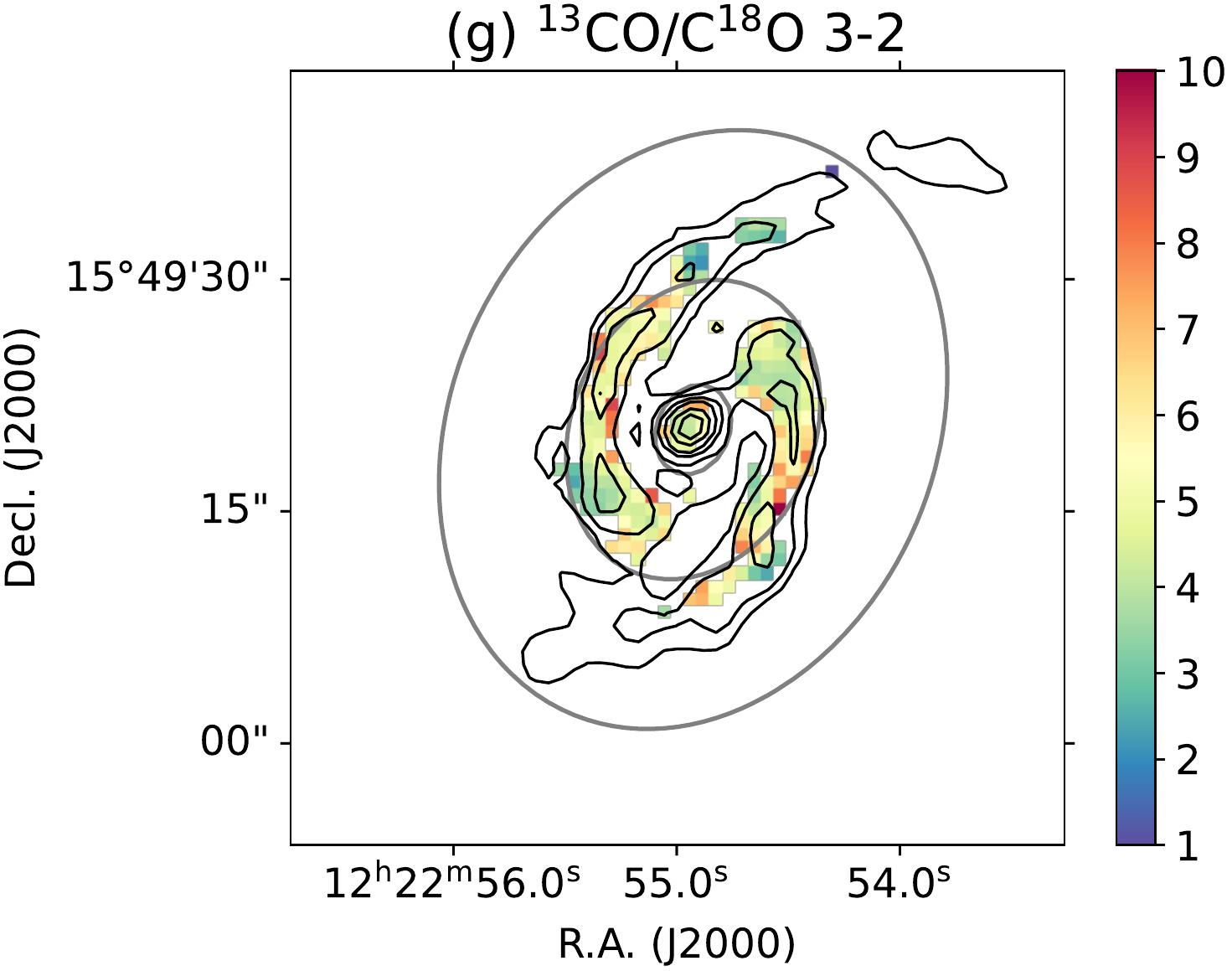}
\caption{Line ratio maps of NGC~4321. Any region with $< 70\%$ flux recovered rate or $< 3\sigma$ detection in either relevant line is masked out in each panel. The gray contours represent the projected galactocentric radii and the black contours show the CO 2--1 integrated intensity, both of which are the same as in Figure~\ref{fig:def_regions}. (a)--(c) show the primarily temperature-sensitive line ratios, and (d)--(g) show the line ratios primarily sensitive to isotopologue abundances or optical depths.}
\label{fig:ratio_4321}
\end{figure*}

\begin{figure*} 
\centering
\includegraphics[width=0.32\linewidth]{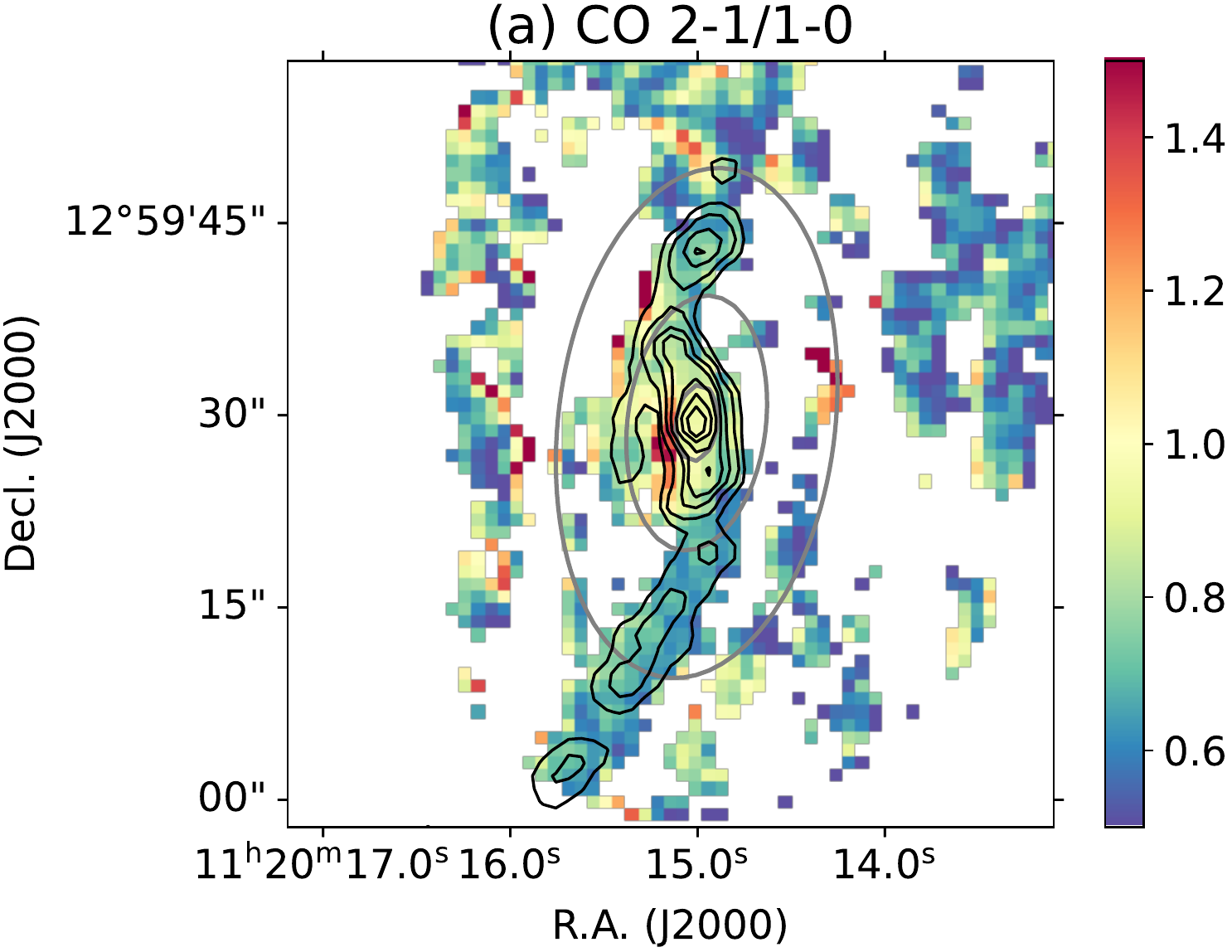} \includegraphics[width=0.32\linewidth]{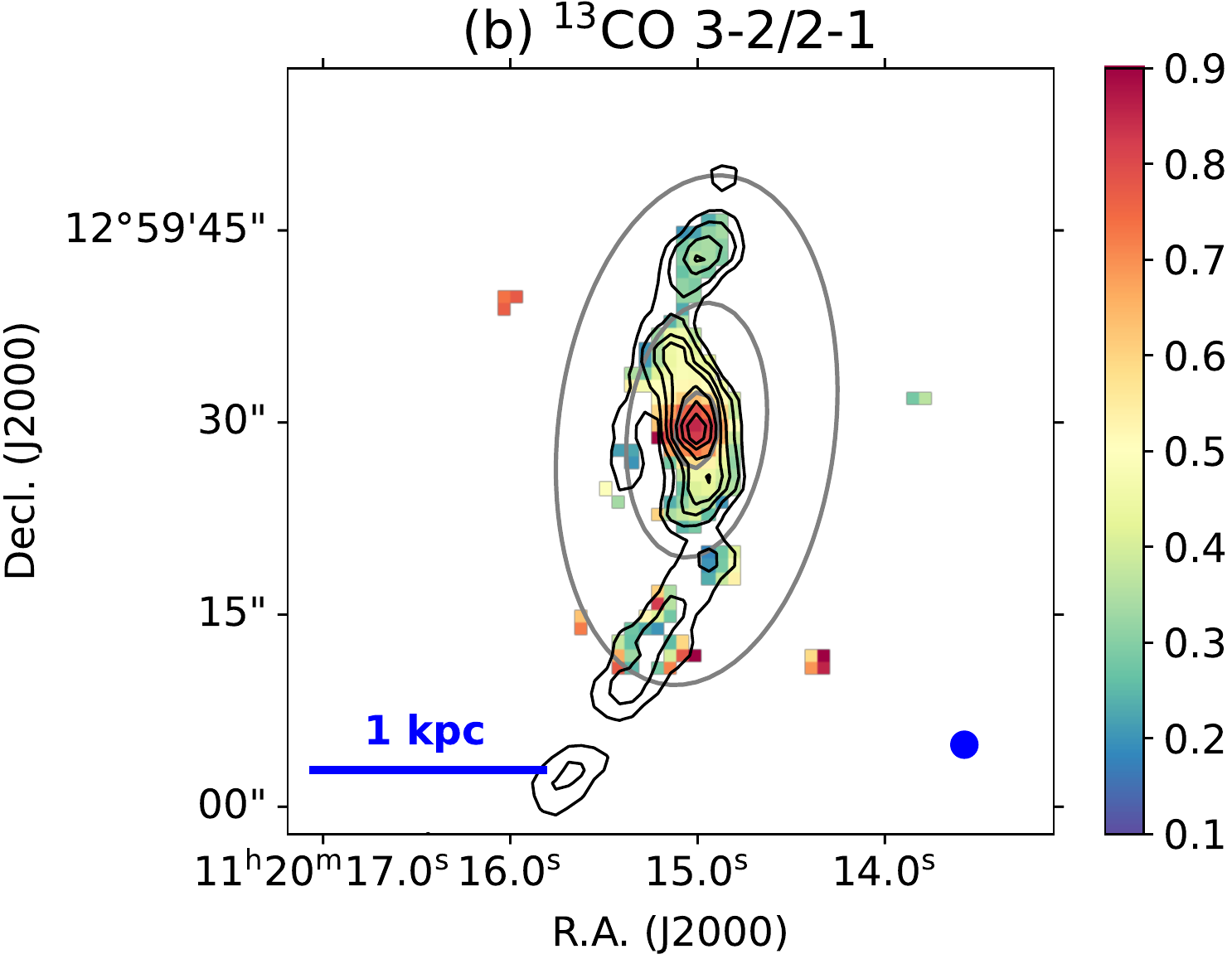} \includegraphics[width=0.32\linewidth]{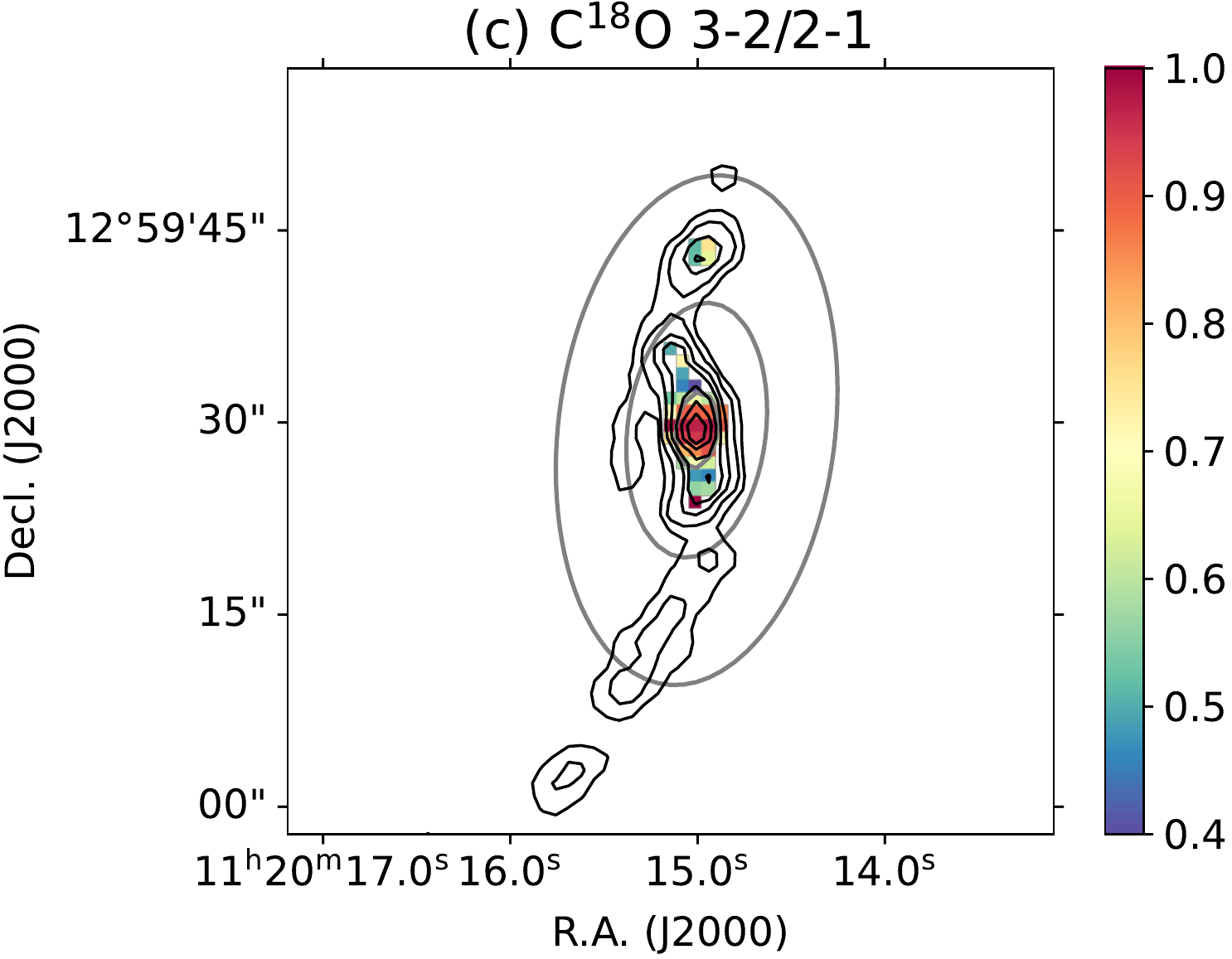} \\
\includegraphics[width=0.33\linewidth]{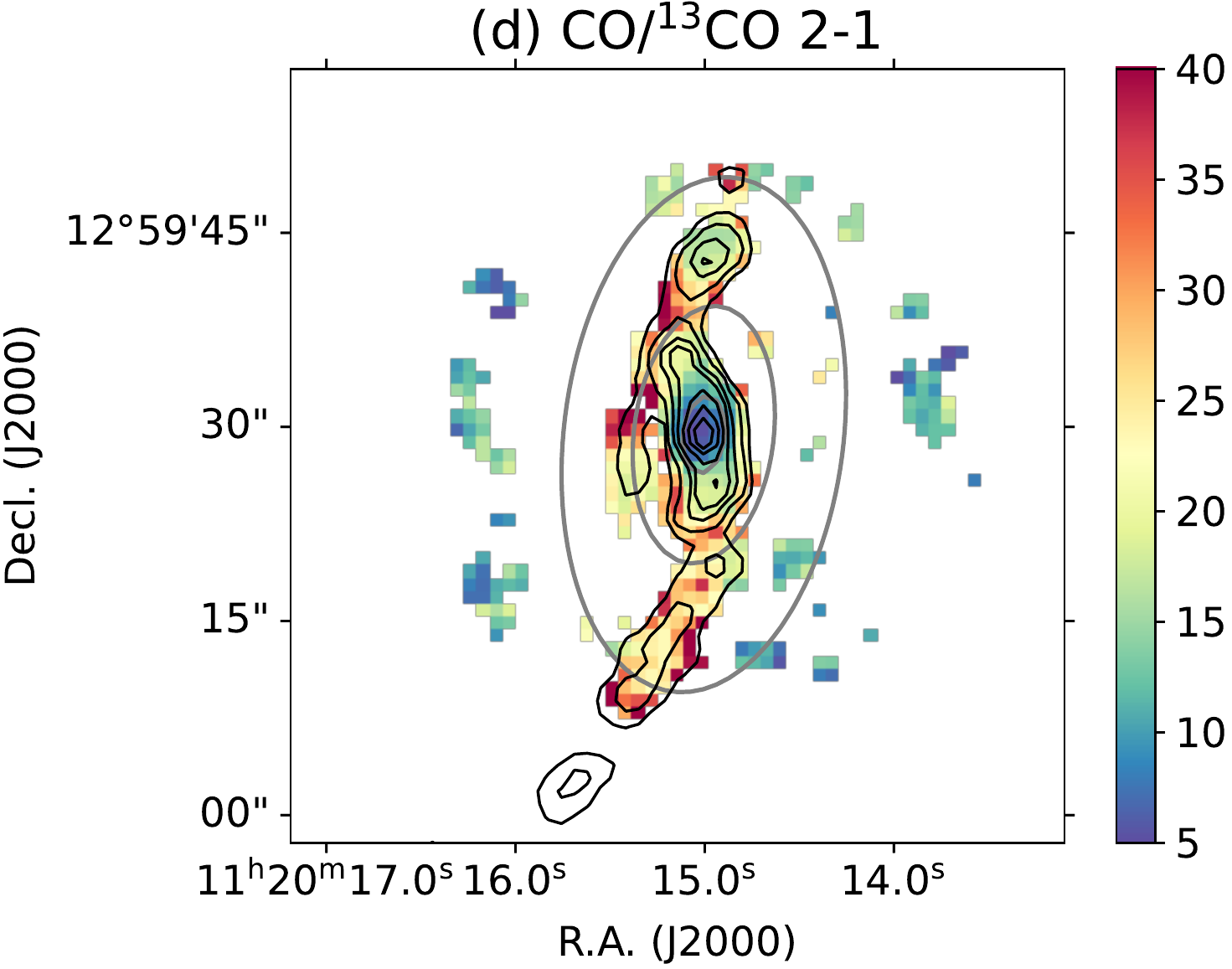} 
\includegraphics[width=0.33\linewidth]{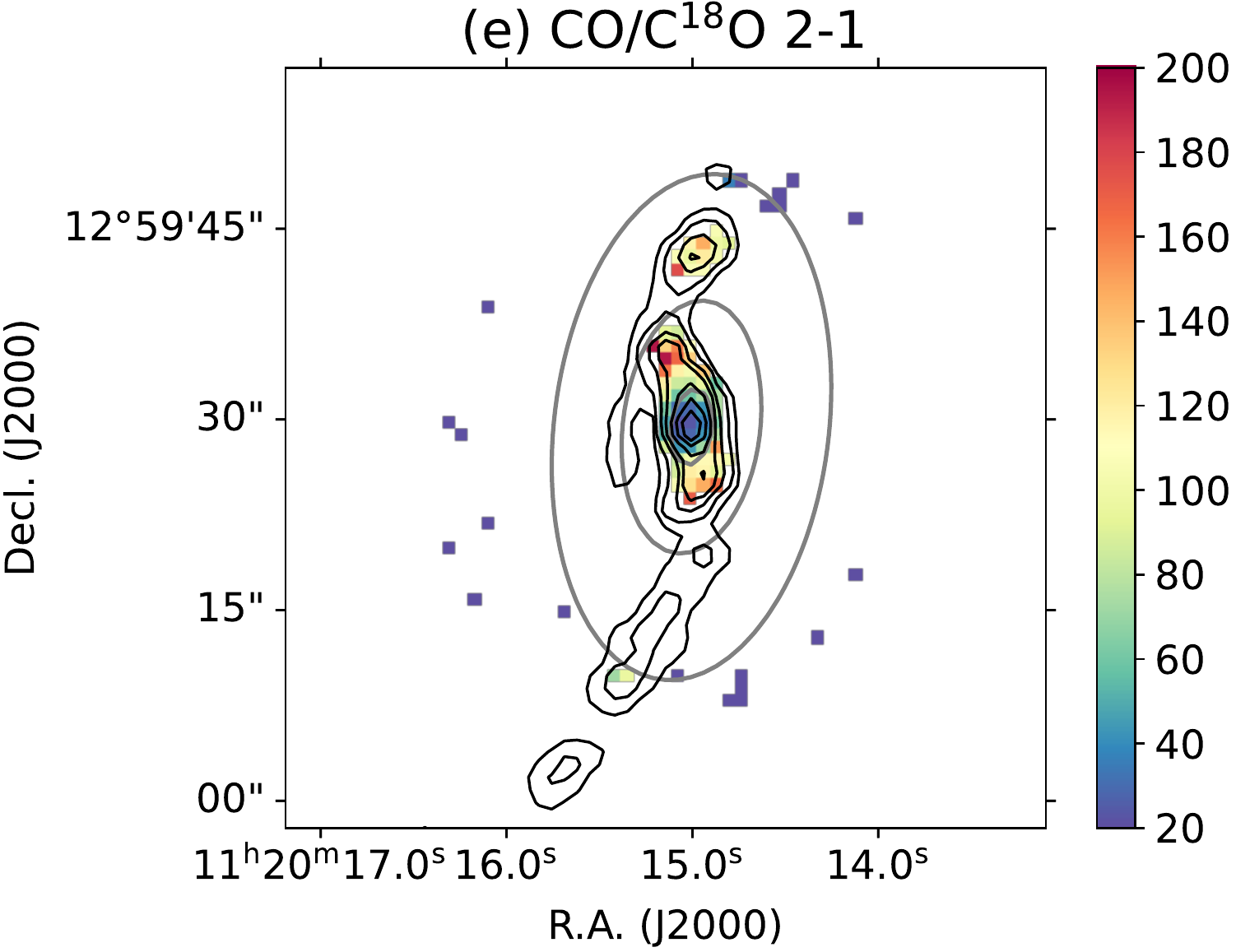} \\ \includegraphics[width=0.33\linewidth]{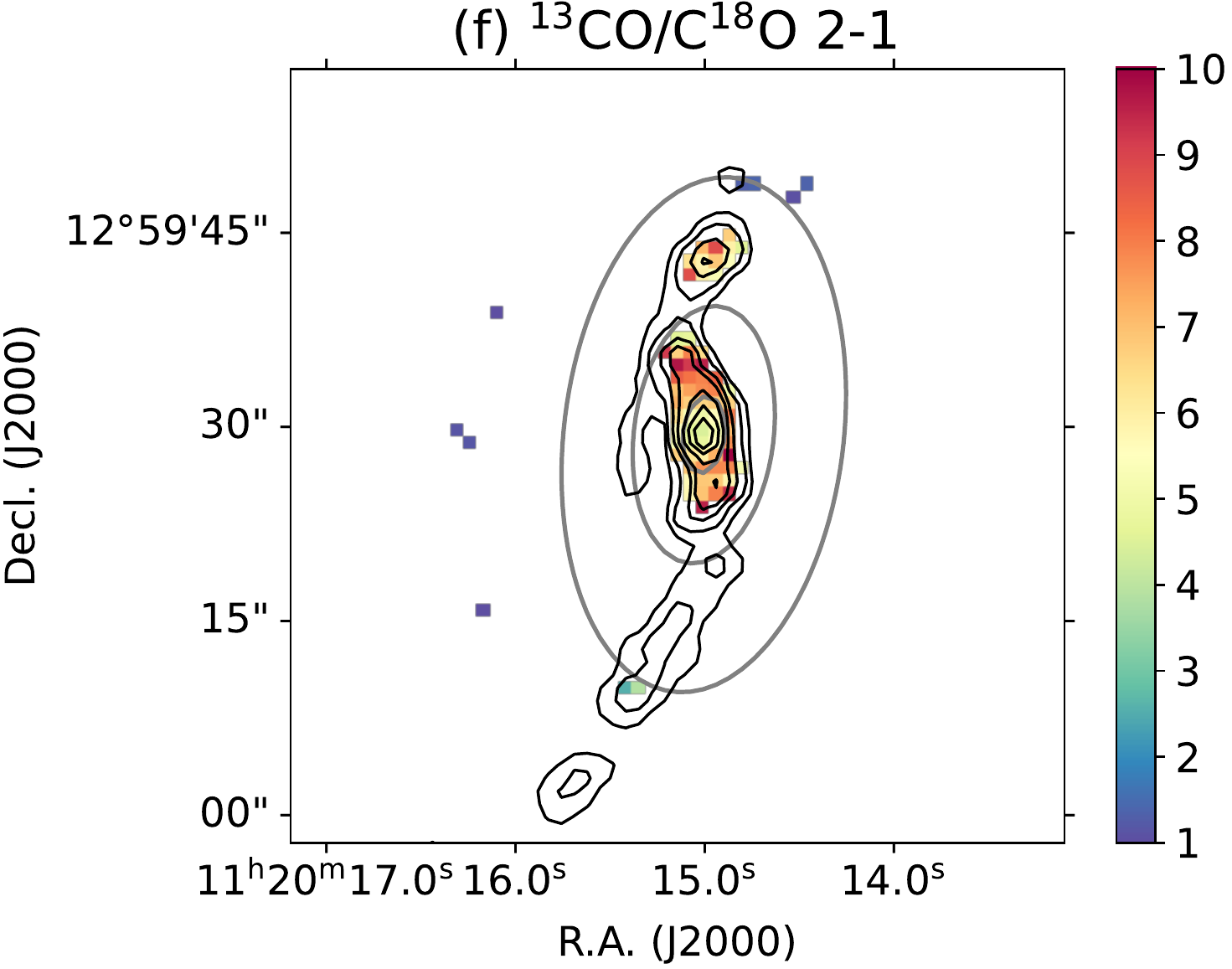} \includegraphics[width=0.33\linewidth]{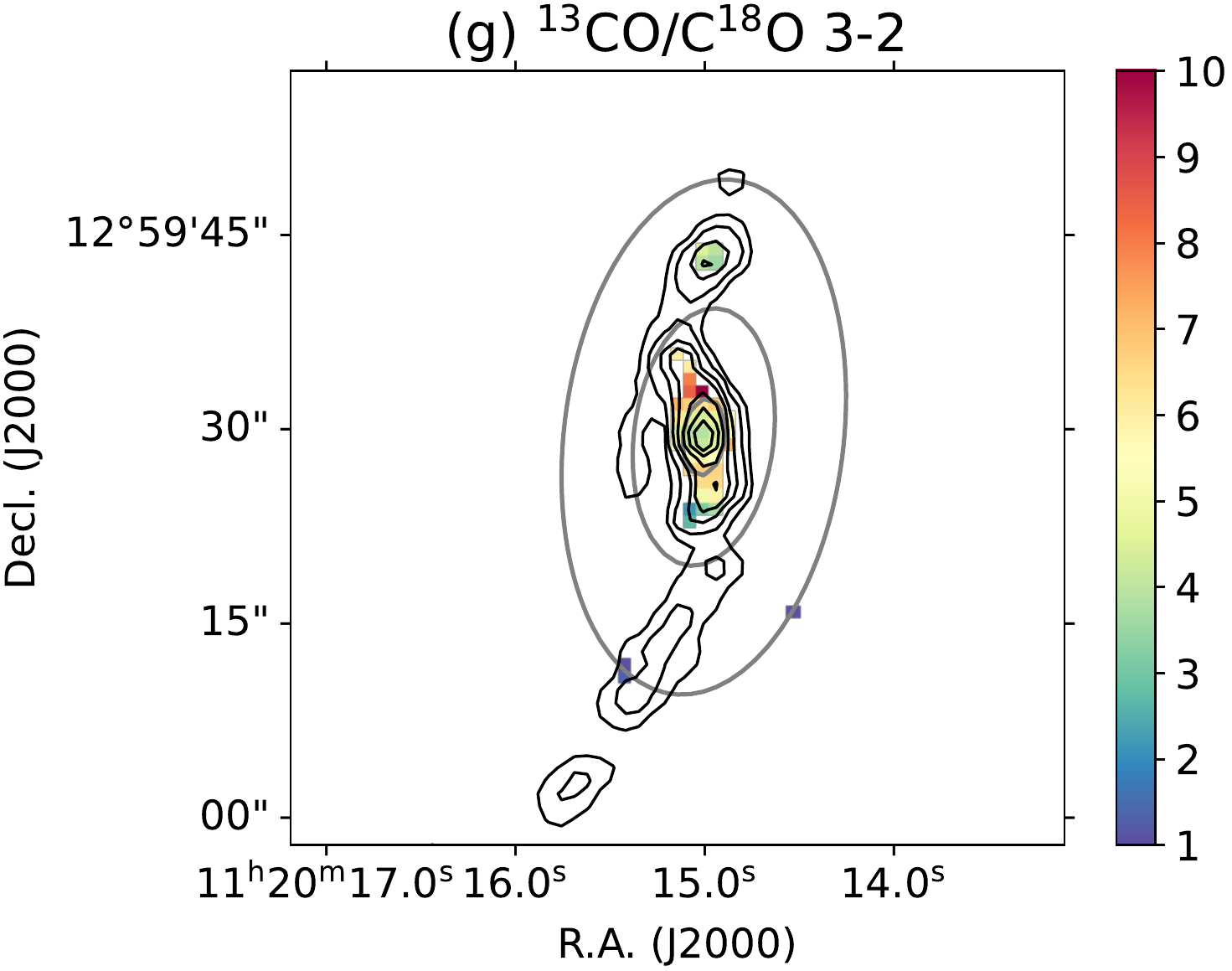}
\caption{Line ratio maps of NGC~3627. Contour levels represent the CO 2--1 integrated intensity at $I_{\rm CO(2-1)} = 50, 100, 200, 300, 500, 700, 900$ K~km~s$^{-1}$. See the caption of Figure~\ref{fig:ratio_4321} for more information.}
\label{fig:ratio_3627}
\end{figure*}

\begin{figure*} 
\begin{minipage}{.335\linewidth}
\centering
\includegraphics[width=\linewidth]{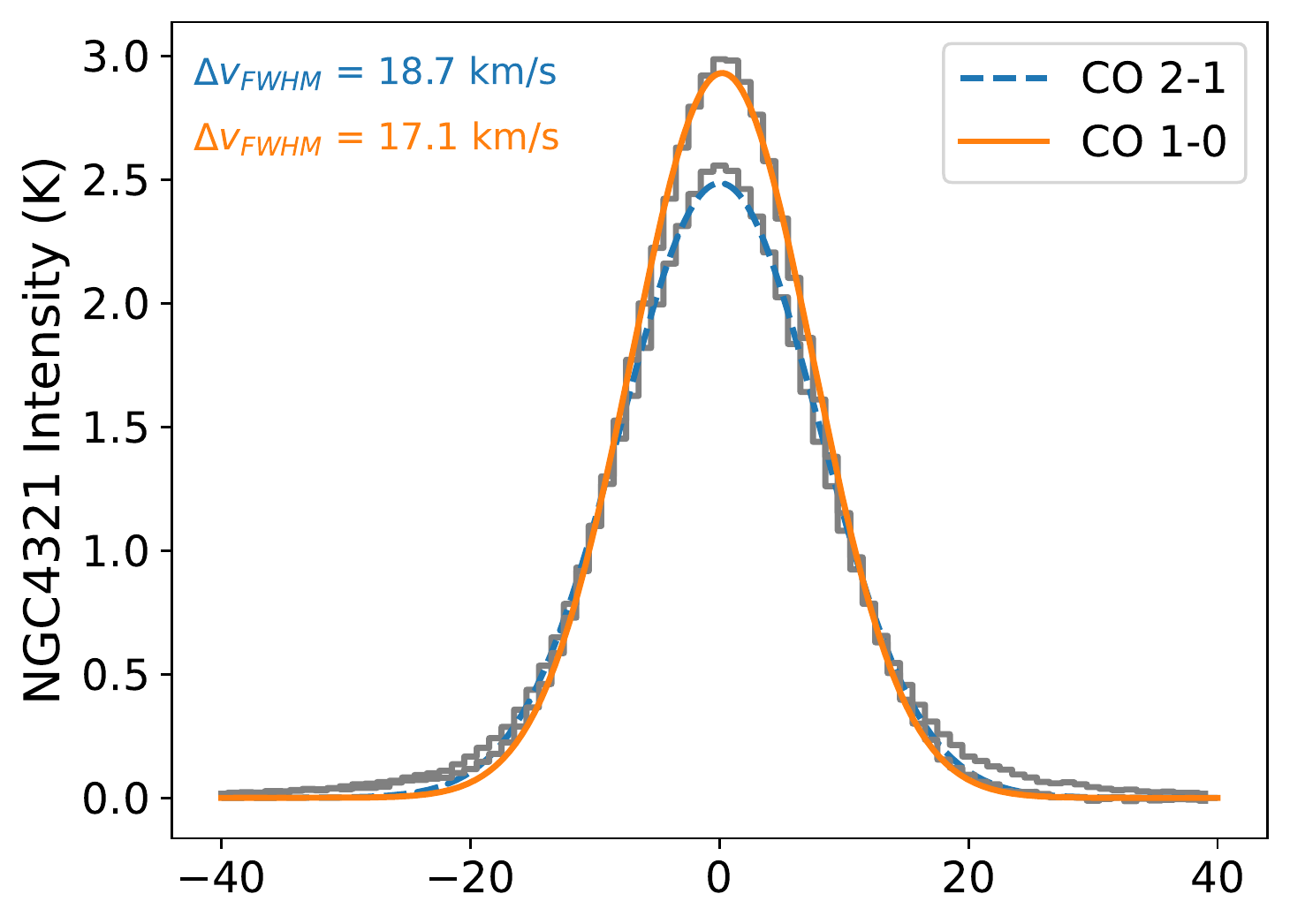}
\end{minipage}
\begin{minipage}{.33\linewidth}
\centering
\includegraphics[width=\linewidth]{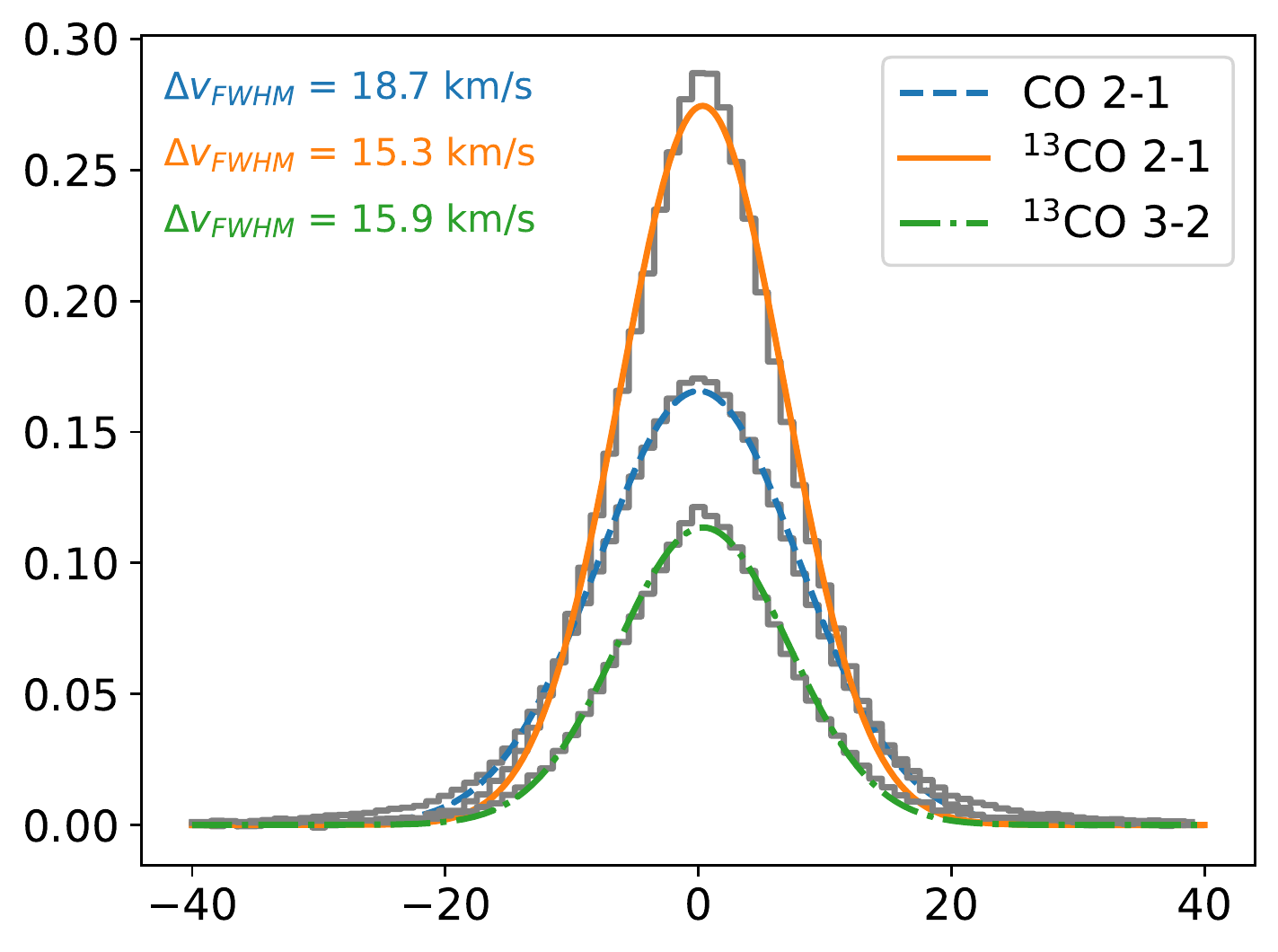}
\end{minipage}
\begin{minipage}{.33\linewidth}
\centering
\includegraphics[width=\linewidth]{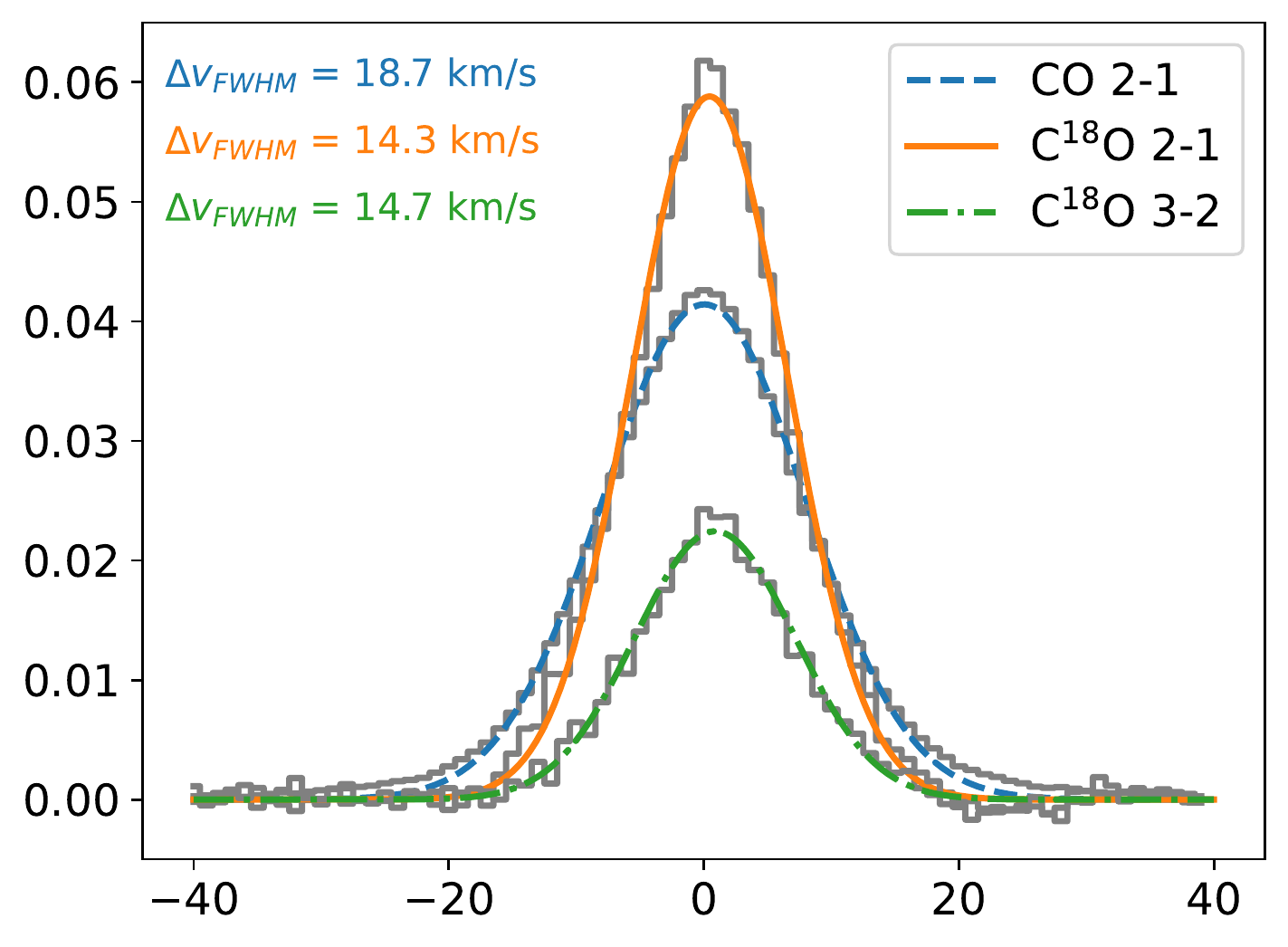}
\end{minipage} \\
\begin{minipage}{.335\linewidth}
\centering
\includegraphics[width=\linewidth]{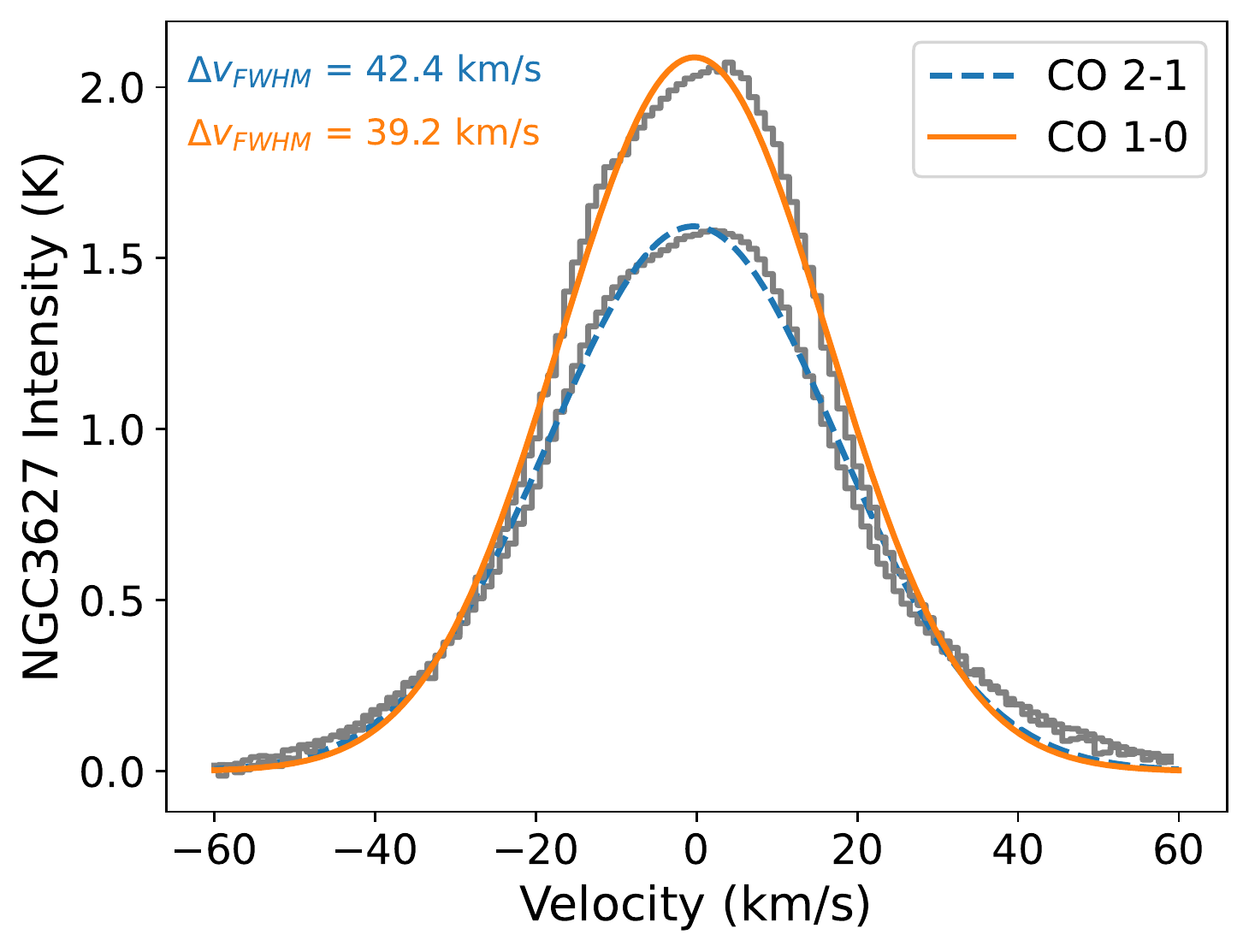} 
\end{minipage}
\begin{minipage}{.33\linewidth}
\centering
\includegraphics[width=\linewidth]{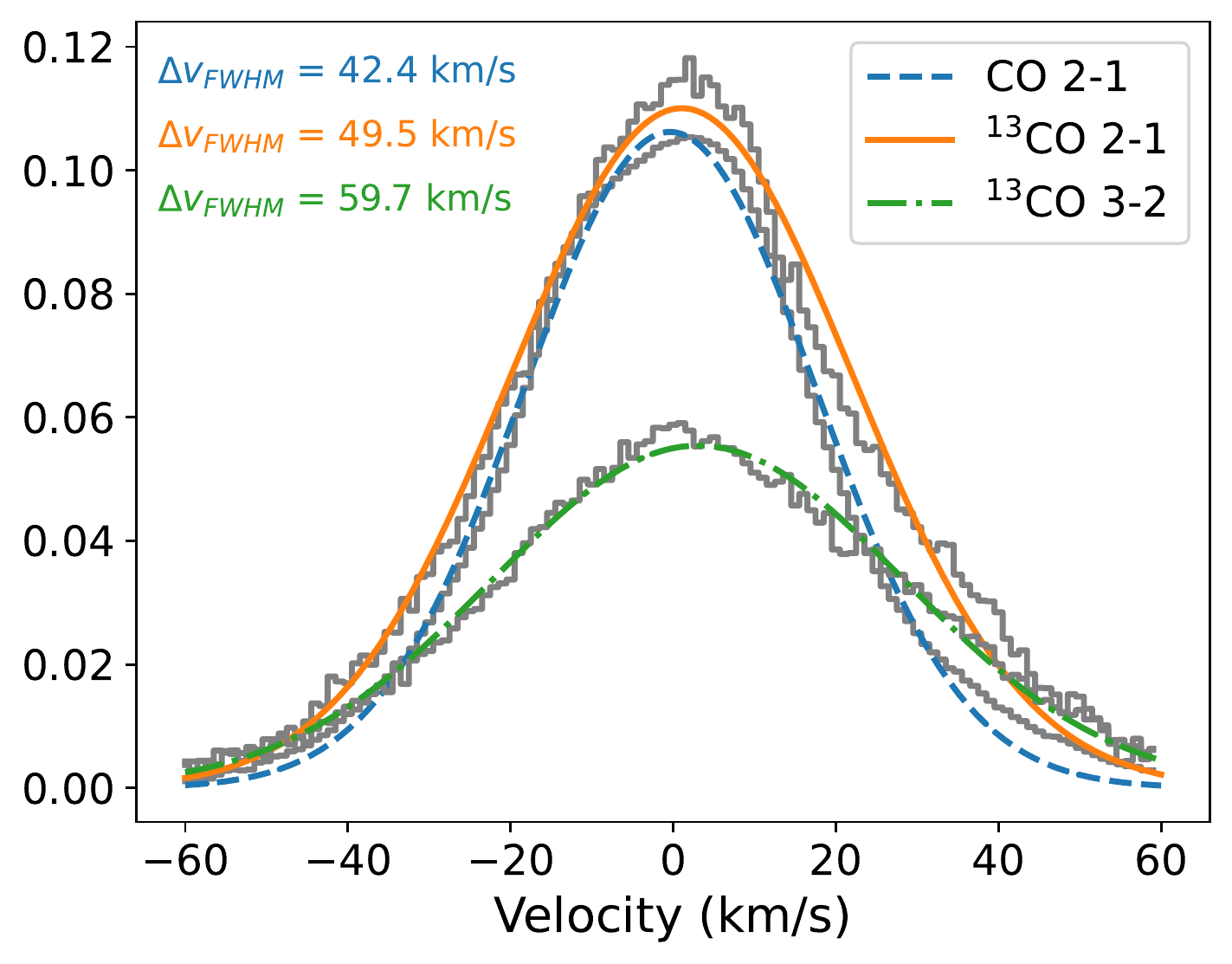} 
\end{minipage}
\begin{minipage}{.34\linewidth}
\centering
\includegraphics[width=\linewidth]{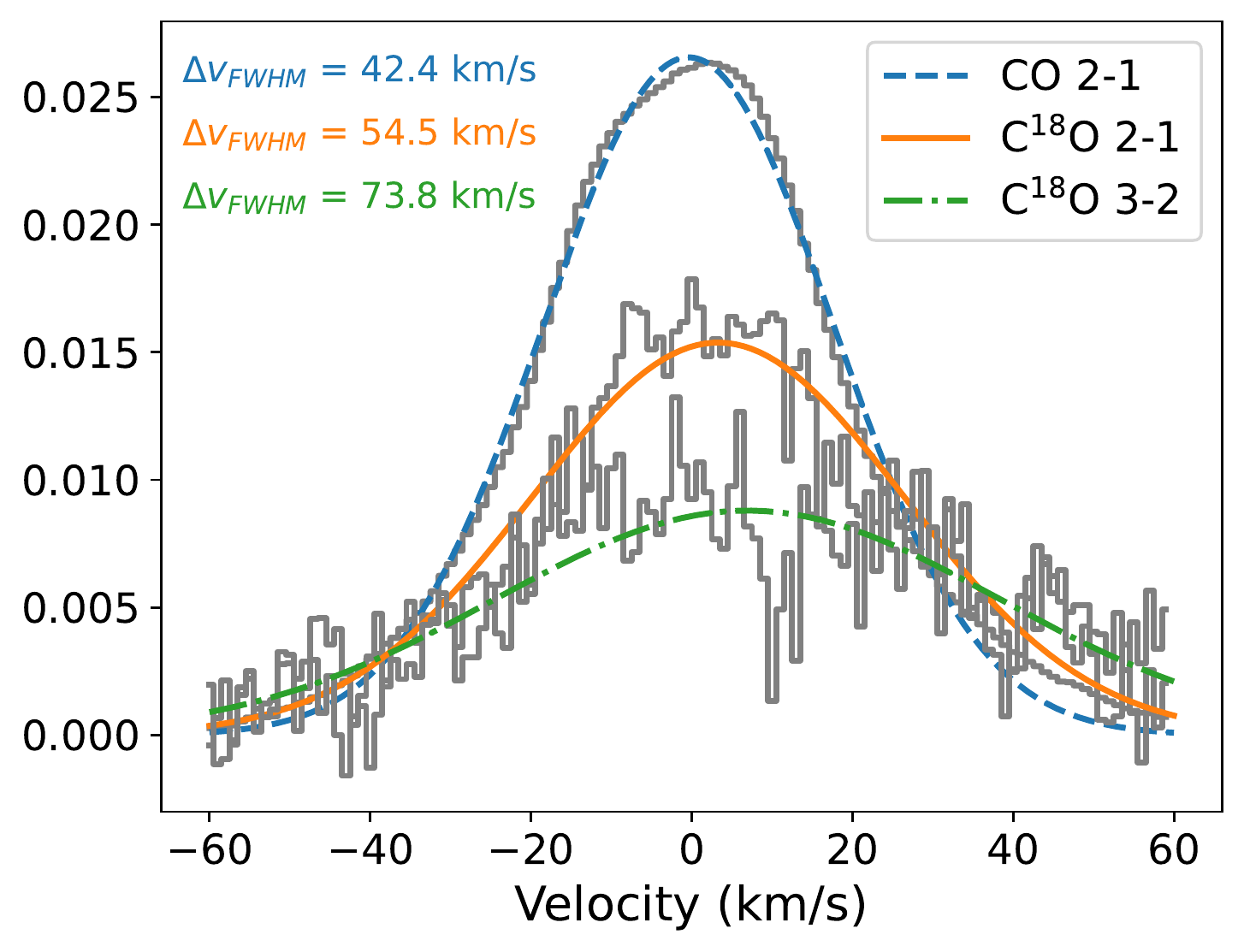} 
\end{minipage}
\caption{Shifted and averaged spectra over the whole kpc regions of NGC~4321 (top row) and NGC~3627 (bottom row), using the moment~1 of CO 2--1 as the fiducial velocity. The intensity of CO 2--1 is scaled down by a factor of 15 (60) in the middle (right) column. The best-fit Gaussian is overlaid, and the upper left corner of each panel lists the fitted FWHM line widths corresponding to the lines in the legend. In both galaxy centers, the overall line width agrees among all six lines within 30--40\%.}
\label{fig:spec_avg}
\end{figure*}

\begin{table*}
%\centering
\begin{center}
\caption{Regional Line Ratios and CO Line Width in NGC~4321. \label{tab:ratio_4321}}
\begin{tabular}{lcccccccccc}
  \hline\hline
  Region & &CO $\frac{2-1}{1-0}$ &$\rm{}^{13}CO$ $\frac{3-2}{2-1}$ &$\rm{C}^{18}O$ $\frac{3-2}{2-1}$ &$\rm \frac{CO}{{}^{13}CO}$ 2--1 &$\rm \frac{CO}{C^{18}O}$ 2--1 &$\rm \frac{{}^{13}CO}{C^{18}O}$ 2--1 &$\rm \frac{{}^{13}CO}{C^{18}O}$ 3--2 &$\Delta v_\mathrm{CO}$ \\
  & & & & & & & & &\,[km~s$^{-1}$] \\
  \hline
  Whole %&Median &0.92  &0.38 &0.46 &16.97 &62.53 &5.46 &5.00\\
  &Mean &0.92 &0.38 &0.46 &16.97 &62.53 &5.46 &5.00 &17.7\\
  &Std.~Dev. &0.22 &0.12 &0.16 &11.02 &34.89 &1.57 &1.44 &6.2\\
  &Integrated Mean &0.92 &0.43 &0.46 &11.51 &49.93 &4.80 &4.69 &---\\
  \hline
  Nucleus %&Median &1.22 &0.48 &0.58 &19.76 &106.68 &5.51 &5.25\\ 
  &Mean &1.22 &0.51 &0.58 &18.97 &101.86 &5.43 &5.25 &34.4\\
  &Std.~Dev. &0.14 &0.15 &0.12 &2.58 &43.60 &1.61 &1.18 &9.1\\
  &Integrated Mean &1.22 &0.61 &0.61 &17.95 &81.38 &4.58 &4.68 &---\\
  \hline
  Inner arms %&Median &1.08 &0.42 &0.44 &10.38 &45.38 &4.91 &5.15\\
  &Mean &0.99 &0.44 &0.44 &9.16 &43.67 &4.82 &5.17 &17.2\\ 
  &Std.~Dev. &0.13 &0.08 &0.12 &2.50 &17.61 &0.94 &1.33 &5.2\\ 
  &Integrated Mean &0.96 &0.46 &0.44 &8.51 &37.72 &4.47 &4.77 &---\\
  \hline
  Outer arms %&Median &0.84 &0.29 &--- &18.59 &90.58 &6.64 &---\\
  &Mean &0.82 &0.29 &--- &18.56 &92.01 &6.68 &--- &16.2\\ 
  &Std.~Dev. &0.14 &0.09 &--- &9.89 &33.60 &1.58 &--- &3.2\\
  &Integrated Mean &0.81 &0.31 &--- &14.60 &80.39 &6.18 &--- &---\\
  \hline
\end{tabular} 
\end{center}
\tablecomments{Line ratios are calculated using the moment~0 maps in units of K~km~$\mathrm{s^{-1}}$. The integrated means are calculated by first averaging the integrated intensities in each region and then dividing them to obtain the ratios, while the mean and standard deviation are for the individual pixels of the map. $\Delta v_\mathrm{CO}$ represents the effective line width$^1$ of CO, not the FWHM. All statistics only take into account the pixels selected for analysis (i.e., S/N $> 3$ in $^{13}$CO, flux recovery rate $> 70\%$, and $I_\mathrm{CO(2-1)} > 50$ K~km~$\mathrm{s^{-1}}$). Due to poor detection of C$^{18}$O 3--2 in the outer arms, the statistics for C$^{18}$O 3--2/2--1 and $^{13}$CO/C$^{18}$O 3--2 in that region are not provided.}
\end{table*} 

\begin{table*}
%\centering
\begin{center}
\caption{Regional Line Ratios and CO Line Width in NGC~3627. \label{tab:ratio_3627}}
\begin{tabular}{lcccccccccc}
  \hline\hline
  Region & &CO $\frac{2-1}{1-0}$ &$\rm{}^{13}CO$ $\frac{3-2}{2-1}$ &$\rm{C}^{18}O$ $\frac{3-2}{2-1}$ &$\rm \frac{CO}{{}^{13}CO}$ 2--1 &$\rm \frac{CO}{C^{18}O}$ 2--1 &$\rm \frac{{}^{13}CO}{C^{18}O}$ 2--1 &$\rm \frac{{}^{13}CO}{C^{18}O}$ 3--2 &$\Delta v_\mathrm{CO}$ \\
  & & & & & & & & &\,[km~s$^{-1}$] \\
  \hline
  Whole &Mean &0.75 &0.46 &0.73 &19.88 &87.84 &6.13 &5.16 &34.1\\
  &Std.~Dev. &0.18 &0.19 &0.19 &9.57 &46.56 &2.28 &1.78 &12.3\\
  &Integrated Mean &0.81 &0.63 &0.85 &12.55 &55.51 &5.67 &4.60 &---\\
  \hline
  Nucleus &Mean &0.93 &0.74 &0.86 &7.51 &44.75 &5.69 &4.94 &60.9\\ 
  &Std.~Dev. &0.03 &0.08 &0.12 &2.22 &21.41 &0.97 &0.94 &8.2\\
  &Integrated Mean &0.93 &0.78 &0.91 &6.65 &34.29 &5.16 &4.40 &---\\
  \hline
  Inner arms &Mean &0.85 &0.43 &0.65 &21.06 &109.64 &7.29 &5.86 &38.9\\
  &Std.~Dev. &0.18 &0.13 &0.19 &8.63 &37.60 &1.52 &1.82 &10.4\\ 
  &Integrated Mean &0.84 &0.49 &0.64 &15.78 &98.93 &7.12 &5.89 &---\\
  \hline
  Outer arms &Mean &0.72 &0.35 &--- &27.20 &--- &--- &--- &26.2\\
  &Std.~Dev. &0.09 &0.15 &--- &8.43 &--- &--- &--- &4.5\\
  &Integrated Mean &0.70 &0.32 &--- &23.15 &--- &--- &--- &---\\
  \hline
\end{tabular} 
\end{center}
\tablecomments{Due to poor detection of both C$^{18}$O lines in the outer arms, the statistics for C$^{18}$O 3--2/2--1, CO/C$^{18}$O 2--1 and $^{13}$CO/C$^{18}$O 2--1 and 3--2 in that region are not provided. See Table~\ref{tab:ratio_4321} notes for more information.}
\end{table*}

Figure~\ref{fig:ratio_4321} and~\ref{fig:ratio_3627} show the line ratio maps of CO 2--1/1--0, $^{13}$CO 3--2/2--1, C$^{18}$O 3--2/2--1, CO/$^{13}$CO 2--1, CO/C$^{18}$O 2--1, $^{13}$CO/C$^{18}$O 2--1, and $^{13}$CO/C$^{18}$O 3--2, which are generated from the moment 0 maps in units of K~km~$\mathrm{s^{-1}}$. These line ratio maps reveal clear variation among our defined regions. All the same-species ratios, which are primarily sensitive to temperature (panels (a)--(c) in Figures~\ref{fig:ratio_4321} and~\ref{fig:ratio_3627}), show clear enhancement in the nucleus of NGC~4321 and 3627, suggesting warmer and/or denser gas toward both galactic nuclei. However, the two galaxies show different trends in the same-transition ratios which are mostly sensitive to abundance/opacity. In NGC~4321, the CO/$^{13}$CO and CO/C$^{18}$O ratios in the inner arms are ${\sim}2$ times lower than in the nucleus, while the $^{13}$CO/C$^{18}$O line ratios are similar between the arms and the nucleus. On the other hand, NGC~3627 shows that all four abundance/opacity sensitive ratios are lower in the nucleus than in the rest of the regions. This likely indicates different variations in optical depths and/or CO isotopologue abundances in these galaxy centers, which will be addressed via our modeling (Section~\ref{sec:model}).

The regional statistics of the observed line ratios are listed in Table~\ref{tab:ratio_4321} and~\ref{tab:ratio_3627}. The means and standard deviations are calculated from the ensemble of pixel-by-pixel measurements in the relevant region, while the integrated means are determined by dividing between the regionally integrated intensities. Since the C$^{18}$O line(s) are not commonly detected in the outer arms, the relevant line ratios for those regions are not listed. The CO 2--1/1--0 ratio averaged over the whole field of view is 0.9 for NGC~4321 and 0.8 for NGC~3627, and it is even higher in their inner 300~pc nuclei. This is consistent with recent line ratio studies at kpc resolution \citep{2021MNRAS.504.3221D,2021PASJ...73..257Y,2022ApJ...927..149L}. 

In Figure~\ref{fig:spec_avg}, we present the averaged spectra over the entire region defined in Figure~\ref{fig:def_regions} for all six lines. The spectra are obtained by applying the stacking technique \citep{stacking} and using the CO 2--1 moment~1 (see maps in Appendix~\ref{sec:vel_maps}) as the fiducial velocity centroid. Except for the poorly detected C$^{18}$O 3--2 line in NGC~3627, the averaged spectra of all the lines in each galaxy show similar line widths within 30--40\%. This means that the velocity dispersion among different observed lines are overall in good agreement, and we also see agreement on pixel-by-pixel scales via a thorough check of each individual line of sight (e.g., Appendix~\ref{sec:multi_comp}). Comparing to NGC~4321, we notice that the averaged spectra for NGC~3627 show some level of discrepancy with the best-fit Gaussian function. The discrepancy is possibly due to a larger fraction of area in NGC~3627 having multi-component gas along the same lines of sight, and this will be further discussed in Appendix~\ref{sec:multi_comp}.
The regional statistics of the CO 2--1 effective line widths are also listed in Table~\ref{tab:ratio_4321} and~\ref{tab:ratio_3627}, and their maps can be found in Appendix~\ref{sec:vel_maps}. The line widths in CO 1--0 are consistent with CO 2--1 within $10\%$.

To investigate the physical implications of these line ratio and line width variations, we determine the gas physical conditions pixel by pixel in these galaxy centers in Section~\ref{sec:model}, using multi-line radiative transfer modeling without assuming local thermodynamic equilibrium (LTE).

\section{Multi-line Bayesian Modeling} \label{sec:model}

\subsection{Modeling Setup} \label{subsec:model_setup}

\begin{table}
%\centering
\caption{RADEX Input Parameters \label{tab:radex}}
\begin{tabular}{lcc}
  \hline\hline
  Parameter & Range & Step Size \\
  \hline
  $\log(n_\mathrm{H_2}\,[\mathrm{cm}^{-3}])$ & $2.0{-}5.0$ & $0.2$~dex \\
  $\log(T_\mathrm{k}\,[\mathrm{K}])$ & $1.0{-}2.7$ & $0.1$~dex \\
  $\log(N_\mathrm{CO}\,[\mathrm{cm}^{-2}])$ & $15.0{-}20.0$ & $0.2$~dex \\
  $X_{12/13}$ & $10{-}200$ & $10$ \\
  $X_{13/18}$ & $2{-}20$ & $1$ \\
  $\log(\Phi_\mathrm{bf})$ & $-1.3{-}0$ & $0.1$~dex \\
  $\Delta v\,[\mathrm{km~s}^{-1}]$ & $15.0$ & --- \\
  \hline
\end{tabular} 
\tablecomments{The fixed $\Delta v$ of 15 km~s$^{-1}$ is only a fiducial value for the model grid. The parameter of interest is $N_\mathrm{CO}/\Delta v$.}
\end{table}

To constrain the physical conditions and $\alpha_\mathrm{CO}$ in different sub-regions of the galaxy centers, we run a non-LTE radiative transfer code, RADEX \citep{radex}, to construct a one-component model and fit it with our observations at $\sim$100~pc scales. 
RADEX assumes a homogeneous medium and uses radiative transfer equations based on the escape probability formalism to find a converged solution for the excitation temperature and level population. On a pixel-by-pixel basis, we model the integrated intensities of the six CO, $^{13}$CO, and C$^{18}$O lines under various combinations of $\mathrm{H_2}$ volume density ($n_\mathrm{H_2}$), kinetic temperature ($T_\mathrm{k}$), CO column density per line width ($N_\mathrm{CO}/\Delta v$), CO/$^{13}$CO ($X_{12/13}$) and $^{13}$CO/C$^{18}$O ($X_{13/18}$) abundance ratios, and the beam-filling factor ($\Phi_\mathrm{bf}$). 

This model assumes the same beam-filling factor for all six observed lines. We note that earlier studies on some barred galaxy centers found high CO/$^{13}$CO line ratios in bar regions, which may be explained by the existence of diffuse molecular components that lead to differences in the beam-filling factor of CO and $^{13}$CO lines \citep{2000A&A...363...93H,2011MNRAS.411.1409W}. However, those studies worked at near-kpc resolutions, and the near-GMC resolution used in this work should reduce the possible beam-filling factor mismatch. While $\Phi_\mathrm{bf}$ could still be lower for emission from higher transitions or less abundant isotopologues, investigating how much $\Phi_\mathrm{bf}$ differs between the lines in each region requires more sophisticated modeling or simulation that includes $\Phi_\mathrm{bf}$ as a variable. We briefly describe the modeling setup below and note that the modeling approach is the same as that adopted in \citet{2022ApJ...925...72T}, where readers can find more details about our RADEX implementation and model construction. We also release all the source code and parameters in a GitHub repository\footnote{\url{https://github.com/ElthaTeng/multiline-bayesian-modeling}}. 

We build a six-dimensional RADEX model grid with $\log(n_\mathrm{H_2}\,[\mathrm{cm}^{-3}])$ varied from $2$ to $5$ in steps of $0.2$~dex, $T_\mathrm{k}$ from $10$ to $500$~K in steps of $0.1$~dex, $N_\mathrm{CO}/\Delta v$ from $10^{15}/15$ to $10^{20}/15$~$\mathrm{cm^{-2}}\ \mathrm{(km~s^{-1})^{-1}}$ in steps of $0.2$~dex, $X_{12/13}$ from $10$ to $200$ in steps of~$10$, $X_{13/18}$ from $2$ to $20$ in steps of~$1$, and $\log(\Phi_\mathrm{bf})$ from $-1.3$ to $0$ in steps of~$0.1$~dex (see Table~\ref{tab:radex}). 
While the $N_\mathrm{CO}$ and $\Delta v = 15\mathrm{km\ s^{-1}}$ listed in Table~\ref{tab:radex} are input separately to RADEX\footnote{In RADEX calculation, the line width should be input in FWHM, and thus we converted our effective line width $\Delta v$ to FWHM by a factor of 2.35 in this step.}, it is important to note that the radiative transfer calculation in RADEX depends only on their ratio $N_\mathrm{CO}/\Delta v$ \citep[][see also \citealt{2012ApJ...753...70K,2020ApJ...893...63T,2022ApJ...925...72T}]{radex}. This means that we are essentially fitting $N_\mathrm{CO}/\Delta v$, and thus variation of $\Delta v$ across the observed regions would not affect our results as long as we ensure that $N_\mathrm{CO}/\Delta v$ is unchanged when we derive $N_\mathrm{CO}$ using the observed $\Delta v$.
We set the upper limit of $T_\mathrm{k}$ to $\sim$500~K due to low reliability to distinguish a higher $T_\mathrm{k}$ with $J$=3--2 as the highest transition in our setting. The parameter ranges were determined by ensuring well-covered probability density functions (PDFs) in representative nucleus and arm regions. We note that RADEX fails to converge at several grid points where $T_\mathrm{k} > 200$~K and $N_\mathrm{CO}/\Delta v \ge 10^{19}/15$~$\mathrm{cm^{-2}}\ \mathrm{(km~s^{-1})^{-1}}$, and thus we exclude those solutions in our modeling. We will show that such conditions tend to result in unreasonably large line-of-sight path length which will also be excluded by our line-of-sight prior, so the lack of these models does not impact our analysis. 

Following \citet{2022ApJ...925...72T}, we study pixel by pixel the marginalized PDFs of each parameter using a Bayesian likelihood analysis. With the marginalized 1D PDFs, we will determine the peak parameter values as the ``1DMax'' solutions and the 50th percentile values as the ``median'' solutions. The ``best-fit'' solution which corresponds to the global minimum $\chi^2$ value of the full 6D grid is also derived. In contrast to a single best-fit solution representing the gas physical properties, the PDFs are descriptive of the local variations over the full parameter space, and the 1DMax/median solutions from the PDFs reflect a more complete characterization of the parameter distributions. Therefore, we will focus on the 1DMax/median solutions throughout our analysis, while we also show that 1DMax, median, and best-fit solutions agree well in many cases. 
In our $\chi^2$ calculation, we include the measurement uncertainty and an estimated flux calibration uncertainty of 10\% for Band~3 and 20\% for Band~6 or~7, respectively \citep{2017ApJ...840....8S,2019MNRAS.485.1188B}. 
For regions with C$^{18}$O detection $< 3\sigma$ (e.g., outer arms), we still include the C$^{18}$O intensity with its (higher) associated uncertainty in our modeling.
However, those lines are excluded from our fitting if the pixel has negative C$^{18}$O integrated intensity below $1\sigma$. Thus, the solutions for some pixels can be constrained by less than six lines, although this situation only occurs in the outer arms of NGC~3627.        

To avoid solutions that result in unrealistically large line-of-sight path lengths ($\ell_\mathrm{los}$), we also set a prior by requiring
\begin{equation}
\ell_\mathrm{los} = N_\mathrm{CO} (\sqrt{\Phi_\mathrm{bf}}\ n_\mathrm{H_2}\,x_\mathrm{CO})^{-1} < 200\ \mathrm{pc}.
\label{eqn_los}
\end{equation}
where $x_\mathrm{CO}$ is the CO/H$_2$ abundance ratio that is normally found or adopted as $3\times10^{-4}$ in active star-forming regions \citep{1994ApJ...428L..69L,2003ApJ...587..171W,2014ApJ...796L..15S}.
This $200$~pc constraint considers the typical molecular gas scale height of ${\sim}100$~pc for our Galaxy and nearby disk galaxies \citep{2014AJ....148..127Y,2015ARA&A..53..583H}, as well as a tolerance of a factor-of-two increase due to galaxy inclination. Since $n_\mathrm{H_2}$ and $\Phi_\mathrm{bf}$ are both our modeled parameters, and $N_\mathrm{CO}$ can be obtained by multiplying the modeled $N_\mathrm{CO}/\Delta v$ with the observed line width, this line-of-sight prior can be easily implemented by excluding all the grid points (i.e., parameter combinations) that give $\ell_\mathrm{los} > 200$~pc and setting their probability to zero. As shown by Equation~\ref{eqn_los}, the prior tends to rule out solutions with high column densities of $> 10^{19}/15$~$\mathrm{cm^{-2}}\ \mathrm{(km~s^{-1})^{-1}}$ and low volume densities of $< 300$~$\mathrm{cm^{-3}}$. This also means that most of the conditions where RADEX fails to converge are excluded by the line-of-sight constraint due to high CO column densities.

\subsection{Molecular Gas Physical Conditions} \label{subsec:model_result}

\begin{figure*}
\begin{minipage}{.5\linewidth}
\centering
\includegraphics[width=\linewidth]{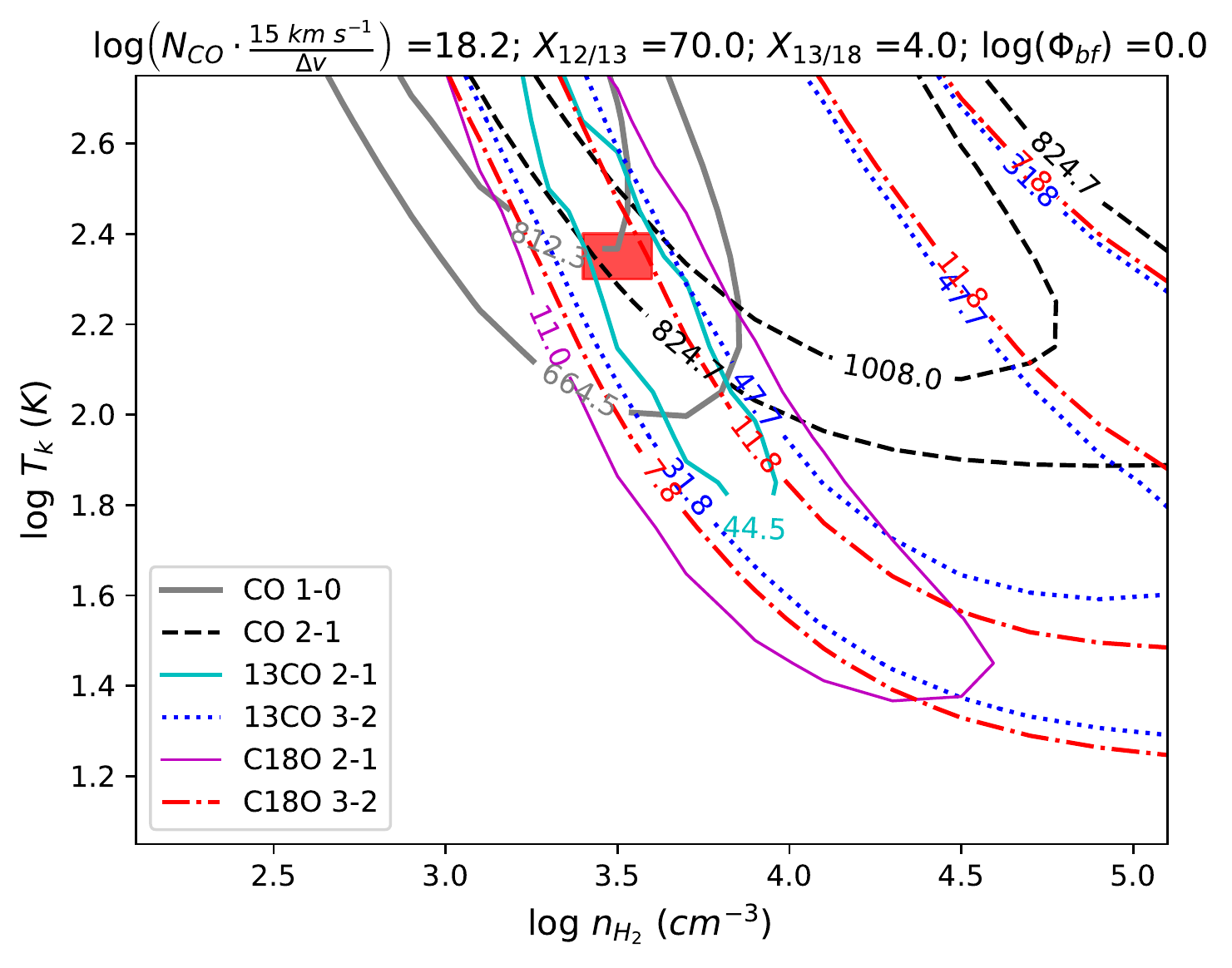}\\
(a)
\end{minipage}
%\hfill
\begin{minipage}{.5\linewidth}
\centering
\includegraphics[width=\linewidth]{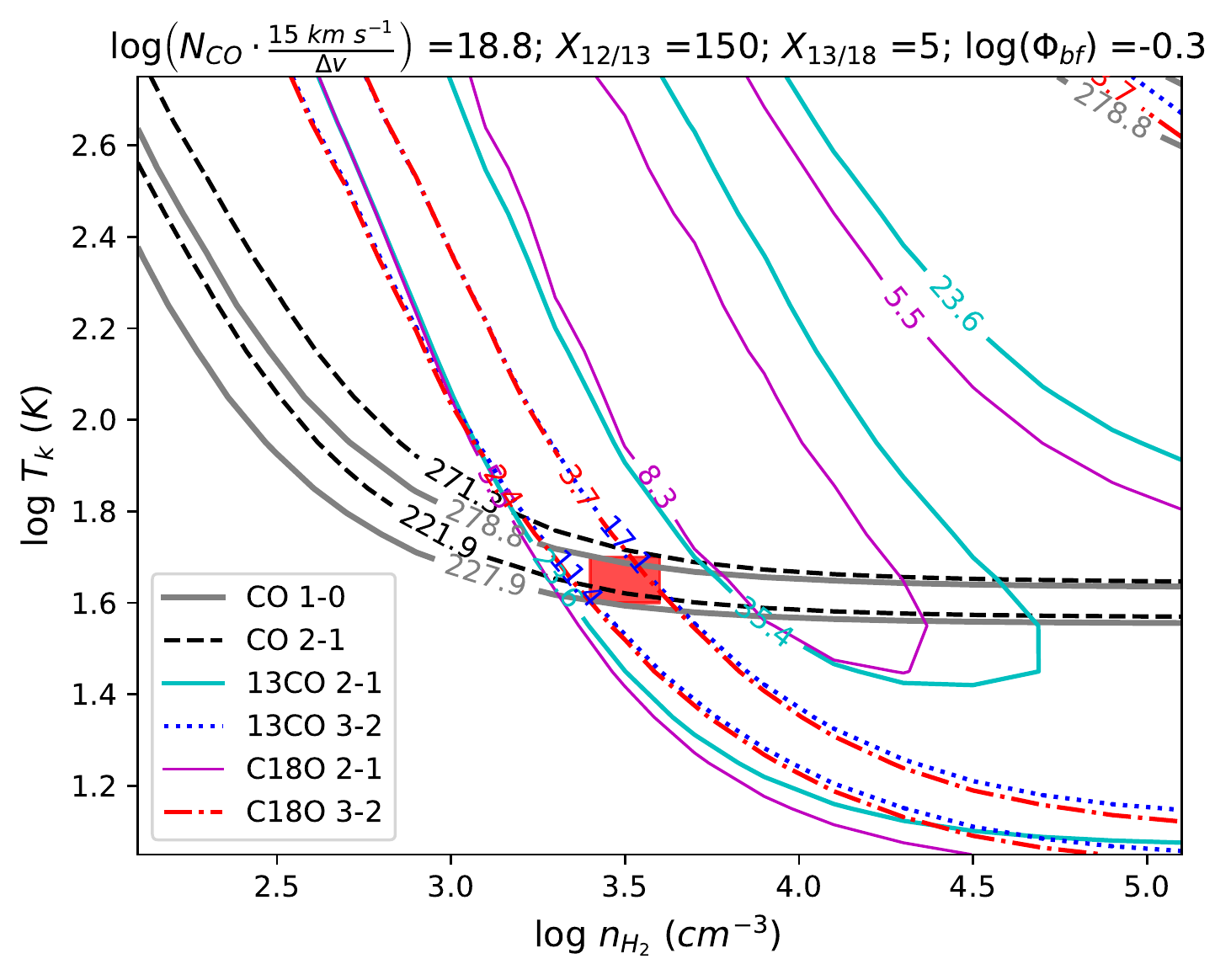}\\
(b)
\end{minipage}
\caption{Best-fit (i.e. lowest $\chi^2$) constraints from the six observed line fluxes at (a) the central pixel and (b) a pixel in the northern, inner arm of NGC~4321. Contours show the ranges of observed line intensities $\pm1\sigma$ uncertainties, including the measurement and calibration uncertainties. Red boxes represent the best-fit solutions. Note that these are the solutions with the lowest $\chi^2$ value in the full grid, not the 1DMax solutions based on the marginalized PDFs, and thus the $X_{12/13}$ values here may deviate from the lower $X_{12/13}$ suggested by 1DMax solutions. Except for $X_{12/13}$, other parameters are similar to the 1DMax solutions as their 1D likelihoods are single-peaked and well-constrained (see Figure~\ref{fig:corner_center_4321} and related discussion).}
\label{fig:flux_contour_4321}
\end{figure*}

\begin{figure*}
\centering
\includegraphics[width=\linewidth]{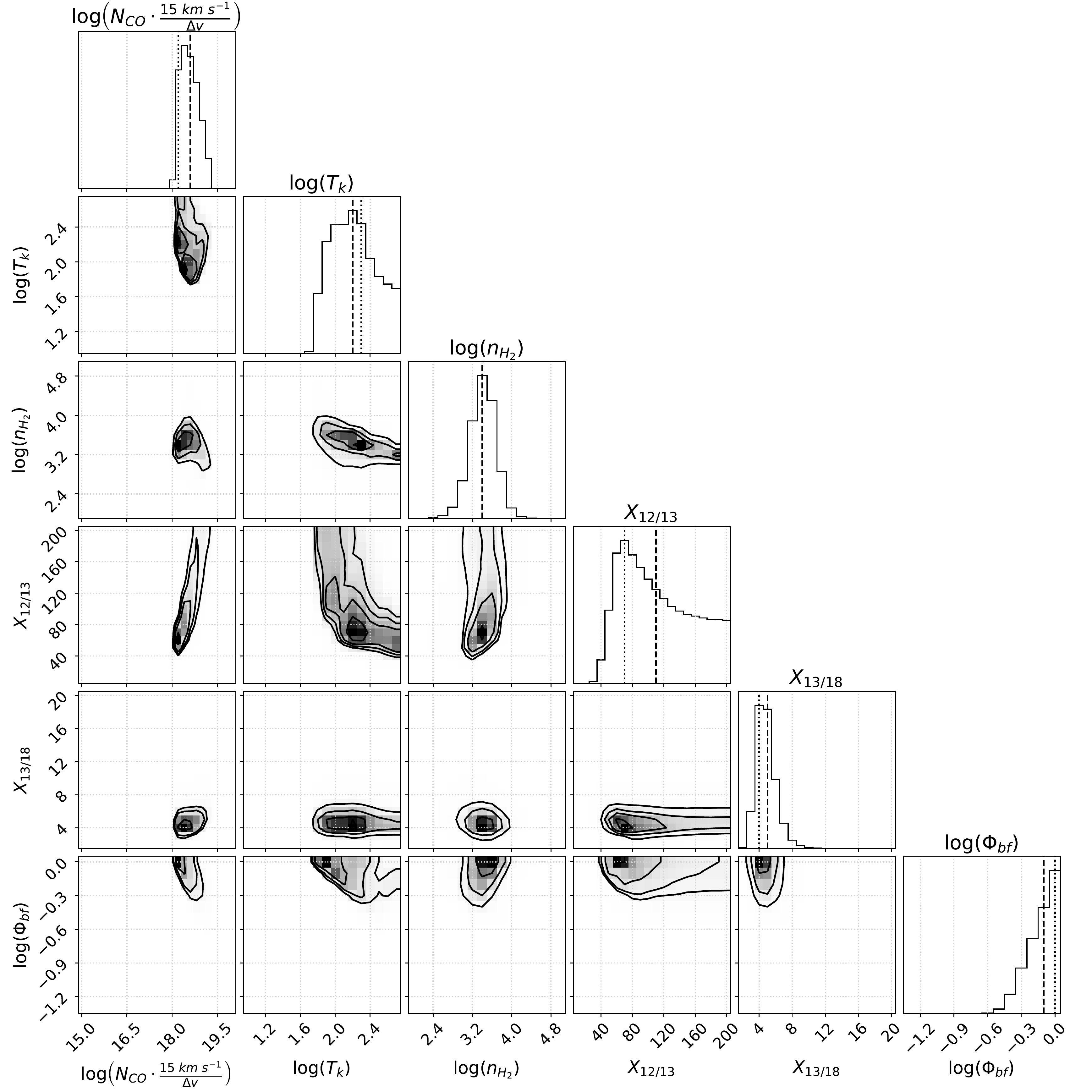}
\caption{Marginalized 1D and 2D probability distributions of the central pixel of NGC~4321. In the panels of 1D PDFs (on the diagonal), the dashed lines represent the 50th percentile (i.e., median) values of the cumulative 1D PDFs, and the dotted lines label the best-fit solution as in Figure~\ref{fig:flux_contour_4321}(a). Except for the PDF of $X_{12/13}$ which is less constrained, the median values of other parameters are closely aligned with the best-fit and 1DMax solutions.}
\label{fig:corner_center_4321}
\end{figure*}

By fitting the line observations with the constructed RADEX models at each pixel, we obtain well-constrained solutions for most of the physical parameters. Figure~\ref{fig:flux_contour_4321} demonstrates how the best-fit solutions are constrained by the six observed line intensities ($\pm 1\sigma$, including measurement and calibration uncertainties) at the central pixel and another pixel in the inner arm region of NGC~4321. In general, we find the best-fit solution of each pixel well within the constraints given by all the observed lines, although the number or species of lines that give crucial constraints varies from pixel to pixel (see Section~\ref{subsec:line_constraint} for further discussion). 

Overall, we find the marginalized PDFs for each parameter to be single-peaked and well-covered by the parameter space. However, we notice that the PDFs of $X_{12/13}$ tend to be broader than other parameters, implying that the $X_{12/13}$ abundance ratio is generally less constrained by the model based on the observed lines. This was also seen in similar modeling toward other galaxy centers or (U)LIRGs \citep{2014ApJ...796L..15S,2017ApJ...840....8S,2022ApJ...925...72T}.
In Figure~\ref{fig:corner_center_4321}, we show the marginalized 1D and 2D PDFs for the central pixel of NGC~4321. The vertical dashed lines on the 1D PDFs represent the 50th percentile values (median), which generally agree with the 1D PDF peaks (1DMax) as well as the best-fit solutions shown in Figure~\ref{fig:flux_contour_4321}(a). In this pixel, the 1DMax solution of $X_{12/13}$ matches the best-fit solution, but it is inconsistent with the median due to the broader and asymmetric PDF of $X_{12/13}$. More examples of the PDFs and/or best-fit solutions for other pixels in NGC~4321 and NGC~3627 are presented in Appendix~\ref{sec:pixel_solutions}. Over the entire observed regions, we find that the best-fit and 1DMax solutions of $N_\mathrm{CO}/\Delta v$, $n_\mathrm{H_2}$, $T_\mathrm{k}$, $X_{13/18}$ and $\Phi_\mathrm{bf}$ are mostly consistent, while the 1DMax $X_{12/13}$ can deviate from the best-fit or median solutions in some regions due to less constrained PDFs.

\begin{figure*} \centering
\includegraphics[width=\linewidth]{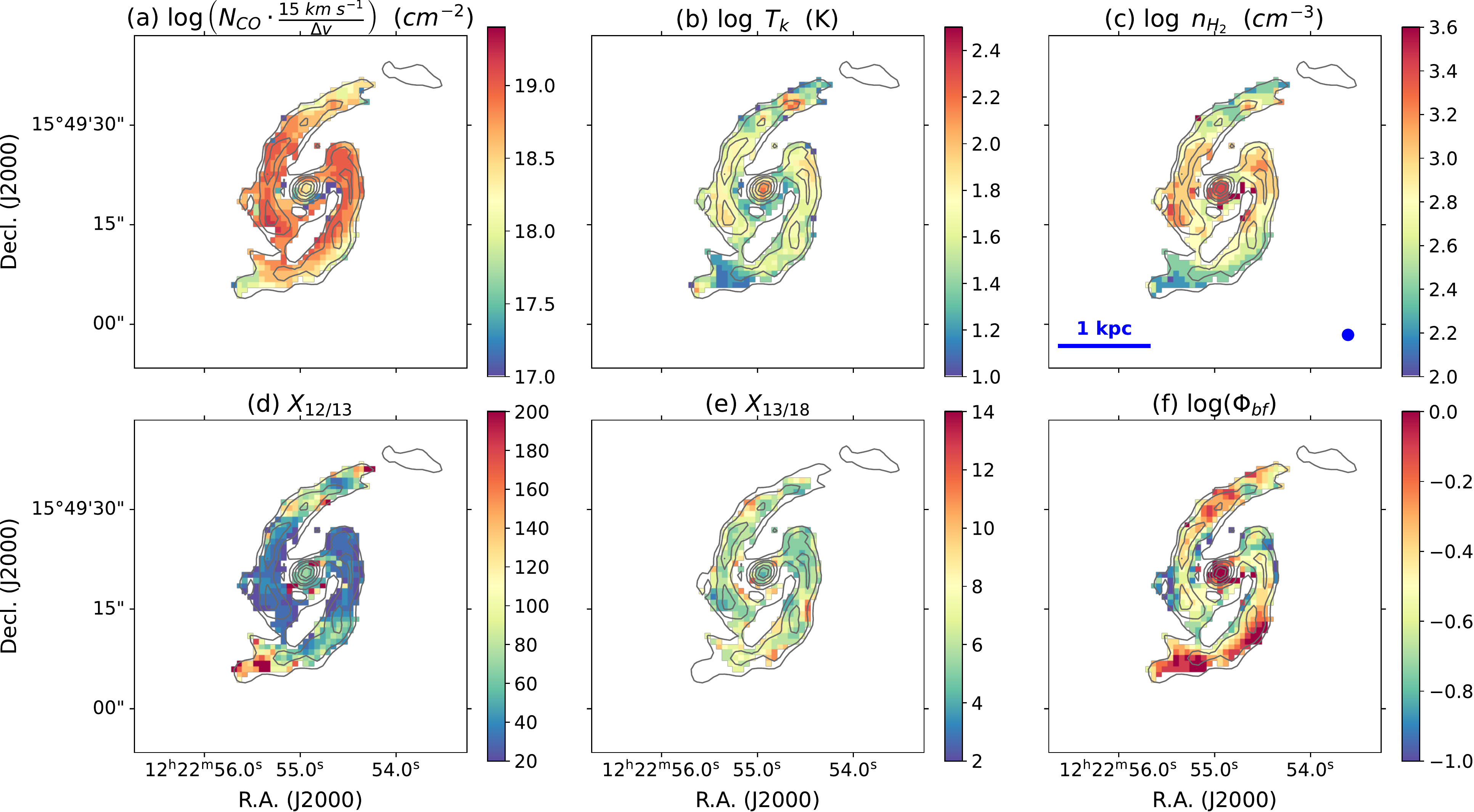} 
\caption{Maps of the 1DMax physical conditions derived from the modeling for NGC~4321. Panel~(a) shows $\log(N_\mathrm{CO})$ normalized to a fiducial line width of $15$~km~s$^{-1}$ over the whole region. Contours represent the CO 2--1 emission shown in Figure~\ref{fig:mom0_4321}(b). A $3\sigma$ mask of the $\rm C^{18}O$ 2--1 image is applied to (e).}
\label{fig:1dmax_4321}
\end{figure*}

\begin{figure*} \centering
\includegraphics[width=\linewidth]{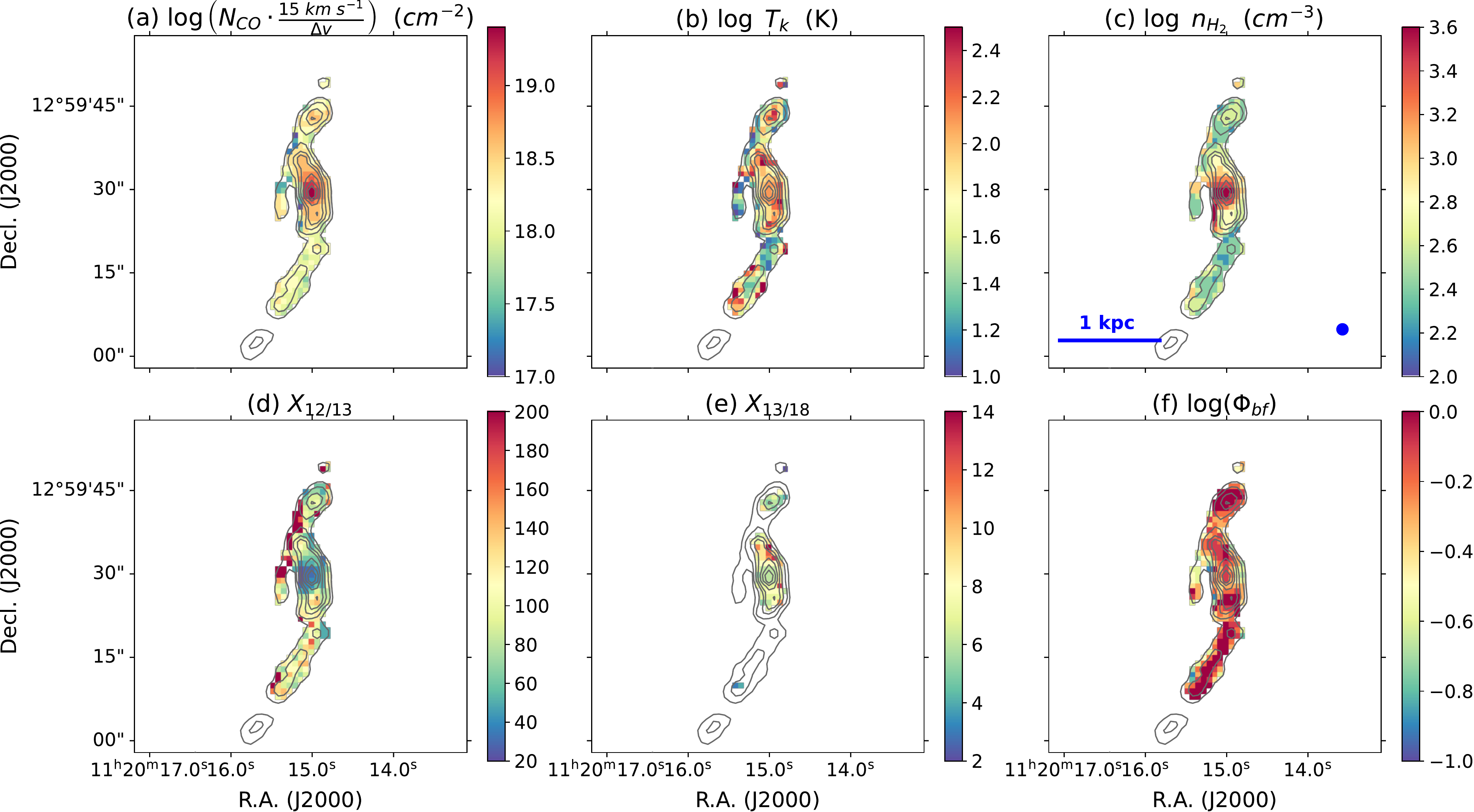} 
\caption{Same as Figure~\ref{fig:1dmax_4321} but for NGC~3627.}
\label{fig:1dmax_3627}
\end{figure*}

\begin{table*}
\begin{center}
\caption{Regional Averages and Standard Deviations of the 1D PDF Solutions \label{tab:stat_prob1d}}
\begin{tabular}{lccccccc}
  \hline\hline
  Region & &$\log \left(N_\mathrm{CO}\,\frac{\mathrm{15~km~s^{-1}}}{\Delta v}\right)$ &$\log T_\mathrm{k}$ &$\log n_\mathrm{H_2}$ &$X_{12/13}$ &$X_{13/18}$ &$\log\ \Phi_\mathrm{bf}$ \\
  & &($\rm cm^{-2}$) &(K) &($\rm cm^{-3}$) & & & \\
  \hline
\textbf{NGC~4321}\\
  Whole &1DMax &$18.60 \pm 0.47$ &$1.57 \pm 0.24$ &$2.78 \pm 0.31$ &$60.95 \pm 44.77$ &$6.43 \pm 1.71$ &$-0.46 \pm 0.27$ \\ 
  ($N_\mathrm{pix}=412$) &Median &$18.33 \pm 0.35$ &$1.60 \pm 0.19$ &$2.86 \pm 0.28$ &$91.58 \pm 22.87$ &$6.84 \pm 2.06$ &$-0.49 \pm 0.20$ \\ 
  \hline
  Nucleus &1DMax  &$18.16 \pm 0.62$ &$1.69 \pm 0.31$ &$3.19 \pm 0.36$ &$76.96 \pm 46.94$ &$6.78 \pm 2.30$ &$-0.13 \pm 0.16$ \\ 
  ($N_\mathrm{pix}=23$) &Median &$18.05 \pm 0.47$ &$1.79 \pm 0.24$ &$3.21 \pm 0.27$ &$95.51 \pm 18.71$ &$7.06 \pm 2.63$ &$-0.27 \pm 0.10$ \\ 
  \hline
  Inner arms &1DMax  &$18.79 \pm 0.43$ &$1.63 \pm 0.19$ &$2.95 \pm 0.19$ &$37.88 \pm 28.18$ &$6.02 \pm 1.49$ &$-0.59 \pm 0.21$ \\ 
  ($N_\mathrm{pix}=203$) &Median &$18.49 \pm 0.32$ &$1.63 \pm 0.13$ &$3.00 \pm 0.20$ &$81.12 \pm 16.03$ &$6.25 \pm 1.76$ &$-0.59 \pm 0.15$ \\ 
  \hline
  Outer arms &1DMax  &$18.44 \pm 0.38$ &$1.48 \pm 0.25$ &$2.55 \pm 0.24$ &$84.14 \pm 46.37$ &$7.07 \pm 1.74$ &$-0.35 \pm 0.25$ \\ 
  ($N_\mathrm{pix}=186$) &Median &$18.20 \pm 0.28$ &$1.54 \pm 0.21$ &$2.66 \pm 0.21$ &$102.50 \pm 24.37$ &$7.78 \pm 2.07$ &$-0.40 \pm 0.19$ \\ 
  \hline
\textbf{NGC~3627}\\  
  Whole &1DMax &$18.24 \pm 0.41$ &$1.75 \pm 0.40$ &$2.70 \pm 0.36$ &$100.51 \pm 47.23$ &$7.31 \pm 2.18$ &$-0.23 \pm 0.22$ \\ 
  ($N_\mathrm{pix}=214$) &Median &$18.07 \pm 0.35$ &$1.78 \pm 0.26$ &$2.71 \pm 0.32$ &$110.52 \pm 23.94$ &$7.84 \pm 2.17$ &$-0.32 \pm 0.16$ \\  
  \hline
  Nucleus &1DMax  &$19.07 \pm 0.23$ &$2.05 \pm 0.07$ &$3.43 \pm 0.14$ &$41.33 \pm 10.87$ &$6.60 \pm 0.80$ &$-0.23 \pm 0.09$ \\ 
  ($N_\mathrm{pix}=15$) &Median &$18.88 \pm 0.19$ &$2.06 \pm 0.05$ &$3.32 \pm 0.12$ &$89.69 \pm 9.26$ &$6.66 \pm 0.86$ &$-0.27 \pm 0.06$ \\ 
  \hline
  Inner arms &1DMax  &$18.27 \pm 0.38$ &$1.76 \pm 0.37$ &$2.79 \pm 0.34$ &$93.37 \pm 42.43$ &$8.24 \pm 2.11$ &$-0.27 \pm 0.22$ \\ 
  ($N_\mathrm{pix}=92$) &Median &$18.09 \pm 0.31$ &$1.82 \pm 0.24$ &$2.82 \pm 0.32$ &$106.62 \pm 25.46$ &$8.81 \pm 2.06$ &$-0.35 \pm 0.16$ \\ 
  \hline
  Outer arms &1DMax  &$18.10 \pm 0.29$ &$1.69 \pm 0.44$ &$2.51 \pm 0.19$ &$114.95 \pm 46.41$ &$5.40 \pm 1.74$ &$-0.19 \pm 0.23$ \\ 
  ($N_\mathrm{pix}=107$) &Median &$17.95 \pm 0.20$ &$1.72 \pm 0.26$ &$2.54 \pm 0.16$ &$116.79 \pm 21.54$ &$6.30 \pm 1.91$ &$-0.31 \pm 0.17$ \\ 
  \hline
\end{tabular} 
\end{center}
\tablecomments{The averages are determined from the ensemble of 1DMax and median solutions from the 1D PDFs. The listed uncertainties represent the standard deviations across pixels in each region, while the deviations between 1DMax and medians can reflect the uncertainties in individual PDFs. $N_\mathrm{pix}$ indicates the number of pixels used for each region.}
\end{table*} 

We present the 1DMax solution maps for each parameter in Figures~\ref{fig:1dmax_4321} and~\ref{fig:1dmax_3627}. The regional statistics of the 1DMax and median solutions are listed in Table~\ref{tab:stat_prob1d}. 
Both NGC~4321 and NGC~3627 show clear trends of increasing $T_\mathrm{k}$ and $n_\mathrm{H_2}$ from the outer arms to the nucleus. $N_\mathrm{CO}/\Delta v$, which reflects the optical depth, also increases toward the centers, except in the nucleus of NGC~4321. Despite having a 0.2--0.5~dex lower $N_\mathrm{CO}/\Delta v$, the nucleus of NGC~4321 has similar $N_\mathrm{CO}$ as the arm regions. This is because the line width at the NGC~4321 nucleus is $> 30$~$\mathrm{km\ s^{-1}}$, which is larger than the $10-20$~$\mathrm{km\ s^{-1}}$ line widths in the inner/outer arms by more than a factor of two (see Table~\ref{tab:ratio_4321}). 
We note that the line width at the nucleus of NGC~3627 is also $> 2\times$ larger than that in its arm regions, and thus the central enhancement of $N_\mathrm{CO}$ is even more dramatic than that shown by the $N_\mathrm{CO}/\Delta v$ enhancement in Table~\ref{tab:stat_prob1d}. 
With a typical line width of ${\sim}60\ \mathrm{km\ s^{-1}}$, the mean $N_\mathrm{CO}$ in the NGC~3627 nucleus exceeds $3\times10^{19}$~$\mathrm{cm^{-2}}$. 

High temperature and volume density are also found in the nucleus of NGC~3627, with mean $T_\mathrm{k}$ and $n_\mathrm{H_2}$ reaching $> 100$~K and $3 \times 10^{3}$~$\mathrm{cm^{-3}}$, respectively. The nucleus of NGC~4321 also shows $T_\mathrm{k} \sim$100~K and $n_\mathrm{H_2} > 10^{3}$~$\mathrm{cm^{-3}}$, higher than the average conditions of the arm regions in both galaxies. We note that the nuclear type of NGC~4321 is mostly classified as a low-ionization nuclear emission region (LINER) and NGC~3627 as either a LINER or a Seyfert~2 AGN \citep{1997ApJS..112..315H,2000ApJS..129...93F,2010ApJS..190..233M,2022A&A...659A..26B}. As the inner $\sim$300~pc regions may be impacted by nuclear activity, it is reasonable to find much more excited gas conditions in these regions.

From Figures~\ref{fig:1dmax_4321} and~\ref{fig:1dmax_3627}, we find a consistent 1DMax solution of $X_{12/13} \sim$40 across the inner arms of NGC~4321 and the nucleus of NGC~3627, though the median solutions imply higher $X_{12/13}$ of 80--90 (see Table~\ref{tab:stat_prob1d} and Appendix~\ref{sec:pixel_solutions}). On the other hand, the 1DMax and medians in the inner/outer arms of NGC~3627 agree well, suggesting $X_{12/13} \sim$100. Both the 1DMax and median solutions in the nucleus and outer arms of NGC~4321 also imply higher $X_{12/13}$ of 80--100. Similar to the $X_{12/13}$ distribution, we also derive the lowest $X_{13/18}$ ($\sim$6) in the inner arms of NGC~4321 and the nucleus of NGC~3627. The decrease of both $X_{12/13}$ and $X_{13/18}$ in those regions may indicate $^{13}$C and $^{18}$O enrichment from enhanced star formation. We note that the derived $X_{13/18}$ across both galaxy centers are well-constrained at a range of 6--8 which is similar to the Galactic Center value \citep{2018A&A...612A.117A}. On the other hand, our derived $X_{12/13}$ values are higher than $X_{12/13} \sim$25 found in our Galactic Center \citep{1994ARA&A..32..191W, 2005ApJ...634.1126M,2023A&A...670A..98Y} as well as the central kpc of NGC~3351 \citep{2022ApJ...925...72T}. This is in line with the higher $X_{12/13}$ values varying from $\sim$40 to $>$100 that have been commonly found in other starburst galaxy centers or (U)LIRGs, likely due to higher inflow rates and/or stellar nucleosynthesis enrichment \citep{2014A&A...565A...3H,2014ApJ...796L..15S,2017ApJ...840....8S,2019A&A...629A...6T}.

\subsection{CO-to-H$_2$ Conversion Factors} \label{subsec:alpha_CO}

\begin{figure*}
\begin{minipage}{.46\linewidth}
\centering
\includegraphics[width=\linewidth]{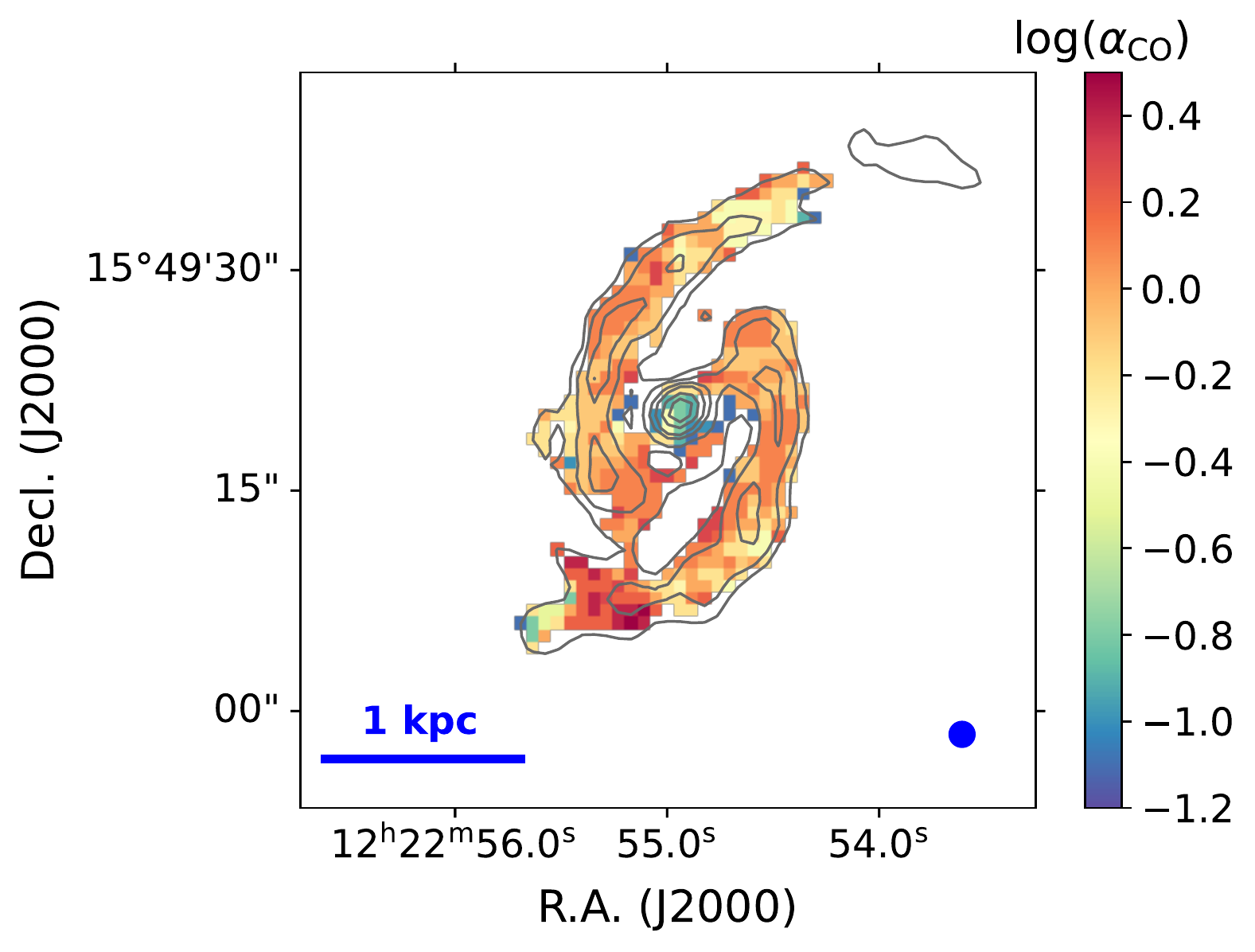}\\
(a)
\end{minipage}
\hfill
\begin{minipage}{.52\linewidth}
\centering
\includegraphics[width=\linewidth]{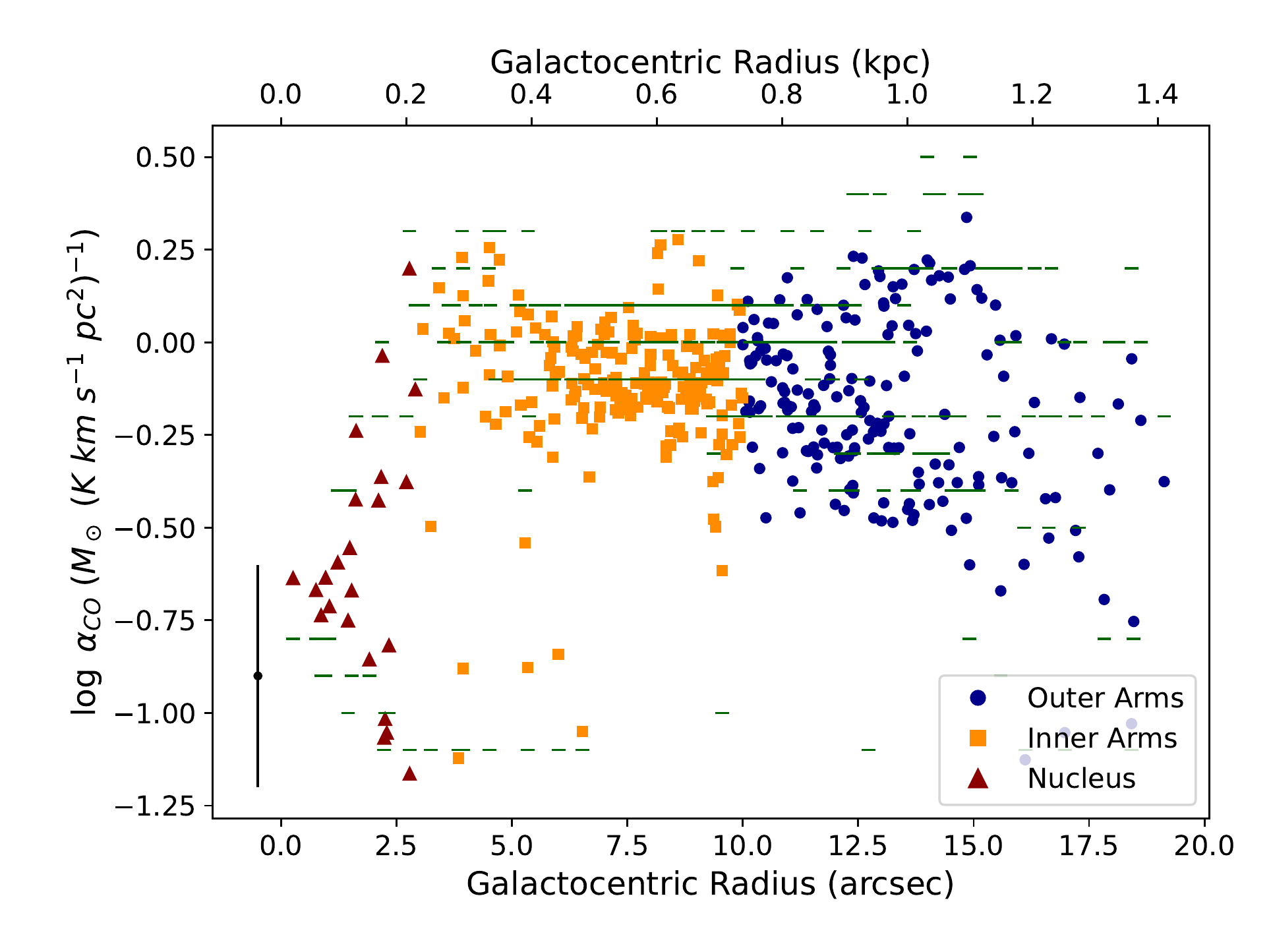}\\
(b)
\end{minipage}
\caption{Spatial variation of $\alpha_\mathrm{CO}$ in NGC~4321. (a) 1DMax $\log (\alpha_\mathrm{CO})$ map in units of $\mathrm{M_\odot\ (K~km~s^{-1}~pc^2)^{-1}}$; the contours represent the moment 0 of CO 2--1. 
(b) Relation between the modeled $\alpha_\mathrm{CO}$ and galactocentric radius. The colored points correspond to the median $\alpha_\mathrm{CO}$ in different regions, and the green horizontal lines present the 1DMax solutions which are similar to the medians. A typical error bar of $\sigma = \pm 0.3$~dex is shown in the lower left corner.
All $\alpha_\mathrm{CO}$ values are below the Galactic disk average of $\log(\alpha_\mathrm{CO}) = 0.64$ or $\alpha_\mathrm{CO} = 4.4$~$\mathrm{M_\odot\ (K~km~s^{-1}~pc^2)^{-1}}$. In the nucleus, $\alpha_\mathrm{CO}$ is a factor of 3--5 lower than in the arms. Furthermore, $\alpha_\mathrm{CO}$ in the outer arms shows a decreasing trend with galactocentric radius (see Section~\ref{subsec:alpha_env} for further discussion).}
\label{fig:alpha_4321}
\end{figure*}

\begin{figure*}
\begin{minipage}{.47\linewidth}
\centering
\includegraphics[width=\linewidth]{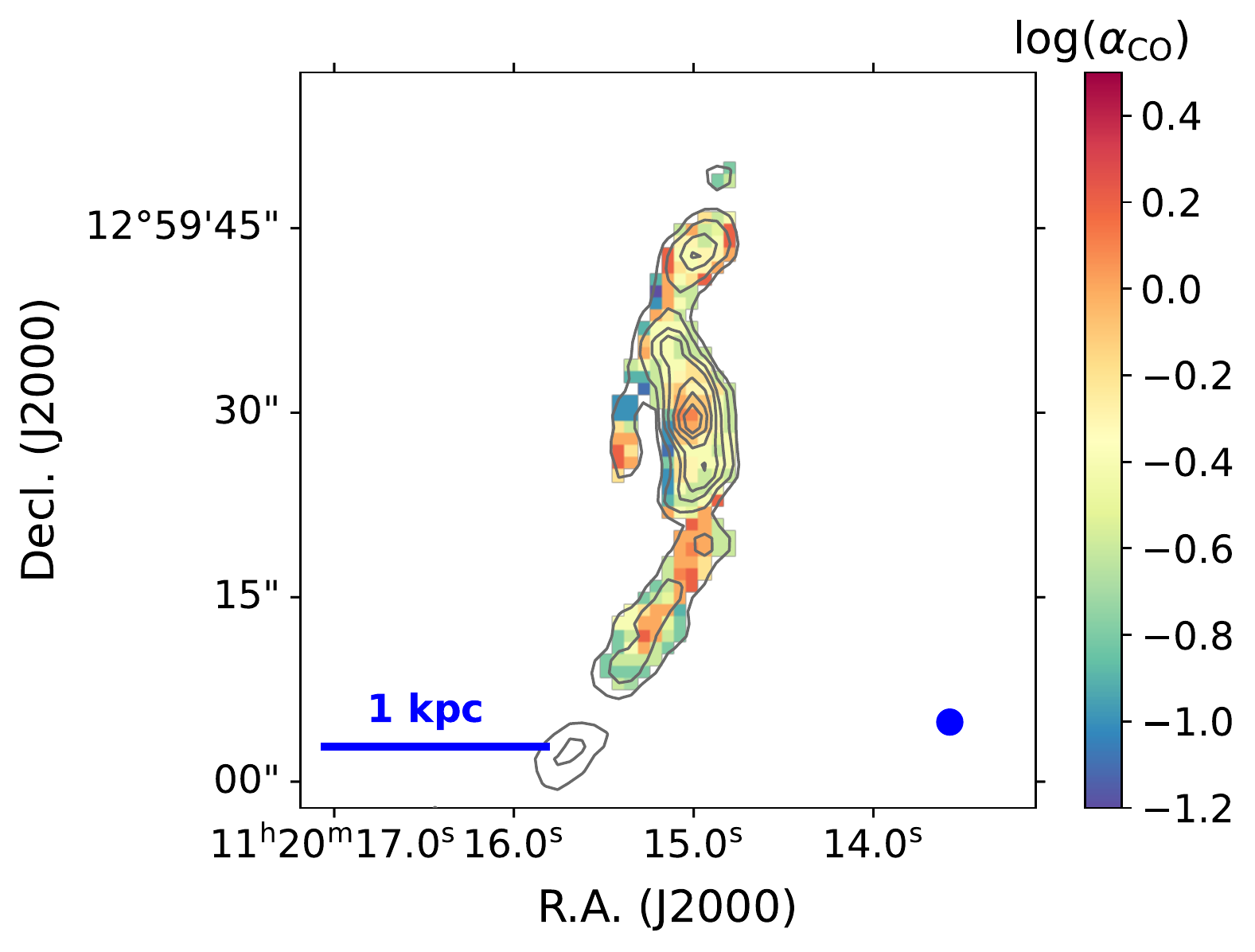}\\
(a)
\end{minipage}
\hfill
\begin{minipage}{.51\linewidth}
\centering
\includegraphics[width=\linewidth]{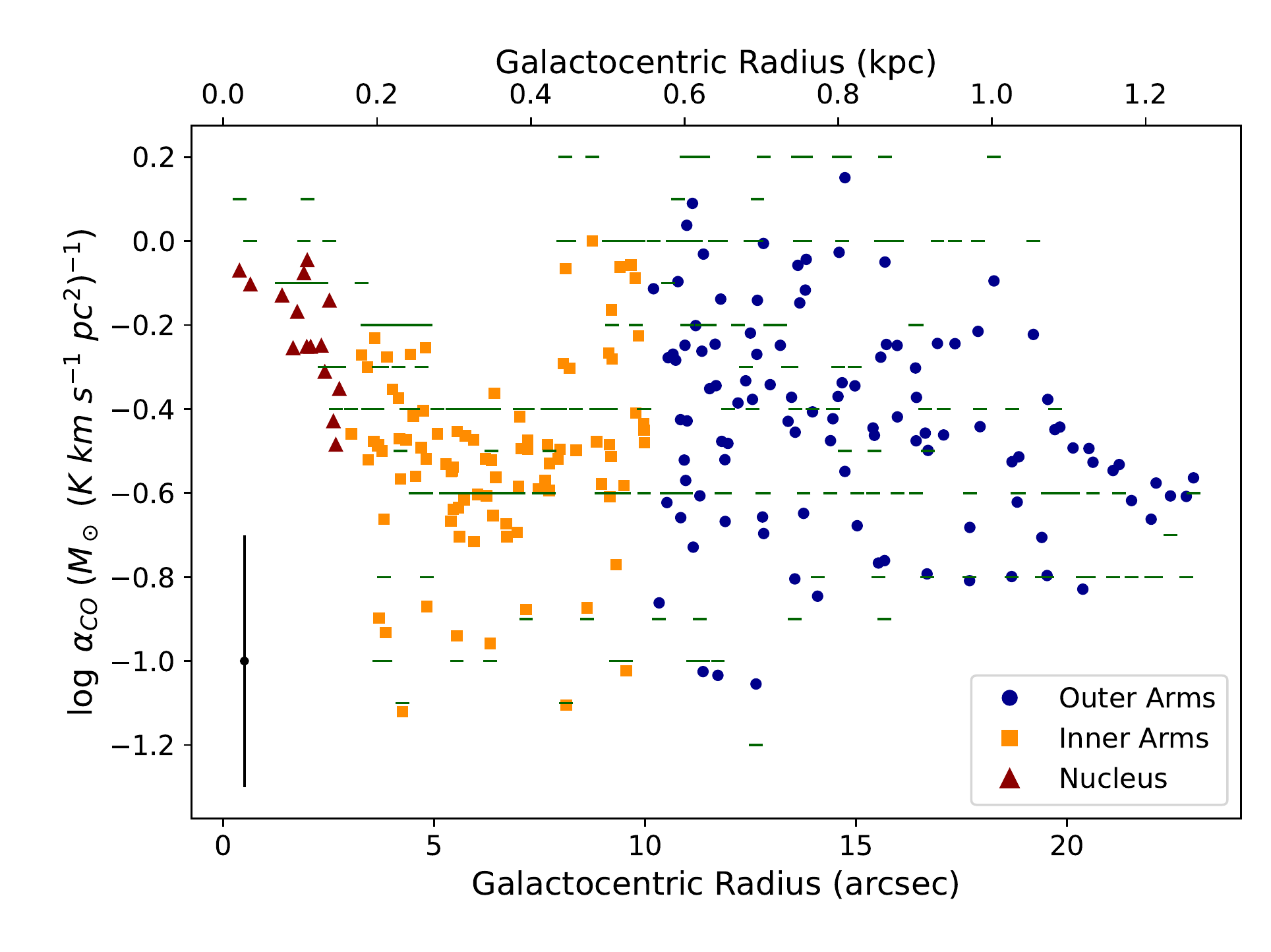}\\
(b)
\end{minipage}
\caption{Spatial variation of $\alpha_\mathrm{CO}$ in NGC~3627. See the caption of Figure~\ref{fig:alpha_4321} for more information. The derived $\alpha_\mathrm{CO}$ values are generally lower than NGC~4321 and substantially lower than the Galactic disk average. $\alpha_\mathrm{CO}$ decreases sharply from the nucleus to the inner arms, while the outer arms show a larger scatter likely due to limited constraints from the $^{13}$CO 3--2 and C$^{18}$O observations.}
\label{fig:alpha_3627}
\end{figure*}

The CO-to-H$_2$ conversion factor (see Equation~\ref{def_alpha_intro} for definition) can be expressed as a function of $N_\mathrm{CO}$, $\Phi_\mathrm{bf}$, and the CO 1--0 intensity $I_\mathrm{CO(1-0)}$:
\begin{equation}\label{eqn_alpha}
\begin{aligned}
\alpha_\mathrm{CO} &=\frac{M_\mathrm{mol}}{L_\mathrm{CO(1-0)}}\,\left[\frac{\mathrm{M_\odot}} {\mathrm{K~km~s^{-1}~pc^2}}\right] \\
&= \frac{1.36\,m_\mathrm{H_2}\,[\mathrm{M}_\odot]\,N_\mathrm{CO}\,[\mathrm{cm}^{-2}]\,A\,[\mathrm{cm}^2]\,\Phi_\mathrm{bf}} {x_\mathrm{CO}\,I_\mathrm{CO(1-0)}\,[\mathrm{K~km~s}^{-1}]\,A\,[\mathrm{pc}^2]} \\
&= \frac{1}{4.5\times10^{19}} \cdot \frac{N_\mathrm{CO}\,[\mathrm{cm}^{-2}]\,\Phi_\mathrm{bf}} {x_\mathrm{CO}\,I_\mathrm{CO(1-0)}\,[\mathrm{K~km~s^{-1}]}}~,
\end{aligned}
\end{equation}
where $x_\mathrm{CO}$ is the CO/H$_2$ abundance ratio.
In the second step of the above equation, the factor of  $1.36$ is to include the mass contribution from helium, $m_\mathrm{H_2}$ is the mass of a hydrogen molecule, and $A$ is the area relevant to the conversion between $I_\mathrm{CO}$ and $L_\mathrm{CO}$. All of these factors are reduced to the constant in the final step of Equation~\ref{eqn_alpha}. We note that galaxy inclinations do not affect the result of $\alpha_\mathrm{CO}$, because the inclination correction on $N_\mathrm{CO}$ and $I_\mathrm{CO(1-0)}$ (which includes $\Delta v$) would cancel out in Equation~\ref{eqn_alpha}.

Since our modeling directly constrains $N_\mathrm{CO}$ and $\Phi_\mathrm{bf}$ and provides a prediction of the $I_\mathrm{CO(1-0)}$ values that matches the observed one, we can derive the spatial distribution of $\alpha_\mathrm{CO}$ from the modeling with an assumption of $x_\mathrm{CO}$. 
While $N_\mathrm{CO}$ can be determined by multiplying $N_\mathrm{CO}/\Delta v$ with $\Delta v$, we caution that the $\Delta v$ should be consistent with the line width of the observed $I_\mathrm{CO(1-0)}$. This is different from the $\alpha_\mathrm{CO}$ calculation in \citet{2022ApJ...925...72T}, where the line widths were not consistent. As we will compare their result on NGC~3351 with ours in Section~\ref{sec:discussion}, we list the updated $\alpha_\mathrm{CO}$ values of NGC~3351 in Appendix~\ref{sec:alpha_3351} for self-consistency. We note that the key conclusions in \citet{2022ApJ...925...72T} are unchanged, but the updated $\alpha_\mathrm{CO}$ values are overall lowered by a factor of two to three (see Appendix~\ref{sec:alpha_3351} for more details).        

Throughout our analysis, we assume $x_\mathrm{CO} = 3 \times 10^{-4}$, which is supported by measurements of warm/dense star-forming clouds \citep[e.g.,][]{1994ApJ...428L..69L,2004ApJ...605..272S,2008ApJ...687.1075S} and commonly adopted in various starburst regions \citep[e.g.,][]{2012ApJ...753...70K,2014ApJ...795..174K,2014ApJ...796L..15S,2017ApJ...840....8S}. As this value assumes that most carbon is in the form of CO, which is not necessarily true in some galaxy centers \citep[e.g.,][]{2022arXiv221209661L,2023ApJ...944L..19L}, the uncertainty in our $\alpha_\mathrm{CO}$ values could be at the factor of 2--3 level because of this assumption. In addition, while we do not expect that elemental abundance variations of C and O are large enough to drive $x_\mathrm{CO}$ variations on sub-kpc scales, there are other mechanisms which may destroy the CO molecule and lower the CO/H$_2$ abundance, such as photodissociation by FUV radiation and cosmic rays in starburst or AGN environments \citep{2018ApJ...858...16G,2021MNRAS.502.2701B,2022arXiv221209661L}. The effect of photodissociation is the strongest in optically thin and CO-faint regions where the shielding of FUV radiation is weak, such as interarm and outer galaxy regions.
However, we emphasize that a better prediction of the $x_\mathrm{CO}$ value is not feasible with current dataset and analysis. 
Thus, it is important to note that our derived $\alpha_\mathrm{CO}$ values depend inversely on $x_\mathrm{CO}$, i.e., $\alpha_\mathrm{CO}^\mathrm{true} = \alpha_\mathrm{CO}^\mathrm{derived} \times (3\times10^{-4} / x_\mathrm{CO})$.  

With Equation~\ref{eqn_alpha} and following the procedure described in \citet[Section~4.4]{2022ApJ...925...72T}, we create a grid of $\log(\alpha_\mathrm{CO})$ from -2.5 to 2.5 with a step size of 0.1 and obtain marginalized PDFs of $\alpha_\mathrm{CO}$ for each pixel. Then, we extract the 1DMax/median $\alpha_\mathrm{CO}$ solutions from the PDFs. With this method, the derived $\alpha_\mathrm{CO}$ does not depend on the best-fit/1DMax/median solutions of $N_\mathrm{CO}$ and $\Phi_\mathrm{bf}$ determined in Section~\ref{subsec:model_result}, since those parameters are fit simultaneously within the full grid before marginalization. We refer readers to \citet{2022ApJ...925...72T} for more details. 

Figures~\ref{fig:alpha_4321} and~\ref{fig:alpha_3627} show the spatial variations of $\alpha_\mathrm{CO}$ across the observed regions. The 1DMax and median $\alpha_\mathrm{CO}$ solutions are similar and have consistent trends with the galactocentric radius. We will mainly present the median solutions hereafter due to their continuity across the $\alpha_\mathrm{CO}$ parameter space resulting from interpolation. As shown in Figure~\ref{fig:alpha_4321}(b), the arm regions of NGC~4321 have a roughly constant $\log(\alpha_\mathrm{CO})$ around -0.1 (i.e., $\alpha_\mathrm{CO} \approx 0.8$~$\mathrm{M_\odot\ (K~km~s^{-1}~pc^2)^{-1}}$), and there is a decreasing trend toward the outer arms. On the other hand, the nucleus region shows $\log(\alpha_\mathrm{CO}) \approx -0.7$ (or $\alpha_\mathrm{CO} \approx 0.2$~$\mathrm{M_\odot\ (K~km~s^{-1}~pc^2)^{-1}}$), which is a factor of 3--5 lower than in the arms. In NGC~3627, the general $\alpha_\mathrm{CO}$ values are even a factor of 2--3 lower than in NGC~4321. Furthermore, Figure~\ref{fig:alpha_3627}(b) shows that $\alpha_\mathrm{CO}$ decreases sharply from the nucleus to the inner arms, while the outer arms show a large scatter of $\alpha_\mathrm{CO}$ which likely results from the lower S/N of $^{13}$CO 3--2 and C$^{18}$O data in this region.

We note that trends of decreasing $\alpha_\mathrm{CO}$ with radius are seen in the inner/outer arm regions of both NGC~3627 and NGC~4321. We will discuss these $\alpha_\mathrm{CO}$ trends seen in barred galaxy centers in Section~\ref{subsec:alpha_env}. Moreover, our modeling results show that all pixels across the observed regions in both galaxy centers have $\alpha_\mathrm{CO}$ that is 4--15 times below the Galactic disk average of $4.4$~$\mathrm{M_\odot\ (K~km~s^{-1}~pc^2)^{-1}}$. This range of lower $\alpha_\mathrm{CO}$ is consistent with previous kpc-scale estimations toward galaxy centers using independent techniques \citep{2004A&A...422L..47S,2013ApJ...777....5S,2020A&A...635A.131I,2023arXiv230203044D}. In Section~\ref{subsec:alpha_literature}, we will compare our kpc-averaged $\alpha_\mathrm{CO}$ with those studies which included NGC~3627 and NGC~4321.

\section{Discussion} \label{sec:discussion}

\subsection{$\alpha_\mathrm{CO}$ Distribution and Environmental Dependence} \label{subsec:alpha_env}

\begin{figure*}
\begin{minipage}{.49\linewidth}
\centering
\includegraphics[width=\linewidth]{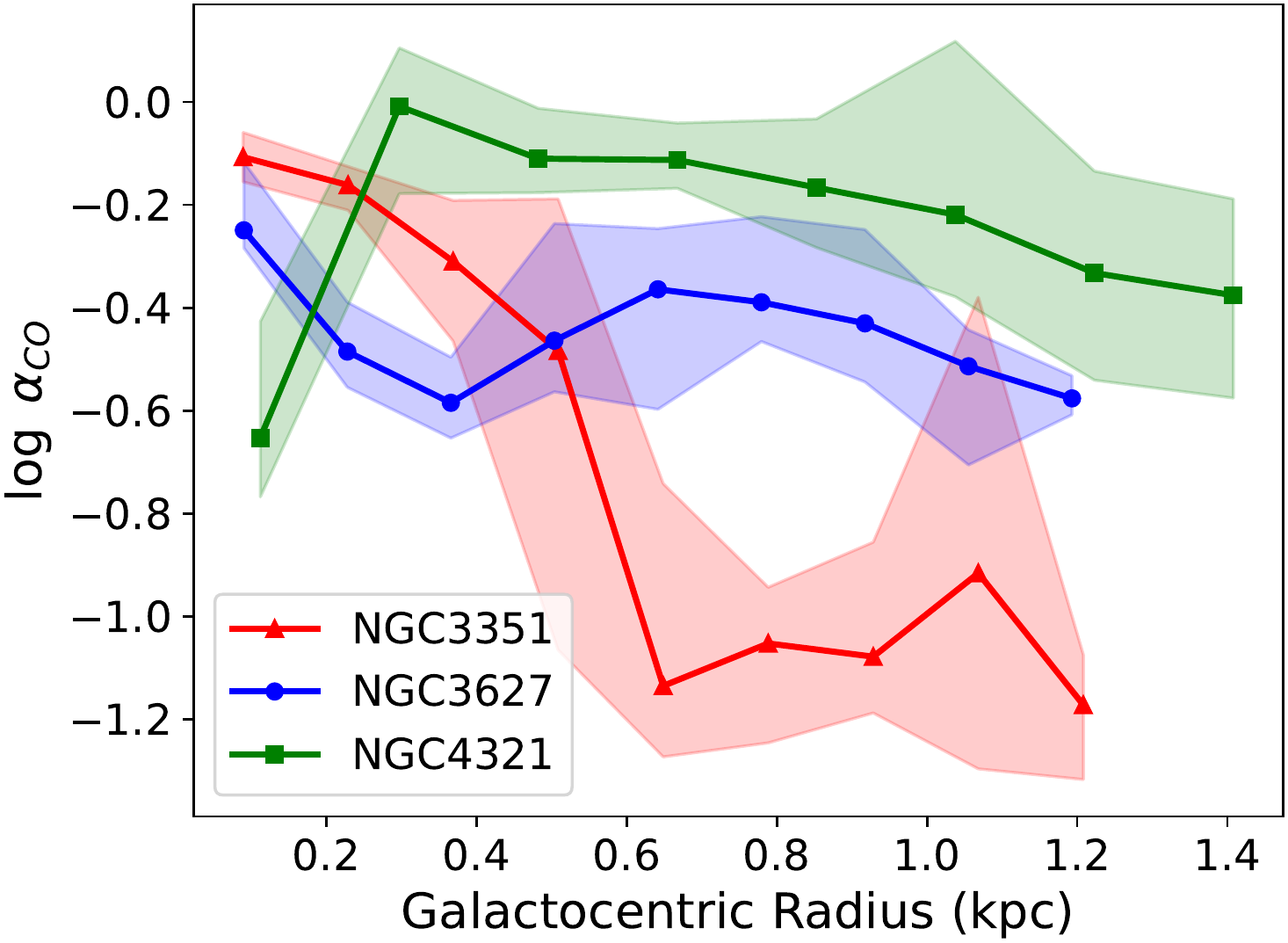}\\
(a)
\end{minipage}
\hfill
\begin{minipage}{.5\linewidth}
\centering
\includegraphics[width=\linewidth]{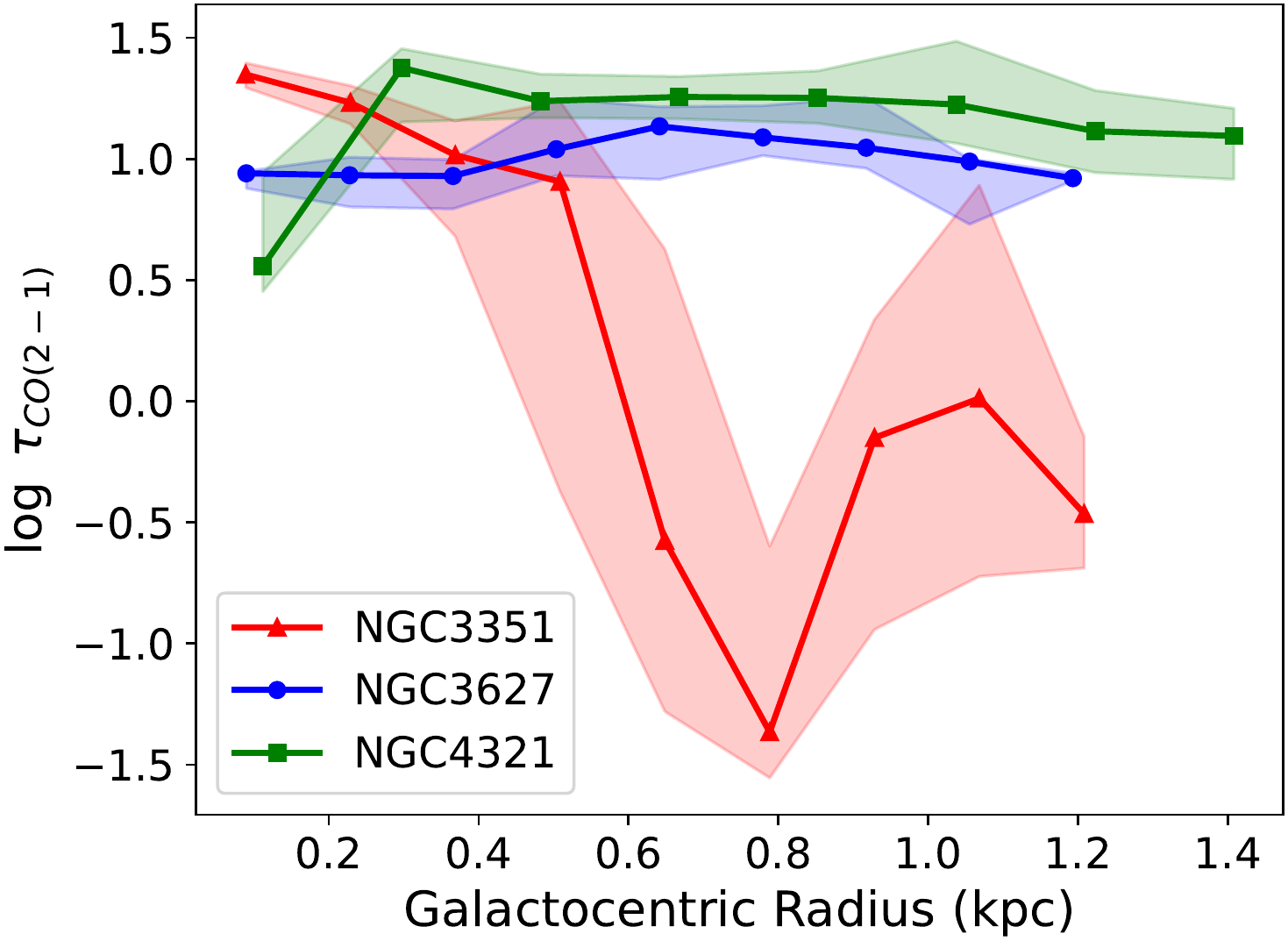}\\
(b)
\end{minipage}
\caption{Medians of the modeled (a) $\alpha_\mathrm{CO}$ in units of $\mathrm{M_\odot\ (K~km~s^{-1}~pc^2)^{-1}}$ and (b) line center $\tau_\mathrm{CO(2-1)}$ within $\sim$100~pc galactocentric radii bins in the centers of NGC~3351 (red), NGC~3627 (blue), and NGC~4321 (green). Shaded areas span the 25th and 75th percentile ranges. All regions show $\alpha_\mathrm{CO}$ at least a factor of four lower than the Galactic value of $\log(\alpha_\mathrm{CO}) = 0.64$. The radial trend of $\alpha_\mathrm{CO}$ is mostly consistent with that of $\tau_\mathrm{CO(2-1)}$ in all three galaxy centers except in NGC~3627's nucleus.}
\label{fig:cross_radius}
\end{figure*}

\begin{figure*}
\begin{minipage}{.49\linewidth}
\centering
\vspace{3ex}
\includegraphics[width=\linewidth]{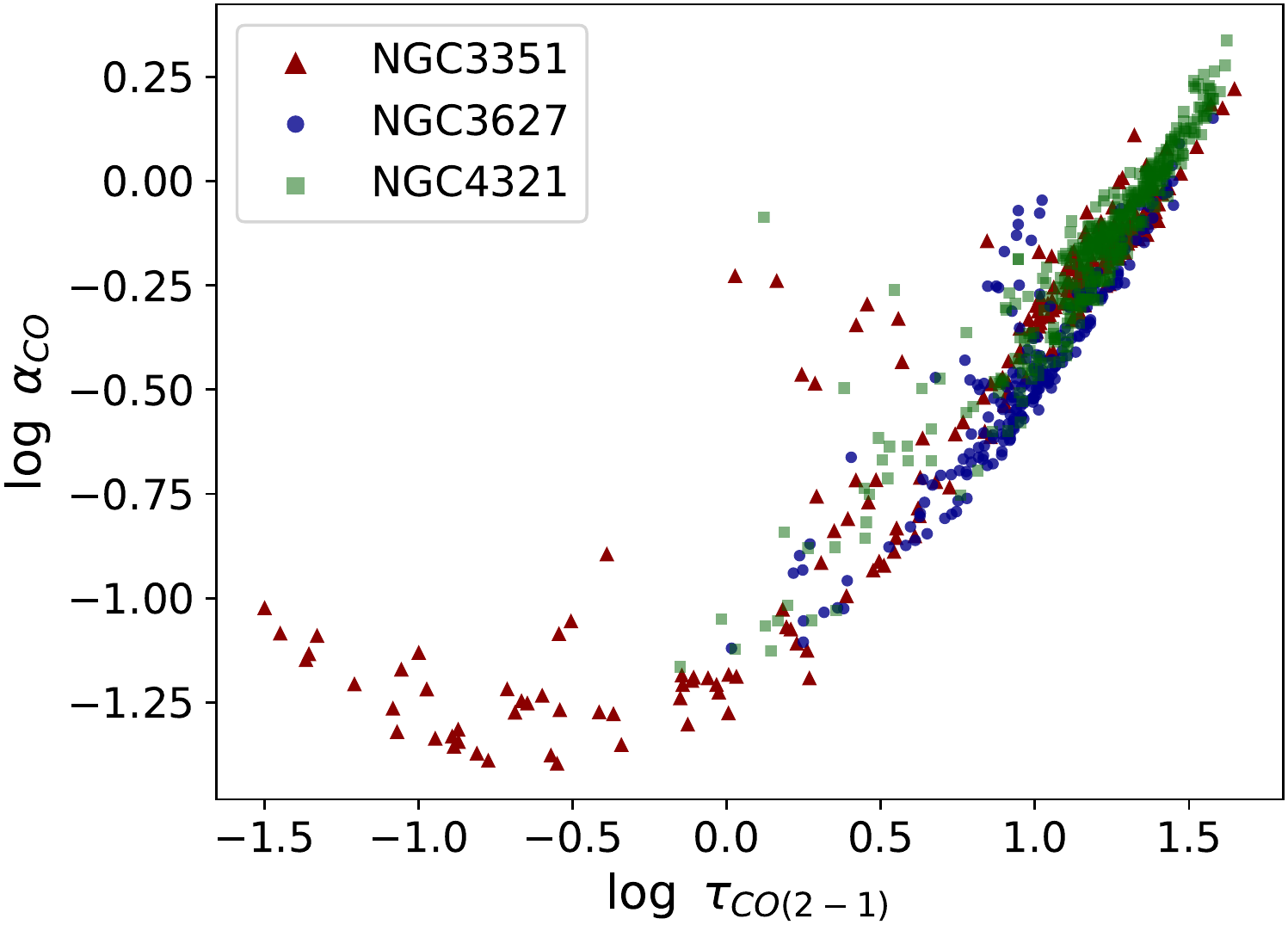}\\
(a)
\end{minipage}
\hfill
\begin{minipage}{.48\linewidth}
\centering
\includegraphics[width=\linewidth]{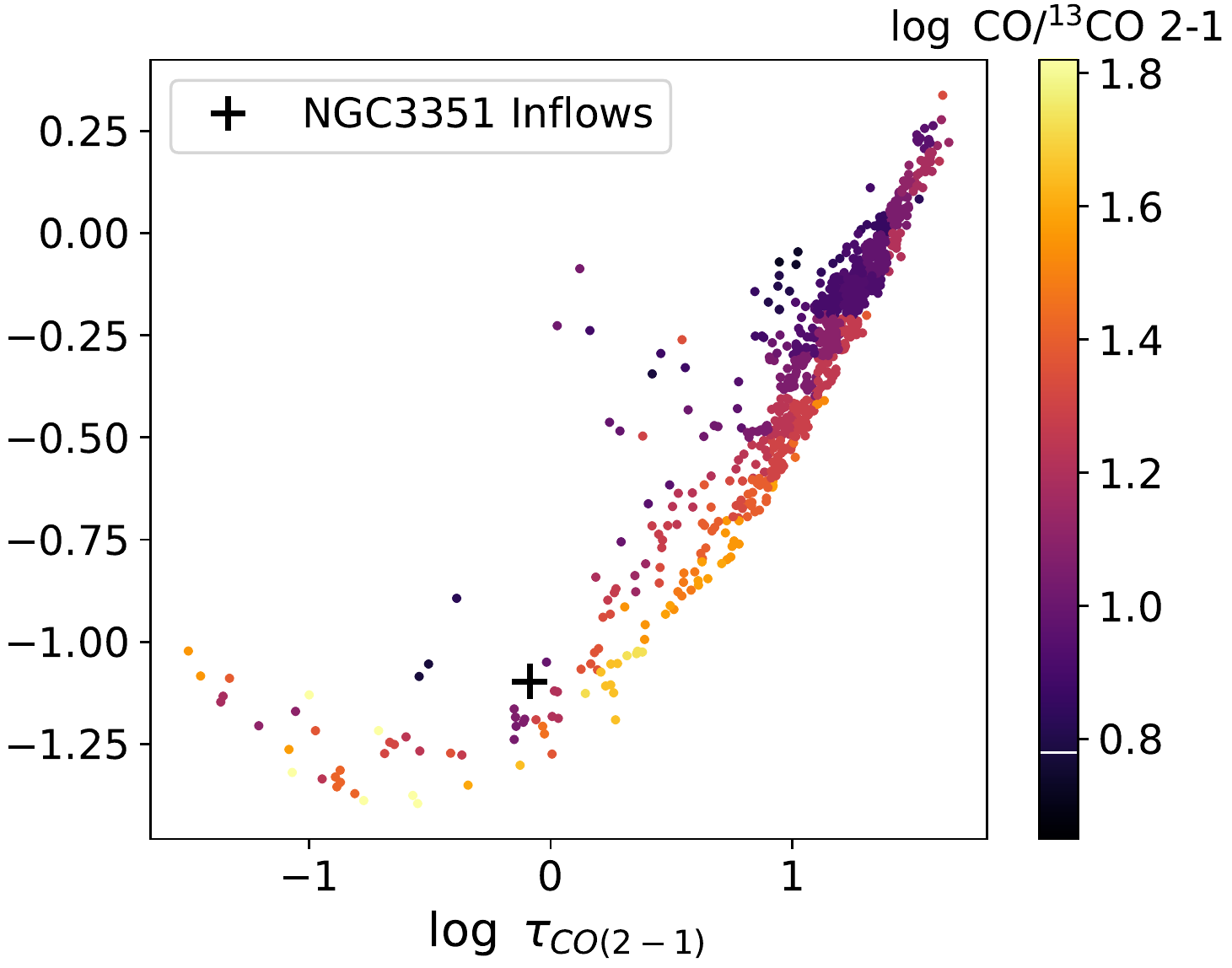}\\
(b)
\end{minipage}
\caption{Modeled $\alpha_\mathrm{CO}$ and $\tau_\mathrm{CO(2-1)}$, color-coded by (a) three galaxies and (b) 2D-binned medians of the observed CO/$^{13}$CO 2--1 ratios. 
In panel (b), the black cross sign represents spectral stacking result of the NGC~3351 inflows (see \citealt{2022ApJ...925...72T} and Appendix~\ref{sec:alpha_3351}), and the white line on the color bar indicates the typical ratio found in the disks of Milky Way or other nearby galaxies. In the optically thick regime ($\tau_\mathrm{CO} > 1$), a positive correlation of $\alpha_\mathrm{CO}$ with $\tau_\mathrm{CO}$ is constantly seen in the three galaxy centers, and the CO/$^{13}$CO 2--1 ratio generally reflects the $\tau_\mathrm{CO}$ variation in optically thick regions.}
\label{fig:alpha_tau_scatter}
\end{figure*}

\begin{figure*}
\begin{minipage}{.485\linewidth}
\centering
\includegraphics[width=\linewidth]{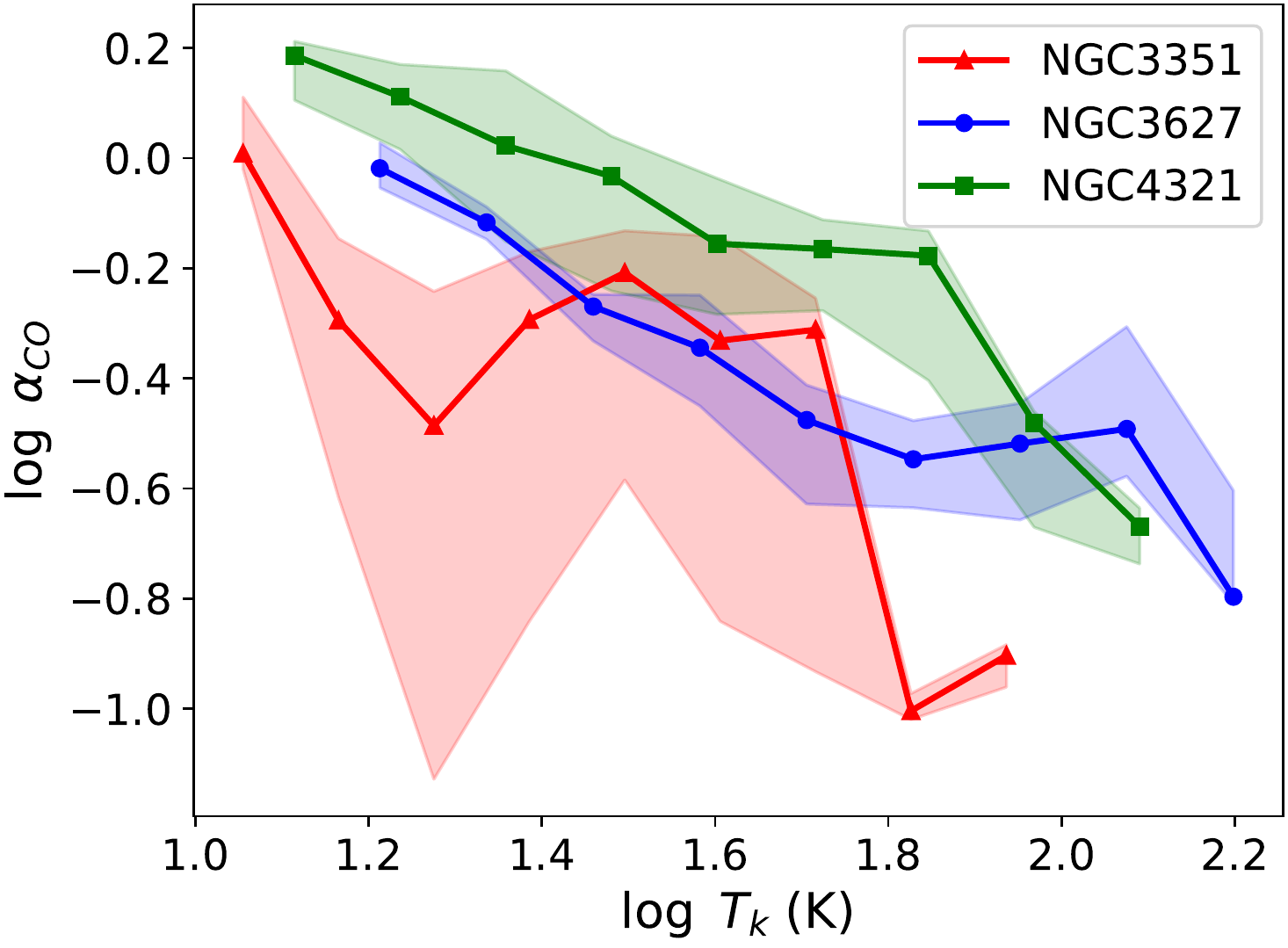}\\
(a)
\end{minipage}
\hfill
\begin{minipage}{.5\linewidth}
\centering
\includegraphics[width=\linewidth]{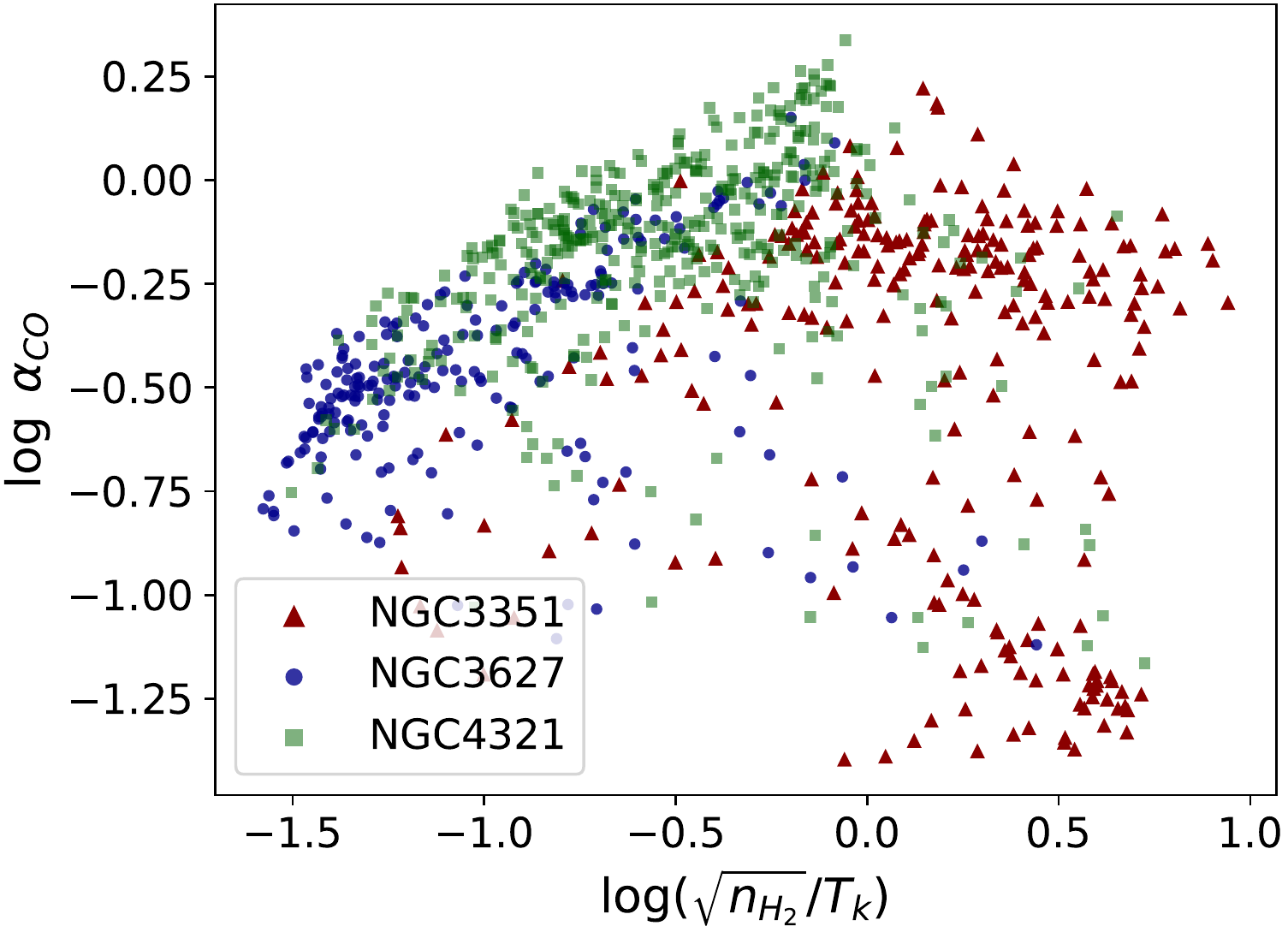}\\
(b)
\end{minipage}
\caption{Relation of the modeled $\log(\alpha_\mathrm{CO})$ with (a) $\log(T_\mathrm{k})$ and (b) $\log(\sqrt{n_\mathrm{H_2}}$/$T_\mathrm{k})$ for NGC~3351 (red), NGC~3627 (blue), and NGC~4321 (green). Both NGC~3627 and NGC~4321 show clear correlations of $\alpha_\mathrm{CO}$ decreasing with $T_\mathrm{k}$ and increasing with $\sqrt{n_\mathrm{H_2}}$/$T_\mathrm{k}$. In NGC~3351, there is no strong dependence on $T_\mathrm{k}$, and its lower $\alpha_\mathrm{CO}$ than the virial balance assumed $\sqrt{n_\mathrm{H_2}}$/$T_\mathrm{k}$ trend may indicate super-virial gas in the center of NGC~3351.}
\label{fig:alpha_n_T}
\end{figure*}

To study the spatial variation of $\alpha_\mathrm{CO}$ at $\sim$100~pc scales in barred, star-forming galaxy centers, we present a cross comparison among the results from \citet{2022ApJ...925...72T} on NGC~3351 and this work on NGC~3627 and NGC~4321. As the non-LTE radiative transfer modeling also predicts the optical depth for each line, we derive full PDFs of the CO optical depths (in both 1--0 and 2--1) as well as $\sqrt{n_\mathrm{H_2}}/T_\mathrm{k}$, using the same technique for determining the $\alpha_\mathrm{CO}$ solutions in Section~\ref{subsec:alpha_CO}. It is interesting to compare our $\alpha_\mathrm{CO}$ with CO optical depth ($\tau_\mathrm{CO}$) and gas temperature ($T_\mathrm{k}$), as they together determine the amount of escaped CO emission that can change $\alpha_\mathrm{CO}$ \citep{2012ApJ...751...10P,2022ApJ...925...72T}. We also derive $\sqrt{n_\mathrm{H_2}}/T_\mathrm{k}$ because $\alpha_\mathrm{CO}$ is expected to be approximately proportional to this quantity for isolated and virialized clouds when CO lines are optically thick and thermalized \citep{co-to-h2,2020ApJ...903..142G}. 
The grids for both $\log(\tau_\mathrm{CO})$ and $\log(\sqrt{n_\mathrm{H_2}}/T_\mathrm{k})$ range from -2 to 2 with a step size of 0.1.  

Figure~\ref{fig:cross_radius}(a) shows the radially-binned medians and the 25--75th percentile ranges of the median $\alpha_\mathrm{CO}$ solutions presented in Figure~\ref{fig:alpha_4321}(b) and~\ref{fig:alpha_3627}(b), together with the NGC~3351 results (see Appendix~\ref{sec:alpha_3351}). The bin size is $\sim$150~pc in galactocentric radius. A similar plot showing the median solutions of $\tau_\mathrm{CO(2-1)}$ is provided in Figure~\ref{fig:cross_radius}(b).    
It is clear that all regions in the three galaxy centers have $\alpha_\mathrm{CO}$ at least four times lower than the Galactic value of $\log(\alpha_\mathrm{CO}) \sim$0.64. In addition, all three galaxies show a globally decreasing $\alpha_\mathrm{CO}$ trend until a radius of $\sim$1.5~kpc, and $\alpha_\mathrm{CO}$ in NGC~3351 inflow regions \citep[i.e., beyond a radius of $\sim$0.5~kpc;][]{2022ApJ...925...72T} drops substantially. Excluding the NGC~3351 inflow regions, $\alpha_\mathrm{CO}$ in the three galaxy centers vary between 0.2--1.5 $\mathrm{M_\odot\ (K~km~s^{-1}~pc^2)^{-1}}$. 
By comparing between Figure~\ref{fig:cross_radius}(a) and (b), it is clear that for each galaxy the radial variation of $\alpha_\mathrm{CO}$ and $\tau_\mathrm{CO(2-1)}$ are overall similar. We note that the spatial variations are consistent between $\tau_\mathrm{CO(2-1)}$ and $\tau_\mathrm{CO(1-0)}$, except that the values of $\tau_\mathrm{CO(2-1)}$ are generally higher than $\tau_\mathrm{CO(1-0)}$.
The higher $\tau_\mathrm{CO(2-1)}$ than $\tau_\mathrm{CO(1-0)}$ in our galaxy centers is likely caused by the higher density/temperature that efficiently excites CO to upper-$J$ and thus depopulates the lower-$J$ levels, and it is consistent with theoretical predictions for gas with $N_\mathrm{CO} \gtrsim 10^{17}$~cm$^{-2}$ \citep[e.g.,][]{2022ApJ...931...28H}.

Motivated by the resemblance of Figure~\ref{fig:cross_radius}(a) and (b), we further investigate the correlation between $\alpha_\mathrm{CO}$ and $\tau_\mathrm{CO}$. Figure~\ref{fig:alpha_tau_scatter} presents the pixel-based median solutions of $\alpha_\mathrm{CO}$ and $\tau_\mathrm{CO(2-1)}$ from all three galaxies, where we can see a tight, positive trend between $\alpha_\mathrm{CO}$ and $\tau_\mathrm{CO}$ in optically thick regions. On the other hand, the optically thin gas from the bar-driven inflows of NGC~3351 show substantially lower $\alpha_\mathrm{CO}$ with little dependence on $\tau_\mathrm{CO}$, which matches the expectation for relatively diffuse ($n_\mathrm{H_2} < 300$~cm$^{-2}$) gas in simulations \citep{2018ApJ...858...16G,2020ApJ...903..142G}. The positive correlation between $\alpha_\mathrm{CO}$ and $\tau_\mathrm{CO}$ agrees well with theoretical predictions for thermalized emission, where $\alpha_\mathrm{CO} \propto \tau/[1-\exp(-\tau)] \approx \tau$ is expected for optically thick emission with $\tau \gg 1$ \citep{2012ApJ...751...10P}. Since $\tau_\mathrm{CO}$ is by definition proportional to $N_\mathrm{CO}$ and $\Delta v$, this means that gas concentration toward galaxy centers (which increases $N_\mathrm{CO}$) and turbulence/shear effects (which changes $\Delta v$) play important roles in setting $\alpha_\mathrm{CO}$ in the central kpc of these barred galaxies. It is thus possible that the overall higher velocity dispersion in galaxy centers can lower the optical depth and lead to systematically lower $\alpha_\mathrm{CO}$ than the Galactic disk value across our maps (see Section~\ref{subsubsec:alpha_spec} for further discussion). 

While there is a strong $\alpha_\mathrm{CO}$ dependence on $\tau_\mathrm{CO}$, we also notice diverging $\alpha_\mathrm{CO}$ toward the nucleus ($r \lesssim 300$~pc) of NGC~3627, where $\alpha_\mathrm{CO}$ is increasing while $\tau_\mathrm{CO(2-1)}$ remains unchanged. This means that the $\alpha_\mathrm{CO}$ variation cannot be solely explained by $\tau_\mathrm{CO}$, and thus there must be additional factors at play. Theoretical studies have suggested that $\alpha_\mathrm{CO}$ may decrease with temperature as the optically thick CO 1--0 intensity increases with temperature \citep{2012MNRAS.421.3127N,co-to-h2,2022ApJ...931...28H}. We present the relation between our modeled $\log(\alpha_\mathrm{CO})$ and $\log(T_\mathrm{k})$ in Figure~\ref{fig:alpha_n_T}(a). While NGC~3351 does not show strong evidence for $\alpha_\mathrm{CO}$ varying with $T_\mathrm{k}$ \citep[see also][]{2022ApJ...925...72T}, we find a clear decrease of $\alpha_\mathrm{CO}$ with $T_\mathrm{k}$ in NGC~3627 and NGC~4321. Notably, the local peak of $\alpha_\mathrm{CO}$ for NGC~3627 (blue curve) near $\log(T_\mathrm{k}) = 2.1$ corresponds to the $\alpha_\mathrm{CO}$ increase in NGC~3627's nucleus, and $\alpha_\mathrm{CO}$ continues to drop in regions with even higher temperature. Though the nucleus in NGC~3627 already has high $T_\mathrm{k} \gtrsim 100$~K, the highest $T_\mathrm{k}$ actually occurs in regions surrounding the nucleus (see Figure~\ref{fig:1dmax_3627}(b)). This temperature drop toward the nucleus could explain why $\alpha_\mathrm{CO}$ rises while $\tau_\mathrm{CO}$ stays flat in Figure~\ref{fig:cross_radius}. 

Based on NGC~3627 and NGC~4321, the 25th--75th percentile scatter of $\alpha_\mathrm{CO}$ in the $\alpha_\mathrm{CO}$--$T_\mathrm{k}$ relation is $\sim$0.4~dex, which is larger than the $\sim$0.1~dex scatter in the $\alpha_\mathrm{CO}$--$\tau_\mathrm{CO}$ relation shown in Figure~\ref{fig:alpha_tau_scatter}. This suggests that optical depth and gas temperature effects contribute $\sim$80\% and 20\% of the change in the derived $\alpha_\mathrm{CO}$, respectively, assuming they are independent and no other factors are at play. In that case, $\tau_\mathrm{CO}$ is likely the main driver of $\alpha_\mathrm{CO}$ variation in these galaxy centers, while $T_\mathrm{k}$ plays a secondary role in changing $\alpha_\mathrm{CO}$. Using the results from all three galaxies but excluding the optically thin inflow regions of NGC~3351, we fit the $\alpha_\mathrm{CO}$, $\tau_\mathrm{CO}$, and $T_\mathrm{k}$ relation with a power law and obtain
\begin{equation}
\begin{aligned}
&\log\frac{\alpha_\mathrm{CO}}{\mathrm{M_\odot\ (K~km~s^{-1}~pc^2)^{-1}}} \\
&\quad = 0.78\ \log \tau_\mathrm{CO(2-1)} - 0.18\ \log \frac{T_\mathrm{k}}{\mathrm{K}} - 0.84~.
\end{aligned}
\label{eqn_alpha_fit}
\end{equation}     
By performing bootstrapping and refitting 1000 times, we determine an uncertainty of $\pm 0.03$ for the slopes with respect to either $\log\,\tau_\mathrm{CO(2-1)}$ or $\log\,T_\mathrm{k}$, and $\pm 0.08$ for the intercept. 

\begin{figure}
\centering
\includegraphics[width=\linewidth]{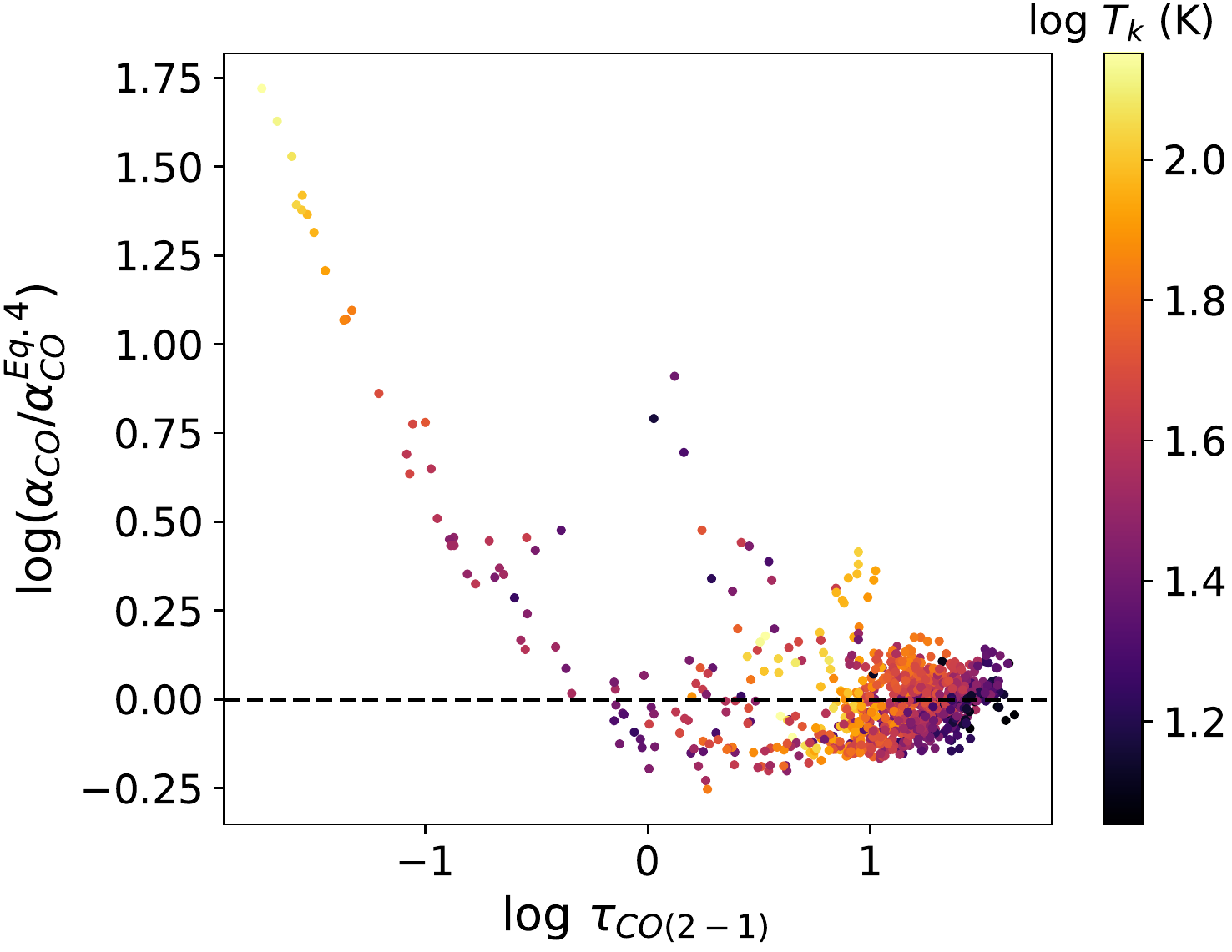}
\caption{Ratio of the modeled and fitting-predicted $\alpha_\mathrm{CO}$ (by Equation~\ref{eqn_alpha_fit}) versus the CO optical depth, colorcoded by the modeled gas temperature. The dashed line indicates perfect agreement between the modeled and predicted $\alpha_\mathrm{CO}$. The fitted Equation~\ref{eqn_alpha_fit} should be limited to optically thick regions, where the 25--75th percentile scatter along the y-axis is 0.12~dex.}
\label{fig:alpha_tau_T}
\end{figure}

Figure~\ref{fig:alpha_tau_T} illustrates how the ratio of $\alpha_\mathrm{CO}$ measured from our modeling and predicted by Equation~\ref{eqn_alpha_fit} varies with $\tau_\mathrm{CO}$ and $T_\mathrm{k}$. In the optically thick regime, the 25--75th percentile scatter is 0.12~dex, which is similar to that seen in the $\alpha_\mathrm{CO}$--$\tau_\mathrm{CO}$ relation (see Figure~\ref{fig:alpha_tau_scatter}). A rough inverse trend can be seen between $\tau_\mathrm{CO}$ and $T_\mathrm{k}$, which is expected as higher temperature can increase level population in high-$J$ transitions and decrease the optical depth of low-$J$ line emission. It is also clear that the power-law fit underestimates $\alpha_\mathrm{CO}$ in the optically-thin inflow regions of NGC~3351. Therefore, we emphasize that Equation~\ref{eqn_alpha_fit} should only be applied to optically thick regions.

We further remind readers that this paper focuses on disentangling the emissivity-related drivers of $\alpha_\mathrm{CO}$, and thus $x_\mathrm{CO}$ is assumed constant at a starburst value of $3\times10^{-4}$ over the entire region. This means that Equation~\ref{eqn_alpha_fit} should either be limited to starburst-like environments with higher $x_\mathrm{CO}$, or be adjusted by multiplying a factor of $3\times10^{-4}/x_\mathrm{CO}$. For instance, molecular clouds in the Milky Way disk can have $\gtrsim 3$ times higher $\alpha_\mathrm{CO}$ values than that predicted by Equation~\ref{eqn_alpha_fit}, as they normally have $x_\mathrm{CO} \lesssim 10^{-4}$ \citep{1982ApJ...262..590F,1987ApJ...315..621B,2002PhDT........28K,2008ApJ...687.1075S,2019MNRAS.484..305P}.

The spatial variation of $x_\mathrm{CO}$ may also affect the derived $\alpha_\mathrm{CO}$ variation. If $x_\mathrm{CO}$ in the arms is lower than in the nucleus, as expected from increasing CO-dark H$_2$ fraction with galactocentric radius due to decreasing gas surface density \citep[e.g.,][]{2014MNRAS.441.1628S}, then $\alpha_\mathrm{CO}$ in the arms of NGC~3627 would become similar to the nucleus having higher $\alpha_\mathrm{CO}$ values. Alternatively, increasing $x_\mathrm{CO}$ in the NGC~3627 nucleus is also possible via the enrichment of $^{12}$C through stellar nucleosynthesis from intermediate or high-mass stars. While it is typically expected that stronger cosmic ray ionization would decrease $x_\mathrm{CO}$ in starburst or AGN-host galaxy centers \citep{2020ApJ...903..142G,2021MNRAS.502.2701B,2022arXiv221209661L}, exceptions have been found in places reaching high gas temperature of $\sim$100~K due to the trigger of OH formation that further increases $x_\mathrm{CO}$ \citep{2017ApJ...839...90B}. Therefore, with the modeled $T_\mathrm{k} > 100$~K near the nucleus of NGC~3627, the potential rise of $x_\mathrm{CO}$ could also lead us to overestimate $\alpha_\mathrm{CO}$, implying that $\alpha_\mathrm{CO}$ in the nucleus may not be distinctly higher than the arm regions. However, we emphasize again that our modeling cannot constrain the absolute $x_\mathrm{CO}$ values, and thus the net change of $x_\mathrm{CO}$ is still to be studied in more detail with the comprehensive effects mentioned above.

To investigate whether the theoretical expectation of $\alpha_\mathrm{CO} \propto \sqrt{n_\mathrm{H_2}}$/$T_\mathrm{k}$ under the virial assumption also holds in the three galaxy centers, Figure~\ref{fig:alpha_n_T}(b) shows the relation between the modeled $\log(\alpha_\mathrm{CO})$ and $\log(\sqrt{n_\mathrm{H_2}}$/$T_\mathrm{k})$. It is clear that both NGC~3627 and NGC~4321 show a positive correlation of $\alpha_\mathrm{CO}$ with $\sqrt{n_\mathrm{H_2}}/T_\mathrm{k}$, which may indicate that the molecular clouds are overall close to virial balance or have a similar virial parameter. However, a similarly high virial parameter should be more likely in our case, given that previous studies already reported high virial parameters of 2--10 for GMCs in both galaxy centers assuming a Galactic-like or metallicity-dependent $\alpha_\mathrm{CO}$ \citep{2017ApJ...839..133P,2021MNRAS.502.1218R}. The high virial parameter in these galaxy centers may indicate unbound molecular clouds that could suppress star formation \citep[e.g.,][]{2012PASJ...64...51S,2013MNRAS.429.2175N}. On the other hand, NGC~3351 shows the highest $\sqrt{n_\mathrm{H_2}}/T_\mathrm{k}$ values due to generally lower $T_\mathrm{k}$ and higher $n_\mathrm{H_2}$, and the NGC~3351 inflows (data points in the bottom-right corner) are strong dynamical feature with optically thin CO emission \citep{2022ApJ...925...72T} and thus do not match the assumption for $\sqrt{n_\mathrm{H_2}}/T_\mathrm{k}$ dependence \citep{co-to-h2,2020ApJ...903..142G}. We do not see a clear correlation in the center of NGC~3351, which shows a roughly constant $\alpha_\mathrm{CO}$ that is lower than the positive trend formed by the other two galaxies. This lower $\alpha_\mathrm{CO}$ could be explained by the increased turbulence and shear near the NGC~3351 inflows, making the clouds there super-virial.  

In summary, we find a strong, positive $\alpha_\mathrm{CO}$ dependence on $\tau_\mathrm{CO}$ after combining the modeling results of the central kpc of NGC~3351, NGC~3627, and NGC~4321. This correlation is in line with theoretical expectations for thermalized and optically thick clouds, and it can explain most of the $\alpha_\mathrm{CO}$ variations found in the three galaxy centers. Additionally, an anti-correlation between $\alpha_\mathrm{CO}$ and $T_\mathrm{k}$ is clearly seen in NGC~3627 and~4321, suggesting $T_\mathrm{k}$ as a secondary driver of $\alpha_\mathrm{CO}$ variation after $\tau_\mathrm{CO}$. The $\alpha_\mathrm{CO}$ in NGC~3627 and~4321 also shows a positive but weaker correlation with $\sqrt{n_\mathrm{H_2}}/T_\mathrm{k}$, which suggests that the molecular clouds in those regions have similar (likely high) virial parameters.

\subsection{Comparison to $\alpha_\mathrm{CO}$ Measurements in Literature} \label{subsec:alpha_literature}

To compare our $\alpha_\mathrm{CO}$ results with previous measurements on kpc scales \citep{2013ApJ...777....5S,2020A&A...635A.131I}, we calculate the intensity-weighted mean $\alpha_\mathrm{CO}$ over the observed regions. Based on Equation~\ref{eqn_alpha}, we compute the average and standard deviation of 2000 likelihood-weighted random draws of $N_\mathrm{CO}$, $\Phi_\mathrm{bf}$, and $I_\mathrm{CO}$ from the full model grid for each pixel. The procedure is described in \citet[Section~5.1]{2022ApJ...925...72T} in greater detail. 
Since the $\alpha_\mathrm{CO}$ values in \citet{2013ApJ...777....5S} were derived from CO 2--1 intensities assuming a constant $R_\mathrm{21}$ of 0.7, we will directly compare the $\alpha_\mathrm{CO(2-1)}$ values to avoid uncertainties originating from $R_\mathrm{21}$. This means that $I_\mathrm{CO(1-0)}$ in Equation~\ref{eqn_alpha} will be replaced with $I_\mathrm{CO(2-1)}$ when we derive the intensity-weighted $\alpha_\mathrm{CO}$ for comparison to the dust-based results.

The intensity-weighted mean $\alpha_\mathrm{CO(2-1)}$ is $0.93 \pm 0.04$ and $0.62 \pm 0.04$ $\mathrm{M_\odot\ (K~km~s^{-1}~pc^2)^{-1}}$ over the central $\sim$kpc of NGC~4321 and NGC~3627 included in our analysis. 
Using dust modeling and CO 2--1 observations, \citet{2013ApJ...777....5S} derived $\alpha_\mathrm{CO(2-1)} = 0.9^{+0.4}_{-0.3}$ and $0.8^{+0.3}_{-0.1}$~$\mathrm{M_\odot\ (K~km~s^{-1}~pc^2)^{-1}}$ in the central 2.6~kpc and 1.7~kpc region of NGC~4321 and NGC~3627, respectively. 
For NGC~4321, our $\alpha_\mathrm{CO(2-1)}$ value is consistent with their dust-based estimate, as well as the carbon budget-based estimate of $\alpha_\mathrm{CO(2-1)} \sim$0.96 by \citet{2020A&A...635A.131I} towards the central $22''$ region after applying our integrated mean $R_\mathrm{21} = 0.92$ in Table~\ref{tab:ratio_4321}. We note that \citet{2020A&A...635A.131I} also reported $T_\mathrm{k} > 100$~K with $X_{12/13}=80$ for the center of NGC~4321 using a two-component model. Over the central kpc region of NGC~3351, the mean $\alpha_\mathrm{CO(2-1)}$ of 0.75 is also consistent with the \citet{2013ApJ...777....5S} estimate of $1.0^{+0.4}_{-0.3}$ (see Appendix~\ref{sec:alpha_3351}).
  
For NGC~3627, our derived mean $\alpha_\mathrm{CO(2-1)}$ of 0.62 is slightly lower than 0.8 from \citet{2013ApJ...777....5S}, while it is higher than $\alpha_\mathrm{CO(2-1)} \sim$0.43 determined by \citet{2020A&A...635A.131I} applying our integrated mean $R_\mathrm{21}$ of 0.81 in Table~\ref{tab:ratio_3627}.
With assumptions on dust-to-gas ratios and applying our $R_\mathrm{21} = 0.81$, a recent work by \citet{2021MNRAS.504.2360J} also suggests $\alpha_\mathrm{CO(2-1)} = 0.99 \pm 0.37$, which overlaps with the solutions from this work and \citet{2013ApJ...777....5S}.
One potential explanation for the discrepancy between \citet{2013ApJ...777....5S} and our result is a calibration issue of the HERACLES CO 2--1 data used by \citet{2013ApJ...777....5S}. 
As shown in \citet[Appendix C]{2021MNRAS.504.3221D}, the HERACLES data of NGC~3627 has been found to have significant calibration uncertainties with up to a factor of two lower intensity than the PHANGS--ALMA data we use. This implies that the $\alpha_\mathrm{CO}$ solution determined by \citet{2013ApJ...777....5S} could be overestimated using the HERACLES data with fainter CO emission. 

Moreover, it is also possible that our modeling overestimates $\alpha_\mathrm{CO}$ in NGC~3627 due to the underestimation of $T_\mathrm{k}$. Figure~\ref{fig:1dmax_3627}(b) shows that many regions in NGC~3627 have high $T_\mathrm{k}$ that exceed few hundreds K, potentially due to the AGN in its nucleus.
Since our modeled lines only include transitions up to $J$=3--2, such line combination may not be sufficient to reveal temperatures above a few hundred K. We have also tested regions with $T_\mathrm{k} > 100$~K using the two-component model constructed by \citet{2022ApJ...925...72T}, and still find $\gtrsim$100~K for the dominant component. Additionally, we have checked the spectral line energy distribution (SLED) of CO in all three galaxy centers using Herschel SPIRE/FTS data at 40$\arcsec$ resolution covering up to CO $J$=9--8 (A.~Crocker, private communication). We find that the SLED of NGC~3627 is peaked in higher-$J$ lines than the other two galaxy centers, which also supports the scenario of higher $T_\mathrm{k}$ in the center of NGC~3627. Thus, our modeling could have underestimated $T_\mathrm{k}$ in NGC~3627, and higher-$J$ CO lines may be needed to accurately constrain such high $T_\mathrm{k}$. If the center of NGC~3627 in fact has higher $T_\mathrm{k}$ than what we derived, this could lead to overestimation of $\alpha_\mathrm{CO}$ as long as $n_\mathrm{H_2}$ does not deviate much from our modeling result \citep{2012ApJ...751...10P,co-to-h2}.
The $\alpha_\mathrm{CO}$ estimate from \citet{2020A&A...635A.131I} also has the issue of lacking high-$J$ CO lines, and the author reported $T_\mathrm{k} \lesssim 60$~K in the center of NGC~3627 which is even lower than our results and inconsistent with the bright emission seen in high-$J$ transitions. 

We conclude that the overall $\alpha_\mathrm{CO}$ in the central kpc of NGC~3627 is likely a factor of 5--10 lower than the Galactic $\alpha_\mathrm{CO}$, while the actual value is still uncertain as seen from the inconsistency among \citet{2013ApJ...777....5S}, \citet{2020A&A...635A.131I}, \citet{2021MNRAS.504.2360J}, and this work. High-resolution observations toward high-$J$ CO transitions will be essential to securely measure the environmental conditions and $\alpha_\mathrm{CO}$ in this region. On the other hand, the $\alpha_\mathrm{CO}$ values in the center of NGC~4321 match perfectly well among these studies using independent methods, which increases the reliability of the NGC~4321 results.

\subsection{Observational Tracers for $\alpha_\mathrm{CO}$ Variations} \label{subsec:alpha_obs}

\begin{figure*}
\begin{minipage}{.5\linewidth}
\centering
\includegraphics[width=\linewidth]{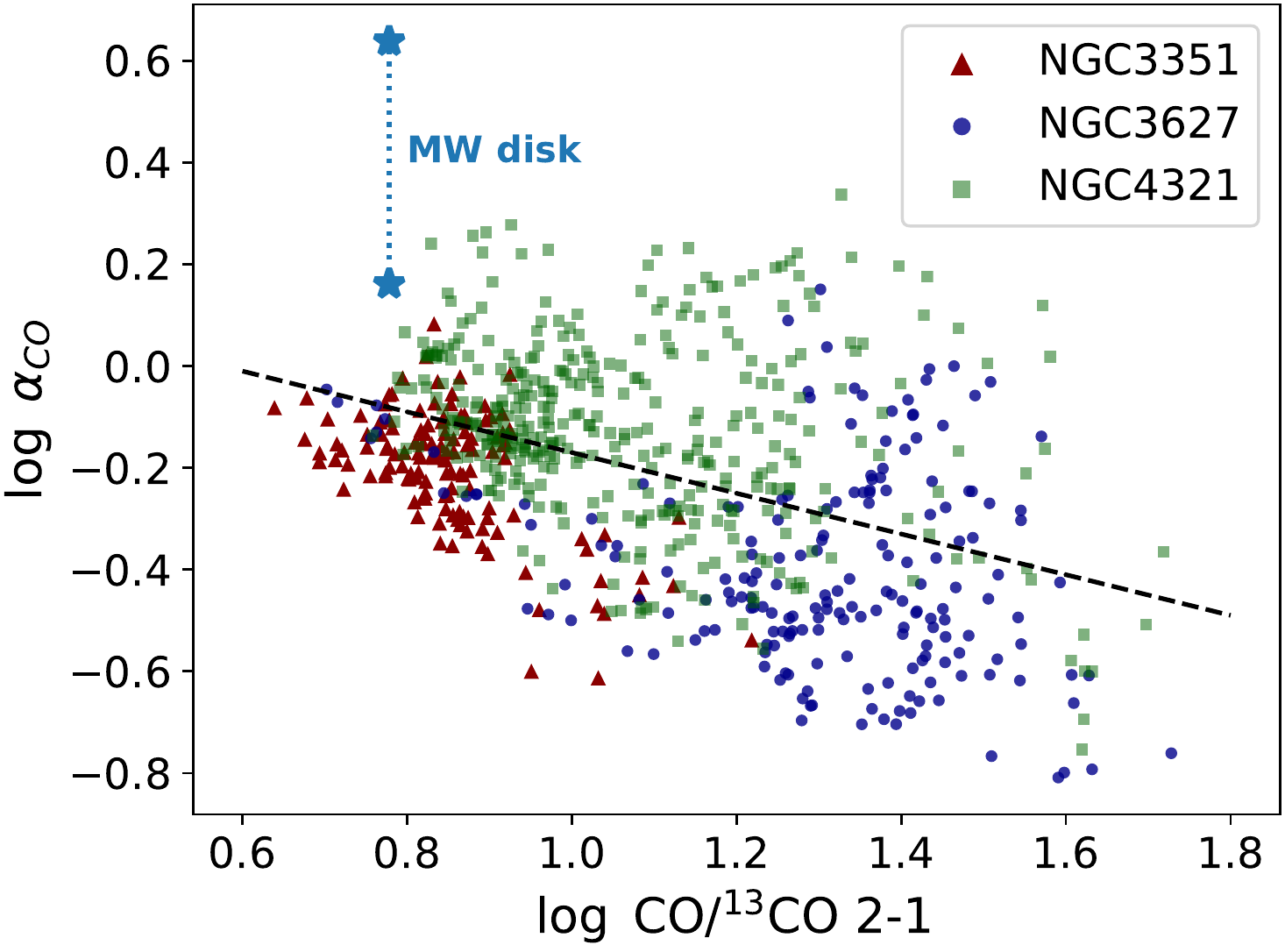}\\
(a)
\end{minipage}
\hfill
\begin{minipage}{.485\linewidth}
\centering
\includegraphics[width=\linewidth]{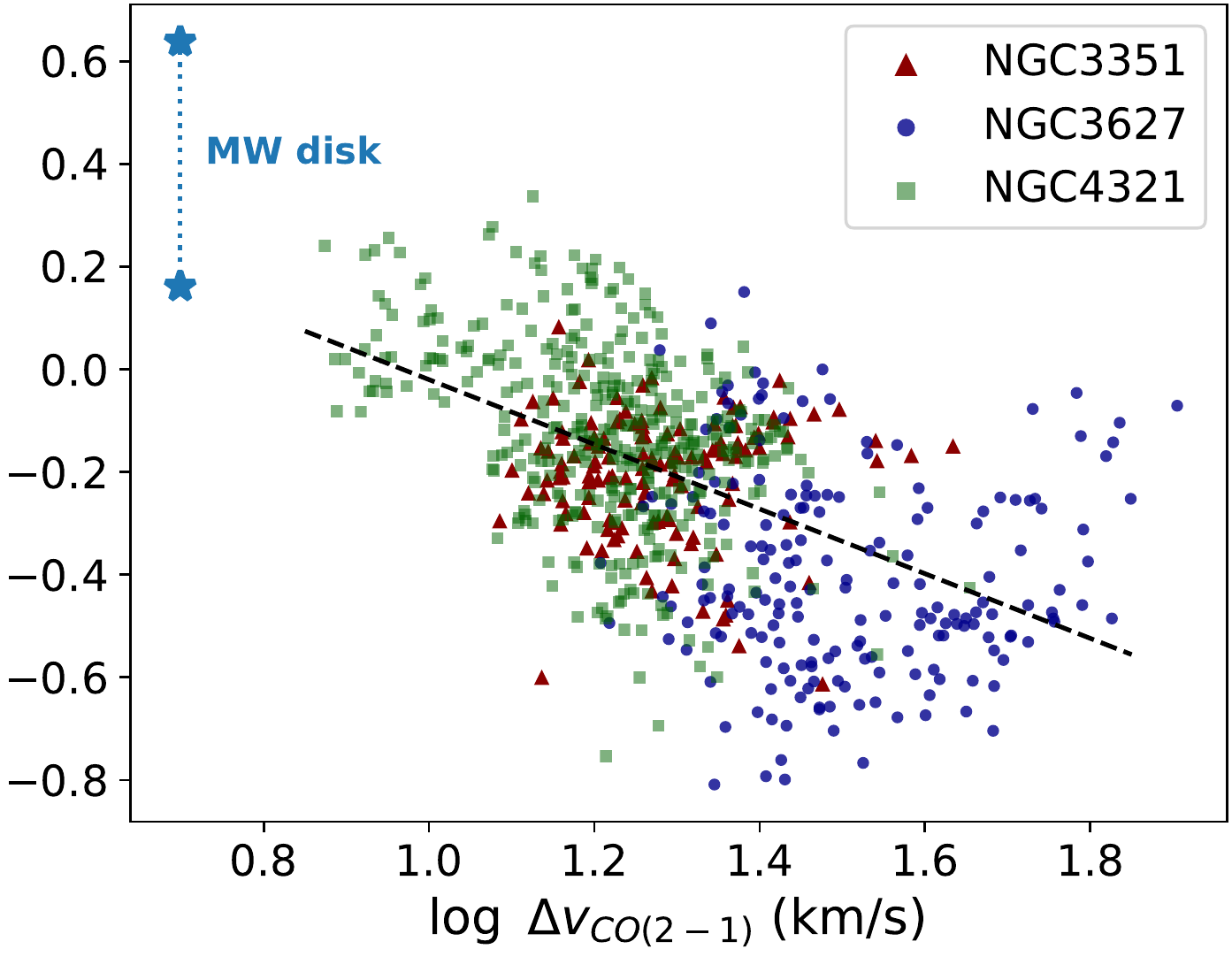}\\
(b)
\end{minipage}
\caption{Relation of the modeled $\log(\alpha_\mathrm{CO})$ with the observed (a) CO/$^{13}$CO 2--1 line ratio and (b) CO 2--1 effective line width in optically thick regions with $\tau_\mathrm{CO(2-1)} > 5$. The dashed lines represent the best-fit power law relations (Equations~\ref{eqn_ratio21_fit} and~\ref{eqn_ew21_fit}). The typical Milky Way disk values with $x_\mathrm{CO}$ ranging from ${\sim} 10^{-4}$ (higher $\alpha_\mathrm{CO}$) to $3\times10^{-4}$ (lower $\alpha_\mathrm{CO}$) are labeled by the blue stars. The data points cover all regions except the inflow regions of NGC~3351. Despite a $\sim$0.4~dex scatter in both relations, there is a clear trend of $\alpha_\mathrm{CO}$ decreasing with the line ratio and CO line width, suggesting these observable properties as potential tracers for $\alpha_\mathrm{CO}$ variations in galaxy centers.} 
\label{fig:alpha_tracer}
\end{figure*}

\subsubsection{The $\mathrm{CO/^{13}CO}$ Line Ratios} \label{subsubsec:alpha_ratio}

As shown in Figure~\ref{fig:alpha_tau_scatter}, all three galaxy centers show a clear correlation between $\alpha_\mathrm{CO}$ and $\tau_\mathrm{CO}$.
In Figure~\ref{fig:alpha_tau_scatter}(b), the colors represent the median CO/$^{13}$CO 2--1 ratios within each (two-dimensional) bin of $\alpha_\mathrm{CO}$ and $\tau_\mathrm{CO}$. The adopted bin size is 0.1 in both $\log(\alpha_\mathrm{CO})$ and $\log(\tau_\mathrm{CO})$ dimensions, which is consistent with the bin size we adopted when deriving the PDFs and solutions for $\log(\alpha_\mathrm{CO})$ and $\log(\tau_\mathrm{CO})$. The median-filtered line ratios shown by the color gradient is a visualization choice to emphasize the overall trend of the line ratio with $\tau_\mathrm{CO}$ or $\alpha_\mathrm{CO}$. We find that the color-coded CO/$^{13}$CO 2--1 ratios form a gradient across the parameter space when $\tau_\mathrm{CO} > 1$, suggesting an anti-correlation of the line ratio with $\tau_\mathrm{CO}$ or $\alpha_\mathrm{CO}$.  
It is also clear that the CO/$^{13}$CO ratio in these galaxy centers is higher than the Galactic disk-like ratio of $\sim$6 \citep{1995A&A...300..369A,2016ApJ...818..144R}, which is consistent with the finding of elevated CO/$^{13}$CO ratios in LIRGs or central starburst regions \citep{1995A&A...300..369A,2010A&A...522A..59A,2012ApJ...753...46S,2014ApJ...796L..15S,2017ApJ...840....8S}.

The inverse relation between $\alpha_\mathrm{CO}$ and the CO/$^{13}$CO 2--1 ratio is also clearly demonstrated by Figure~\ref{fig:alpha_tracer}(a). Here we only include optically thick regions with $\tau_\mathrm{CO(2-1)} > 5$, where $\alpha_\mathrm{CO}$ strongly depends on $\tau_\mathrm{CO}$. 
With $\tau_\mathrm{CO} < 30$ and $X_{12/13} > 40$ across our measurements, we obtain $^{13}$CO optical depth that is solidly in the optically thin regime. Therefore, the correlations suggested by Figures~\ref{fig:alpha_tau_scatter}(b) and \ref{fig:alpha_tracer}(a) agree with the interpretation that the observed CO/$^{13}$CO 2--1 ratio is generally tracing $\tau_\mathrm{CO(2-1)}$ variations inversely in the three galaxy centers. This is because the decrease of $\tau_\mathrm{CO}$ can lead to more escaped CO emission and thus increasing the CO/$^{13}$CO ratio when CO is optically thick and $^{13}$CO is optically thin. 
Since there is also a strong correlation between $\alpha_\mathrm{CO}$ and $\tau_\mathrm{CO}$, this implies that the CO/$^{13}$CO ratio may be used as an observational tracer for $\alpha_\mathrm{CO}$ variation. As indicated by the dashed line on Figure~\ref{fig:alpha_tracer}(a), we conduct a power law fit to the data points and find
\begin{equation}
\log\frac{\alpha_\mathrm{CO}}{\mathrm{M_\odot\ (K~km~s^{-1}~pc^2)^{-1}}} = -0.40\ \log R_\mathrm{12/13} + 0.23~,
\label{eqn_ratio21_fit}
\end{equation}
where $R_\mathrm{12/13}$ is the observed CO/$^{13}$CO 2--1 line ratio, and both the fitted slope and intercept have an uncertainty of $\pm 0.03$. Similar to Equation~\ref{eqn_alpha_fit}, this fitted relation is only appropriate for starburst-like regions with a higher CO abundance $x_\mathrm{CO}$, unless the predicted value is further scaled by a factor of $3\times10^{-4}/x_\mathrm{CO}$. This scaling of $x_\mathrm{CO}$ can explain the factor of 3--4 discrepancy between the fit (with $x_\mathrm{CO} = 3\times10^{-4}$) and the typical Galactic disk $\alpha_\mathrm{CO}$ value (with $x_\mathrm{CO} \lesssim 10^{-4}$) as shown in Figure~\ref{fig:alpha_tracer}(a). 
There is a dispersion of $\sigma \sim 0.2$~dex between the modeled and fitting-predicted $\alpha_\mathrm{CO}$, which likely originates from the uncertainty in $X_{12/13}$ variation as well as the exclusion of temperature effects.  

The CO/$^{13}$CO ratio should also vary with the molecular abundance $X_{12/13}$, which is one of our directly modeled parameters. As presented in Section~\ref{subsec:model_result}, most regions show 1DMax $X_{12/13}$ solutions consistent with the best-fit solutions at $X_{12/13} \sim$80--100 (e.g., compare Figure~\ref{fig:flux_contour_4321}(a) with Figure~\ref{fig:corner_center_4321}). Even in several regions with 1DMax $X_{12/13} {\sim} 40$, their median $X_{12/13}$ also show higher $X_{12/13} \sim$~80--100 that is similar to their best-fit solutions. Thus, even though the $X_{12/13}$ PDFs are generally not as well constrained as other parameters, it is likely that most regions have $X_{12/13} \sim$80--100 based on the match between the 1DMax/median and best-fit solutions. Moreover, the $X_{13/18}$ abundances are roughly constant and well-constrained at 6--8 over both galaxies, which implies that significant spatial variations in $X_{12/13}$ is unlikely from a nucleosynthesis perspective as enrichment of both $^{12}$C and $^{18}$O would be expected from massive stars. Therefore, $X_{12/13}$ may not be the main driver of the CO/$^{13}$CO line ratio variations. The roughly constant $X_{12/13}$ and varying CO optical depths in these galaxy centers can explain why the CO/$^{13}$CO 2--1 line ratio is overall reflecting $\tau_\mathrm{CO}$ variations in Figures~\ref{fig:alpha_tau_scatter}(b) and \ref{fig:alpha_tracer}(a). 

In \citet{2022ApJ...925...72T}, the bar-driven inflows of NGC~3351 shows enhanced CO/$^{13}$CO 2--1 ratio with nearly optically thin CO emission, which also suggests the inverse relation between $\tau_\mathrm{CO}$ and CO/$^{13}$CO line ratio. Notably, their stacking result for the inflow regions revealed well-constrained $X_{12/13}$ PDFs showing 1DMax and median $X_{12/13} \sim$30. Since the value is similar to that found in the central regions of NGC~3351, it provided evidence for $\tau_\mathrm{CO}$ being the main driver of the CO/$^{13}$CO line ratio. 
Furthermore, \citet{2018MNRAS.475.3909C} also reported anti-correlations of $\alpha_\mathrm{CO}$ with CO/$^{13}$CO 1--0 ratio across the disks of several galaxies, using dust-based $\alpha_\mathrm{CO}$ (from \citealt{2013ApJ...777....5S}) with single-dish CO observations at $\sim$1.5~kpc resolutions. Interestingly, such an anti-correlation was only seen in the three galaxies hosting starburst-dominated nuclei in their sample, but not in the other five normal star-forming galaxies. This can be explained by the increased optical depth variation in starburst environments. Similarly, barred galaxy centers tend to have variable gas dynamics and conditions due to higher excitation, turbulence, shear, and gas concentration, which altogether can lead to even more significant $\tau_\mathrm{CO}$ variations. Within the three barred galaxy centers presented in this work, we find that $\alpha_\mathrm{CO}$ is positively correlated with $\tau_\mathrm{CO}$, and that the CO/$^{13}$CO 2--1 line ratio mainly traces the $\tau_\mathrm{CO}$ variation. These results suggest that the CO/$^{13}$CO ratio can be a useful observational tracer for $\alpha_\mathrm{CO}$ variation, particularly in galaxy centers where optical depth is generally high and spans a wide dynamic range.

\subsubsection{Spectral Line Widths and Peak Temperatures} \label{subsubsec:alpha_spec}

\begin{figure*}
\begin{minipage}{.5\linewidth}
\centering
\includegraphics[width=\linewidth]{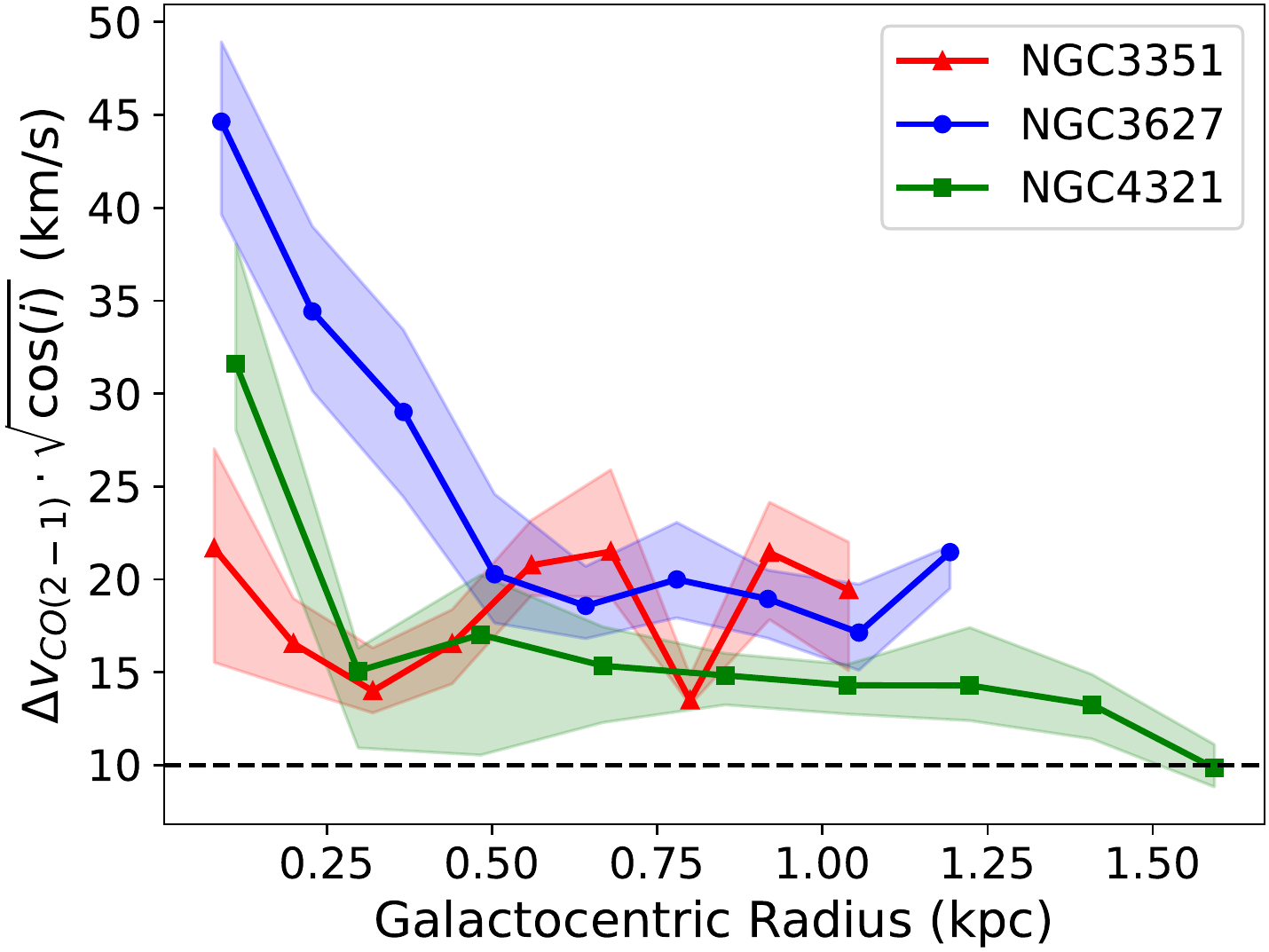}\\
(a)
\end{minipage}
\hfill
\begin{minipage}{.485\linewidth}
\centering
\includegraphics[width=\linewidth]{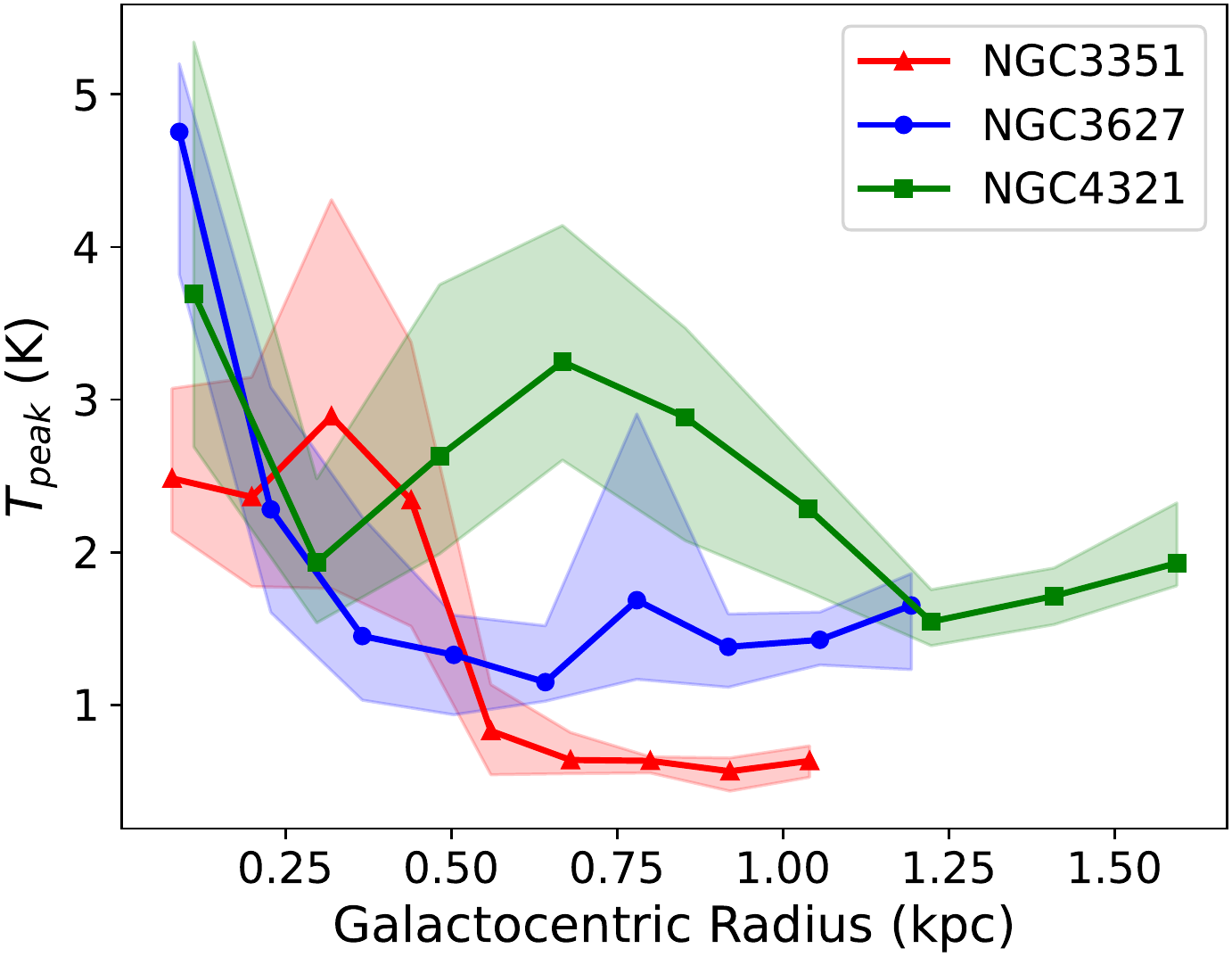}\\
(b)
\end{minipage}
\caption{Galactocentric radial profiles of the CO 2--1 (a) effective line widths with a $\sqrt{\cos(i)}$ inclination correction and (b) spectral peak intensities for NGC~3351 (red), NGC~3627 (blue), and NGC~4321 (green). Colored lines show the radial-binned medians and shaded areas show the 25th and 75th percentile ranges. The horizontal dashed line represents an approximate boundary of velocity dispersion that distinguishes between gas in barred centers and in disks or unbarred centers \citep{2020ApJ...901L...8S}.}
\label{fig:spec_radius}
\end{figure*}

With spectroscopic observations, the line width ($\Delta v$) and brightness temperature at the line peak ($T_\mathrm{peak}$) provide two direct observables that may contain information about gas properties. For optically thick lines like CO, $T_\mathrm{peak}$ can be a probe of the excitation temperature ($T_\mathrm{ex}$) if the beam-filling factor is known or fixed. On the other hand, line width represents the one-dimensional velocity dispersion, which is indicative of turbulent motions. Since the line center optical depth is a function of surface density and velocity dispersion, variations in line widths may also provide hints for optical depth changes.

As we have already shown the strong $\alpha_\mathrm{CO}$ dependence on $\tau_\mathrm{CO}$, it is likely that the observed $\Delta v$ can also trace the $\alpha_\mathrm{CO}$ variations. Figure~\ref{fig:alpha_tracer}(b) presents a scatter plot of the modeled $\alpha_\mathrm{CO}$ versus the observed CO 2--1 line width for regions with $\tau_\mathrm{CO(2-1)} > 5$. It is clear that $\alpha_\mathrm{CO}$ decreases with $\Delta v$, consistent with the expectation of $\alpha_\mathrm{CO}$ increasing with $\tau_\mathrm{CO}$. The power law fit is presented by the dashed line, indicating 
\begin{equation}
\log\frac{\alpha_\mathrm{CO}}{\mathrm{M_\odot\ (K~km~s^{-1}~pc^2)^{-1}}} = -0.63\ \log \frac{\Delta v_\mathrm{CO}}{\mathrm{km\ s^{-1}}} + 0.61~,
\label{eqn_ew21_fit}
\end{equation}
where the uncertainty is $\pm 0.04$ and $\pm 0.05$ for the fitted slope and intercept, respectively. The dispersion with respect to this prediction is $\sigma \sim 0.2$~dex, which is reasonable as the surface density term in the optical depth and the temperature are also included as parameters in our modeled $\alpha_\mathrm{CO}$. The stars in Figure~\ref{fig:alpha_tracer}(b) indicate the Galactic $\alpha_\mathrm{CO}$ values at $\Delta v = 5$~$\mathrm{km\ s^{-1}}$, which is typical for Galactic disk clouds with size of $\sim$100~pc \citep[e.g.,][]{2015ARA&A..53..583H}. We emphasize that the line-of-sight $\Delta v$ is what relates directly to $\tau_\mathrm{CO}$, and thus there is no need to correct $\Delta v$ for inclination effects among different galaxies.

Figure~\ref{fig:spec_radius} shows the radial profiles of the observed CO 2--1 line width and $T_\mathrm{peak}$ in NGC~3351, NGC~3627, and NGC~4321. Here we multiply the observed line width by a $\sqrt{\cos(i)}$ factor to eliminate the line width dependency on galaxy inclination, following the empirical correction found by \citet{2022AJ....164...43S} based on data with similar resolution of $150$~pc. This correction is only applied here to bring out the line width effects from small-scale turbulence or large-scale dynamical processes, ensuring a fair comparison among different galaxies. With the inclination correction, $\Delta v$ in the three galaxies becomes roughly aligned at radii beyond 500~pc. We find a significant increase in line width toward the nuclei ($r \lesssim 200$~pc) of NGC~3627 and NGC~4321, which is consistent with increased velocity dispersion being the cause of lower optical depths seen in Figure~\ref{fig:cross_radius}(b). The increase of line width in the inflow regions of NGC~3351 ($r \gtrsim 500$~pc) is also notable, reaching comparable values to its central nucleus and being higher than in the other two galaxies. Contrary to most situations where $T_\mathrm{peak}$ dominates the integrated intensity variation \citep[e.g.,][]{2022ApJ...935...64E}, $T_\mathrm{peak}$ is consistently low in the NGC~3351 inflows, and thus the enhanced velocity dispersion plays a more important role in the observed CO emission of this region.

Interestingly, all the mentioned regions with enhanced velocity dispersion are places where abrupt changes in $\alpha_\mathrm{CO}$ are found (see Figure~\ref{fig:cross_radius}(a)). Furthermore, \citet{2020ApJ...901L...8S} found that molecular gas in barred galaxy centers tends to have higher velocity dispersion than in galaxy disks or non-barred galaxy centers, and they can be distinguished by an approximate boundary of $\Delta v = 10$~km/s (without inclination correction). As shown in Figure~\ref{fig:spec_radius}(a), almost all pixels in the three galaxy centers have $\Delta v > 10$~km/s even after inclination correction. The average line width in our galaxy centers (see Tables~\ref{tab:ratio_4321} and~\ref{tab:ratio_3627}) are also 3--5 times higher than that of the galaxy disk sample in \citet{2020ApJ...901L...8S}. Such higher velocity dispersion can lead to lower optical depths in galaxy centers, and may explain the overall lower-than-disk $\alpha_\mathrm{CO}$ across the whole central kpc regions \citep{2013ApJ...777....5S}.
Notably, this scenario of higher velocity dispersion lowering $\alpha_\mathrm{CO}$ in galaxy centers is compatible with the kpc-scale $\alpha_\mathrm{CO}$ dependence on stellar mass surface density found by \citet{co-to-h2} and I-D. Chiang et al.~(in preparation). This is because the stellar mass surface density can track additional external pressure from the ISM that sets the high velocity dispersion in galaxy centers. 
Therefore, in addition to the CO/$^{13}$CO line ratio, the observed line width may also be useful in predicting $\alpha_\mathrm{CO}$ changes due to its relation with optical depth. 

\begin{figure*}
\begin{minipage}{.34\linewidth}
\centering
\includegraphics[width=\linewidth]{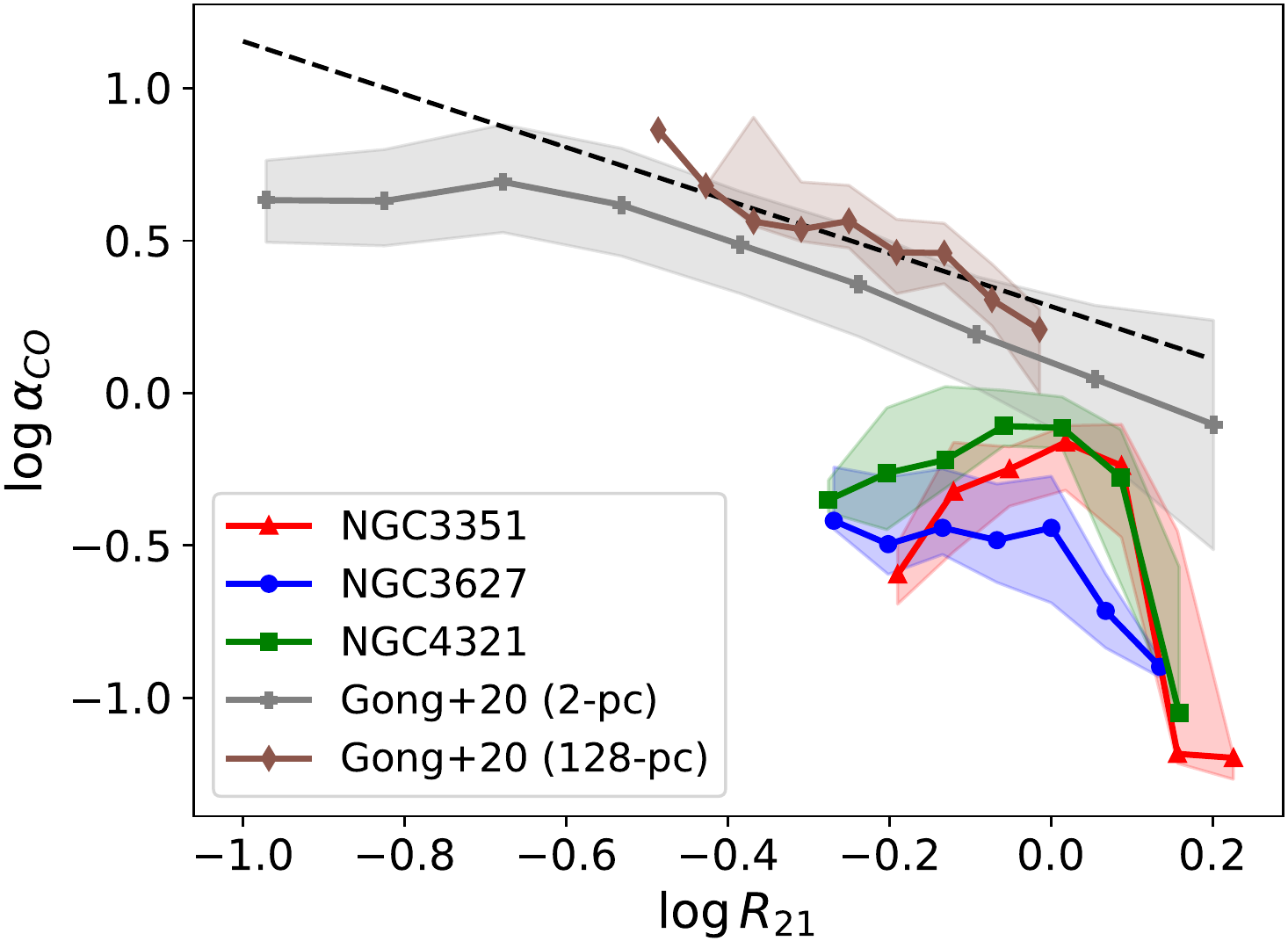} \\
(a)
\end{minipage} 
\begin{minipage}{.325\linewidth}
\centering
\includegraphics[width=\linewidth]{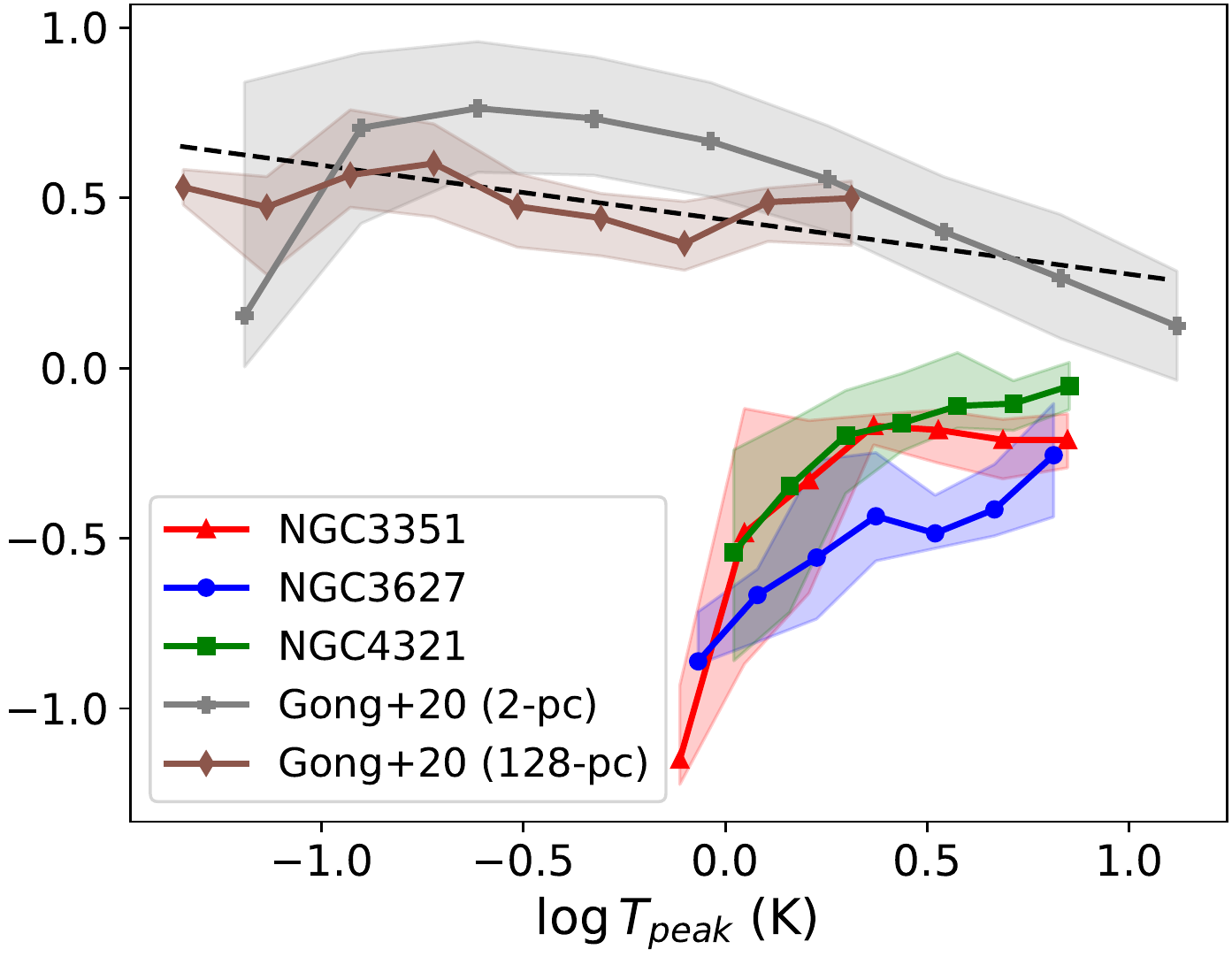} \\
(b)
\end{minipage}
\begin{minipage}{.325\linewidth}
\centering
\includegraphics[width=\linewidth]{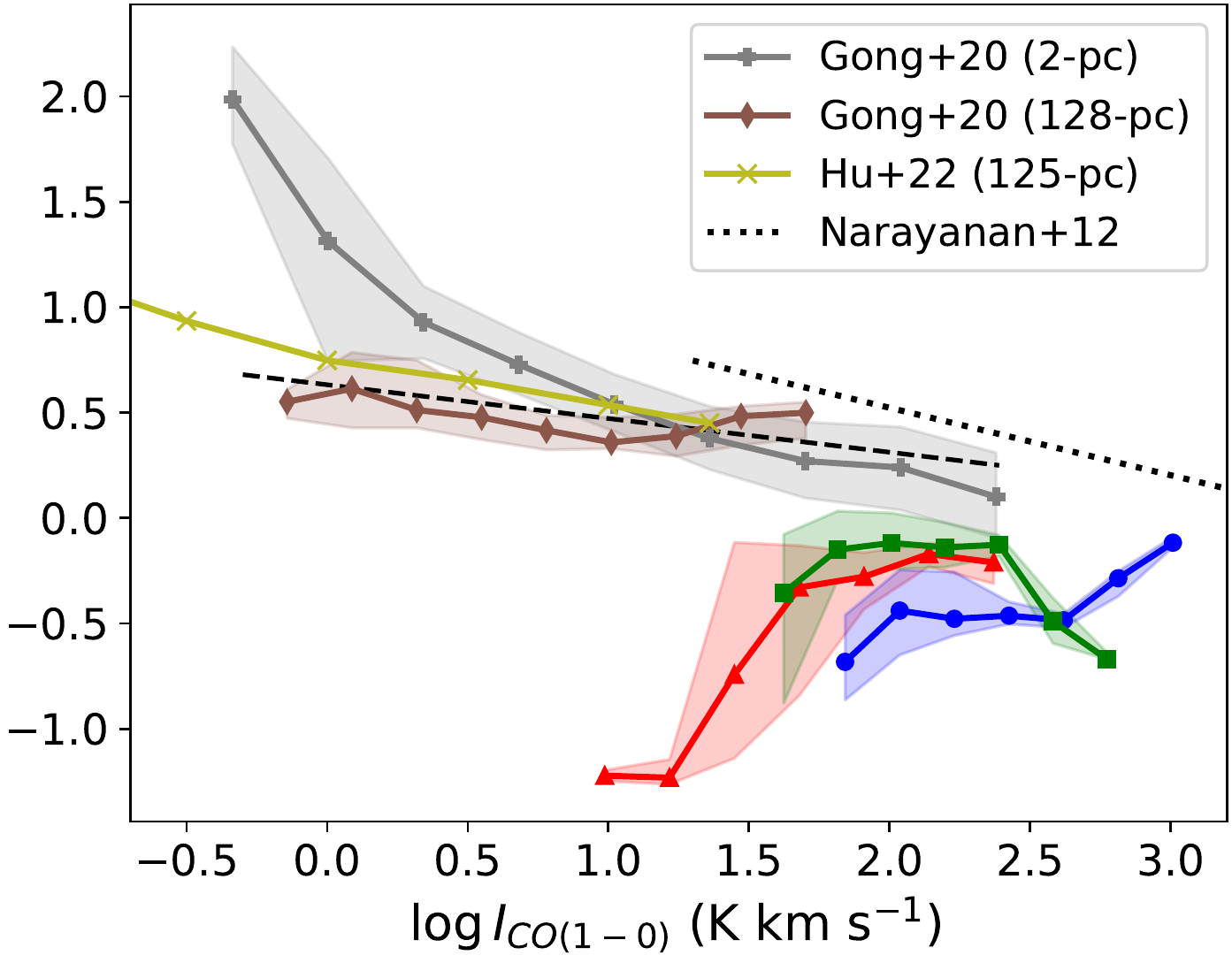} \\
(c)
\end{minipage}
\caption{$\alpha_\mathrm{CO}$ relations with the observed (a) CO 2--1/1--0 ratio, (b) peak temperature of CO 1--0, and (c) CO 1--0 integrated intensity, comparing our observations (colored lines as before) with the simulations. The solid lines and shaded areas indicate the binned medians and 25th/75th percentile ranges. The brown (gray), solid lines show the simulated data at 128-pc (2-pc) resolution from \citet{2020ApJ...903..142G}, and the dashed lines represent their suggested prescriptions with a 100~pc beam size. In panel (c), the dotted line and the yellow line show the simulation-based predictions suggested by \citet{2012MNRAS.421.3127N} and \citet{2022ApJ...931...28H}, respectively. The mismatch of gas conditions such as CO excitation and velocity dispersion between the observation and simulations may explain why $\alpha_\mathrm{CO}$ values in the galaxy centers are lower than that predicted by simulations probing Galactic disk-like environments.}
\label{fig:alpha_prediction}
\end{figure*}

Another potential observational tracer for $\alpha_\mathrm{CO}$ is $T_\mathrm{peak}$, which can be indicative of the excitation temperature $T_\mathrm{ex}$ as well as the total integrated intensity when CO is optically thick. However, we do not find the observed $T_\mathrm{peak}$ tracking the modeled $\alpha_\mathrm{CO}$, even though $\alpha_\mathrm{CO}$ is found to anti-correlate with $T_\mathrm{k}$ (see Section~\ref{subsec:alpha_env}). This means that $T_\mathrm{peak}$ is not a good indicator of $T_\mathrm{k}$ in our case, which can be due to deviation from LTE in most regions as we generally find $T_\mathrm{ex} < T_\mathrm{k}$. As shown in Figure~\ref{fig:spec_radius}(b), the radial variation of the observed $T_\mathrm{peak}$ differs from that of $\alpha_\mathrm{CO}$ in Figure~\ref{fig:cross_radius}(a). The relation between $\alpha_\mathrm{CO}$ and the observed $T_\mathrm{peak}$ is presented in Figure~\ref{fig:alpha_prediction}(b) in Section~\ref{subsubsec:alpha_prescription}.

In this work, we find that the sub-kpc scale $\alpha_\mathrm{CO}$ variation in galaxy centers is dominated by $\tau_\mathrm{CO}$ variations, which can be reflected by the observed CO/$^{13}$CO line ratio as well as the CO line width. While we also find a secondary effect of $T_\mathrm{k}$ on $\alpha_\mathrm{CO}$, the observed $T_\mathrm{peak}$ do not trace the $\alpha_\mathrm{CO}$ variation well given the non-LTE conditions as well as density or optical depth variations in barred galaxy centers. In Section~\ref{subsubsec:alpha_prescription}, we will show that simulations also predict only a mild $\alpha_\mathrm{CO}$ dependence on $T_\mathrm{peak}$, and that dependence can be washed out if observed with a $\sim$100~pc beam size.

\subsection{Comparison with Existing $\alpha_\mathrm{CO}$ Prescriptions} \label{subsubsec:alpha_prescription}

Recent simulation studies have developed predictions for $\alpha_\mathrm{CO}$ in terms of metallicity, CO line ratios, and/or CO integrated intensities \citep{2012MNRAS.421.3127N, 2017MNRAS.464.3315A,2017MNRAS.470.4750A,2020ApJ...903..142G,2022ApJ...931...28H}. Such predictions have the potential to greatly improve the assessment of molecular gas content in galaxies, and therefore testing them is critical. However, most simulations focus on low-metallicity or Galactic disk-like environments, which do not capture the dense, turbulent conditions or gas inflows that are common in galaxy centers. 
To test if the current $\alpha_\mathrm{CO}$ predictions can be applied to star-forming galaxy centers, we compare our $\alpha_\mathrm{CO}$ results with the established prescriptions and discuss the consistency/discrepancy.

Based on magneto-hydrodynamic simulations of the ISM on kpc-sized chunks of galactic disks down to 2~pc resolution, \citet{2020ApJ...903..142G} proposed three different $\alpha_\mathrm{CO}$ prescriptions as a function of metallicity, beam size, as well as CO line related properties: $R_\mathrm{21}$, $T_\mathrm{peak}$, and CO integrated intensity $I_\mathrm{CO(1-0)}$, respectively. The prescriptions are cautioned to be only applicable to disk-like environment with $I_\mathrm{CO(1-0)} < 200$~K~km~s$^{-1}$, which is the maximum intensity of their simulated data at 2~pc resolution. Their native 2~pc resolution data also span a range of $2 > R_\mathrm{21} > 0.1$ and $20 > T_\mathrm{peak} > 0.1$~K.
Figure~\ref{fig:alpha_prediction} compares our modeled $\alpha_\mathrm{CO}$ with the \citet{2020ApJ...903..142G} simulated data (brown curves) and prescriptions (black dashed lines) at solar metallicity and a $\sim$100~pc beam size. 
We also show their simulated data at 2~pc resolution (gray curves), which seems to extend to denser/hotter regions that are more consistent with the main sample of our observations. However, we note that their 2-pc data should be resolving individual molecular clouds, while our results at $\sim$100~pc resolutions are sampling beam-averaged, unresolved gas.  

Overall, it is clear that our observational data have higher $R_\mathrm{21}$, $T_\mathrm{peak}$, and $I_\mathrm{CO(1-0)}$ than the simulated data at $\gtrsim$100~pc scales, and that our $\alpha_\mathrm{CO}$ results are systematically lower than the extrapolated predictions. The $R_\mathrm{21}$-dependent prediction has the most potential to match our data within a factor of three discrepancy, while the $T_\mathrm{peak}$- and $I_\mathrm{CO(1-0)}$-dependent predictions show deviations with a factor of 3--10. This is in line with the suggestion by \citet{2020ApJ...903..142G} to adopt the $R_\mathrm{21}$ prescription for larger ($> 100$~pc) beam sizes, as $R_\mathrm{21}$ can better reflect CO excitation and suffers less from beam dilution. As shown in Figure~\ref{fig:alpha_prediction}, the predicted $\alpha_\mathrm{CO}$ correlations with $T_\mathrm{peak}$ and $I_\mathrm{CO(1-0)}$ are fairly weak, likely due to significant beam-averaging over temperature and density at $\gtrsim$100~pc resolutions. 

In addition, we find that the $\gtrsim$100~pc simulated data have some overlap with all three galaxies in the dynamic range of $T_\mathrm{peak}$ but almost no overlap in that of $I_\mathrm{CO(1-0)}$, implying that the line width is generally broader in our case (see also Section~\ref{subsubsec:alpha_spec}). This can be a vital reason for the discrepancy on $\alpha_\mathrm{CO}$ between the simulations and observations, as the enhanced velocity dispersion due to strong dynamical effects in these galaxy centers cannot be captured by such simulations, where gas inflows and central starbursts were not taken into account. 
Another possible reason for the difference with simulations is our assumption of constant CO abundance, $x_\mathrm{CO}$. This could be important in some regions with low $T_\mathrm{peak}$ and $I_\mathrm{CO}$, where the optical depth is low and the photodissociation may lower $x_\mathrm{CO}$, leading to possible underestimation of $\alpha_\mathrm{CO}$ in our modeling. However, the difference in $x_\mathrm{CO}$ cannot explain the overall lower $\alpha_\mathrm{CO}$ seen in the majority of our observed data with $n_\mathrm{H_2} > 300$~cm$^{-3}$, since most of the simulated data from \citet{2020ApJ...903..142G} in this regime reaches their maximum $x_\mathrm{CO}$ of $3.2\times 10^{-4}$, which is similar to our assumption. 

It is important to note that the simulation by \citet{2020ApJ...903..142G} represents a different regime of physical conditions than our measurements, as nearly half of the simulated data are optically thin and sub-thermally excited. However, they have also explored $\alpha_\mathrm{CO}$ dependence in the optically thick and thermally excited regime, which is closer to the conditions of our data and may explain the $\tau_\mathrm{CO}$ correlation we observe. 
Compared to \citet{2020ApJ...903..142G}, our three galaxy centers lie beyond the ``high-density'' regime ($n_\mathrm{H_2} \gtrsim 300$~cm$^{-3}$) where they found saturated CO emission with growing $N_\mathrm{H_2}$ due to increased optical depths. This saturated level corresponds to $N_\mathrm{H_2} \gtrsim 5 \times 10^{21}$~cm$^{-2}$ or $N_\mathrm{CO} \gtrsim 1.5 \times 10^{18}$~cm$^{-2}$ (assuming consistent $x_\mathrm{CO}$ of $3\times10^{-4}$), the value of which agrees with our $N_\mathrm{CO}$ solutions.
In the sub-thermal regime where CO intensity is not yet saturated, \citet{2020ApJ...903..142G} reported decreasing $\alpha_\mathrm{CO}$ with $n_\mathrm{H_2}$, which can be explained by increasing excitation temperature and CO abundance. Meanwhile, they also found that $\alpha_\mathrm{CO}$ starts to increase with $n_\mathrm{H_2}$ when entering the thermal regime where CO becomes fully optically thick. This ``turnover'' trend of $\alpha_\mathrm{CO}$ suggests that the impact from optical depth effects can take over in dense, optically thick regions like galaxy centers, which potentially explains why optical depth effects dominate the $\alpha_\mathrm{CO}$ trend in our results (see Section~\ref{subsec:alpha_env} and~\ref{subsubsec:alpha_ratio}). 

In addition to \citet{2020ApJ...903..142G}, other hydrodynamic simulations also suggested $\alpha_\mathrm{CO}$ as a multivariate function of metallicity, CO integrated intensity, and/or beam size \citep{2012MNRAS.421.3127N,2022ApJ...931...28H}. As shown in Figure~\ref{fig:alpha_prediction}(c), the prediction by \citet{2012MNRAS.421.3127N} is within 0.2~dex higher than that by \citet{2020ApJ...903..142G} at $I_\mathrm{CO(1-0)} \gtrsim 30$~K~km~s$^{-1}$ and solar metallicity. We also overplot the simulated data at 125-pc resolution from a recent study by \citet{2022ApJ...931...28H}, which predicts a similar $\alpha_\mathrm{CO}$ trend to \citet{2020ApJ...903..142G} and reaches a maximum $x_\mathrm{CO}$ of $2.8\times 10^{-4}$. We find our modeled $N_\mathrm{CO}$ generally higher than the predicted relations between CO optical depth and column density at solar metallicity in \citet{2022ApJ...931...28H}.

By assembling previous observations at $>$kpc scales including nearby disks \citep{2013ApJ...777....5S} and (U)LIRGs \citep{1998ApJ...507..615D,2012ApJ...751...10P}, \citet{co-to-h2} also suggested a prescription of $\alpha_\mathrm{CO}$ as a function of metallicity ($Z'$, normalized to the solar value), characteristic giant molecular cloud surface density ($\Sigma_\mathrm{GMC}$), and the total (gas + star) surface density ($\Sigma_\mathrm{tot}$):
\begin{equation}
\alpha_\mathrm{CO} = 2.9 \exp\left(\frac{40}{Z'\ \Sigma_\mathrm{GMC}}\right) \left(\frac{\Sigma_\mathrm{tot}}{100\ \mathrm{M_\odot\ pc^{-2}}}\right)^{-\gamma}
\label{eqn_alpha_B13}
\end{equation}
$\mathrm{M_\odot\ (K~km~s^{-1}~pc^2)^{-1}}$, where $\gamma = 0.5$ if $\Sigma_\mathrm{tot} > 100$ $\mathrm{M_\odot}$~pc$^{-2}$ or $\gamma = 0$ otherwise.
To compare our results with this kpc-based prescription, we calculate the stellar mass surface densities $\Sigma_*$ using the PHANGS--MUSE data at a native resolution of ${\sim}1.5\arcsec$ \citep{2022A&A...659A.191E}. We weight the $\Sigma_*$ with our observed $I_\mathrm{CO(1-0)}$ and then average over the entire region covered in our analysis. Similarly, we derive the average molecular gas mass surface density ($\Sigma_\mathrm{mol}$) by multiplying the $I_\mathrm{CO(1-0)}$ maps with our modeled $\alpha_\mathrm{CO}$ and then calculating the intensity-weighted mean across the maps.

The resulting $\Sigma_*$ ($\Sigma_\mathrm{mol}$) for the centers of NGC~3351, 3627, and 4321 are approximately 5000 (63), 4500 (115), and 2100 (94) $\mathrm{M_\odot\,pc^{-2}}$. It is clear that $\Sigma_\mathrm{tot}$ is dominated by $\Sigma_*$ in all three galaxy centers, and the derived $\Sigma_\mathrm{mol}$ is similar to the $\Sigma_\mathrm{GMC} = 100$~$\mathrm{M_\odot\,pc^{-2}}$ adopted in \citet{co-to-h2}.   
Finally, we correct the derived surface densities with their galaxy inclinations by a cosine factor and then substitute into Equation~\ref{eqn_alpha_B13}, assuming $\Sigma_\mathrm{GMC} = 100$~$\mathrm{M_\odot\,pc^{-2}}$ at solar metallicity \footnote{For the \citet{co-to-h2} prescription, the solar metallicity condition is suggested to be paired with a fixed GMC surface density of 100~$\mathrm{M_\odot\,pc^{-2}}$. This is because the exponential term (see Equation~\ref{eqn_alpha_B13}) can easily lead to unrealistic $\alpha_\mathrm{CO}$ values even with small variations in the adopted GMC surface density \citep[see also][]{2023ApJ...945L..19S}. We therefore use the suggested value of 100 $\mathrm{M_\odot\,pc^{-2}}$ here to avoid such issues, and we also show that the intensity-weighted $\Sigma_\mathrm{mol}$ at kpc scales for our galaxy centers roughly agrees with that value.}. 

With the corrected surface density, Equation~\ref{eqn_alpha_B13} predicts $\log(\alpha_\mathrm{CO})$ of -0.22, -0.20, and -0.04 $\mathrm{M_\odot\ (K~km~s^{-1}~pc^2)^{-1}}$ over the centers of NGC~3351, 3627, and 4321, respectively. Our modeled $\alpha_\mathrm{CO}$ distributions show $\log(\alpha_\mathrm{CO})$ of $-0.22^{+0.12}_{-0.39}$, $-0.46^{+0.24}_{-0.21}$, and $-0.14^{+0.17}_{-0.24}$~$\mathrm{M_\odot\ (K~km~s^{-1}~pc^2)^{-1}}$ in the central $r < 20\arcsec$ of these galaxies. The intensity-weighted mean $\alpha_\mathrm{CO(2-1)}$ derived in Section~\ref{subsec:alpha_literature} is equivalent to $\log(\alpha_\mathrm{CO})$ of -0.12, -0.30, and -0.07 $\mathrm{M_\odot\ (K~km~s^{-1}~pc^2)^{-1}}$ if we convert the CO 2--1 intensity back to CO 1--0 via the integrated mean $R_{21}$ \citep[see Tables~\ref{tab:ratio_4321},~\ref{tab:ratio_3627}, and Table~2 of][]{2022ApJ...925...72T}. 
Both results overlap well with the predicted values from Equation~\ref{eqn_alpha_B13}, assuming a reasonable 0.2~dex uncertainty of the prediction. Notably, the range of our $\Sigma_\mathrm{tot}$ (dominated by $\Sigma_*$) is also similar to the (U)LIRG samples used in \citet{co-to-h2} to develop the prescription.

We conclude that on kpc scales, our $\alpha_\mathrm{CO}$ results are compatible with the \citet{co-to-h2} prescription. On sub-kpc scales, the existing simulation-based prescriptions may overestimate $\alpha_\mathrm{CO}$ when being applied to galaxy centers with higher surface density, CO intensity, and velocity dispersion. Future simulations capturing gas inflows and local turbulence will be needed to develop a better $\alpha_\mathrm{CO}$ prescription appropriate for galaxy centers or other extreme environments.

\subsection{Multi-line Constraints in the Modeling} \label{subsec:line_constraint}

\begin{figure*}
\begin{minipage}{.35\linewidth}
\centering
Quantities determined by 4 lines:\\
CO 1--0, CO 2--1, $^{13}$CO 2--1, $^{13}$CO 3--2
\includegraphics[width=\linewidth]{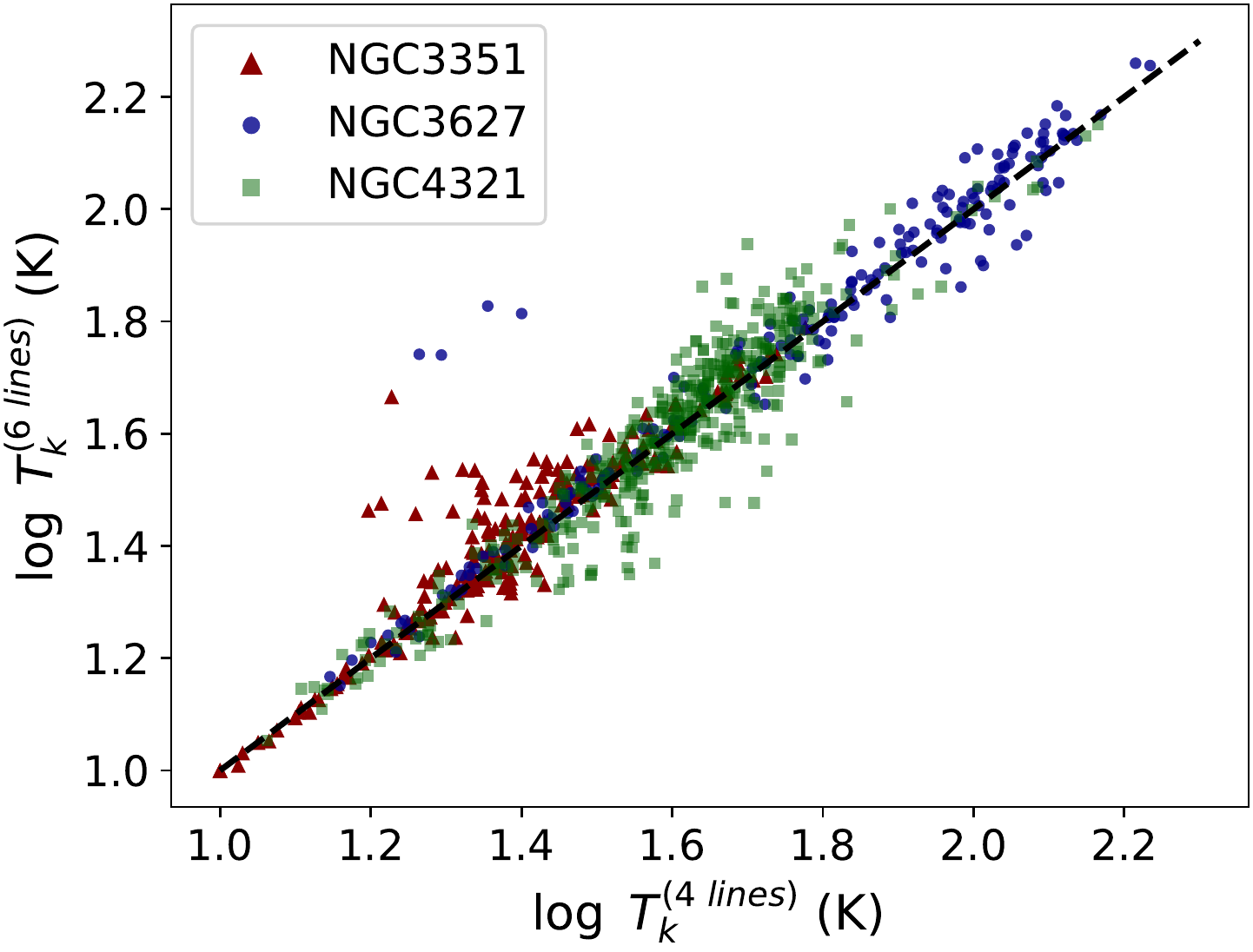}
\end{minipage}
\hfill
\begin{minipage}{.31\linewidth}
\centering
Quantities determined by 3 lines:\\
CO 2--1, $^{13}$CO 2--1, $^{13}$CO 3--2
\includegraphics[width=\linewidth]{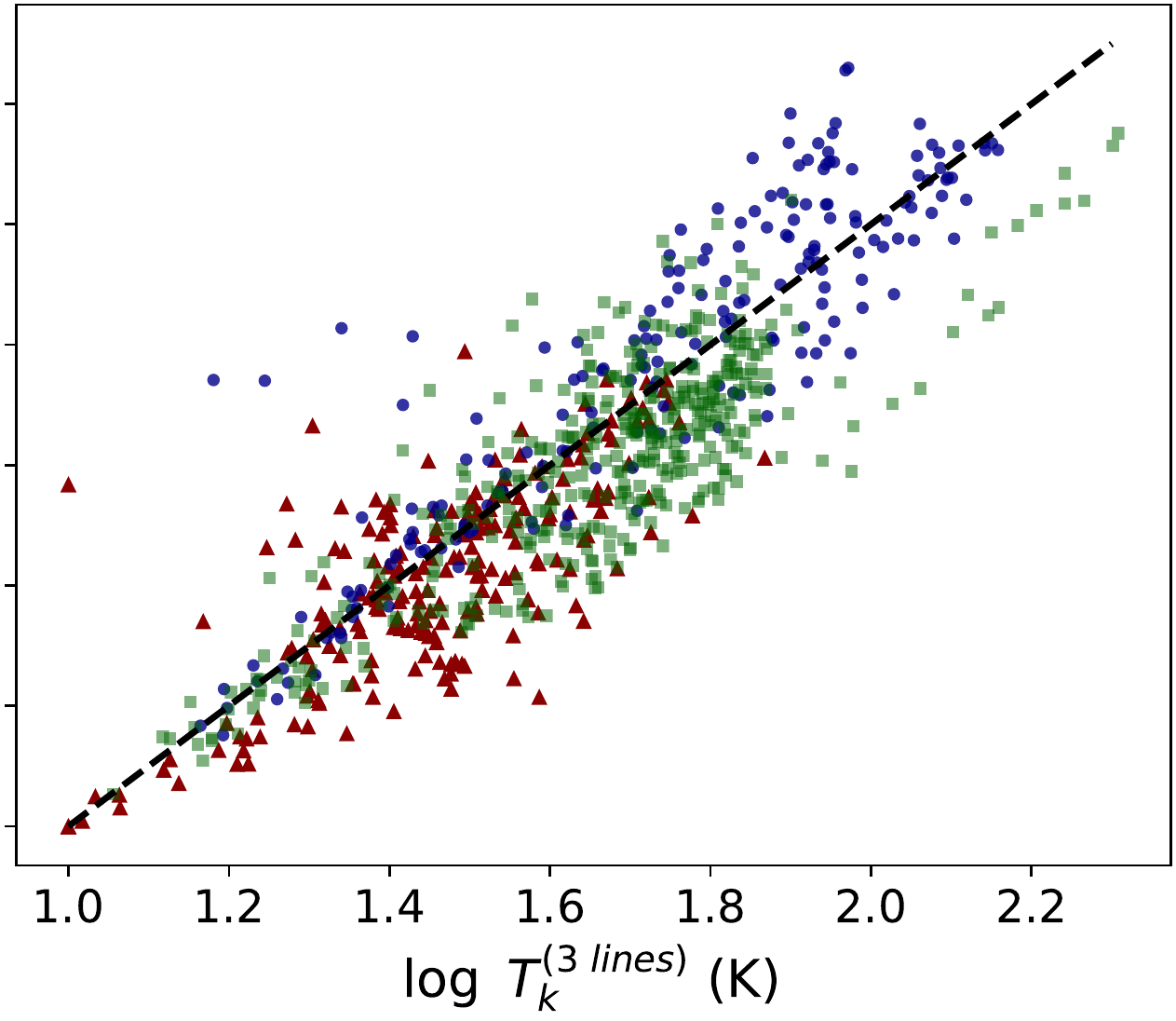}
\end{minipage}
\hfill
\begin{minipage}{.31\linewidth}
\centering
Quantities determined by 3 lines:\\
CO 1--0, CO 2--1, $^{13}$CO 2--1
\includegraphics[width=\linewidth]{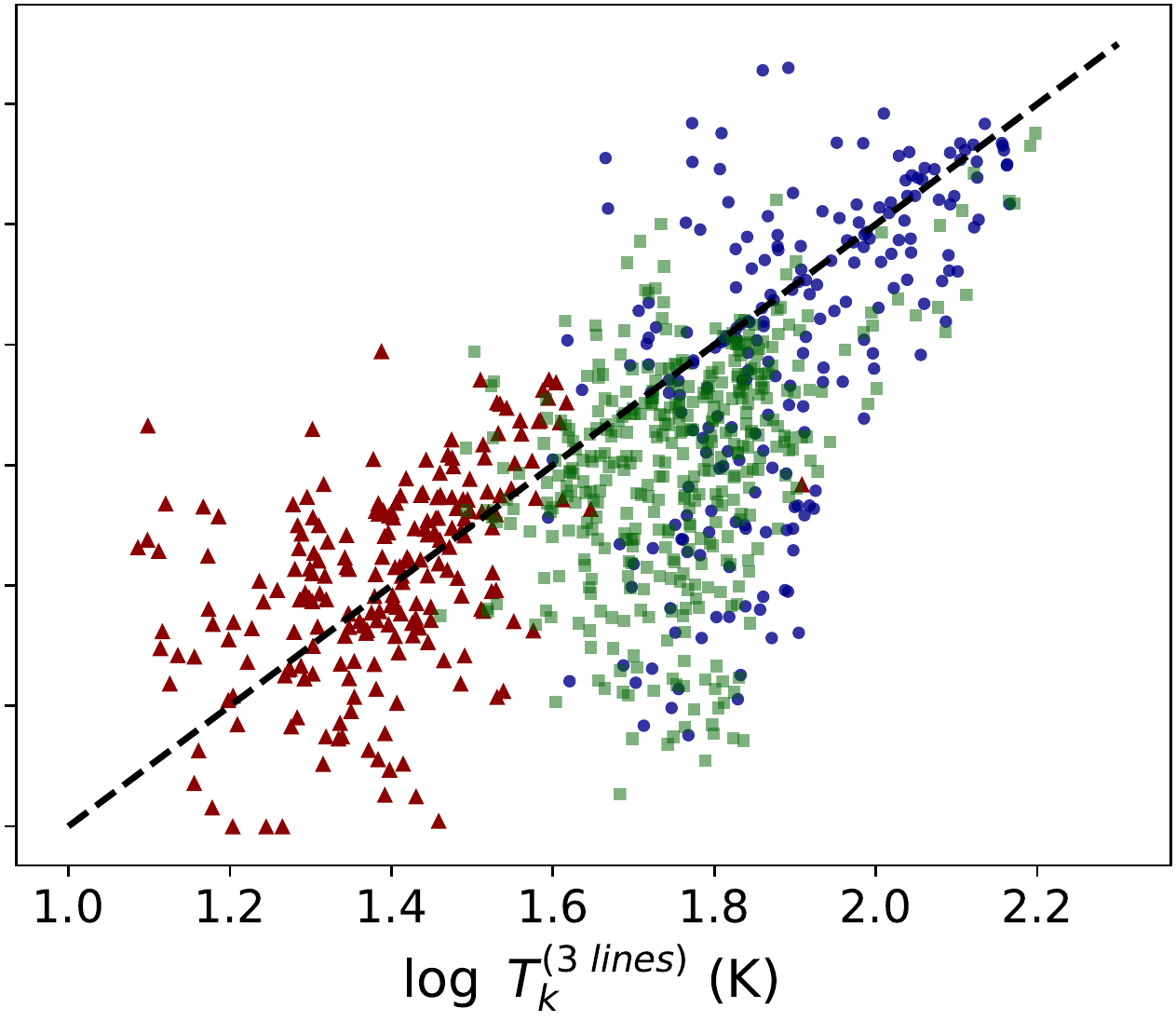}
\end{minipage} 
\\
\begin{minipage}{.35\linewidth}
\centering
\includegraphics[width=\linewidth]{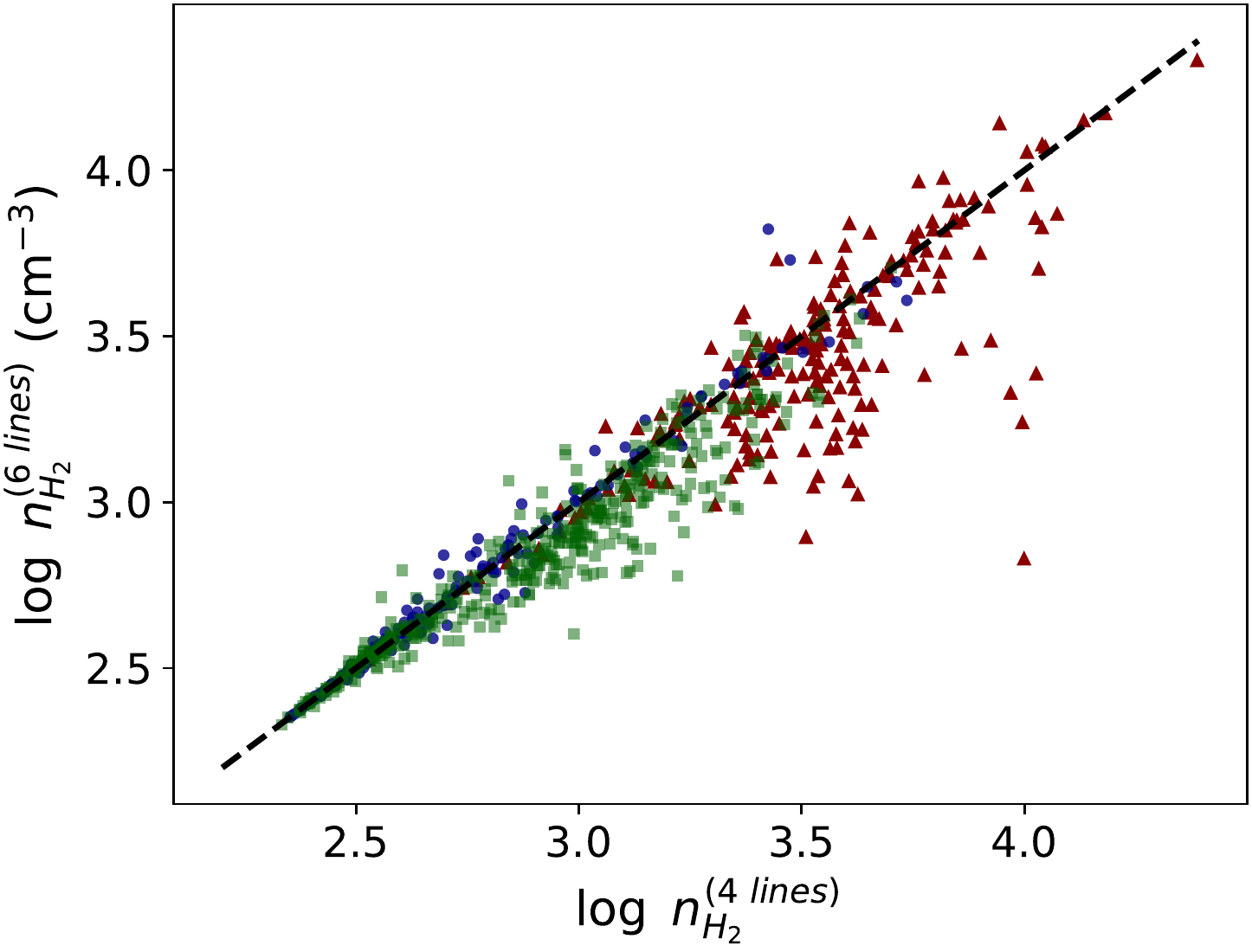}
\end{minipage}
%\hfill
\begin{minipage}{.31\linewidth}
\centering
\includegraphics[width=\linewidth]{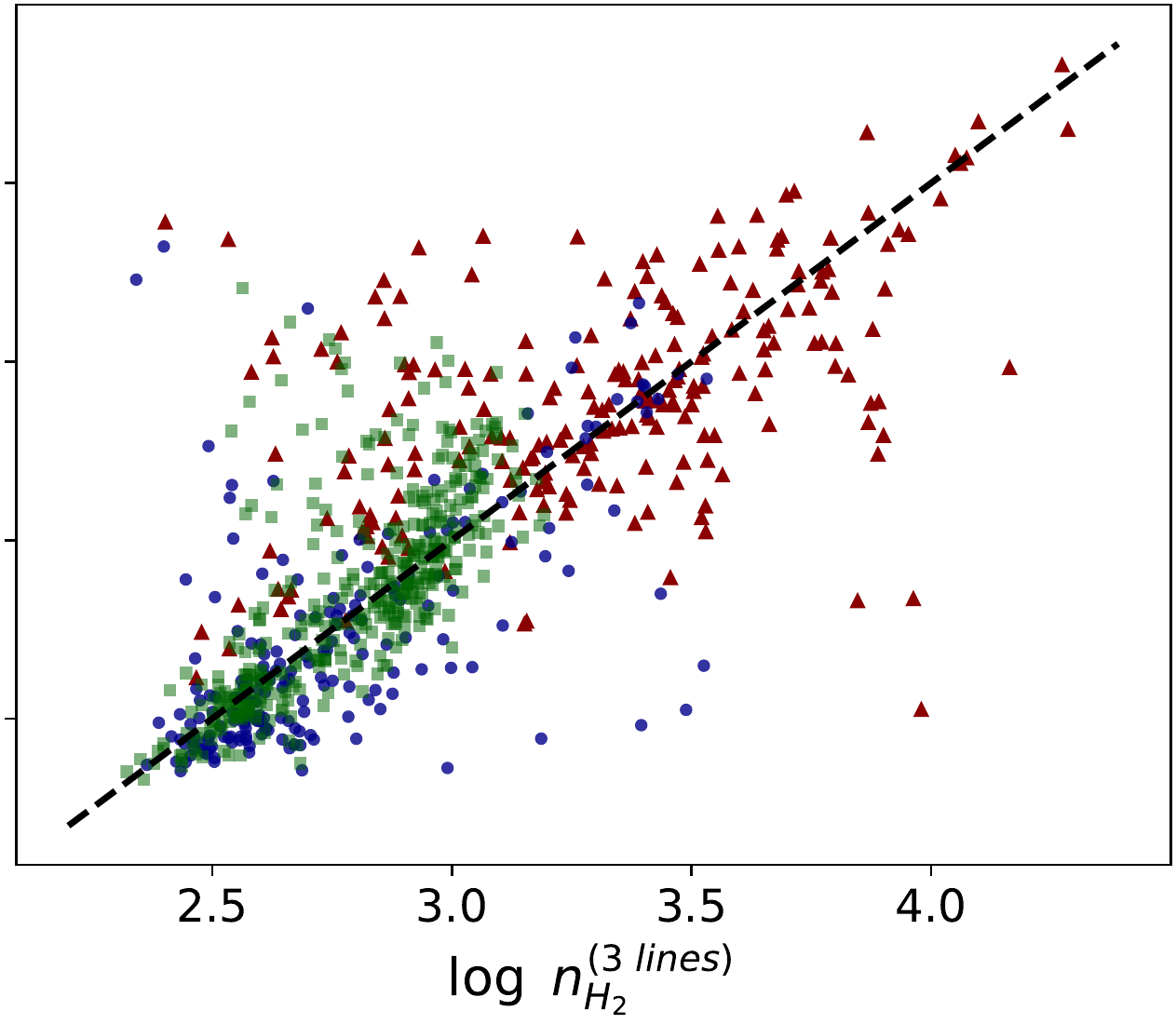}
\end{minipage}
%\hfill
\begin{minipage}{.31\linewidth}
\centering
\includegraphics[width=\linewidth]{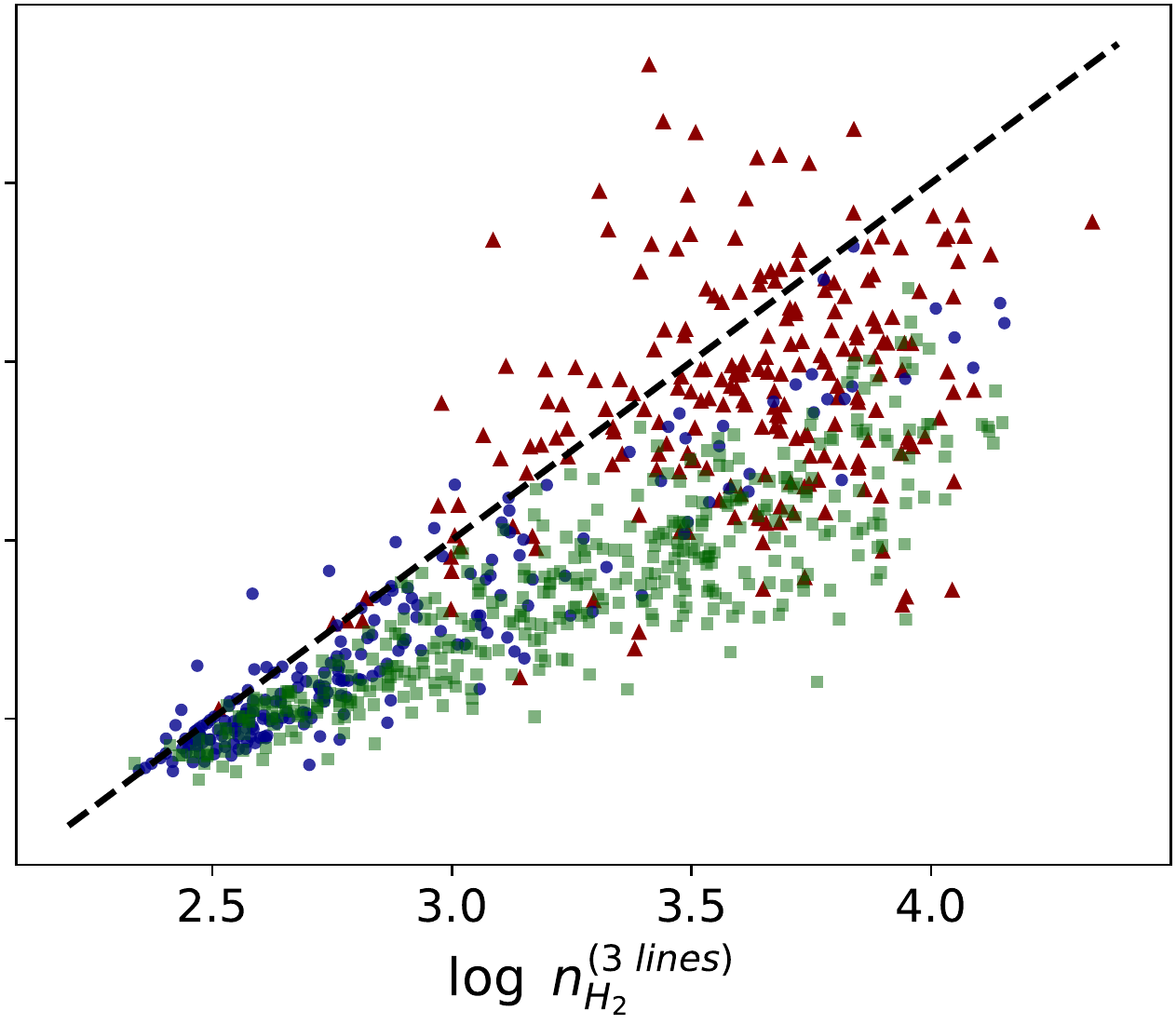}
\end{minipage}
\\
\begin{minipage}{.36\linewidth}
\centering
\includegraphics[width=\linewidth]{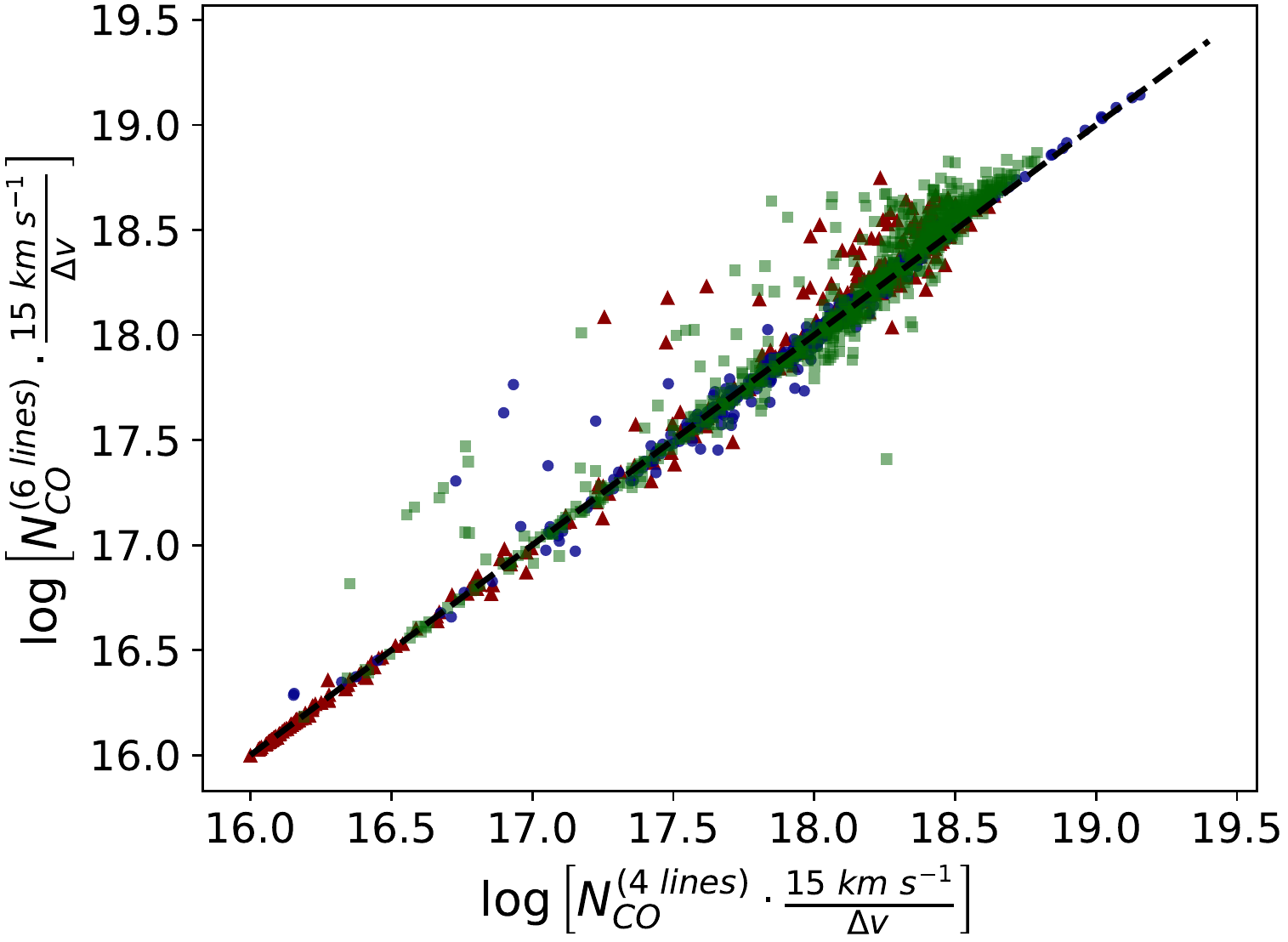}
\end{minipage}
%\hfill
\begin{minipage}{.31\linewidth}
\centering
\includegraphics[width=\linewidth]{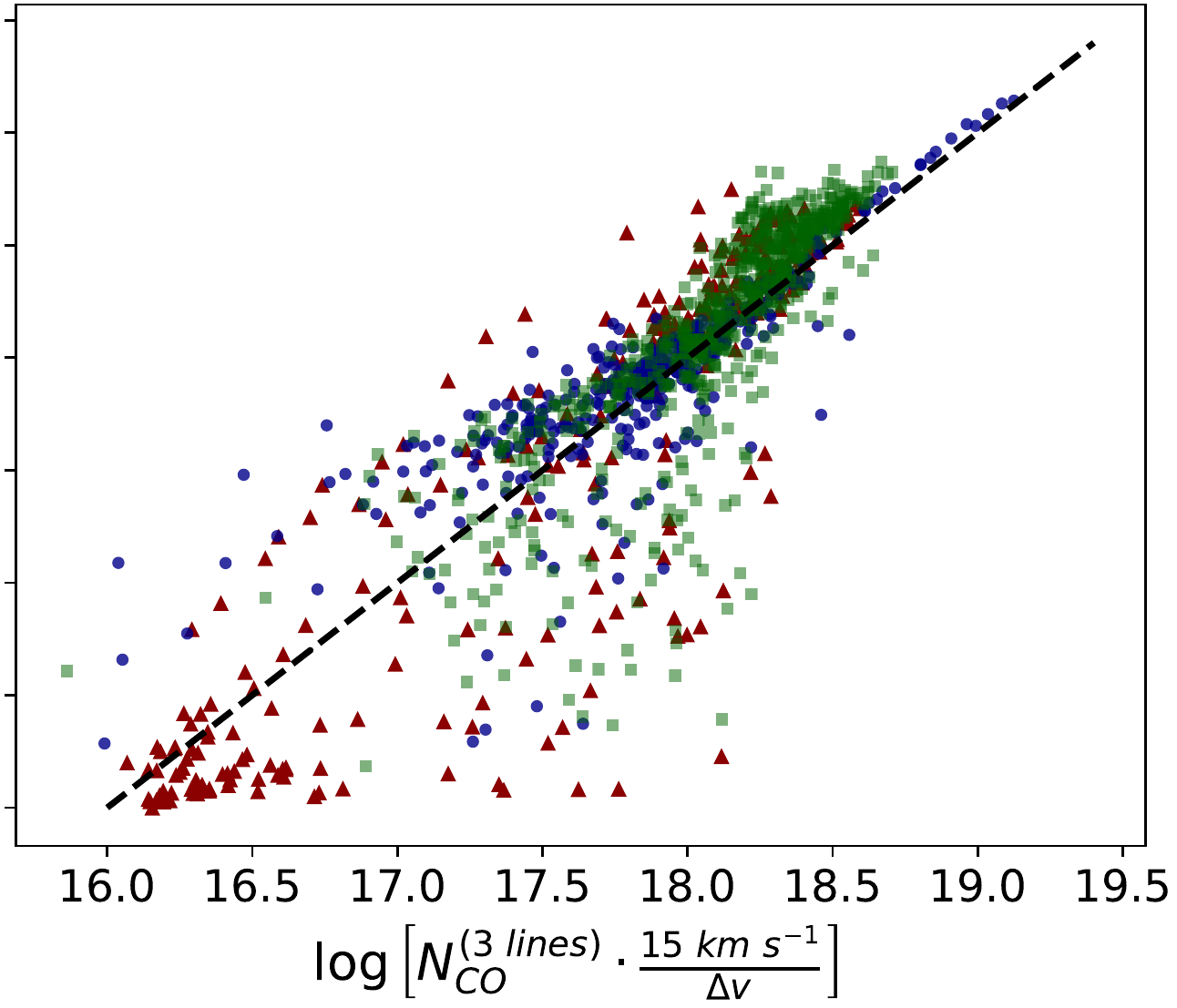}
\end{minipage}
%\hfill
\begin{minipage}{.31\linewidth}
\centering
\includegraphics[width=\linewidth]{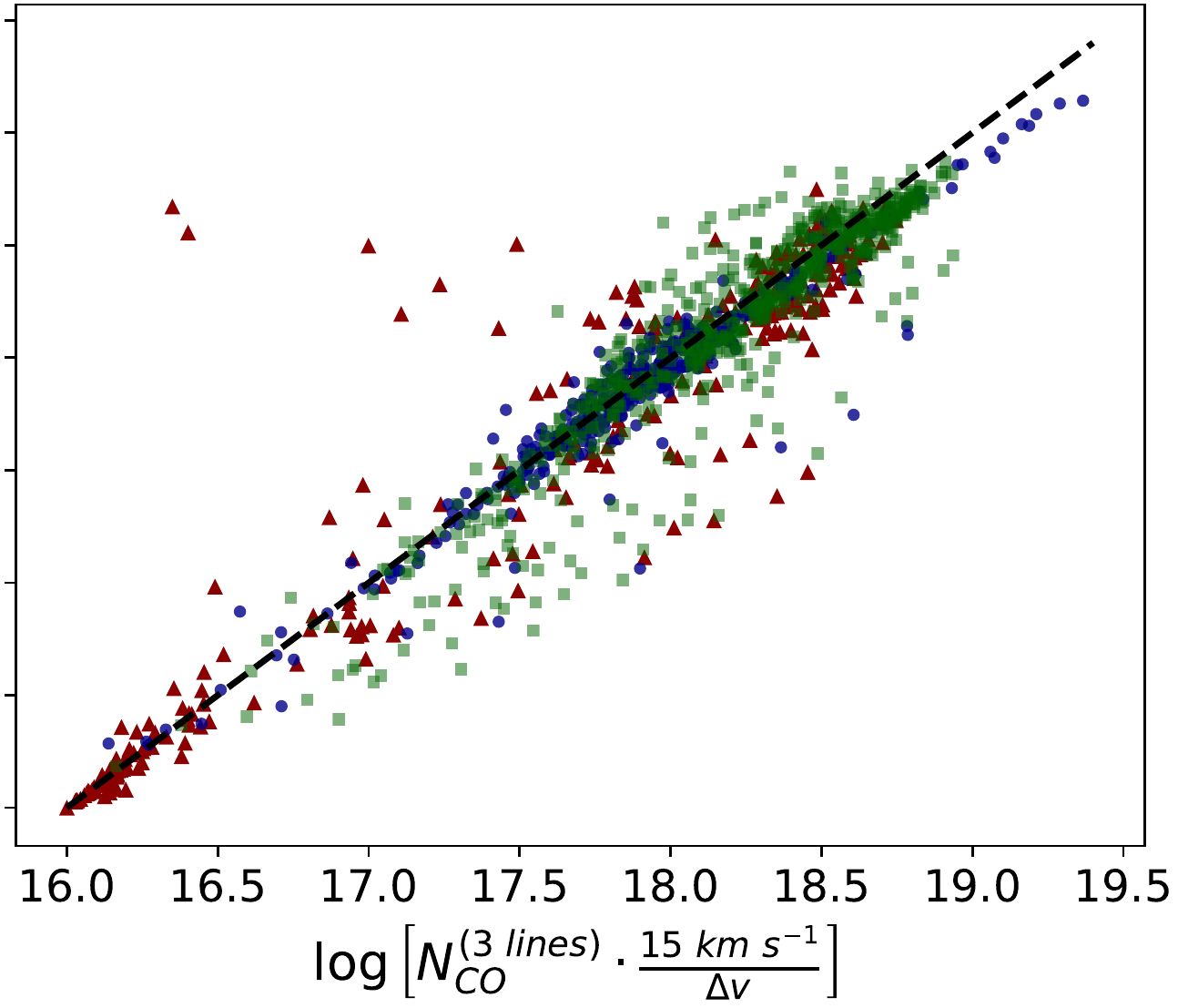}
\end{minipage}
\caption{Median solutions of $T_\mathrm{k}$ (first row), $n_\mathrm{H_2}$ (middle row), and $N_\mathrm{CO}/\Delta v$ (bottom row) determined by multi-line modeling with different sets of emission lines in the central kpc regions of NGC~3351 (red), NGC~3627 (blue), and NGC~4321 (green). The y-axes represent the environmental parameters constrained by all six lines, which are compared to the x-axes showing those constrained by only a subset of lines. \textit{Left column:} removing the two C$^{18}$O lines still reproduce consistent solutions with those determined by the six-line modeling. \textit{Middle column:} CO 2--1, $^{13}$CO 2--1, and $^{13}$CO 3--2 are key constraints, while the addition of CO 1--0 (left column) is crucial for reducing the scatter and constraining $N_\mathrm{CO}/\Delta v$. \textit{Right column:} without the high-$J$ constraint from $^{13}$CO 3--2, the derived $T_\mathrm{k}$ and $n_\mathrm{H_2}$ would deviate from the six-line modeling results.}
\label{fig:line_constraint}
\end{figure*}

Our multi-line modeling jointly analyzes six low-$J$ transitions of CO, $^{13}$CO, and C$^{18}$O. In this subsection, we discuss how modeling solutions would change if different subsets of lines were used. We will compare our solutions from the six-line modeling (Section~\ref{subsec:model_result}) to those determined from various combinations of lines, and identify the most critical measurements that enable good constraints on the parameters. In addition to the modeling of NGC~3627 and NGC~4321, we will also include the modeling results of NGC~3351 from \citet{2022ApJ...925...72T} as the observed lines and modeling approach are the same.    

From Figure~\ref{fig:flux_contour_4321}, we find that the constraints given by the $^{13}$CO 2--1 and C$^{18}$O 2--1 are almost identical, and the same applies to $^{13}$CO 3--2 and C$^{18}$O 3--2 (see also Figure~\ref{fig:flux_contour_3627} in the Appendix). This means that the best-fit solutions would remain the same even if we remove the constraints from both C$^{18}$O lines. \citet{2022ApJ...925...72T} also reported that the C$^{18}$O line constraints are not critical to the results for the inflow regions in NGC~3351, though they did not further examine other regions. To test how the removal of C$^{18}$O would affect the modeling solutions, we present in the left column of Figure~\ref{fig:line_constraint} the pixel-by-pixel solutions of $T_\mathrm{k}$, $n_\mathrm{H_2}$, and $N_\mathrm{CO}/\Delta v$ modeled with and without the two C$^{18}$O lines on the y-axes and x-axes, respectively. It is clear that solutions obtained from both scenarios are consistent with only a ${\sim}0.2$~dex scatter for all quantities. We note that C$^{18}$O emission is weak in the arms of NGC~3627, but it is well detected in NGC~4321 and the central $\sim$kpc of NGC~3351, so the low S/N of the C$^{18}$O measurements is not the main reason for such consistency. We conclude that the combination of CO 1--0, CO 2--1, $^{13}$CO 2--1, and $^{13}$CO 3--2 can already provide strong constraints on the gas properties, while C$^{18}$O 2--1 and 3--2 tend to give constraints similar to $^{13}$CO and thus do not provide much additional information. 

The middle and right columns of Figure~\ref{fig:line_constraint} compare the six-line modeling solutions with the solutions determined by only three of the lines. In the middle column panels, we present the case with CO 2--1, $^{13}$CO 2--1, and $^{13}$CO 3--2, and the rightmost panels show the case with CO 1--0, CO 2--1, and $^{13}$CO 2--1. We include CO 2--1 and $^{13}$CO 2--1 in both cases because the detection of these lines were used to define our analyzed regions (i.e., CO 2--1 flux recovery $> 70\%$ and $^{13}$CO 2--1 S/N $> 3$). We find that the inclusion of $^{13}$CO 3--2 (middle panels) is critical to obtaining accurate solutions for $T_\mathrm{k}$ and $n_\mathrm{H_2}$, as removing that line leads to much larger scatter and/or bias in the reconstructed $T_\mathrm{k}$ and $n_\mathrm{H_2}$ even with the inclusion of CO 1--0 (right panels). This is likely because $^{13}$CO 3--2 is the only high-$J$ transition in the set of lines, and thus it provides critical constraints on density and temperature in addition to the lower-$J$ 1--0 or 2--1 transitions. On the other hand, including both CO 1--0 and 2--1 can significantly reduce the scatter in $N_\mathrm{CO}/\Delta v$ since the $^{12}$CO emission is highly dependent on optical depth (see bottom panels of Figure~\ref{fig:line_constraint}). 

In summary, CO 2--1, $^{13}$CO 2--1, and $^{13}$CO 3--2 can be an efficient combination to measure gas temperature and volume density via multi-line modeling, while the addition of CO 1--0 would be important to obtain more reliable and precise estimates for optical depth. Since $^{13}$CO has a slightly higher effective critical density and much lower optical depth than CO, the inclusion of a $^{13}$CO line ratio can better constrain regions with higher volume/surface density or optical depth such as galaxy centers. Notably, the C$^{18}$O lines give degenerate but lower quality information to the $^{13}$CO lines. Since it is observationally expensive to securely detect the faint C$^{18}$O lines, this result may help reduce observing time for similar studies in the future or over larger area. 
However, we emphasize that the conclusion is simply drawn from the modeling results toward the central kpc of NGC~3351, NGC~3627, and NGC~4321, which are all barred centers with starburst or AGN signatures. It is likely that different transitions or species are needed to constrain regions such as galaxy disks or unbarred galaxy centers.

\section{Conclusions} \label{sec:conclusion}

We present ALMA observations of six low-$J$ CO, $^{13}$CO, and C$^{18}$O lines toward the inner 2--3~kpc regions of NGC~3627 and NGC~4321 at ${\sim}100$~pc resolution. Using non-LTE radiative transfer modeling with Bayesian likelihood analysis, we constrain molecular gas properties including density, temperature, and CO isotopologue abundances on a pixel-by-pixel basis. With the modeling, we further derive $\alpha_\mathrm{CO}$ and correlate with parameters such as optical depth, temperature, velocity dispersion, and line ratios to discuss the physical drivers and observational tracers of $\alpha_\mathrm{CO}$ variations in barred galaxy centers. The results on NGC~3351 from \citet{2022ApJ...925...72T} are incorporated in our discussion for a more comprehensive view. We also compare the results with existing $\alpha_\mathrm{CO}$ estimates and predictions. Our main findings and conclusions are as follows: 

\begin{enumerate}

\item The moment 0 images of all six lines reveal a bright nucleus with size of $\sim$300~pc in diameter in both NGC~3627 and NGC~4321. The nuclei are connected with inner spiral arms or bar lanes, which are observed in all the lines for NGC~4321 but not securely detected in C$^{18}$O for NGC~3627.
The temperature sensitive line ratios are significantly higher in both nuclei with an integrated mean $R_\mathrm{21}$ of 0.9 for NGC~3627 and 1.2 for NGC~4321, suggesting high excitation and thermalized gas. The integrated mean $R_\mathrm{21}$ over the entire central region of NGC~3627 and~4321 is 0.8 and 0.9, respectively, which are consistent with previous observations on kpc scales.  

\item Our modeling results in well-constrained solutions for most physical parameters. Both galaxies show increasing kinetic temperature ($T_\mathrm{k}$) and H$_2$ volume density ($n_\mathrm{H_2}$) trends toward the centers, with both nuclei reaching $T_\mathrm{k} \gtrsim 100$~K and $n_\mathrm{H_2} > 10^{3}$~$\mathrm{cm^{-3}}$. We find that the $^{13}$CO/C$^{18}$O abundance ratio ($X_\mathrm{13/18}$) varies in the range 6--8, which is similar to the Galactic Center values. The $^{12}$CO/$^{13}$CO abundance ratio ($X_\mathrm{12/13}$) ranges from 80--100 for most regions, despite being less constrained than other parameters.  

\item Assuming the CO/H$_2$ abundance ratio $x_\mathrm{CO} = 3\times10^{-4}$, all the pixels in both galaxy centers show lower CO conversion factor ($\alpha_\mathrm{CO}$) than the standard Galactic value by a factor of 4 to 15. We find that most regions have $\alpha_\mathrm{CO} < 1\,(3\times10^{-4}/x_\mathrm{CO}$)~$\mathrm{M_\odot\ (K~km~s^{-1}~pc^2)^{-1}}$, and it generally decreases with galactocentric radius till a radius of 1.5~kpc. This decreasing $\alpha_\mathrm{CO}$ trend with similar values was also seen within the inner 1~kpc nuclear ring of NGC~3351 \citep{2022ApJ...925...72T}.  

\item We derive intensity-weighted mean $\alpha_\mathrm{CO(2-1)}$ of $0.62\pm0.04$ and $0.93\pm0.04$ over the central $\sim$2~kpc regions for NGC~3627 and NGC~4321, respectively. The result for NGC~4321 matches well with previous dust-based and carbon budget-based studies at lower resolutions \citep{2013ApJ...777....5S,2020A&A...635A.131I}. However, our $\alpha_\mathrm{CO}$ value for NGC~3627 is in between those studies with a $\sim$0.2~dex discrepancy. The disagreement may be related to calibration issues for NGC~3627 in previous CO mapping. Another possibility may be that the high temperatures in NGC~3627 are not well measured in our observations with lines only up to $J$=3--2 . 

\item Based on the modeling results on three barred galaxy centers (including NGC~3351), we find a strong, positive $\alpha_\mathrm{CO}$ dependence with CO optical depth ($\tau_\mathrm{CO}$) that is responsible for $\sim$80\% of the changes in $\alpha_\mathrm{CO}$. The rest of the $\alpha_\mathrm{CO}$ variation is driven by $T_\mathrm{k}$, which varies inversely with $\alpha_\mathrm{CO}$ and can explain the $\alpha_\mathrm{CO}$ variation in local regions where the $\alpha_\mathrm{CO}$ and $\tau_\mathrm{CO}$ trends do not match. This suggests that emissivity-related terms are critical in driving $\alpha_\mathrm{CO}$ in barred galaxy centers, and that optical depth is likely a more dominant driver of $\alpha_\mathrm{CO}$ than gas temperature in this regime. 

\item The observed $^{12}$CO/$^{13}$CO 2--1 ratio and line width generally trace the $\tau_\mathrm{CO}$ variation inversely in all three galaxy centers. With the tight correlation seen in $\alpha_\mathrm{CO}$ and $\tau_\mathrm{CO}$, this indicates that both the line ratio and line width can be good observational tracers for predicting $\alpha_\mathrm{CO}$ variations in galaxy centers, where optical depth effects are dominant. We find the velocity dispersion in the centers of the barred galaxies studied here is higher than the typical values in galaxy disks or non-barred centers by a factor of 3--5, which may explain the overall lower-than-Galactic disk $\alpha_\mathrm{CO}$.  

\item We have tested current $\alpha_\mathrm{CO}$ prescriptions based on observations and simulations. The \citet{co-to-h2} prescription matches our average $\alpha_\mathrm{CO}$ values across the three galaxy centers, given the high total surface density of gas and stars.
On the other hand, current simulation-based prescriptions do not probe similar physical conditions to our galaxy centers, and their extrapolation into this regime tends to overpredict $\alpha_\mathrm{CO}$. 
Future simulations that capture gas inflows and local turbulence have the potential to provide better $\alpha_\mathrm{CO}$ predictions for more extreme environments such as in galaxy centers or (U)LIRGs.

\item We also test our multi-line modeling by varying input combinations of observed molecular lines and comparing the solutions with those from modeling all six lines (i.e., CO 1--0 and 2--1, $^{13}$CO 2--1 and 3--2, and C$^{18}$O 2--1 and 3--2). Combining the results of three galaxy centers, we find that CO 2--1, $^{13}$CO 2--1, and $^{13}$CO 3--2 are the most essential constraints that lead to the six-line solutions, while CO 1--0 also plays a significant role in constraining the CO column density per line width ($N_\mathrm{CO}/\Delta v$). The addition of both C$^{18}$O lines is not crucial as they often duplicate the constraints provided by the $^{13}$CO lines. However, the well-determined $X_{13/18}$ abundances derived from the C$^{18}$O lines can be useful information in particular when the $X_{12/13}$ solutions are uncertain.  

\end{enumerate}

In general, our results suggest that CO optical depth is the dominant driver for $\alpha_\mathrm{CO}$ variations in the central kpc of barred galaxy centers, which can cover a compact nucleus and its surrounding bar lanes or inner spiral arms. To the second order, the increase/decrease of gas temperature in local regions can further lower/raise the $\alpha_\mathrm{CO}$ values.
The lower-than-Galactic disk $\alpha_\mathrm{CO}$ in these barred centers can be explained by the overall enhanced velocity dispersion that lowers the opacity. As we find the CO/$^{13}$CO 2--1 ratio and CO line width mainly reflecting the changes of CO optical depth, these observables may be useful in predicting $\alpha_\mathrm{CO}$ variation in other galaxy centers or similar environments.

\appendix

\section{Effects of Multiple Velocity Components} 
\label{sec:multi_comp}

\begin{figure*}
\begin{minipage}{.33\linewidth}
\centering
\includegraphics[width=\linewidth]{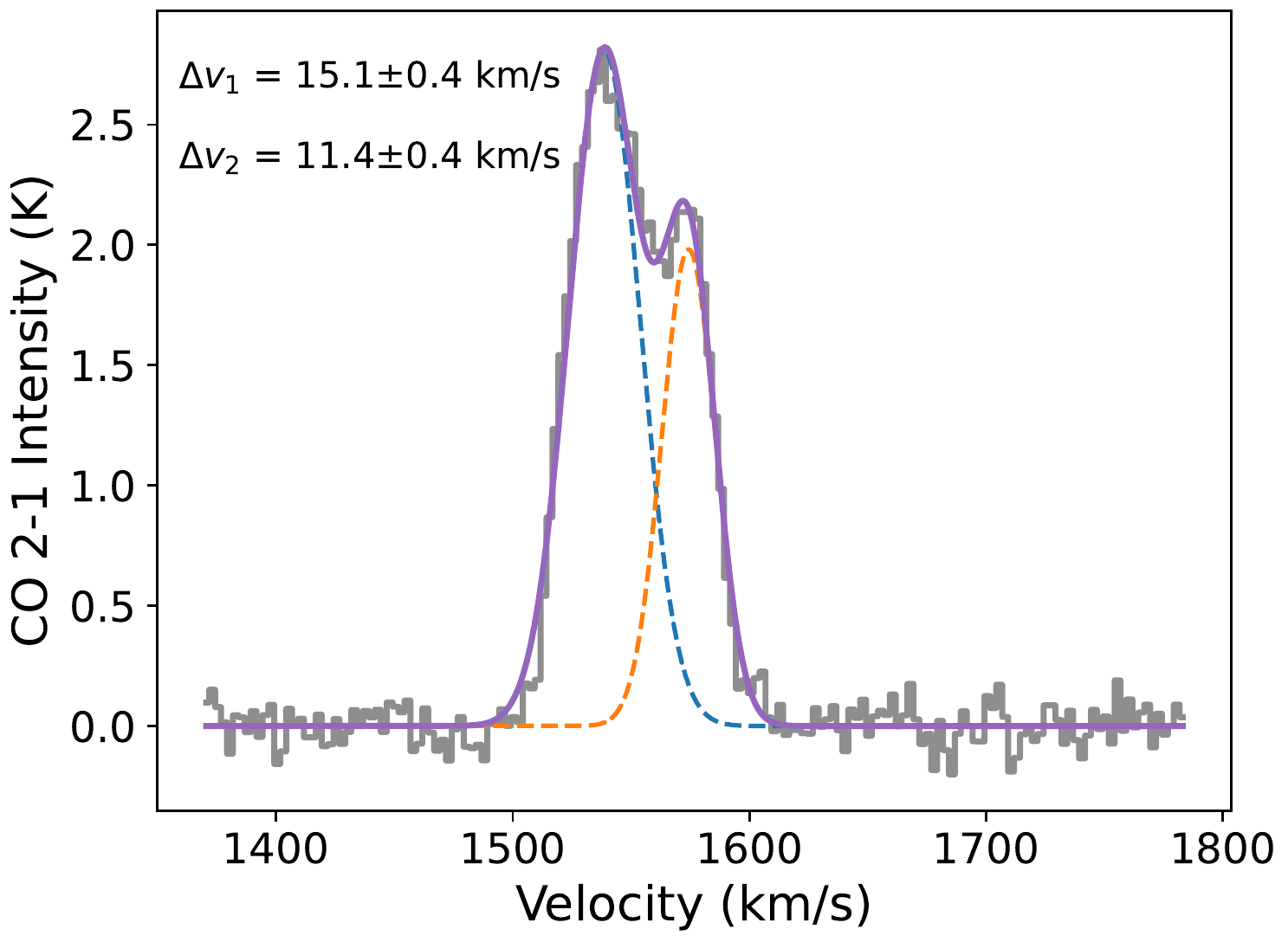}
\end{minipage}
\begin{minipage}{.34\linewidth}
\centering
\includegraphics[width=\linewidth]{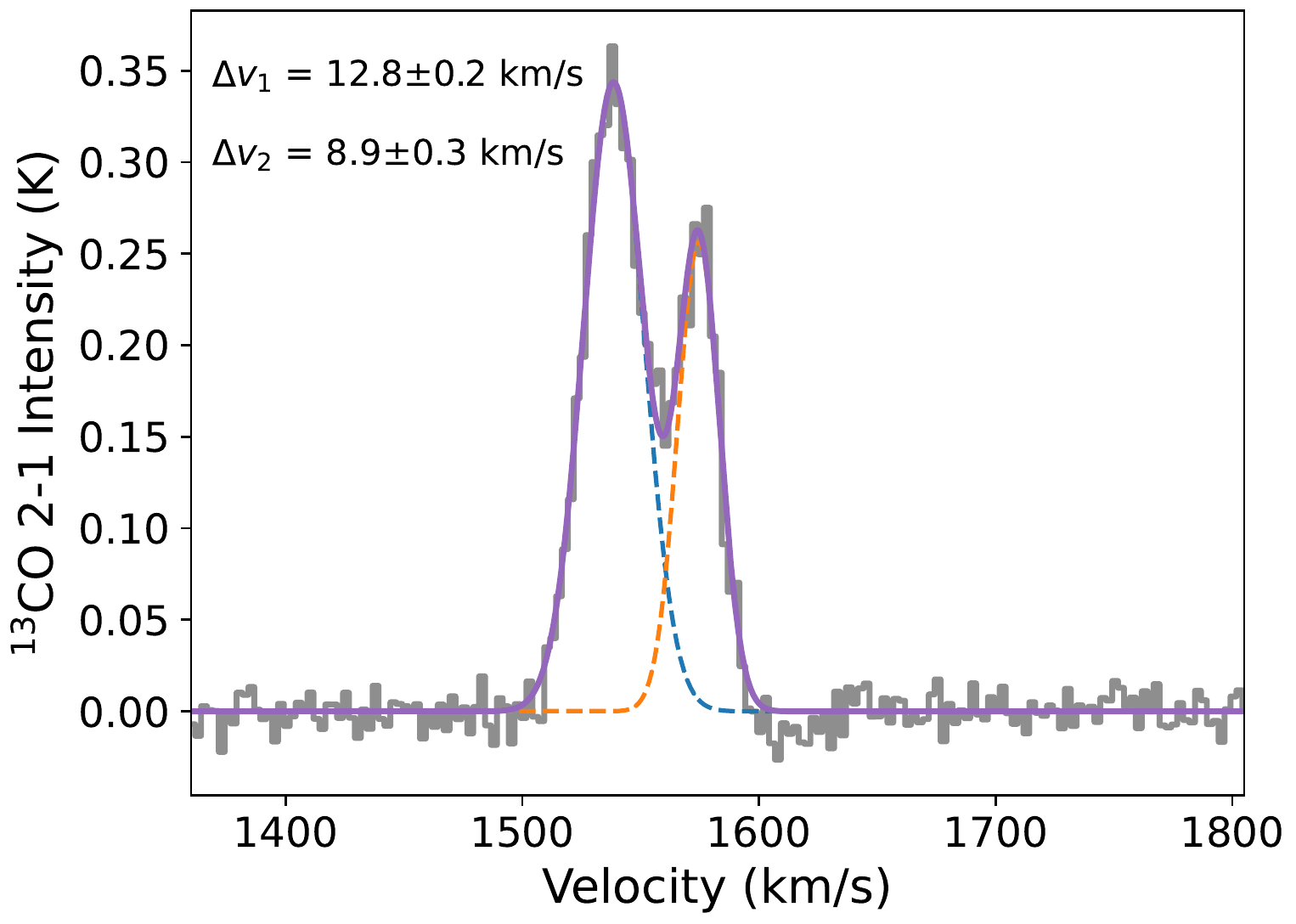}
\end{minipage}
\begin{minipage}{.325\linewidth}
\centering
\includegraphics[width=\linewidth]{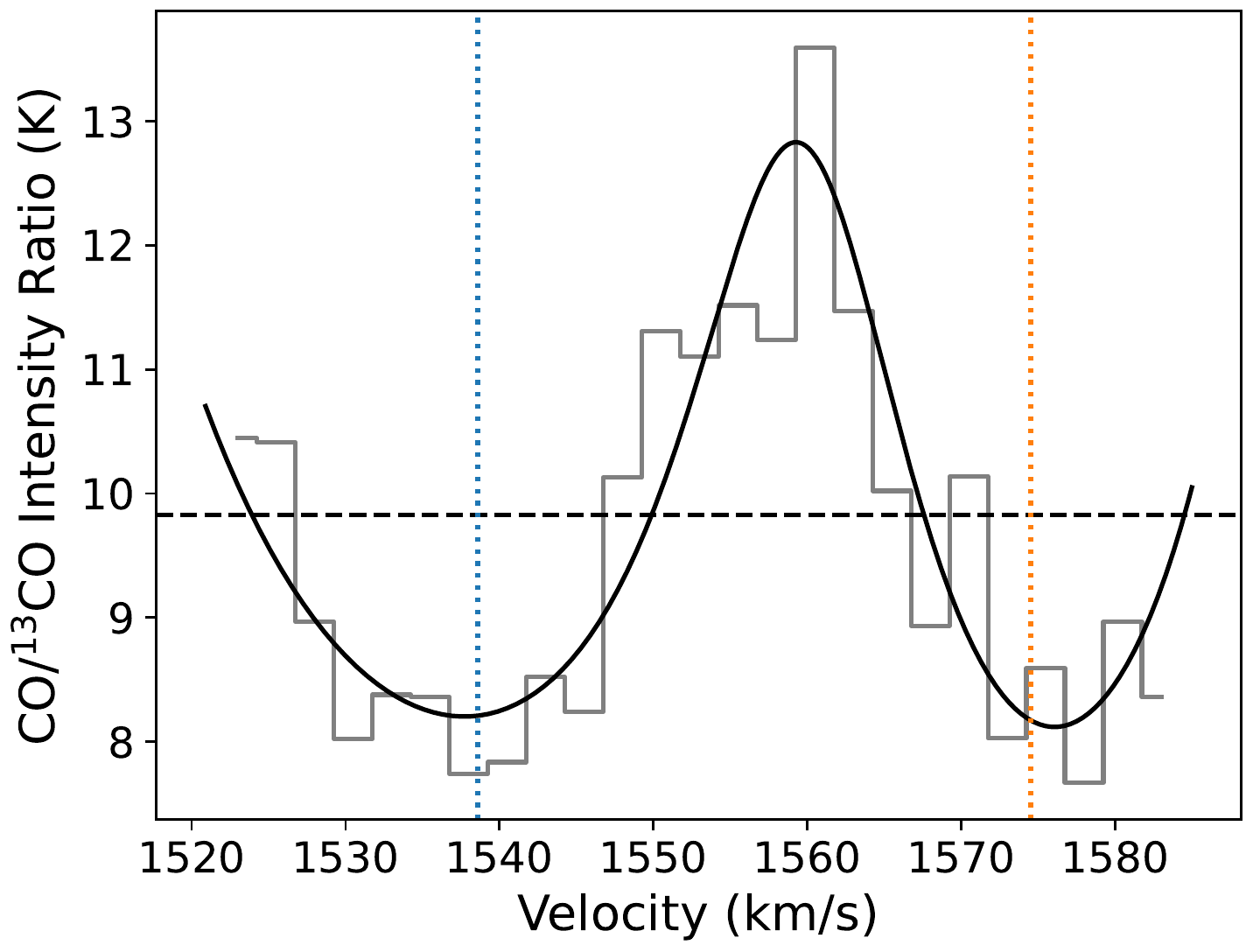}
\end{minipage} 
\\
\begin{minipage}{.33\linewidth}
\centering
\includegraphics[width=\linewidth]{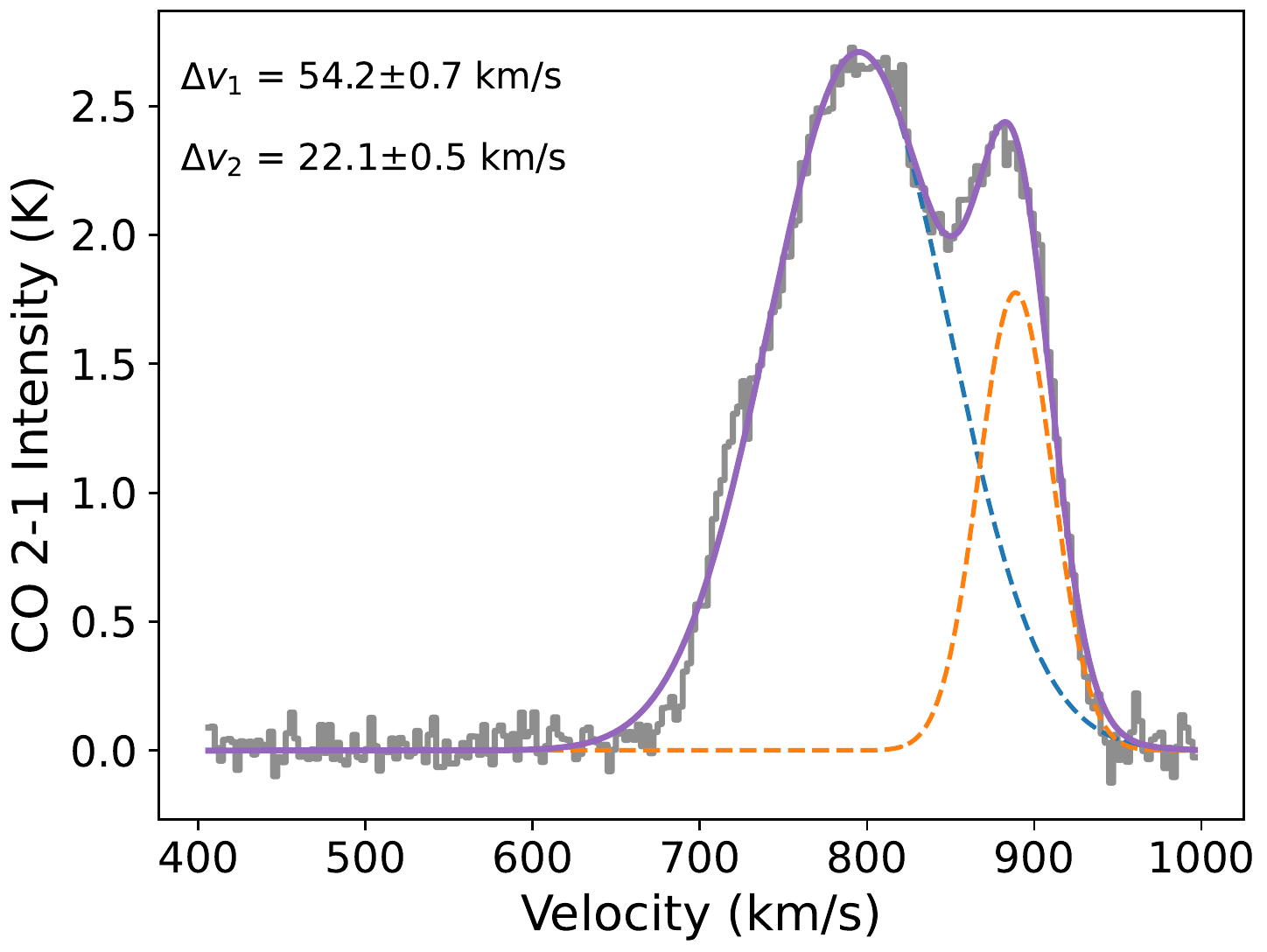}
\end{minipage}
\begin{minipage}{.34\linewidth}
\centering
\includegraphics[width=\linewidth]{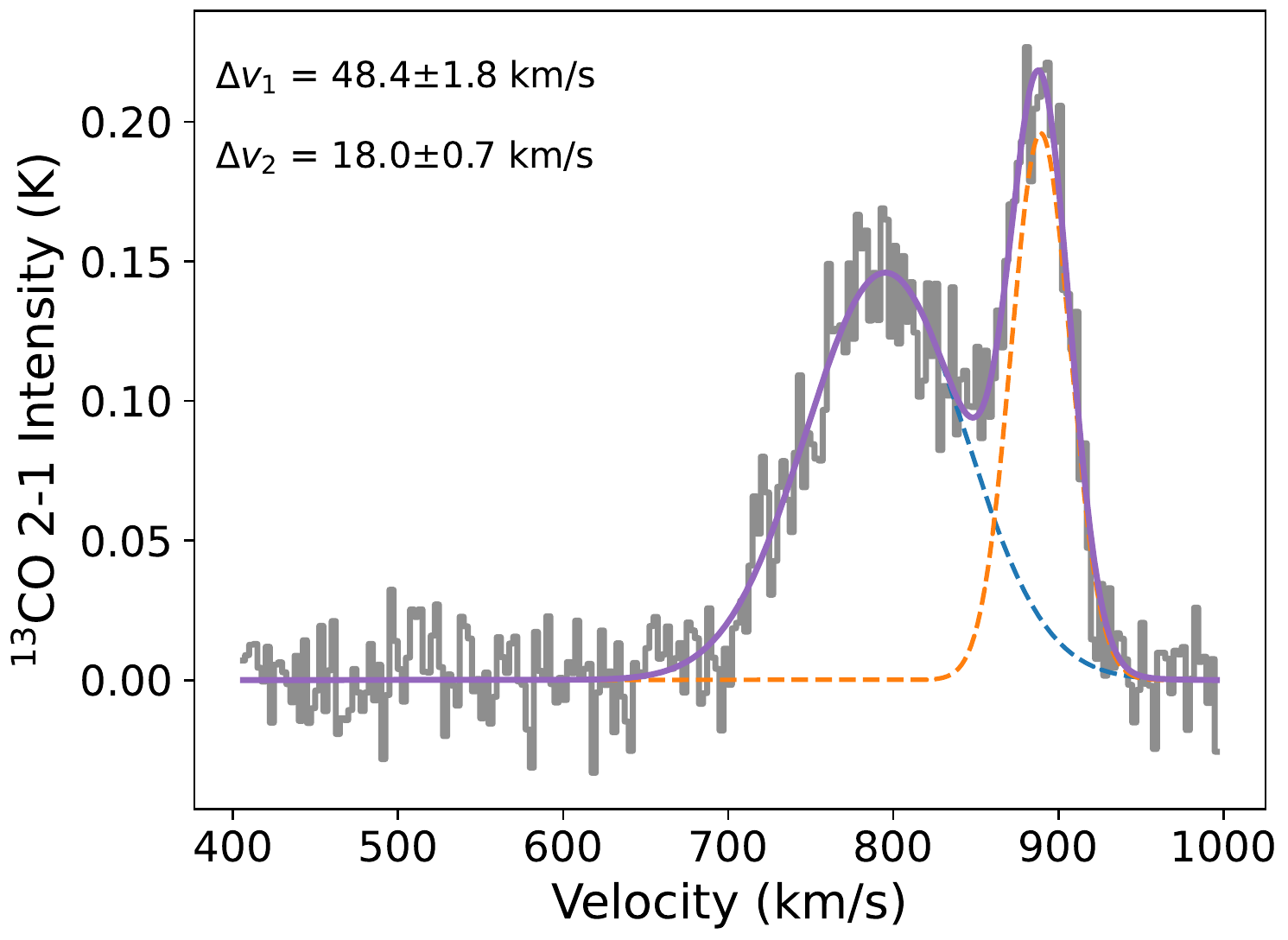}
\end{minipage}
\begin{minipage}{.33\linewidth}
\centering
\includegraphics[width=\linewidth]{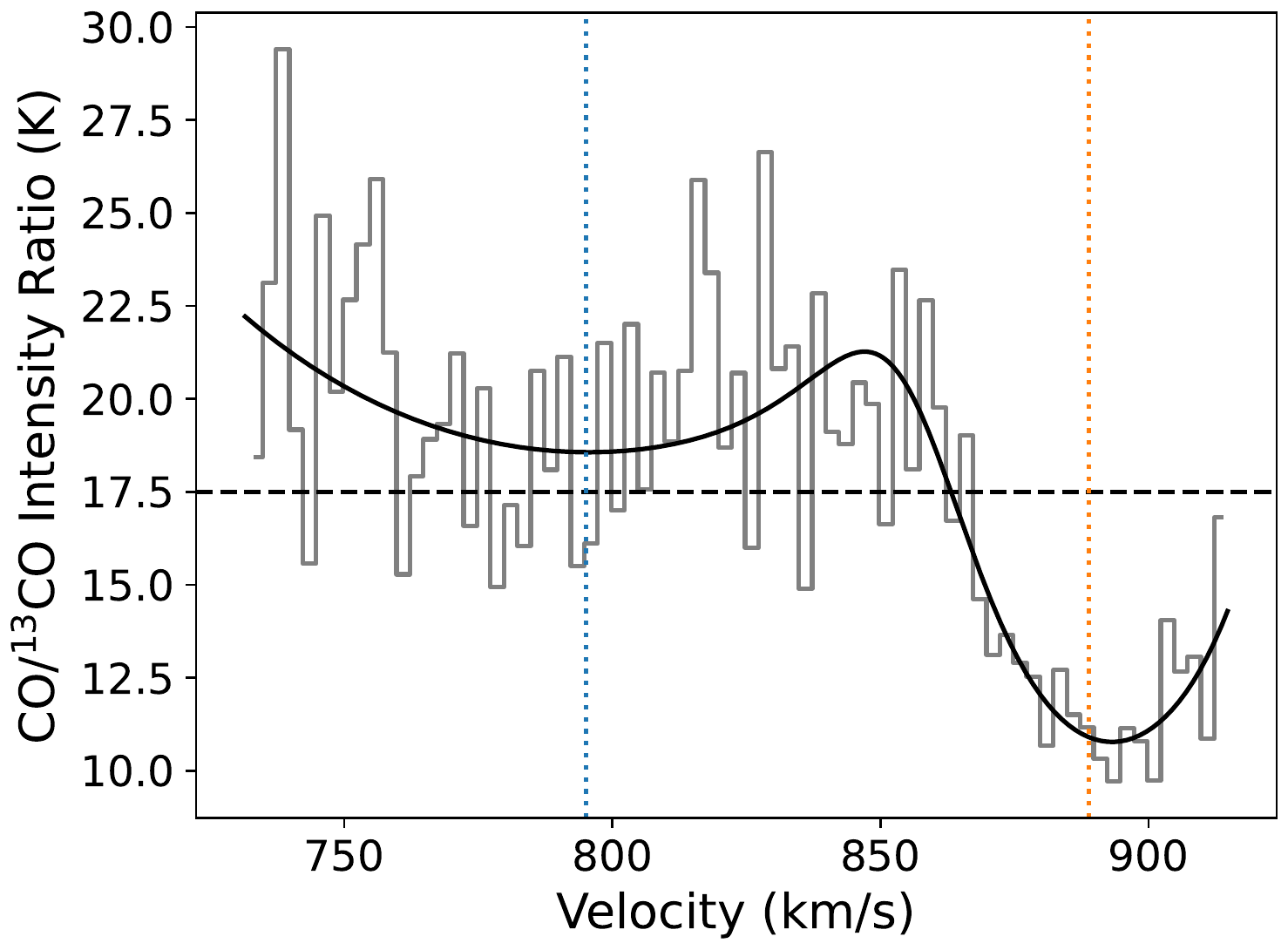}
\end{minipage}
\caption{Example of double-peaked line profiles in CO 2--1 (left column) and $^{13}$CO 2--1 (middle column), together with their intensity ratios (right column). The top and bottom rows showcase pixels from the western inner arm of NGC~4321 and the southern inner arm of NGC~3627, respectively. The line profiles can be well described by the sum (solid, purple lines) of two Gaussian functions (dashed lines), and the fitted line widths for both components are shown in the upper left corners. In the right column, the horizontal dashed line represents the integrated line intensity ratio, and the vertical dotted lines label the peak velocities of the two Gaussian fits. The bottom row demonstrates an extreme case with opposite trends of relative intensities in the two components, which leads to almost a factor of two difference in the CO/$^{13}$CO line ratio.}
\label{fig:multi_comp}
\end{figure*}

While our modeling and analysis assume one-component gas structure along each sightline, it is likely that a few regions have multi-component gas along the same sightlines.
To investigate whether there is multi-component gas present, we check each individual spectrum in the cube. 

By inspecting the spectra of all the pixels included in our analysis, we find that $\sim$8\% (28\%) of sightlines in NGC~4321 (NGC~3627) shows double-peaked line profiles. 
The higher fraction of multi-component sightlines in NGC~3627 may be partially due to the higher inclination of NGC~3627 compared to NGC~4321, and it is also consistent with previous studies which identified more overlapping giant molecular clouds/associations in the center of NGC~3627 than in NGC~4321 \citep{2017ApJ...839..133P,2021MNRAS.502.1218R}. In Figure~\ref{fig:multi_comp}, we present an example of double-peaked spectra in CO 2--1 and $^{13}$CO 2--1 and show how their intensity ratio varies with velocity. The lower/upper limit of the velocity range in the right panels of Figure~\ref{fig:multi_comp} corresponds to the FWHM line width of the lower/higher velocity component. 

For NGC~4321, multi-component sightlines are mostly found around the middle along both inner arms at similar declinations to the nucleus. As shown by the top-right panel of Figure~\ref{fig:multi_comp}, we find that the double-peaked spectra in NGC~4321 generally have consistent CO/$^{13}$CO ratios in both components, and the ratio is also similar to the integrated line ratio observed from the moment~0 map. In this case, the emissivity properties of both components would agree with what we derived using the single-component model, even if there were separate components along the sightline.

On the other hand, double/multi-peaked line profiles are scattered in both the inner and outer arms of NGC~3627, and the CO/$^{13}$CO ratios can vary between components along the same sightlines. The bottom row of Figure~\ref{fig:multi_comp} presents an extreme case found in NGC~3627 where the two spectral peaks clearly show opposite trends in their relative intensities. We find that the CO/$^{13}$CO ratio in this case can differ by almost a factor of two between both components, and the integrated line ratio only agrees with the dominant component, which has a broader line width. This implies that the optical depth and $\alpha_\mathrm{CO}$ values derived from our one-component modeling could be biased toward one of the components in such regions, and the observed line width of that component would also be overestimated (e.g., $\sim$50~km~s$^{-1}$ instead of 67~km~s$^{-1}$ for the showcased pixel in NGC~3627).

With the extreme case in NGC~3627, we have tested how the results would change by modeling the two components separately using the integrated intensity per Gaussian component for all six lines. We find that the component with a broader line width (which dominates the integrated intensity) has similar gas conditions ($< 0.2$~dex difference in all the modeled parameters) and the same $\alpha_\mathrm{CO}$ as what we obtained with the one-component modeling. On the contrary, the other component shows a different gas condition with higher temperature and lower density, optical depth, and $X_{12/13}$ abundance ratio, which altogether leads to the lower CO/$^{13}$CO line ratio seen in the bottom-right panel of Figure~\ref{fig:multi_comp}. 
We note that the relations in Figure~\ref{fig:alpha_prediction} and Equations~\ref{eqn_ratio21_fit} and~\ref{eqn_ew21_fit} would remain the same even if the difference in CO/$^{13}$CO line ratio (a factor of two at most) solely reflects optical depth changes, as the scatter in those $\alpha_\mathrm{CO}$ correlations are larger than a factor of two. In addition, the scatter of the line width correlation could even be reduced, since many of the high $\Delta v$ points (from NGC~3627) seen in Figure~\ref{fig:alpha_prediction}(b) were overestimated due to the one-component assumption. 

To summarize, our one-component assumption throughout this work should only impact our parameter estimation in a minority of regions. We find that the center of NGC~4321 is dominated by single velocity components, and the impact on radiative transfer calculation is likely small even in the few sightlines with evidence of multiple velocity components. For the center of NGC~3627, the majority of sightlines also show single-component spectra, while our one-component modeling could be biased toward one of the components in some multi-component sightlines. A more comprehensive modeling that covers different components along the same sightline would require a careful channel-by-channel analysis across all the regions. For NGC~3627, this can be done in future works with the support of SCOUSE \citep{2016ascl.soft01003H} and existing GMC catalogues \citep[e.g.,][]{2021MNRAS.502.1218R}.

\section{Updates of the NGC~3351 $\alpha_\mathrm{CO}$ Values} 
\label{sec:alpha_3351}

\begin{figure*}
\begin{minipage}{.51\linewidth}
\centering
\includegraphics[width=\linewidth]{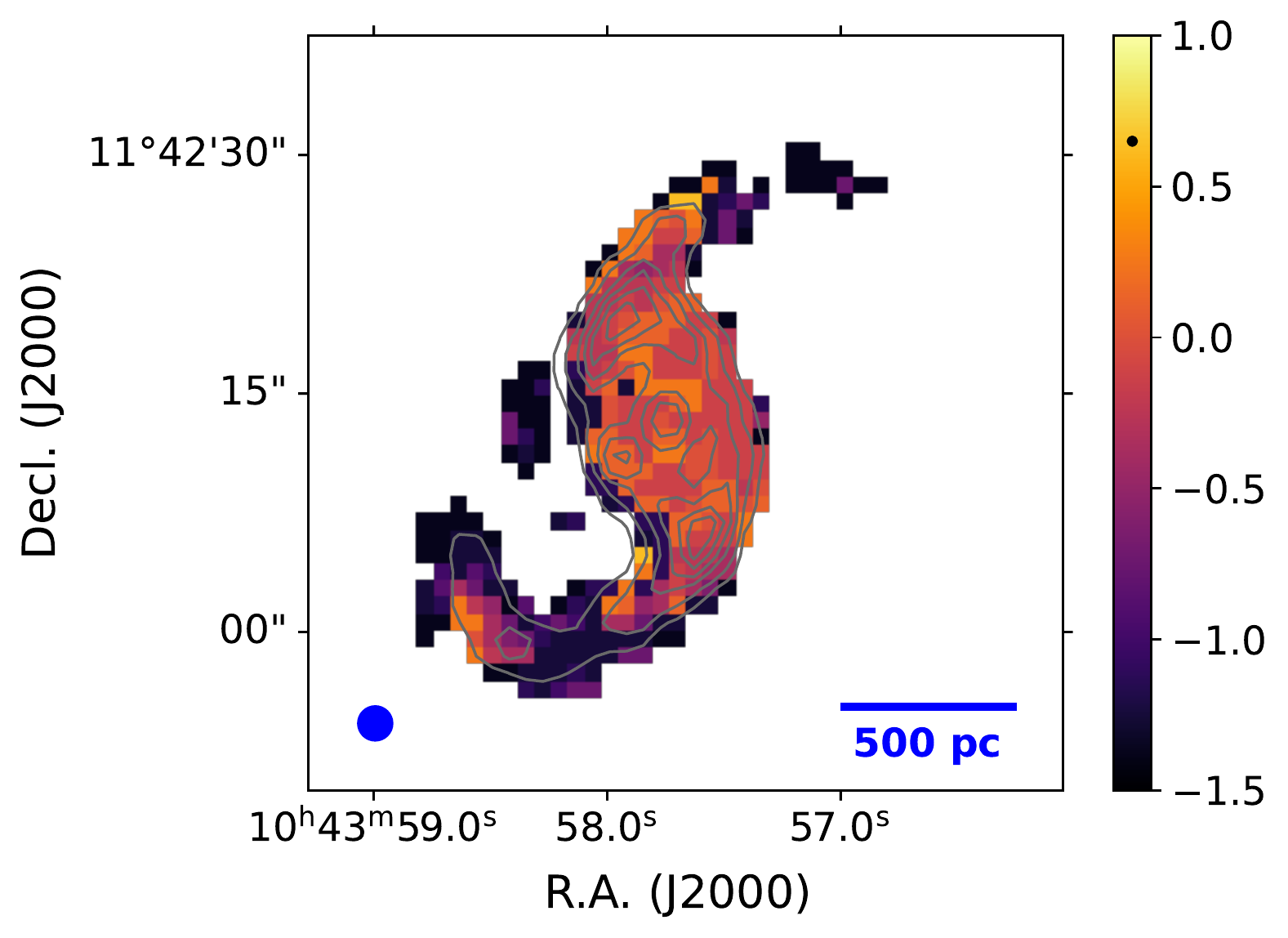}\\
(a)
\end{minipage}
\hfill
\begin{minipage}{.47\linewidth}
\centering
\includegraphics[width=\linewidth]{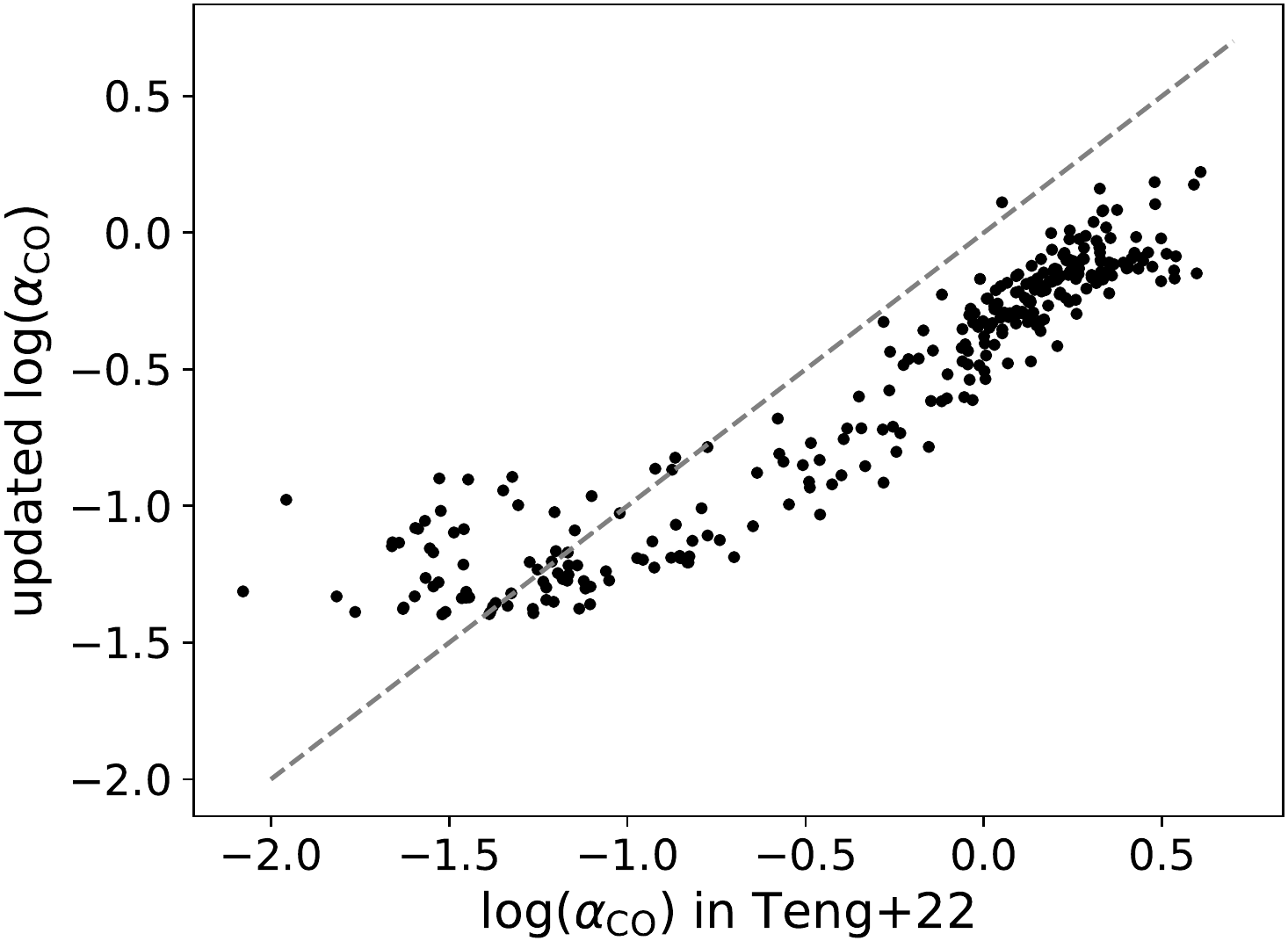}\\
(b)
\end{minipage}
\caption{(a) The updated 1DMax $\alpha_\mathrm{CO}$ map of NGC~3351 as a direct comparison to Figure~10(a) in \citet{2022ApJ...925...72T}. The black dot on the color bar indicates the MW disk $\alpha_\mathrm{CO}$ value, and the contours represent CO 1--0 integrated intensity. The $\alpha_\mathrm{CO}$ distribution is qualitatively unchanged. (b) Relation of the updated $\alpha_\mathrm{CO}$ with those in \citet{2022ApJ...925...72T}. The dashed line indicates equality. The $\alpha_\mathrm{CO}$ values are overall lowered by a factor of two to three.}
\label{fig:alpha_3351}
\end{figure*}

As mentioned in Section~\ref{subsec:alpha_CO}, the $\alpha_\mathrm{CO}$ solutions for NGC~3351 in \citet{2022ApJ...925...72T} should be a factor of 2--3 lower if consistent line widths were adopted when computing the $\alpha_\mathrm{CO}$ model grid. This factor of 2--3 overestimation comes from the observed line width in NGC~3351 being overall 2--3 times higher than the FWHM line width of 15~km~s$^{-1}$ assumed in RADEX. Here we recalculate and update the $\alpha_\mathrm{CO}$ values for NGC~3351 which are used in this work for a self-consistent comparison. We emphasize that this does not change the qualitative results and main findings in \citet{2022ApJ...925...72T}. 

Figure~\ref{fig:alpha_3351}(a) presents the updated 1DMax $\alpha_\mathrm{CO}$ map of NGC~3351. The color bar and scale are set to be the same as \citet[Figure~10a]{2022ApJ...925...72T} for easier comparison. We find no major change in $\alpha_\mathrm{CO}$ in the inflow regions as their observed FWHM line width is close to the RADEX-assumed 15~km~s$^{-1}$. It is also clear that $\alpha_\mathrm{CO}$ in the inflow regions remain substantially lower than the central nuclear ring, even though $\alpha_\mathrm{CO}$ within the nuclear ring becomes 2--3 times lower. The pixel-by-pixel relation between the pre- and post-updated $\alpha_\mathrm{CO}$ values is shown in Figure~\ref{fig:alpha_3351}(a).  

We also report changes in the intensity-weighted mean $\alpha_\mathrm{CO}$ over the entire kpc region ($\langle \alpha_\mathrm{CO} \rangle_\mathrm{kpc}$), as well as the spectrally-stacked $\alpha_\mathrm{CO}$ value of the inflow regions. After correction, $\langle \alpha_\mathrm{CO} \rangle_\mathrm{kpc} = 0.75 \pm 0.04$~$\mathrm{M_\odot\ (K~km~s^{-1}~pc^2)^{-1}}$, which is 2.4 times lower than $1.79 \pm 0.10$ in \citet{2022ApJ...925...72T}. In addition, the spectrally-stacked $\alpha_\mathrm{CO}$ over the inflow arms based on the best-fit solution becomes 0.08~$\mathrm{M_\odot\ (K~km~s^{-1}~pc^2)^{-1}}$, which is still within the range of 0.01--0.1~$\mathrm{M_\odot\ (K~km~s^{-1}~pc^2)^{-1}}$ reported in \citet{2022ApJ...925...72T}.

\section{Additional Figures on the Data and Modeling} 

\subsection{Maps of Moment 1 and Effective Line Width} 
\label{sec:vel_maps}

\begin{figure}
\begin{minipage}{.57\linewidth}
\centering
\includegraphics[width=\linewidth]{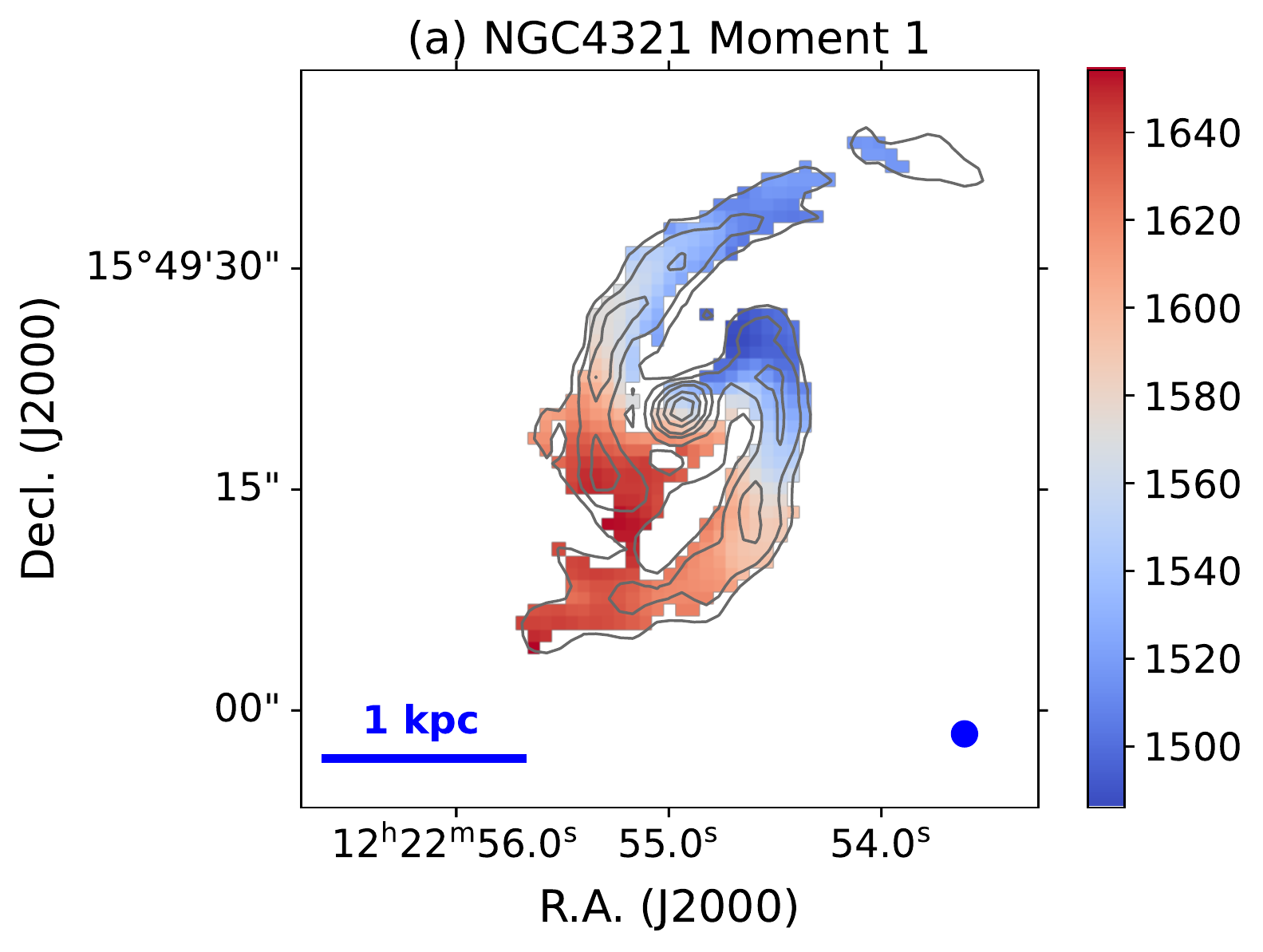}\\
\end{minipage}
\hfill
\begin{minipage}{.42\linewidth}
\centering
\includegraphics[width=\linewidth]{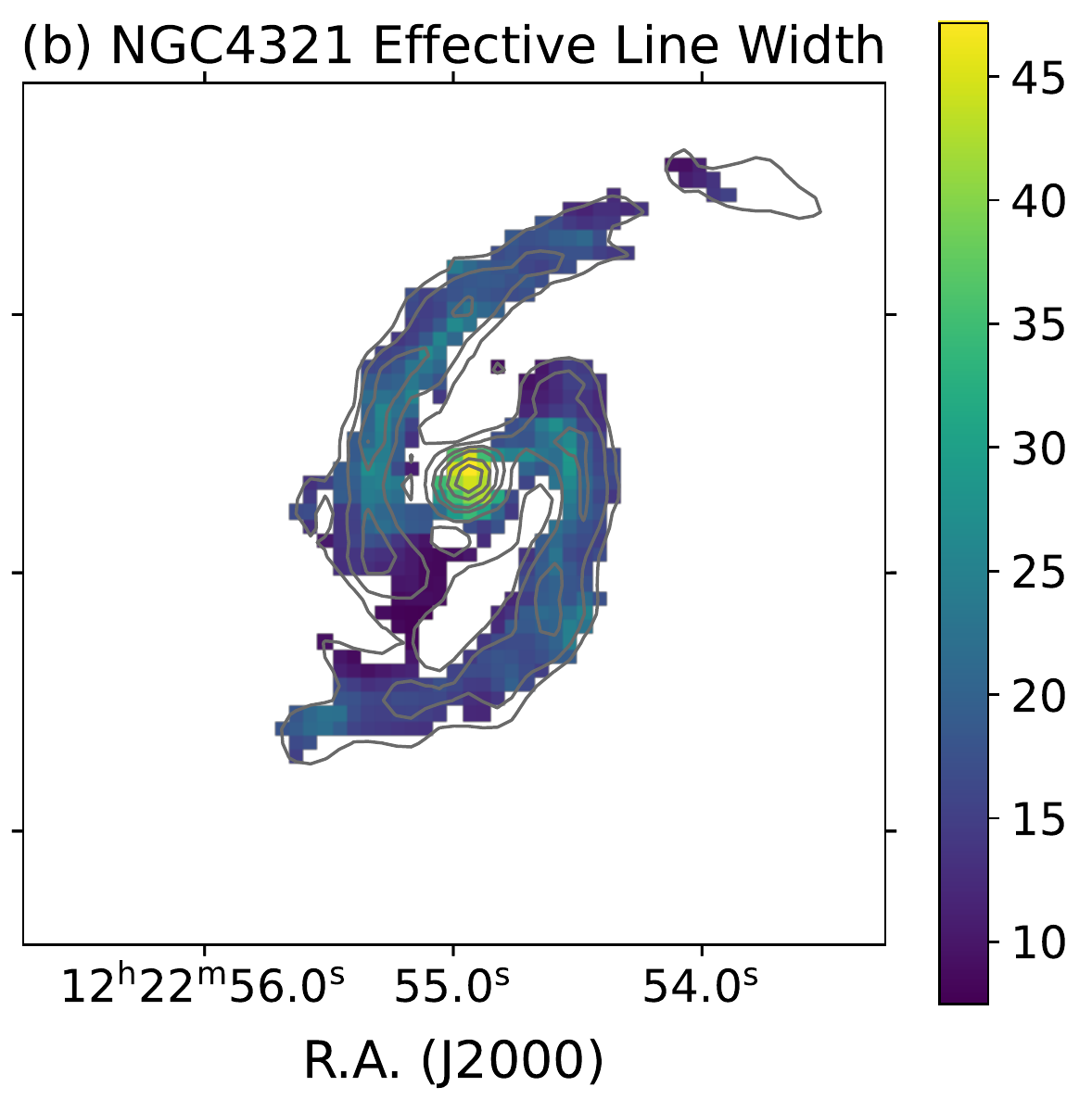}\\
\end{minipage}
\caption{Maps of the CO 2--1 (a) moment 1 and (b) effective line width for NGC~4321, both in units of km~s$^{-1}$.}
\label{fig:vmaps_4321}
\end{figure}

\begin{figure}
\begin{minipage}{.545\linewidth}
\centering
\includegraphics[width=\linewidth]{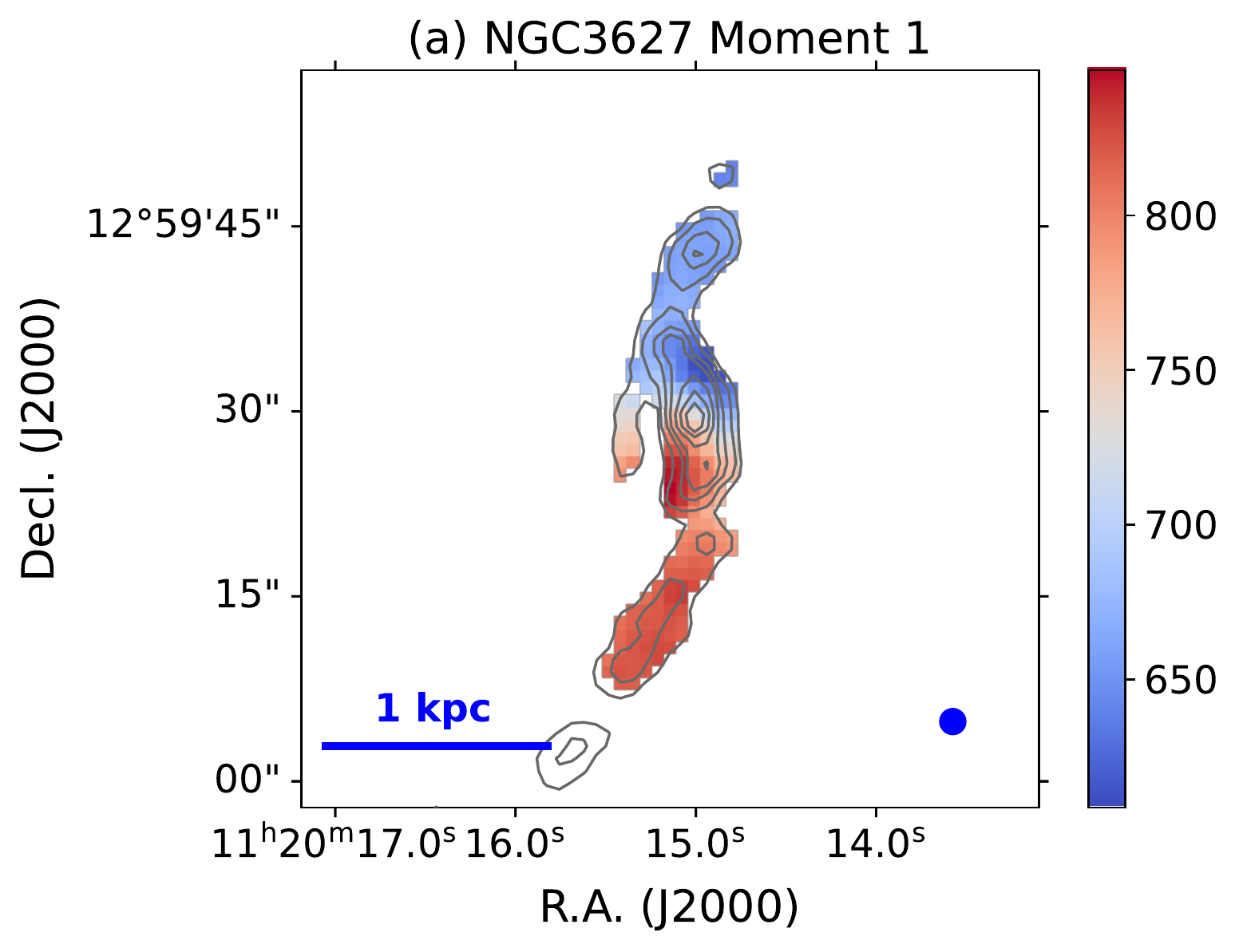}\\
\end{minipage}
\hfill
\begin{minipage}{.45\linewidth}
\centering
\includegraphics[width=\linewidth]{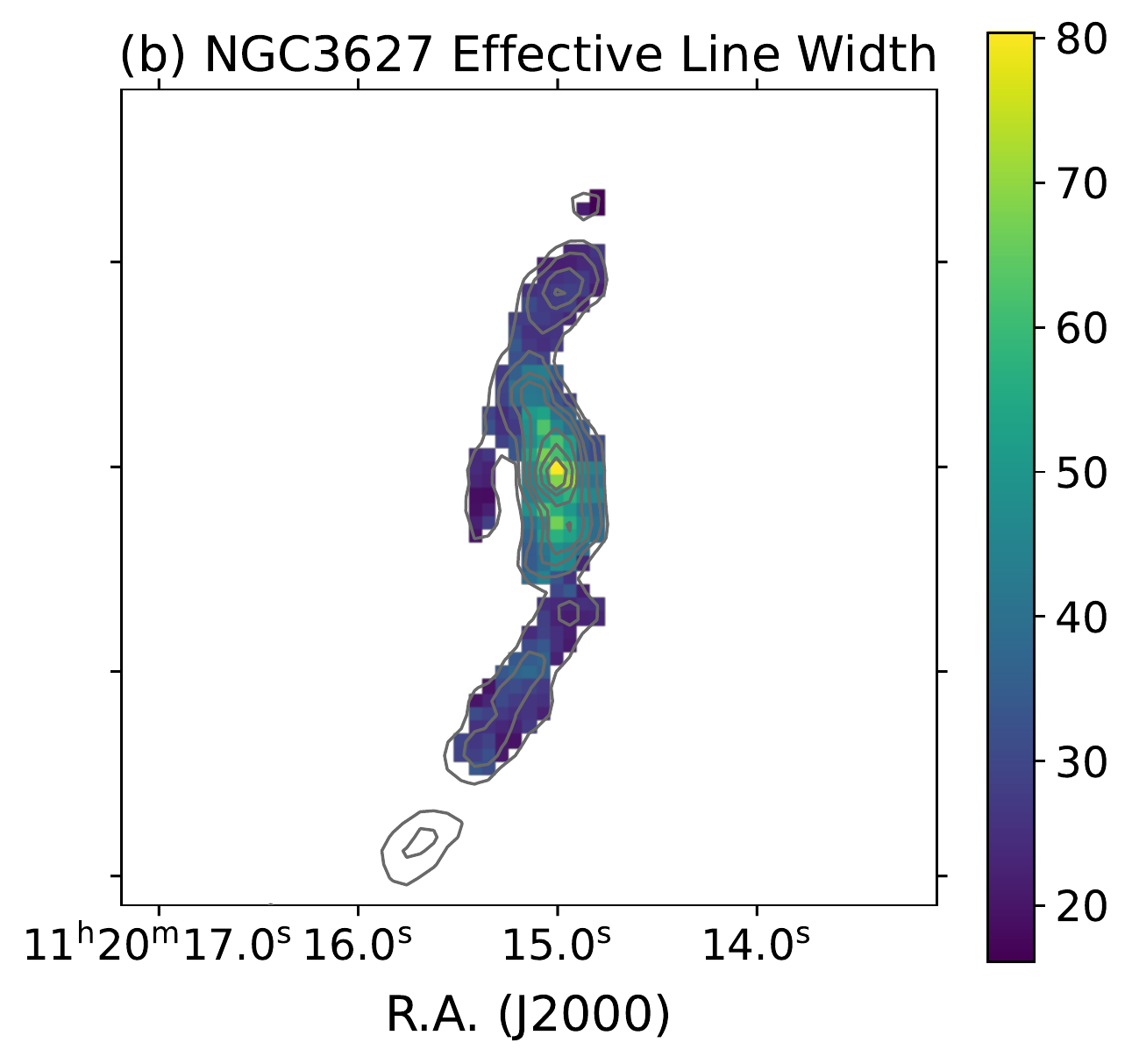}\\
\end{minipage}
\caption{Same as Figure~\ref{fig:vmaps_4321} but for NGC~3627.}
\label{fig:vmaps_3627}
\end{figure}

While this work do not focus on the molecular gas dynamics, we provide here the moment 1 and effective line width maps toward our targets as a reference for future studies. Figures~\ref{fig:vmaps_4321} and~\ref{fig:vmaps_3627} show the CO 2--1 maps for NGC~4321 and NGC~3627, respectively. The moment~1 maps for both galaxies reveal clear signature of counter-clockwise gas rotation, and the effective line widths are highest in the nuclei possibly due to unresolved gas motion within the central beam.

\subsection{Line Constraints and Modeled Probability Distributions}
\label{sec:pixel_solutions}

\begin{figure*}
\centering
\includegraphics[width=\linewidth]{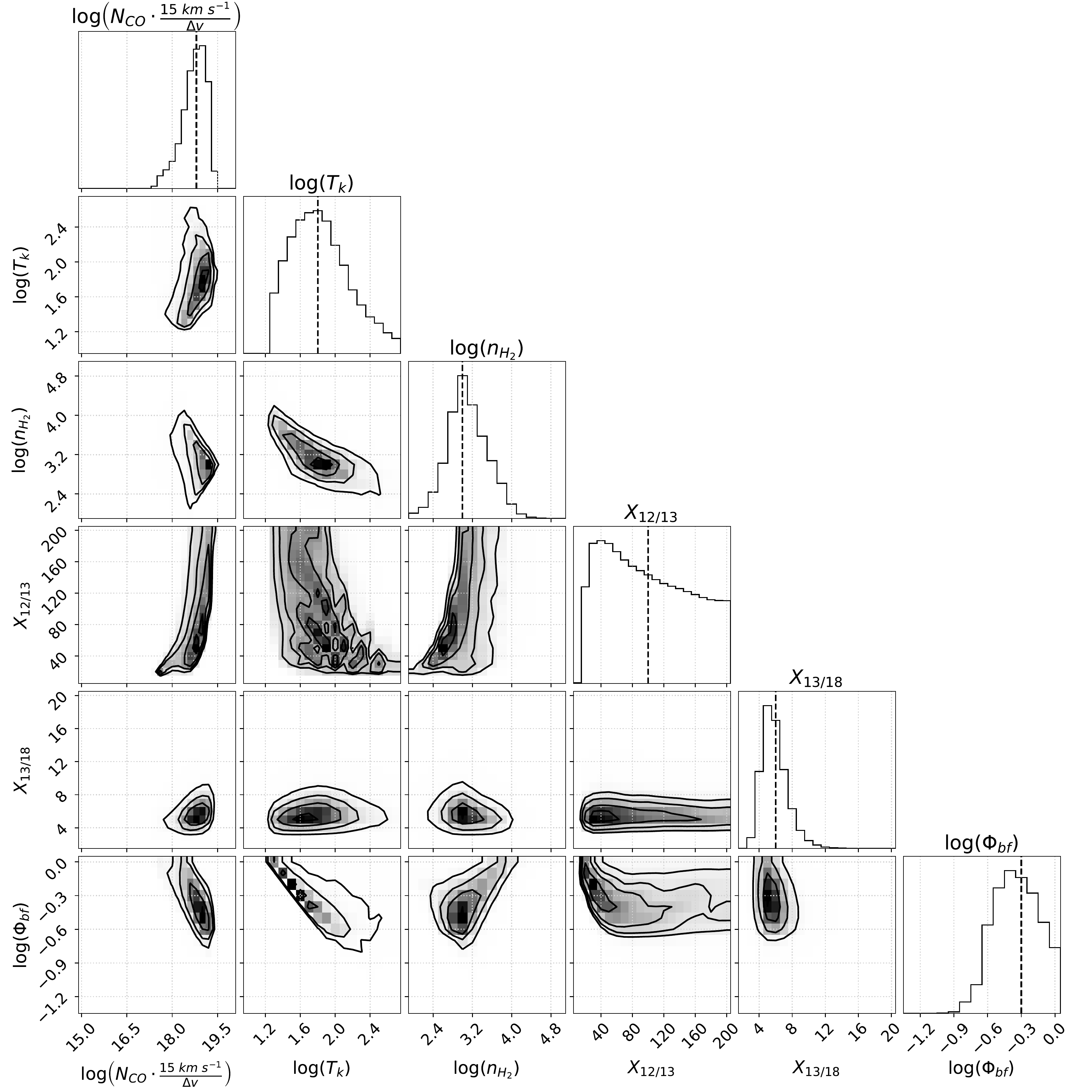}
\caption{Marginalized 1D and 2D likelihood distributions of a pixel in the northern arm of NGC~4321, which is also the same pixel as shown in Figure~\ref{fig:flux_contour_4321}(b).
See the caption of Figure~\ref{fig:corner_center_4321} for more information.}
\label{fig:corner_arms_4321}
\end{figure*}

\begin{figure*}
\begin{minipage}{.49\linewidth}
\centering
\includegraphics[width=\linewidth]{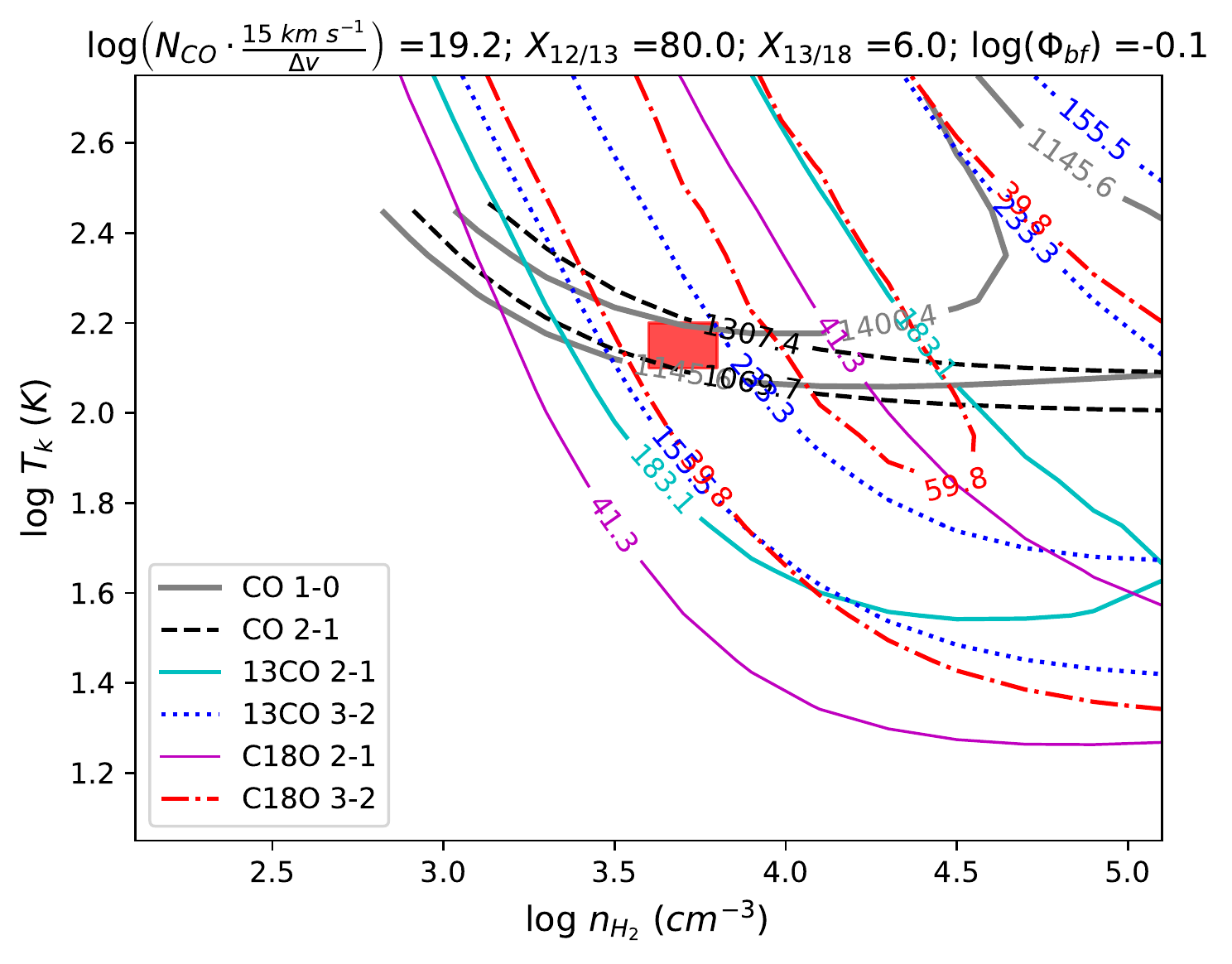}\\
(a)
\end{minipage}
\hfill
\begin{minipage}{.49\linewidth}
\centering
\includegraphics[width=\linewidth]{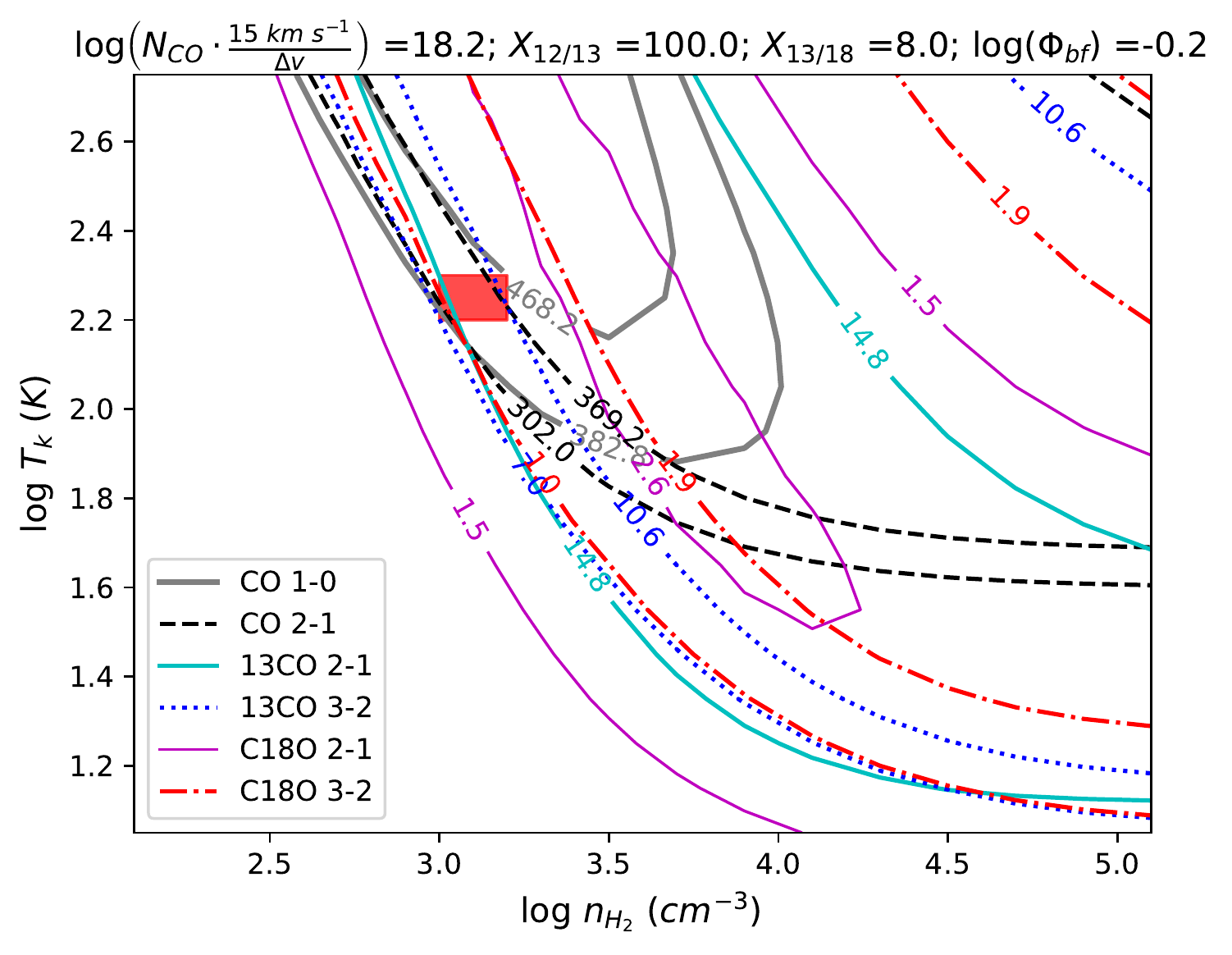}\\
(b)
\end{minipage}
\caption{Same as Figure~\ref{fig:flux_contour_4321} but for (a) the central pixel of NGC~3627 and (b) a pixel in the northern inner arm of NGC~3627. The low-density and high-temperature part in panel (a) is excluded due to violation of the $\ell_\mathrm{los} < 200$~pc constraint.}
\label{fig:flux_contour_3627}
\end{figure*}

\begin{figure*}
\centering
\includegraphics[width=\linewidth]{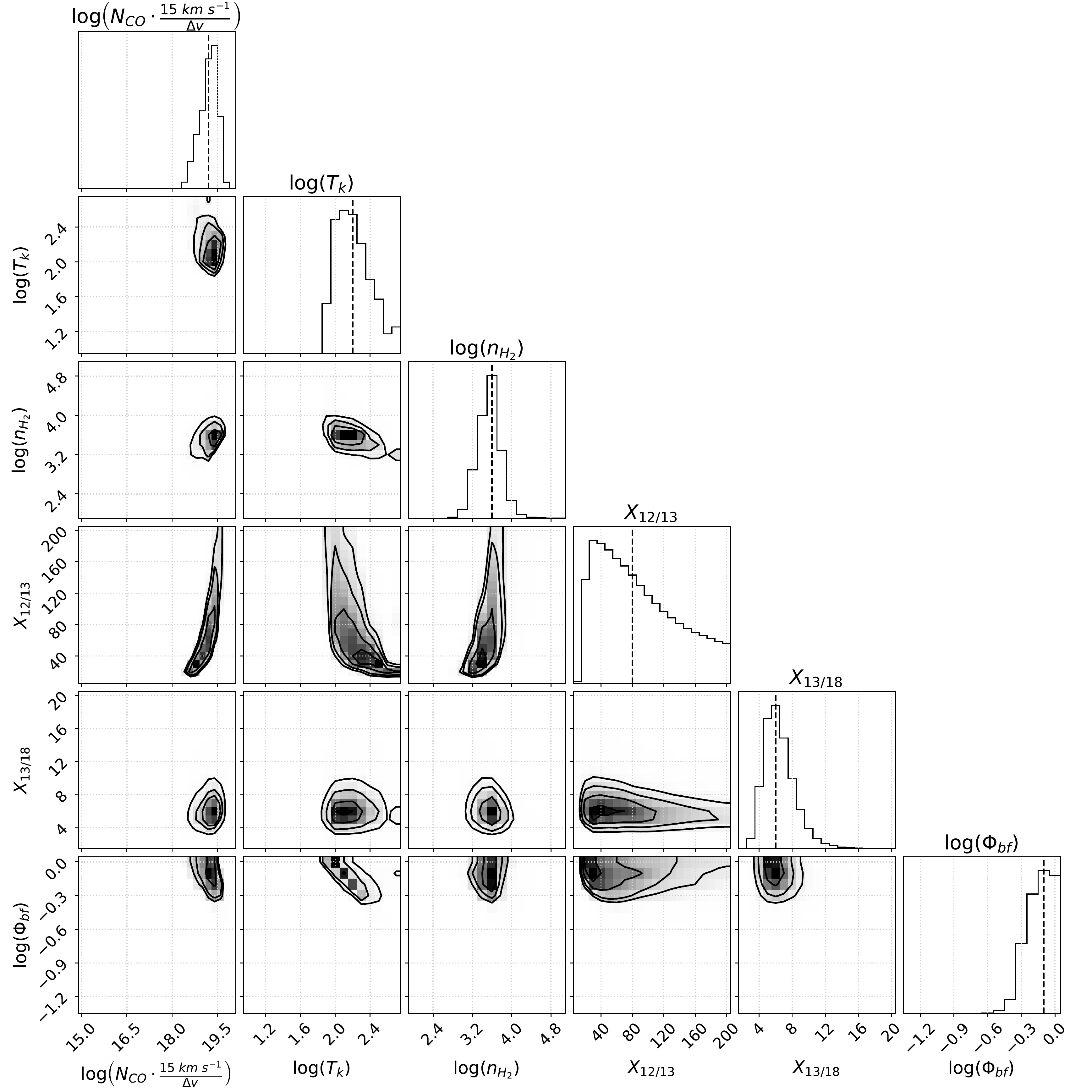}
\caption{Same as Figure~\ref{fig:corner_center_4321} but for the central pixel of NGC~3627, which is also the same pixel as shown in Figure~\ref{fig:flux_contour_3627}(a).}
\label{fig:corner_center_3627}
\end{figure*}

\begin{figure*}
\centering
\includegraphics[width=\linewidth]{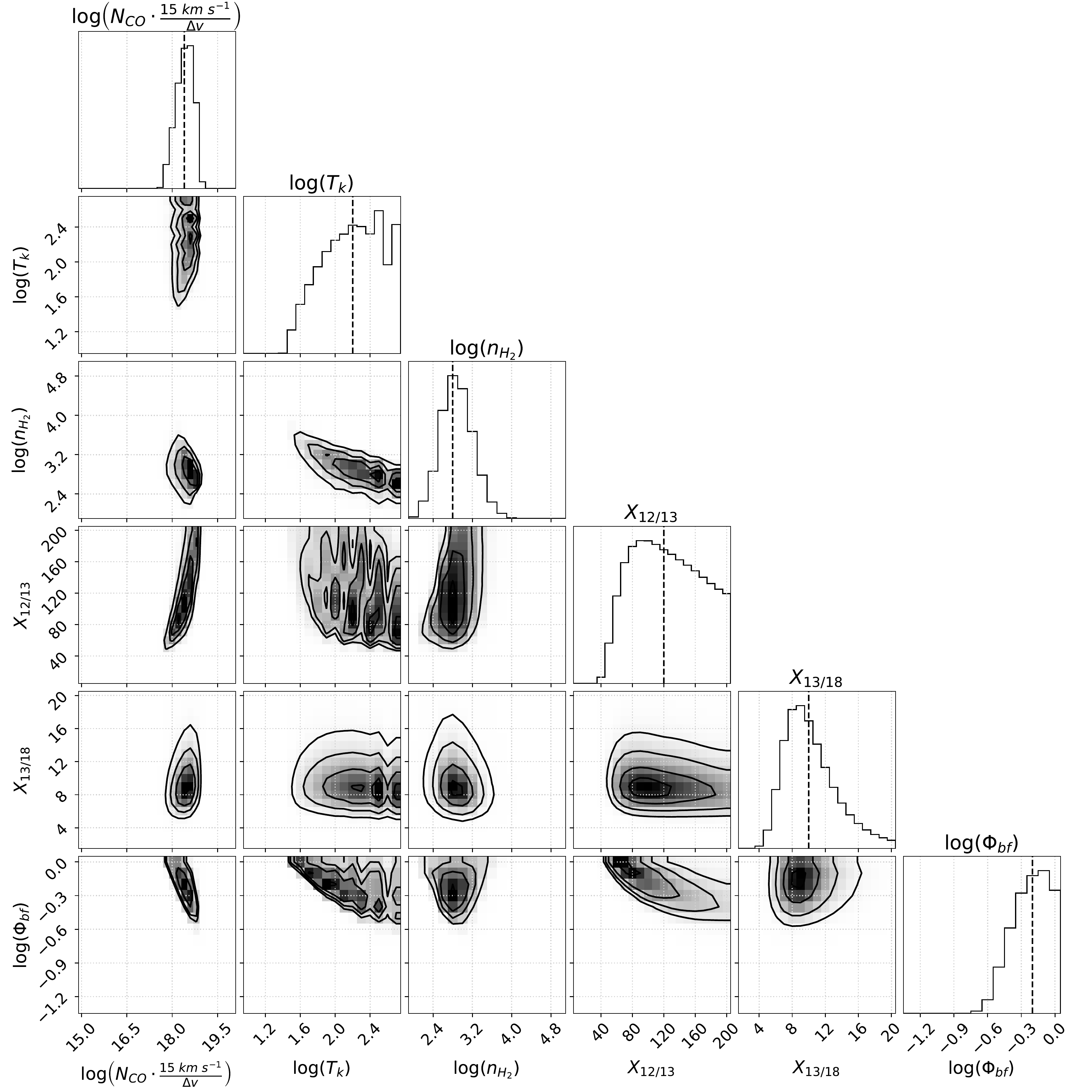}
\caption{Same as Figure~\ref{fig:corner_center_4321} but for a pixel in the northern inner arm of NGC~3627, which is also the same pixel as shown in Figure~\ref{fig:flux_contour_3627}(b).}
\label{fig:corner_arms_3627}
\end{figure*}

Besides the line constraints and modeled PDFs for the NGC~4321 nucleus presented in Figures~\ref{fig:flux_contour_4321} and~\ref{fig:corner_center_4321}, here we include additional figures for other representative regions. Figures~\ref{fig:corner_arms_4321}, \ref{fig:corner_center_3627}, and \ref{fig:corner_arms_3627} demonstrate the PDFs for the inner arms of NGC~4321, the nucleus of NGC~3627, and the inner arms of NGC~3627, respectively. Figure~\ref{fig:flux_contour_3627} shows the line constraints and best-fit solutions for the nucleus and inner arms of NGC~3627. 

\acknowledgments
We thank the referee for insightful comments that helped improve the manuscript. 
Y.-H.T. and K.S. acknowledge funding support from NRAO Student Observing Support Grant SOSPADA-012 and from the National Science Foundation (NSF) under grant No. 2108081.
J.S. acknowledges support by the Natural Sciences and Engineering Research Council of Canada (NSERC) through a Canadian Institute for Theoretical Astrophysics (CITA) National Fellowship. HAP acknowledges support by the National Science and Technology Council of Taiwan under grant 110-2112-M-032-020-MY3.
MQ acknowledges support from the Spanish grant PID2019-106027GA-C44, funded by MCIN/AEI/10.13039/501100011033.
AU acknowledges support from the Spanish grants PGC2018-094671-B-I00, funded by MCIN/AEI/10.13039/501100011033 and by ``ERDF A way of making Europe'', and PID2019-108765GB-I00, funded by MCIN/AEI/10.13039/501100011033. 
RSK and SCOG acknowledge  support  from  the  Deutsche  Forschungsgemeinschaft (DFG) in the Collaborative Research Centre (SFB 881, ID 138713538)  ``The Milky Way System'' (subprojects A1, B1, B2, and B8) and from the Heidelberg Cluster of Excellence (EXC 2181, ID 390900948) ``STRUCTURES'', funded by the German Excellence Strategy. RSK also thanks for funding form the European Research Council in the ERC Synergy Grant ``ECOGAL'' (ID 855130). RSK and SCOG also benefit from computing resources provided by the State of Baden-W\"urttemberg through bwHPC and DFG through grant INST 35/1134-1 FUGG, and from the data storage facility SDS@hd supported through grant INST 35/1314-1 FUGG. FB acknowledges funding from the European Research Council (ERC) under the European Union’s Horizon 2020 research and innovation programme (grant agreement No.726384/Empire).
ES acknowledges funding from the European Research Council (ERC) under the European Union’s Horizon 2020 research and innovation programme (grant agreement No. 694343).
 MCS acknowledges financial support from the Royal Society (URF\textbackslash R1\textbackslash 221118).

This work was carried out as part of the PHANGS collaboration.
This paper makes use of the following ALMA data: ADS/JAO.ALMA\#2013.1.00885.S, 
ADS/JAO.ALMA\#2015.1.00956.S, % PHANGS CO 2-1
ADS/JAO.ALMA\#2015.1.00978.S,  
ADS/JAO.ALMA\#2016.1.00972.S.    
ALMA is a partnership of ESO (representing its member states), NSF (USA), and NINS (Japan), together with NRC (Canada), NSC and ASIAA (Taiwan), and KASI (Republic of Korea), in cooperation with the Republic of Chile. The Joint ALMA Observatory is operated by ESO, AUI/NRAO, and NAOJ. The National Radio Astronomy Observatory is a facility of the National Science Foundation operated under cooperative agreement by Associated Universities, Inc.

We acknowledge the usage of NASA's Astrophysics Data System (\url{http://www.adsabs.harvard.edu}) and \texttt{ds9}, a tool for data visualization supported by the Chandra X-ray Science Center (CXC) and the High Energy Astrophysics Science Archive Center (HEASARC) with support from the JWST Mission office at the Space Telescope Science Institute for 3D visualization.

\vspace{5mm}
\facilities{ALMA}

\software{\texttt{CASA} \citep{2022PASP..134k4501C}, \texttt{ds9} \citep{2000ascl.soft03002S,2003ASPC..295..489J}, \texttt{matplotlib} \citep{Hunter:2007}, \texttt{numpy} \citep{harris2020array}, \texttt{scipy} \citep{Virtanen_2020}, \texttt{astropy} \citep{2013A&A...558A..33A,2018AJ....156..123A}, \texttt{RADEX} \citep{radex}, \texttt{corner} \citep{Foreman-Mackey2016} }

\bibliography{reference}{}
\bibliographystyle{aasjournal}

\end{document}